%% file: PhD_Thesis.tex
\documentclass[12pt]{book} % "draft" serves not to compile figs
\usepackage[top=24mm,text={155mm,225mm},dvips,twoside]{geometry} %
\usepackage[applemac]{inputenc}
\usepackage[british,french]{babel}
\usepackage{ae,aeguill} % for some special fonts
\usepackage{latexsym}
\usepackage{amssymb}
\usepackage{amsmath}
\usepackage{amsfonts}
\usepackage{graphicx}
\usepackage{subfigure}
\usepackage{epsfig} 
\usepackage{feyn} % to draw Feynman diagrams
\usepackage{array,multirow}
\usepackage{psfrag}
\usepackage{bm}
\usepackage{textcomp} % by arXiv
\usepackage[hyperindex]{hyperref} % by arXiv
\usepackage{breakurl} % by arXiv

\usepackage{fancyhdr} % fancy header

\usepackage{verbatim} % by UMB to allow the use of the ambient {comment} 
\usepackage{pst-node}  % to allow the use of the \red  UMB

\newcommand\chap[1]{Chap.~\ref{#1}} % UMB
\newcommand\Sec[1]{Sec.~\ref{#1}} % UMB , "\sec" is for the trigonometric function
\newcommand\fig[1]{Fig.~\ref{#1}} % UMB
\newcommand\bi{\begin{itemize}} % UMB
\newcommand\ei{\end{itemize}} % UMB
\newcommand\ben{\begin{enumerate}} % UMB
\newcommand\een{\end{enumerate}} % UMB
\newcommand\bd{\begin{description}} % UMB
\newcommand\ed{\end{description}} % UMB
\newcommand\ifo{interferometer} % UMB
\newcommand\ifos{interferometers} % UMB
\newcommand\st{scalar-tensor theory} % UMB
\newcommand\sts{scalar-tensor theories} % UMB
\newcommand{\tgv}{three-graviton vertex}

\newcommand{\tfgv}{three- and four-graviton vertex}
\newcommand{\tfgvs}{three- and four-graviton vertices}
\newcommand{\btf}{$\b_3$ and $\b_4$}
\newcommand\q{\quad} % UMB
\newcommand\qq{\qquad} % UMB
\def\l{\lambda}				% UMB
\def\q{\quad}			% UMB
\def\qq{\qquad}			% UMB
\newcommand{\bppn}{\beta_{\rm PPN}}      % Group paper and
\newcommand{\gppn}{\gamma_{\rm PPN}}  % Chapter on Frameworks
\newcommand{\ds}{\displaystyle} % in Ric's Chspter
\newcommand{\pt}[1]{\left(#1\right)}  %  parentesi tonda
\newcommand{\paq}[1]{\left[#1\right]} %  parentesi quadra
\newcommand{\az}{\mathcal{S}}         %

\begin{document}

\include{mydefs} % by UMB to include definitions 

\pagestyle{fancy}
\renewcommand{\chaptermark}[1]{\markboth{#1}{}}
\renewcommand{\sectionmark}[1]{\markright{\thesection\ #1}}
\fancyhf{}
\fancyhead[LE,RO]{\thepage}
\fancyhead[LO]{\leftmark}
\fancyhead[RE]{\rightmark}

%%%%%%%%%%%%%%%%%%%%%%%%%%%%
%%  TITLE  PAGE
%%%%%%%%%%%%%%%%%%%%%%%%%%%%
\flushbottom
\pagenumbering{none}

\begin{titlepage}
\begin{center}

UNIVERSITÉ DE GENÈVE \hfill FACULTÉ DES SCIENCES \\ 
Section de Physique \hfill Professeur Michele Maggiore\\
Département de physique théorique \hfill Docteur Riccardo Sturani \\

\vspace{0.5cm}
\hrulefill

\vspace{3cm}

{\LARGE \bf Effective Field Theory Methods}\\ 
\vspace{0.2cm}
{\LARGE \bf in Gravitational Physics}\\
\vspace{0.2cm}
{\LARGE \bf and Tests of Gravity}\\

\vspace{6cm} 
TH\`ESE\\
présentée à la Faculté des Sciences de l'Université de Genève\\
pour obtenir le grade de Docteur \`es Sciences, mention Physique\\
  
\vspace{1cm}
par\\
{\bf \large Umberto Cannella}\\
de\\
Rome, Italie\\

\vspace{2cm}
Thèse N° 4289 \\
Genève, Fevrier 2011\\
\vspace{0.2cm}
{\small
Atelier d'impression "ReproMail", Université de Genève
}

\end{center}
\end{titlepage}

\setcounter{footnote}{0}
\pagenumbering{roman}
\setcounter{page}{0}

%%%%%%%%%%%%%%%%%%%%%%%%%
% INCLUDE FILES
%%%%%%%%%%%%%%%%%%%%%%%%%
\input{imprimatur.tex}
\cleardoublepage

\input{Merci.tex}

\input{Resume.tex}

\selectlanguage{british}
\tableofcontents
\cleardoublepage

\pagenumbering{arabic}
\setcounter{page}{1}

\input{Introduction.tex}

\input{How_To_Test/How_To_Test.tex}

\input{Chap_EFT/Chap_EFT.tex}

\input{Ric_Paper/Ric_Paper.tex}

\input{Group_Paper/Group_Paper.tex}

\input{Chap_Berti/Chap_Berti.tex}

\input{Conclusions.tex}

\input{abbreviations.tex}

%%%%%%%%%%%%%%%%%%%%%%%%%
%% Bibliography
%%%%%%%%%%%%%%%%%%%%%%%%%

\addcontentsline{toc}{chapter}{Bibliography}
\bibliographystyle{ClasicCite} 
\bibliography{../Bib_Thesis}

\end{document}

%% file: mydefs.tex
% abbrevizioni utili, versione 09/03/09

\renewcommand\({\left(}
\renewcommand\){\right)}
\renewcommand\[{\left[}
\renewcommand\]{\right]}
\newcommand\del{{\mbox {\boldmath $\nabla$}}}
\newcommand\n{{\mbox {\boldmath $\nabla$}}}
\newcommand{\ra}{\rightarrow}

%minore o circa uguale
\def\lsim{\raise 0.4ex\hbox{$<$}\kern -0.8em\lower 0.62
ex\hbox{$\sim$}}

%maggiore o circa uguale
\def\gsim{\raise 0.4ex\hbox{$>$}\kern -0.7em\lower 0.62
ex\hbox{$\sim$}}

\def\lbar{{\hbox{$\lambda$}\kern -0.7em\raise 0.6ex
\hbox{$-$}}}

\newcommand\eq[1]{eq.~(\ref{#1})}
\newcommand\eqs[2]{eqs.~(\ref{#1}) and (\ref{#2})}
\newcommand\Eq[1]{Equation~(\ref{#1})}
\newcommand\Eqs[2]{Equations~(\ref{#1}) and (\ref{#2})}
\newcommand\Eqss[3]{Equations~(\ref{#1}), (\ref{#2}) and (\ref{#3})}
\newcommand\eqss[3]{eqs.~(\ref{#1}), (\ref{#2}) and (\ref{#3})}
\newcommand\eqsss[4]{eqs.~(\ref{#1}), (\ref{#2}), (\ref{#3})
and (\ref{#4})}
\newcommand\eqssss[5]{eqs.~(\ref{#1}), (\ref{#2}), (\ref{#3}),
(\ref{#4}) and (\ref{#5})}
\newcommand\eqst[2]{eqs.~(\ref{#1})--(\ref{#2})}
\newcommand\pa{\partial}
\newcommand\p{\partial}
\newcommand\pdif[2]{\frac{\pa #1}{\pa #2}}
\newcommand\pfun[2]{\frac{\delta #1}{\delta #2}}
\newcommand\Tr{{\rm Tr}\, }

\newcommand\be{\begin{equation}}
\newcommand\ee{\end{equation}}
\def\bea{\begin{array}}
\def\eea{\end{array}} % \def\ea{\end{array}}
\newcommand\bees{\begin{eqnarray}}
\newcommand\ees{\end{eqnarray}}
\def\nn{\nonumber}
\newcommand\sub[1]{_{\rm #1}}
\newcommand\su[1]{^{\rm #1}}

% questo produce lettere greche boldface
\def\v#1{\hbox{\boldmath$#1$}}
\def\vepsilon{\v{\epsilon}}
\def\vPhi{\v{\Phi}}
\def\vomega{\v{\omega}}
\def\vsigma{\v{\sigma}}
\def\vmu{\v{\mu}}
\def\vxi{\v{\xi}}
\def\vpsi{\v{\psi}}
\def\vth{\v{\th}}
\def\vphi{\v{\phi}}
\def\vchi{\v{\chi}}

% abbrevio le lettere greche normali % NB, le seguenti vanno scritte
% cosi perche \o, \l etc confliggono con HyperTex e quindi con arXiv:
\newcommand{\om}{\omega}
\newcommand{\Om}{\Omega}
%NON usare \th , lambda,  Lambda

% queste sotto funzionano senza problemi
\def\f{\phi}
\def\D{\Delta}
\def\a{\alpha}
\def\b{\beta}
\def\ab{\alpha\beta}

\def\s{\sigma}
\def\g{\gamma}
\def\G{\Gamma}
\def\d{\delta}
\def\Si{\Sigma}
\def\eps{\epsilon}
\def\veps{\varepsilon}
\def\Ups{\Upsilon}
\def\Upsun{{\Upsilon}_{\odot}}

\def\dslash{\hspace{-1mm}\not{\hbox{\kern-2pt $\partial$}}}
\def\Dslash{\not{\hbox{\kern-4pt $D$}}}
\def\pslash{\not{\hbox{\kern-2.1pt $p$}}}
\def\kslash{\not{\hbox{\kern-2.3pt $k$}}}
\def\qslash{\not{\hbox{\kern-2.3pt $q$}}}

% libro QFT:

\newcommand{\vac}{|0\rangle}
\newcommand{\cav}{\langle 0|}
\newcommand{\hint}{H_{\rm int}}
\newcommand{\vp}{{\bf p}}
\newcommand{\vq}{{\bf q}}
\newcommand{\vk}{{\bf k}}
\newcommand{\vx}{{\bf x}}
\newcommand{\xp}{{\bf x}_{\perp}}
\newcommand{\vy}{{\bf y}}
\newcommand{\vz}{{\bf z}}
\newcommand{\vu}{{\bf u}}
\newcommand{\bdot}{{\bf\cdot}}

\def\p1{{\bf p}_1}
\def\p2{{\bf p}_2}
\def\k1{{\bf k}_1}
\def\k2{{\bf k}_2}

% abbreviazioni utili in relativita' generale
\newcommand{\emn}{\eta_{\mu\nu}}
\newcommand{\ers}{\eta_{\rho\sigma}}
\newcommand{\emr}{\eta_{\mu\rho}}
\newcommand{\ens}{\eta_{\nu\sigma}}
\newcommand{\ems}{\eta_{\mu\sigma}}
\newcommand{\enr}{\eta_{\nu\rho}}
\newcommand{\eMN}{\eta^{\mu\nu}}
\newcommand{\eRS}{\eta^{\rho\sigma}}
\newcommand{\eMR}{\eta^{\mu\rho}}
\newcommand{\eNS}{\eta^{\nu\sigma}}
\newcommand{\eMS}{\eta^{\mu\sigma}}
\newcommand{\eNR}{\eta^{\nu\rho}}
\newcommand{\ema}{\eta_{\mu\alpha}}
\newcommand{\emb}{\eta_{\mu\beta}}
\newcommand{\ena}{\eta_{\nu\alpha}}
\newcommand{\enb}{\eta_{\nu\beta}}
\newcommand{\eab}{\eta_{\alpha\beta}}
\newcommand{\eAB}{\eta^{\alpha\beta}}

\newcommand{\gmn}{g_{\mu\nu}}
\newcommand{\grs}{g_{\rho\sigma}}
\newcommand{\gmr}{g_{\mu\rho}}
\newcommand{\gns}{g_{\nu\sigma}}
\newcommand{\gms}{g_{\mu\sigma}}
\newcommand{\gnr}{g_{\nu\rho}}
\newcommand{\gsn}{g_{\sigma\nu}}
\newcommand{\gsm}{g_{\sigma\mu}}
\newcommand{\gMN}{g^{\mu\nu}}
\newcommand{\gRS}{g^{\rho\sigma}}
\newcommand{\gMR}{g^{\mu\rho}}
\newcommand{\gNS}{g^{\nu\sigma}}
\newcommand{\gMS}{g^{\mu\sigma}}
\newcommand{\gNR}{g^{\nu\rho}}
\newcommand{\gLR}{g^{\lambda\rho}}
\newcommand{\gSN}{g^{\sigma\nu}}
\newcommand{\gSM}{g^{\sigma\mu}}
\newcommand{\gAB}{g^{\alpha\beta}}
\newcommand{\gab}{g_{\alpha\beta}}

\newcommand{\gBmn}{\bar{g}_{\mu\nu}}
\newcommand{\gBrs}{\bar{g}_{\rho\sigma}}
\newcommand{\gBMN}{\bar{g}^{\mu\nu}}
\newcommand{\gBRS}{\bar{g}^{\rho\sigma}}
\newcommand{\gBMS}{\bar{g}^{\mu\sigma}}
\newcommand{\gBAB}{\bar{g}^{\alpha\beta}}
\newcommand{\gBma}{\bar{g}_{\mu\alpha}}
\newcommand{\gBnb}{\bar{g}_{\nu\beta}}
\newcommand{\gBab}{\bar{g}_{\a\b}}

\newcommand{\hmn}{h_{\mu\nu}}
\newcommand{\hrs}{h_{\rho\sigma}}
\newcommand{\hmr}{h_{\mu\rho}}
\newcommand{\hns}{h_{\nu\sigma}}
\newcommand{\hms}{h_{\mu\sigma}}
\newcommand{\hnr}{h_{\nu\rho}}
\newcommand{\hra}{h_{\rho\alpha}}
\newcommand{\hsb}{h_{\sigma\beta}}
\newcommand{\hma}{h_{\mu\alpha}}
\newcommand{\hna}{h_{\nu\alpha}}
\newcommand{\hmb}{h_{\mu\beta}}
\newcommand{\has}{h_{\alpha\sigma}}
\newcommand{\hab}{h_{\alpha\beta}}
\newcommand{\hnb}{h_{\nu\beta}}
\newcommand{\hcr}{h_{\times}}

\newcommand{\hMN}{h^{\mu\nu}}
\newcommand{\hRS}{h^{\rho\sigma}}
\newcommand{\hMR}{h^{\mu\rho}}
\newcommand{\hNS}{h^{\nu\sigma}}
\newcommand{\hMS}{h^{\mu\sigma}}
\newcommand{\hNR}{h^{\nu\rho}}
\newcommand{\hAB}{h^{\alpha\beta}}
\newcommand{\hij}{h_{ij}}
\newcommand{\hIJ}{h^{ij}}
\newcommand{\hkl}{h_{kl}}
\newcommand{\hTTij}{h_{ij}^{\rm TT}}
\newcommand{\HTTij}{H_{ij}^{\rm TT}}
\newcommand{\dhTTij}{\dot{h}_{ij}^{\rm TT}}
\newcommand{\hTTab}{h_{ab}^{\rm TT}}

\newcommand{\sh}{\mathsf{h}}% h in caratteri sans-serif per il cap 5
\newcommand{\shmn}{\mathsf{h}_{\mu\nu}}
\newcommand{\shrs}{\mathsf{h}_{\rho\sigma}}
\newcommand{\shmr}{\mathsf{h}_{\mu\rho}}
\newcommand{\shns}{\mathsf{h}_{\nu\sigma}}
\newcommand{\shms}{\mathsf{h}_{\mu\sigma}}
\newcommand{\shnr}{\mathsf{h}_{\nu\rho}}
\newcommand{\shra}{\mathsf{h}_{\rho\alpha}}
\newcommand{\shsb}{\mathsf{h}_{\sigma\beta}}
\newcommand{\shma}{\mathsf{h}_{\mu\alpha}}
\newcommand{\shna}{\mathsf{h}_{\nu\alpha}}
\newcommand{\shmb}{\mathsf{h}_{\mu\beta}}
\newcommand{\shas}{\mathsf{h}_{\alpha\sigma}}
\newcommand{\shab}{\mathsf{h}_{\alpha\beta}}
\newcommand{\shnb}{\mathsf{h}_{\nu\beta}}
\newcommand{\shcr}{\mathsf{h}_{\times}}
\newcommand{\shMN}{\mathsf{h}^{\mu\nu}}
\newcommand{\shRS}{\mathsf{h}^{\rho\sigma}}
\newcommand{\shMR}{\mathsf{h}^{\mu\rho}}
\newcommand{\shNS}{\mathsf{h}^{\nu\sigma}}
\newcommand{\shMS}{\mathsf{h}^{\mu\sigma}}
\newcommand{\shNR}{\mathsf{h}^{\nu\rho}}
\newcommand{\shAB}{\mathsf{h}^{\alpha\beta}}
\newcommand{\shij}{\mathsf{h}_{ij}}
\newcommand{\shIJ}{\mathsf{h}^{ij}}
\newcommand{\shkl}{\mathsf{h}_{kl}}
\newcommand{\shTTij}{\mathsf{h}_{ij}^{\rm TT}}
\newcommand{\shTTab}{\mathsf{h}_{ab}^{\rm TT}}

\newcommand{\bhmn}{\bar{h}_{\mu\nu}}
\newcommand{\bhrs}{\bar{h}_{\rho\sigma}}
\newcommand{\bhmr}{\bar{h}_{\mu\rho}}
\newcommand{\bhns}{\bar{h}_{\nu\sigma}}
\newcommand{\bhms}{\bar{h}_{\mu\sigma}}
\newcommand{\bhnr}{\bar{h}_{\nu\rho}}
\newcommand{\bhRS}{\bar{h}^{\rho\sigma}}
\newcommand{\bhMN}{\bar{h}^{\mu\nu}}
\newcommand{\bhNR}{\bar{h}^{\nu\rho}}
\newcommand{\bhMR}{\bar{h}^{\mu\rho}}
\newcommand{\bhAB}{\bar{h}^{\alpha\beta}}

\newcommand{\hax}{h^{\rm ax}}
\newcommand{\haxmn}{h^{\rm ax}_{\mu\nu}}
\newcommand{\hpol}{h^{\rm pol}}
\newcommand{\hpolmn}{h^{\rm pol}_{\mu\nu}}

\newcommand{\dgzz}{{^{(2)}g_{00}}}
\newcommand{\qgzz}{{^{(4)}g_{00}}}
\newcommand{\tgzi}{{^{(3)}g_{0i}}}
\newcommand{\dgij}{{^{(2)}g_{ij}}}
\newcommand{\zTzz}{{^{(0)}T^{00}}}
\newcommand{\dTzz}{{^{(2)}T^{00}}}
\newcommand{\dTii}{{^{(2)}T^{ii}}}
\newcommand{\uTzi}{{^{(1)}T^{0i}}}

\newcommand{\xm}{x^{\mu}}
\newcommand{\xn}{x^{\nu}}
\newcommand{\xr}{x^{\rho}}
\newcommand{\xs}{x^{\sigma}}
\newcommand{\xa}{x^{\a}}
\newcommand{\xb}{x^{\b}}

\newcommand{\hatk}{\hat{\bf k}}
\newcommand{\hatn}{\hat{\bf n}}
\newcommand{\hatx}{\hat{\bf x}}
\newcommand{\haty}{\hat{\bf y}}
\newcommand{\hatz}{\hat{\bf z}}
\newcommand{\hatr}{\hat{\bf r}}
\newcommand{\hatu}{\hat{\bf u}}
\newcommand{\hatv}{\hat{\bf v}}
\newcommand{\xim}{\xi_{\mu}}
\newcommand{\xin}{\xi_{\nu}}
\newcommand{\xia}{\xi_{\a}}
\newcommand{\xib}{\xi_{\b}}
\newcommand{\xiM}{\xi^{\mu}}
\newcommand{\xiN}{\xi^{\nu}}

\newcommand{\tA}{\tilde{\bf A} ({\bf k})}

\newcommand{\pam}{\pa_{\mu}}
\newcommand{\pal}{\pa_{\mu}}
\newcommand{\pan}{\pa_{\nu}}
\newcommand{\parho}{\pa_{\rho}}
\newcommand{\pas}{\pa_{\sigma}}
\newcommand{\paM}{\pa^{\mu}}
\newcommand{\paN}{\pa^{\nu}}
\newcommand{\paR}{\pa^{\rho}}
\newcommand{\paS}{\pa^{\sigma}}
\newcommand{\paa}{\pa_{\alpha}}
\newcommand{\pab}{\pa_{\beta}}
\newcommand{\pat}{\pa_{\theta}}
\newcommand{\paf}{\pa_{\phi}}

\newcommand{\Dam}{D_{\mu}}
\newcommand{\Dan}{D_{\nu}}
\newcommand{\Dar}{D_{\rho}}
\newcommand{\Das}{D_{\sigma}}
\newcommand{\DaM}{D^{\mu}}
\newcommand{\DaN}{D^{\nu}}
\newcommand{\DaR}{D^{\rho}}
\newcommand{\DaS}{D^{\sigma}}
\newcommand{\Daa}{D_{\alpha}}
\newcommand{\Dab}{D_{\beta}}

\newcommand{\DBm}{\bar{D}_{\mu}}
\newcommand{\DBn}{\bar{D}_{\nu}}
\newcommand{\DBr}{\bar{D}_{\rho}}
\newcommand{\DBs}{\bar{D}_{\sigma}}
\newcommand{\DBt}{\bar{D}_{\tau}}
\newcommand{\DBa}{\bar{D}_{\alpha}}
\newcommand{\DBb}{\bar{D}_{\beta}}
\newcommand{\DBM}{\bar{D}^{\mu}}
\newcommand{\DBN}{\bar{D}^{\nu}}
\newcommand{\DBR}{\bar{D}^{\rho}}
\newcommand{\DBS}{\bar{D}^{\sigma}}
\newcommand{\DBA}{\bar{D}^{\alpha}}

\newcommand{\GMnr}{{\Gamma}^{\mu}_{\nu\rho}}
\newcommand{\Glmn}{{\Gamma}^{\lambda}_{\mu\nu}}
\newcommand{\barGMnr}{{\bar{\Gamma}}^{\mu}_{\nu\rho}}
\newcommand{\GMns}{{\Gamma}^{\mu}_{\nu\sigma}}
\newcommand{\GInr}{{\Gamma}^{i}_{\nu\rho}}
\newcommand{\Rmn}{R_{\mu\nu}}
\newcommand{\Rmnrs}{R_{\mu\nu\rho\sigma}}
\newcommand{\RMnrs}{{R^{\mu}}_{\nu\rho\sigma}}
\newcommand{\Tmn}{T_{\mu\nu}}
\newcommand{\Tab}{T_{\a\b}}
\newcommand{\TMN}{T^{\mu\nu}}
\newcommand{\TAB}{T^{\a\b}}
\newcommand{\TBmn}{\bar{T}_{\mu\nu}}
\newcommand{\TBMN}{\bar{T}^{\mu\nu}}
\newcommand{\TRS}{T^{\rho\sigma}}
\newcommand{\tmn}{t_{\mu\nu}}
\newcommand{\tMN}{t^{\mu\nu}}
\newcommand{\RUmn}{R_{\mu\nu}^{(1)}}
\newcommand{\RDmn}{R_{\mu\nu}^{(2)}}
\newcommand{\RTmn}{R_{\mu\nu}^{(3)}}
\newcommand{\RBmn}{\bar{R}_{\mu\nu}}
\newcommand{\RBmr}{\bar{R}_{\mu\rho}}
\newcommand{\RBnr}{\bar{R}_{\nu\rho}}

\newcommand{\dddM}{\kern 0.2em \raise 1.9ex\hbox{$...$}\kern -1.0em \hbox{$M$}}
\newcommand{\dddQ}{\kern 0.2em \raise 1.9ex\hbox{$...$}\kern -1.0em \hbox{$Q$}}
\newcommand{\dddI}{\kern 0.2em \raise 1.9ex\hbox{$...$}\kern -1.0em\hbox{$I$}}
\newcommand{\dddJ}{\kern 0.2em \raise 1.9ex\hbox{$...$}\kern-1.0em
\hbox{$J$}}
\newcommand{\dddcalJ}{\kern 0.2em \raise 1.9ex\hbox{$...$}\kern-1.0em
\hbox{${\cal J}$}}

\newcommand{\dddO}{\kern 0.2em \raise 1.9ex\hbox{$...$}\kern -1.0em
\hbox{${\cal O}$}}
\def\dddz{\raise 1.5ex\hbox{$...$}\kern -0.8em \hbox{$z$}}
\def\dddd{\raise 1.8ex\hbox{$...$}\kern -0.8em \hbox{$d$}}
\def\dddbd{\raise 1.8ex\hbox{$...$}\kern -0.8em \hbox{${\bf d}$}}
\def\ddbd{\raise 1.8ex\hbox{$..$}\kern -0.8em \hbox{${\bf d}$}}
\def\dddx{\raise 1.6ex\hbox{$...$}\kern -0.8em \hbox{$x$}}

\newcommand{\hti}{\tilde{h}}
\newcommand{\hf}{\tilde{h}_{ab}(f)}
\newcommand{\Hti}{\tilde{H}}
\newcommand{\fmin}{f_{\rm min}}
\newcommand{\fmax}{f_{\rm max}}
\newcommand{\frot}{f_{\rm rot}}
\newcommand{\fpol}{f_{\rm pole}}
\newcommand{\omax}{\o_{\rm max}}
\newcommand{\orot}{\o_{\rm rot}}
\newcommand{\op}{\o_{\rm p}}
\newcommand{\tmax}{t_{\rm max}}
\newcommand{\tobs}{t_{\rm obs}}
\newcommand{\fobs}{f_{\rm obs}}
\newcommand{\temis}{t_{\rm emis}}
\newcommand{\DE}{\D E_{\rm rad}}
\newcommand{\DEm}{\D E_{\rm min}}
\newcommand{\msun}{M_{\odot}}
\newcommand{\rsun}{R_{\odot}}
\newcommand{\ogw}{\o_{\rm gw}}
\newcommand{\fgw}{f_{\rm gw}}
\newcommand{\oL}{\o_{\rm L}}
\newcommand{\kL}{k_{\rm L}}
\newcommand{\lL}{\l_{\rm L}}
\newcommand{\mns}{M_{\rm NS}}
\newcommand{\rns}{R_{\rm NS}}
\newcommand{\tret}{t_{\rm ret}}
\newcommand{\Sch}{Schwarzschild }
\newcommand{\rtid}{r_{\rm tidal}}

\newcommand{\ot}{\o_{\rm t}}
\newcommand{\mt}{m_{\rm t}}
\newcommand{\gt}{\g_{\rm t}}
\newcommand{\xit}{\tilde{\xi}}
\newcommand{\xtr}{\xi_{\rm t}}
\newcommand{\xtj}{\xi_{{\rm t},j}}
\newcommand{\dxtj}{\dot{\xi}_{{\rm t},j}}
\newcommand{\ddxtj}{\ddot{\xi}_{{\rm t},j}}
\newcommand{\teff}{T_{\rm eff}}
\newcommand{\samp}{S_{\xi_{\rm t}}^{\rm ampl}}

% defs dal phys rept
\newcommand{\mpl}{M_{\rm Pl}}
\newcommand{\mgut}{M_{\rm GUT}}
\newcommand{\lpl}{l_{\rm Pl}}
\newcommand{\tpl}{t_{\rm Pl}}
\newcommand{\ls}{\lambda_{\rm s}}
\newcommand{\Ogw}{\Omega_{\rm gw}}
\newcommand{\hogw}{h_0^2\Omega_{\rm gw}}
\newcommand{\hn}{h_n(f)}

\newcommand{\sinc}{{\rm sinc}\, }
\newcommand{\Ein}{E_{\rm in}}
\newcommand{\Eout}{E_{\rm out}}
\newcommand{\Et}{E_{\rm t}}
\newcommand{\Er}{E_{\rm refl}}
\newcommand{\lm}{\l_{\rm mod}}

\newcommand{\mrI}{\mathrm{I}}
\newcommand{\mrJ}{\mathrm{J}}
\newcommand{\mrW}{\mathrm{W}}
\newcommand{\mrX}{\mathrm{X}}
\newcommand{\mrY}{\mathrm{Y}}
\newcommand{\mrZ}{\mathrm{Z}}
\newcommand{\mrM}{\mathrm{M}}
\newcommand{\mrS}{\mathrm{S}}

\newcommand{\rmI}{\mathrm{I}}
\newcommand{\rmJ}{\mathrm{J}}
\newcommand{\rmW}{\mathrm{W}}
\newcommand{\rmX}{\mathrm{X}}
\newcommand{\rmY}{\mathrm{Y}}
\newcommand{\rmZ}{\mathrm{Z}}
\newcommand{\rmM}{\mathrm{M}}
\newcommand{\rmS}{\mathrm{S}}
\newcommand{\rmU}{\mathrm{U}}
\newcommand{\rmV}{\mathrm{V}}

% comandi usati nel Vol. 2

\newcommand{\et}{{{\bf e}^t}}
\newcommand{\etm}{{\bf e}^t_{\mu}}
\newcommand{\etn}{{\bf e}^t_{\nu}}

\newcommand{\er}{{{\bf e}^r}}
\newcommand{\erm}{{\bf e}^r_{\mu}}
\newcommand{\ern}{{\bf e}^r_{\nu}}

\newcommand{\hz}{H^{(0)}}
\newcommand{\hu}{H^{(1)}}
\newcommand{\hd}{H^{(2)}}
\newcommand{\thz}{\tilde{H}^{(0)}}
\newcommand{\thu}{\tilde{H}^{(1)}}
\newcommand{\thd}{\tilde{H}^{(2)}}
\newcommand{\tK}{\tilde{K}}
\newcommand{\tZ}{\tilde{Z}}
\newcommand{\tQ}{\tilde{Q}}
\newcommand{\ts}{\tilde{s}}

\newcommand{\inT}{\int_{-\infty}^{\infty}}
\newcommand{\intz}{\int_{0}^{\infty}}
\newcommand{\Dl}{\int{\cal D}\lambda}

\newcommand{\fnl}{f_{\rm NL}}

%% file: imprimatur.tex
%
%\begin{figure}[t]
%\begin{center}
\includegraphics[width=15cm]{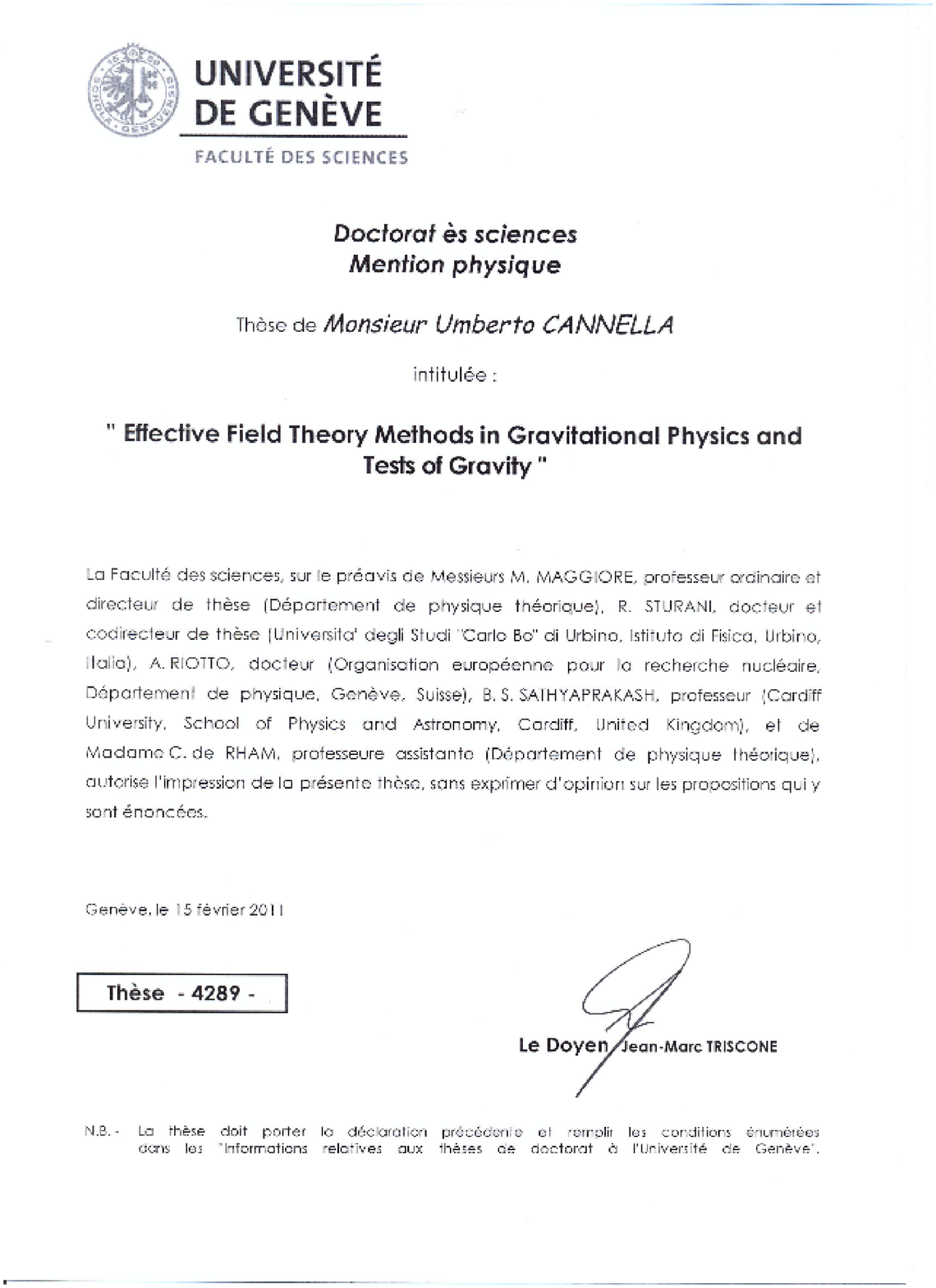} 
%\caption{}
%\end{center}
%\end{figure}
%

\smallskip 
\noindent 
Cette thèse a été publiée sur l'archive télématique arXiv.org avec la référence suivante:
"arXiv:1103.0983 [gr-qc]".  

%% file: Merci.tex
\chapter*{Remerciements}
\chaptermark{}

Je suis très reconnaissant à mon directeur de thèse, le professeur Michele Maggiore, pour m'avoir donné la possibilité de réaliser mes études doctorales à l'Université de Genève, et au docteur Riccardo Sturani pour avoir co-dirigé mon travail pendant ces quatre ans. En suite, c'est un plaisir de remercier Domenico Sapone, un ancien doctorant de l'UniGe, qui a joué le rôle de la chance au moment de ma postulation. Outre Domenico, je me réjouis d'avoir pu partager une partie plus ou moins consistante de mon chemin avec: Filipe, Hillary, Riccardo et Stefano, Antti, Camille, Chiara, Elisa, José, Lukas, Natalia, Rajeev, Marc, Marcus et Max, Clare et Lavinia, Miki et Mona, Dania, Géraldine, Heidi, Iurii, Ivan, Janine, Jian-Li, Kirsten, Marlies, Mathias Albert, Mathias Lunde, Rafa et Simon, Christoph et Sadi, Cyrille, Maher et Philippe; avec tous j'ai partagé de bons moments, avec beaucoup j'ai eu la chance de devenir copain. L'ambiance familiale caractérisant l'École de Physique m'a permis d'avoir des relations très sympathiques avec tout le personnel, notamment avec Andreas, Cécile, Christine, Danièle, Francine, Heriberto, Marisa, Nathalie et Vincenzo, ainsi que avec les professeurs Claudia, Jean-Pierre, Marcos, Markus, Martin Kunz, Martin Pohl, Michel, Ruth, Peter et Werner; en particulier, je me réjouis d'avoir pu devenir ami avec Claudia et Martin Kunz. Parmi les chercheurs qui ont visité le Département de Physique Théorique pendant mon sejour, j'ai trouvé un soutien amicale en Céline Boehm et Daniele Steer. Toujours au sein de l'université, c'était une expérience sympathique de travailler au laboratoire de vulgarisation scientifique, le Physiscope, et d'y rencontrer Adriana, Céline Corthay, Dook, Fanny, Florian, Gaby, Gauthier, Greg, Lidia, Noé, Olivier, Stefano et le professeur Christoph Renner. Au cours de la dernière année de travail, j'ai pu profité de la collaboration precieuse de Emanuele Berti, Ira Rothstein et Manuel Tiglio aux Etats Unis. Une mention à part est pour les membres de mon jury de thèse: Antonio Riotto, Claudia De Rham et Bangalore S. Sathyaprakash. Pour terminer les remerciements du coté universitaire, je veux exprimer ma reconnaisance au Fonds National Suisse et au Département de l'Instruction Publique, qui ont financé mes études. 

Une autre raison pour la quelle j'aimerai me souvenir de Genève est liée aux gens que j'ai rencontré en dehors du lieu de travail: j'éprouve ainsi une grande gratitude pour mes amis Chicca et Alessandro qui m'ont accompagné de maniere vraiment proche; une pensée très amicale va à mes «oncles suisses» Patrizia et Claude; je suis encore très content d'avoir connu Ple et sa cuisine, les Italians de Genéve, les amis qui travaillent pour les Nations Unies et les gens des églises catholiques italienne, espagnole et francophone. Ici, à Genève, j'ai rencontré aussi ma copine Alessandra: à elle va une dédicace spéciale. 
Tout le long de mon experience de vie en Suisse, j'ai pu compté sur la présence fondamentale de mes amis de Rome, qui ont toujours été la pour moi. 
Encore en Italie, un remerciement particulier concerne Emanuele et les amies de Lecceto. 
{\it Last but not least,} mon doctorat n'aurait pas pu être réalisé sans le support et l'amour infinis de ma famille; une pensée particulière est pour mes oncles Dino et Vito, pour tout ce que leurs histoires m'ont appris, et pour mon oncle Carlo pour ces suggestions.

%% file: Resume.tex
%%%%%%%%%%%%%%%%%%%%%%%%%%%%%%%%%%%%%%%%%%%%%%%%%%%%%%%%%%%%%%%
%%  a eliminer
%%%%%%%%%%%%%%%%%%%%%%%%%%%%%%%%%%%%%%%%%%%%%%%%%%%%%%%%%%%%%%%
%\documentclass[a4paper,11pt]{book}
%\usepackage[french]{babel}
%\usepackage[applemac]{inputenc}
%\usepackage{ae,aeguill}
%\usepackage{latexsym}
%\usepackage{amssymb}
%\usepackage{graphicx}
%\usepackage{array,tabularx,multirow,dcolumn,hhline}
%\usepackage{rotate}
%\usepackage{amsmath}
%\usepackage{verbatim} %for the comment envir't
%\pagenumbering{roman}

%
%\begin{document}

%\noindent

%\cleardoublepage

\chapter*{R\'esum\'e}

	L'ère de l'astronomie gravitationnelle a déjà commencé. La première génération d'interféromètres terrestres a été construite et a atteint la sensibilité expérimentale prévue au niveau du projet. Pour l'instant, les ondes gravitationnelles n'ont pas été détectées, mais à un tel niveau de sensibilité, même des résultats négatifs présentent un intérêt physique. 
	Dans ce contexte, deux exemples sont trés remarquables.
\begin{description}
\begin{comment}
\item
Le résultat négatif d'une recherche visant à la détection d'ondes gravitationnelles émises par le pulsar du Crabe a montré que moins de 4\% de l'énergie perdue par ce pulsar provient de l'émission d'ondes gravitationnelles: cette contrainte est plus forte que celle obtenue en étudiant le taux de ralentissement de rotation du pulsar du Crabe.   
\end{comment}
\item
Le résultat négatif d'une recherche visant à la détection d'ondes gravitationnelles d'origine stochastique a permis de dériver une contrainte comparable aux limites cosmologiques obtenues à partir de la nucléosynthèse et du fonds diffus à micro-ondes.  
\item
Le résultat négatif d'une recherche visant à la détection d'ondes gravitationnelles émises par l'étoile à neutrons située dans le rémanent de supernova Cassiopée A a permis, pour la première fois, de placer des limites supérieures sur l'amplitude du mode d'oscillation de type «r» des étoiles à neutrons. 
\end{description}
Il est très important de souligner que ces résultats ont été obtenus par le fait que les interféromètres de première génération ont atteint leurs sensibilités prévues lors de la phase de projet; en d'autres termes: il était attendu que nous puissions explorer des régimes physiques bien précis d'une façon propre aux techniques expérimentales de l'interférométrie gravitationnelle. De même, nous nous attendons aujourd'hui à ce que la deuxième génération d'interféromètres, en voie de réalisation prochaine, puisse inaugurer l'époque de détection d'ondes gravitationnelles de façon routinière. Parmi les sources les plus probables qui seront explorées d'abord, il y a les systèmes binaires d'objets compacts, comme les étoiles à neutrons et les trous noirs: en effet, l'émission d'ondes gravitationnelles est telle que le mouvement orbital descend en spirale et les premiers signaux attendus à être détectés sont émis par les deux objets lors de cette phase. Ce mouvement en spirale est caractérisé par une vitesse relative~$v$ et une séparation relative~$r$, telles que la dynamique orbitale est non-relativiste: par conséquent, ce régime peut être décrit par une série perturbative dans le petit paramètre~$\epsilon$ d'ordre $(v/c)^2$ ou $(m\,G_N / r)$, $c$ étant la vitesse de la lumière, $G_N$ la constante de Newton et $m$ la masse typique des objets; les corrections à l'ordre principale, représentées par la dynamique newtonienne, sont classées selon les puissances du paramètre $\epsilon$ qu'ils contiennent et forment l'expansion dite post-newtonienne.  
	En raison de sa nature perturbative, l'expansion post-newtonienne peut aussi être interprétée d'un point de vue de la théorie des champs et reformulée en termes de diagrammes de Feynman. 
	Au cours des ces dernières années, l'application de ces instruments à la gravité classique a été relancée grâce à l'introduction d'un approche basée sur le théorie {\it effective} des champs. Dans ce contexte, l'étude d'un système binaire est traitée d'une façon similaire au problème des états liés non-relativistes en électrodynamique et cromodynamique quantique: en raison de cette analogie, le traitement d'un système binaire d'objets compacts selon la théorie effective des champs a été appelé «théorie non-relativiste de la Relativité Générale». 
	Dans ma thèse, j'explore différentes utilisations des méthodes typiques de la physique de particules empruntées par la théorie non-relativiste de la Relativité Générale. 
	No\-tamment j'applique de façon étendue sa série perturbative des diagrammes de Feynman. Outre leur convenance au niveau des calculs, la technique des diagrammes peut être adoptée pour faire des investigations autour des théories alternatives de la gravité et pour considérer les informations provenant des expériences d'un point de vue de la théorie des champs. Dans le premier travail original que je présente dans cette thèse, j'ai étudié une extension de la théorie non-relativiste de la Relativité Générale au cas d'une théorie tenseur-scalaire de la gravitation. 
	%\begin{comment}%%%%%%%%%%%%%%%%%%%%%%%%%%%%%%%%%
J'y ai traité la dynamique des sources de type ponctuel et à une dimension; je me suis intéressé en particulier à l'étude de la renormalisation du tenseur énergie-impulsion pour les deux types de sources.
%\end{comment}%%%%%%%%%%%%%%%%%%%%%%%%%%%%%%%%%%%%%
	Le deuxième et troisième travail que je décris au cours de cette thèse se situent dans le contexte des tests de la gravitation. La Relativité Générale est une théorie fortement non-linéaire: jusqu'à présent, elle a fait l'objet de plusieurs tests expérimentaux et elle a été confirmée avec succès. 
	Toutefois, ces tests ont exploré des régimes dynamiques correspondant aux premiers ordres de l'expansion post-newtonienne seulement: en particulier, l'ordre le plus élevé ayant pu être investigué est $(v/c)^5$, qui représente le tout premier terme du secteur radiatif en Relativité Générale: à travers la detection des ondes radio émises par le pulsar de Hulse et Taylor, on a mesuré ce terme et donc prouvé l'existence de la radiation gravitationnelle. Par ailleurs, au moyen de l'astronomie gravitationnelle, nous nous attendons à pouvoir tester la Relativité Générale jusqu'à $(v/c)^{12}$! 
Les effets qui correspondent à ces termes sont si faibles que seulement une technique dediée comme l'interféromètrie gravitationnelle permet de les detecter. 
Concernant l'expansion post-newtonienne jusqu'à l'ordre $(v/c)^5$, une grande quantité de travaux dans la littérature a été consacrée à comprendre quelles sont les caractéristiques de la Relativité Générale testées par une expérience et comment ces caractéristiques diffèrent des prédictions faites par des théories alternatives de la gravitation. En vue des premières détections d'ondes gravitationnelles, il est donc important d'envisager des cadres théoriques et phénoménologiques propres aux nouveaux régimes dynamiques. Au cours des deux derniers chapitres de cette thèse, je discute un exemple original d'un de ces cadres. À travers la théorie non-relativiste de la Relativité Générale, les non-linéarités de la gravitation sont représentées par une série perturbative de diagrammes de Feynman dans lesquels les gravitons classiques interagissent avec les sources matérielles ainsi qu'avec eux-mêmes. Nous avons choisi d'étudier les non-linéarités de la Relativité Générale en désignant les vertex d'auto-interaction gravitationnelle par des paramètres: cela nous a permis, par exemple, de traduire la mesure de la diminution de la période orbitale du pulsar de Hulse et Taylor en une contrainte sur le vertex à trois gravitons, contrainte précise au niveau de 0.1\%. 
Cette attitude originale constitue un nouveau cadre phénoménologique dans le contexte des paramétrisations du régime radiatif et de champ fort de la Relativité Générale: un tel cadre est trés utile pour interpréter les experiences gravitationnelles presents et futures.

%% file: Introduction.tex
%%%%%%%%%%%%%%%%%%%%%%%%%%%%%%
\chapter*{Introduction} \label{chap:intro}
\addcontentsline{toc}{chapter}{Introduction}
\chaptermark{Introduction}
%%%%%%%%%%%%%%%%%%%%%%%%%%%%%%

The era of gravitational-wave (GW) astronomy has already begun.
First-generation ground-based \ifos\ have been built and operated at their design sensitivities. 
No GWs have been detected yet but, at this level of sensitivity, even negative results have physical relevance. 
Most notable examples are the following ones.
\bd
\item The negative result of a search for periodic GWs from the Crab pulsar~\cite{LIGO-Crab} showed that no more than 4\% of the energy loss of the pulsar is caused by the emission of GWs: this bound lies below the upper limit inferred from the spin-down rate of the Crab pulsar. 

\item The negative result of a search for stochastic backgrounds of GWs~\cite{GW_bck-Nat} allowed one to derive upper bounds comparable to limits of cosmological origin inferred from the nucleosynthesis~\cite{GW_bck-BBN} and from the cosmic microwave background~\cite{GW_bck_CMB}.

\item The negative result of a search for periodic GWs from the neutron star in the supernova remnant Cassiopeia~A enabled one to put upper limits on the amplitude of r-mode oscillations of a neutron star for the first time~\cite{LIGO-r_modes}.  
%It searches gravitational wave frequencies from 100 to 300 Hz, and covers a wide range of first and second frequency derivatives appropriate for the age of the remnant and for different spin-down mechanisms. No gravitational wave signal was detected. 
%Within the range of search frequencies, we set 95\% confidence upper limits of 0.7--1.2e-24 on the intrinsic gravitational wave strain, 0.4--4e-4 on the equatorial ellipticity of the neutron star, and 0.005--0.14 on the . 
%These direct upper limits beat indirect limits derived from energy conservation and enter the range of theoretical predictions involving crystalline exotic matter or runaway r-modes. 
\ed

\noindent 
It is important to stress that these scientific results have been obtained because first generation  \ifos\ met their design sensitivities; in other words: it was expected that one would explore well determined physical regimes in a way which is unique to the experimental techniques of GW interferometry. 
Analogously, today it is expected that the soon-to-come second-generation ground-based \ifos\ will open the phase in which GWs are routinely detected. 
%With these data it will be possible to explore the dynamics of gravity in its strong-field/radiative regime and, then, to test General Relativity (GR) in unmatched ways. 
%First detections 
Among the most likely sources to be observed first are systems of compact binaries like neutron stars and/or black holes: in fact, the emission of GWs cause the orbital motion to {\it inspiral} and the first signals awaited are the waves emitted by binary systems in this phase. 
The inspiral motion is characterized by a relative velocity~$v$ and a relative separation~$r$ that make the orbital dynamics non-relativistic: this regime can then be described by a perturbative series in the small parameter $\eps\sim (v/c)^2\sim m\,G_N/r$, where $c$ is the speed of light, $G_N$ is Newton's constant and $m$ the mass of one of the objects; corrections to the leading order Newtonian dynamics are labeled by powers of~$\eps$ and form the so-called post-Newtonian (PN) expansion. 
%The inspiral phase is non-relativistic and can be described analytically by a perturbative series, where corrections to the leading order Newtonian dynamics form the so-called post-Newtonian expansion. 

Being perturbative in nature, the PN expansion can also be interpreted from a ﬁeld-theoretical  standpoint and reformulated in terms of Feynman diagrams. 
In the last few years the use of these instruments in classical gravity has revived thanks to the introduction of an effective field theory (EFT) approach~\cite{NRGR_paper}. 
In this context the dynamics of a binary system is studied in a way analogous to non-relativistic bound states in QED and QCD~\cite{Caswell:1985ui}: for such a reason the EFT framework of ref.~\cite{NRGR_paper} has been dubbed non-relativistic General Relativity (NRGR). 
The use of an EFT framework in the two-body problem in GR is motivated by the fact that the binary dynamics is a typical situation in which the disparity of physical scales allows one to separate the relevant degrees of freedom in subsets with decoupled dynamics.
For example, because the extension of the radiating source is much smaller than the wavelength of gravitational perturbations, the modes corresponding to the internal dynamics of the compact object can be {\it integrated out} and the source is {\it effectively} described as a 
point particle. 
By means of NRGR, it has been possible to obtain previously un-computed spin contributions to the source multipoles at next-to-leading order~\cite{Porto:2010zg}: 
this provides the last missing ingredient required to determine the phase evolution to 3PN accuracy including all spin effects.

\smallskip

In this thesis I explore different uses of the EFT methods of NRGR, in particular of its perturbative series of Feynman diagrams. 
Beside being a convenient tool for calculations in the PN regime of GR, diagrammatic techniques can be employed to investigate alternative theories of gravity and to have a different point of view on the physical information that can be extracted from experiments. 
In the first original study I present in this thesis I have extended NRGR to the case of a \st\ and treated the dynamics of point-like and string-like objects; notably, I have addressed the issue of the renormalization of the energy-momentum tensor for both types of sources. 
The second and third study I report in this thesis are set in the context of testing gravity. 
GR is a highly non-linear theory of gravity: 
so far it has been the subject of many experimental tests and it has passed them all with flying colors. 
However, these tests have probed dynamical regimes that correspond only to the first few orders of the PN approximation: notably, the highest such order is $(v/c)^5$ and represents the very first term of the radiative sector in GR. 
The existence of this term has been assessed by timing Hulse-Taylor pulsar and has provided the fundamental result of the reality of gravitational radiation. 
In contrast, by means of GW astronomy, one aims at testing GR up to $(v/c)^{12}$! 
This corresponds to the very high 3.5 next-to-leading order in the PN expansion, which comprises a set of non-linear effects that are very peculiar to GR and cannot be probed by other means than gravity \ifos. 
For what concerns the orders up to $(v/c)^5$, considerable work has been done in the past to understand the peculiar features of GR that are probed by an experiment and to confront them with the predictions of alternative theories of gravity.  
In view of the first GW detections, it is then relevant to envisage other testing frameworks that are appropriate to the new dynamical regime. 
The one explored in the last two chapters of this thesis constitutes a relevant example:   
here, by means of the NRGR description, gravity non-linearities are represented by a perturbative series of Feynman diagrams in which classical gravitons interact with matter sources and among themselves. 
Tagging the self-interaction vertices of gravitons with parameters, it is possible, for example, to translate the measure of the period decay of Hulse-Taylor pulsar in a constraint on the three-graviton vertex at the 0.1\% level. 
What emerges from this attitude is a phenomenological framework that is suitable for parametrizing the strong-ﬁeld/radiative regime of GR.

\smallskip

This thesis begins with an account of the theoretical and phenomenological frameworks that can be employed to test relativistic gravity according to the dynamical regime. 
\chap{chap:How_To} starts with a concrete example of alternative theories of gravity, \sts, which will be a reference point for many of the following discussions.
In the course of discussing the consistency of \sts\ with experiments, I introduce the various testing regimes; moreover, the derivation of the testable predictions of \sts\  is a first example of the use of diagrammatic techniques in classical gravity. 
These particle-physics tools are the {\it leitmotiv} of the present thesis. 
Afterwards, I move to a more detailed treatment of the regimes, each of which is characterized by a parametrized phenomenological framework. 
In the order, I will discuss: 
\bd 
\item the 1PN order accessible by Solar-System experiments and the parametrized post-Newtonian approach, 
\item the onset of the radiative dynamics of GR at ${\cal O}(v/c)^5$ probed by binary pulsars and the parametrized post-Keplerian framework, 
\item the higher PN orders characterizing the GW phasing formula and one of the possible parametrizations for the latter.
\ed 
%The first dynamical regime is the 1PN order accessible by Solar-System experiments: at this level, there are no dissipative contributions to the dynamics. 
%The relevant framework in this context is the parametrized post-Newtonian approach. 
%The onset of the radiative dynamics of GR is at ${\cal O}(v/c)^5$ and can be probed by means of binary pulsars; here I will present the parametrized post-Keplerian framework. 
%In this context, many effects have been measured on top of the period decay so that the resulting consistency of GR with experimental data is multi-faceted. 
\chap{chap:How_To} constitutes the background material for what concerns the NRGR-based investigations presented in this thesis. 

In \chap{chap:EFT}, then, I discuss the EFT approach to inspiraling compact binaries: here I expose the relevant ingredients needed in order to present my original studies. 
The latter are the subject of the remaining chapters.  
\bd 
\item In \chap{chap:Ric} I report the aforementioned investigation of mine in which, together with Riccardo Sturani, I extended NRGR to the case of a scalar-tensor theory: this allowed us to treat the dynamics of point-like and string-like objects and to address the issue of the renormalization of the energy-momentum tensor for both types of sources.  
\item \chap{chap:Group} contains a study I have performed with the gravity group at the University of Geneva. 
In this work we expressed GR non-linearities in terms of the non-Abelian vertices of the theory and we derived the bounds that can be put on them by means of experiments of relativistic gravity: in such a way, we put forward a proposal for a parametrized field-theoretical framework for testing gravity. 
With binary pulsars timing data we found interesting constraints on the vertices at the first two perturbative levels; then, we assessed the possibility of deriving meaningful constraints with GW observations at a first qualitative level. 
\item In \chap{chap:Berti} I address the issue of bounding the vertices with GW tests in a more quantitative fashion: this is done from the point of view of parameter estimation and relies on a  follow-up study that I have taken up in collaboration with Emanuele Berti at the University of Mississippi. 
In this chapter I treat some technical details of the calculations that are common with the investigation of \chap{chap:Group}; moreover, I compare our original parametrized field-theoretical framework for testing gravity with other approaches in the literature. 
%, including the one discussed in the last section of \chap{chap:How_To}. 
In the course of discussion, I present different and complementary strategies of investigation that take advantage of the peculiarities of GW astronomy ... 
an era that has already begun.
\ed

%% file: How_To_Test/How_To_Test.tex
%%%%%%%%%%%%%%%%%%%%%%%%%%%%%%%%%
\chapter{Theoretical and phenomenological frameworks for \\
testing relativistic gravity} 
\chaptermark{Theoretical and phenomenological frameworks for 
testing gravity}
\label{chap:How_To}
%%%%%%%%%%%%%%%%%%%%%%%%%%%%%%%%%

What the numerous tests of relativistic gravity have probed so far is actually the post-Newtonian (PN) regime, i.e. the first orders of a perturbative series that can be used to approximate General Relativity (GR) as well as other theories of gravity;   
 %%As a matter of fact, in every viable theory of gravity one can study 
 non-fully-relativistic systems %non-relativistic
% ~\footnote{The term "non-relativistic" is used 
% in opposition to "fully relativistic" as referred to systems where gravity 
% is so strong that one cannot rely on perturbative schemes for their 
% treatment. It then makes perfect sense to study "non-relativistic" 
% systems to understand relativistic gravity.}
are indeed describable 
 by means of an expansion in a small parameter 
 represented by two inter-related quantities: the velocities with which 
 the objects move and the gravitational fields they generate.
%In this sense then, we have tests on a class of theories of gravity 
%among which GR also seats.

Until present, experiments conducted in the PN regime have confirmed GR and tightly constrained alternative theories. 
In the confrontation between theoretical predictions and experimental tests, it is convenient to pursue two avenues: one which is theoretically minded, the other which is based on a phenomenological attitude.
In the first case, one compares the observable predictions of GR against those of consistent alternative theories of gravity. 
%this line of reasoning has, by definition, the advantage of relying on sound theoretical basis.
In a phenomenological context, the attitude is more oriented towards understanding which features of GR as the theory of gravity are actually tested by means of a specific experiment. 
"This is a bit similar to the well-known psycho-physiological fact that the best way to appreciate a nuance of color is to surround a given patch of color by other patches with slightly different colors"~\cite{Dam_Como}. 
In the case of testing gravity, this is a very convenient way of proceeding. 
In fact, GR is a very peculiar theory where many symmetries imply that otherwise possible effects are strictly zero. 
For instance, GR requires that the inertial and gravitational mass coincide. 
Another example is the effacement of the internal structure of compact objects, i.e. the fact that in GR one can describe the external field of a self-gravitating body by means of a point particle approximation to a high degree of accuracy. 
These are only two examples of the many symmetries of GR that have been confirmed by approaching experiments with 
%Nevertheless, to be able to grasp an indication of a departure from GR as provided by future experiments, it makes sense to ask 
questions like "what would be the observable manifestation of a violation of Lorentz invariance?". 
In other words, one can entertain the possibility of violating GR symmetries even outside the domain of consistent alternative theories of gravity: for the purpose of testing a symmetry, one considers models, instead of theories, that are directly built to exhibit a possible violation.

Both the theoretical and the phenomenological attitudes make use of parameters. 
In the case of alternative theories, the parameters enter the couplings of the action functional; on the other hand, in the set of phenomenological models, the parameters stem from tagging different physical effects. 
While GR predicts a number for some observable, both models and theories have this number multiplied by a parameter, whose range of values usually contains the one predicted by GR. % at least for the effects probed so far.
%A framework like the parametrized post-Newtonian (PPN) can encompass both types of approaches: 
Therefore, in the comparison with experiments, at times one may consider parameters coming from the PN expansion of a theory like GR, at times one may deal with  parameters that belong to a model and have been introduced only to weigh a specific contribution to an observable. 
This last case applies to studies I have conducted personally, where one re-interprets gravity non-linearities in a particle-physics context and tries to constrain the values of the non-Abelian vertices of the theory. 
In this context, one makes use of instruments like Feynman diagrams, which were first applied to classical gravity in 1960~\cite{BP}. 
More recently a particle-physics attitude has been employed with the aim of testing gravity, notably through the confrontation of GR with a specific class of alternative theories. 
In refs.~\cite{Dam_Far-ST_CQG,Dam_Far_2PN_Diags:PRD53} Damour and Esposito-Farese (DEF) have started a thorough investigation of scalar-tensor theories and their observable predictions employing Feynman-like diagrams. 
I will begin this chapter by giving an account of the studies by DEF as an example of the theoretical attitude in testing gravity; moreover, discussing the adoption of diagrammatic techniques in gravity will serve as a basis for introducing my work in the subject. %, discussed in Chapters \ref{chap:Group} and \ref{chap:Berti}.

In describing the observable predictions of \sts, I will introduce the different testing regimes of gravity, defined according to the typical values of the velocities and the gravitational fields at play in a given system. 
A more detailed account of these dynamical regimes will be presented when I discuss the phenomenological frameworks employed in testing gravity.
This discussion begins in \Sec{sec:PPN} with Solar-System tests of the weak-field regime and the parametrized post-Newtonian (PPN) approach. 
In the Solar System gravity is so weak and velocities are so small that only the first PN order has been completely mapped out: 
the conclusion is that the slow-motion/weak-field limit 
of GR is fairly consistent with all the experiments performed so far.

With the discovery and timing of a binary pulsar~\cite{HT}, it has been possible 
to probe GR through 2.5 PN order, i.e. up to corrections of order $(v^5 / c^5)$ included.
At this order, the dynamics of the system is made of both conservative 
and dissipative contributions: therefore, binary-pulsar tests are sensitive to 
both the strong-field and the radiative regime of the theory of gravity. 
I will present these tests in \Sec{sec:PPK}, where I discuss the parametrized post-Keplerian approach.
Most notably, the detection of radio pulses from the Hulse-Taylor pulsar~\cite{HT} has allowed one to measure the decay of the system's orbital period, which is a loss of mechanical energy due to the emission of gravitational radiation. 
This constituted the first proof of the existence of gravitational waves (GWs) and confirmed the GR expectation for this effect at the $0.02 \%$ level: for these reasons, R. A. Hulse and  J. H. Taylor have been awarded the Nobel Prize in 1993.
An even more spectacular laboratory for testing GR is represented by the recently-discovered double pulsar J0737$-$3039~\cite{double_pulsar}.
The uniqueness of this system is due to the fact that both constituent objects are neutron stars pulsating in the direction of the Earth.
Besides being oriented favorably, this binary is so relativistic that it has been possible to measure more parameters than with Hulse-Taylor pulsar and in much less time.

For the sake of testing the PN theory of gravity to orders higher 
than $(v/c)^5$, %the techniques of GW interferometry~\cite{Maggiorebook} 
one needs to detect objects that are more relativistic than binary pulsars, like inspiraling compact binaries: these are systems of neutron stars and/or black holes which are in the very last stages of their orbital motion. 
The high PN orders that govern the dynamics of the inspiral stem from the interplay of very peculiar non-linearities that are intrinsic to the specific theory of gravity. %: this is why detecting and analyzing GW signals will bring new and fundamental information about physics. 
In this context, it is convenient to adopt the same point of view that originates from phenomenological frameworks like the PPN: here GR is thought of as one candidate in the space of parametrized theories of gravity and the confrontation among theories is done at the next-to-leading order. 
It is relevant to extend this phenomenological approach to the high PN orders of the GW regime, where strong-field and radiative effects come into play. 
I will conclude this chapter by discussing one such extension of the PPN: this approach stems from a parametrization of the GW phasing formula and has been studied in refs.~\cite{Arun:2006yw,Arun:2006hn,Mishra:2010tp}. 
Another extension of the PPN framework is discussed in Chapters \ref{chap:Group} and \ref{chap:Berti}: this study is based on the aforementioned particle-physics attitude and constitutes part of the original work presented in this thesis.

%%%%%%%%%%%%%%%%%%%%%%%%%%%%%%%%%
\section[Tests of gravity through diagrammatic techniques - \\
The case of \sts]
    {Tests of gravity through diagrammatic techni\-ques -
    The case of \sts %
  \sectionmark{Tests of gravity through diagrammatic tech\-niques} } 
    \sectionmark{Tests of gravity through diagrammatic tech\-niques}
\label{sec:Dam_Far}

% where:    
% "[]" is for the table of contents
% the subsequent "{}" is for where the section actually is
% 1st "sectionmark" is for the first page containing the section
% 2nd "sectionmark" is for the other pages containing the section

%%%%%%%%%%%%%%%%%%%%%%%%%%%%%%%%%

% cite Dass-Feyn_Quad ? (1982)

The first use of field theoretical diagrammatic techniques in classical gravity dates back to 1960 with the work of Bertotti and Plebanski~\cite{BP} but a more systematic use had to wait until the 90's. 
% We are talking about CLASSICAL gravity so we have to agree w/ what DEF say in Dam_Far_2PN_Diags, in footnote 28 at page 5569.
It is at this moment that DEF have put forward a program to construct the PN limit of \sts\ in order to compare it with that of GR.
This program has been pioneered in ref.~\cite{Dam_Far-ST_CQG} and is well summarized in the recent review from Esposito-Farese~\cite{EF:09} that we will follow in the presentation.

To better address the issue of modified theories of gravity, it is useful to quickly remind which are the basis hypothesis of GR:
\bd   
  \item[$\bullet$] Matter ﬁelds are {\it universally} and {\it minimally} coupled to 
  one {\bf single} metric tensor $g_{\mu\nu}$.
%  \be
%  \label{eq:Sm}
%  {\cal S}_{m} \equiv {\cal S}_{m} [\psi_i , g_{\mu\nu}]
 % \ee
%where the matter field $\psi_i$ stands for a generic non-metric field, therefore including bosonic fields. 
  This enforces the Einstein Equivalence Principle (EEP), which embraces the so-called "universality of free-fall" together with local Lorentz invariance and local position invariance (see Will's review~\cite{Will_LRR:2006} for a comprehensive discussion).
  
  \item[$\bullet$] The metric $g_{\mu\nu}$ propagates as a pure helicity-2 ﬁeld, 
  i.e. its kinetic term is given by the Einstein-Hilbert action~\footnote{In this chapter we adopt the "mostly plus" convention for the metric $\eta_{\mu\nu}$=\text{diag}(--,+,+,+).}
  \be
  \label{eq:SEH}
  {\cal S}_{EH} = \frac{c^3}{16\pi G_N} \int d^4 x \sqrt{-g} R
  \ee
where $G_N$ is Newton's gravitational coupling constant, $g$ is the determinant of the metric and $R$ its Ricci scalar.
  \ed
  
The theory of GR can then be encoded in the following action:
\be
  \label{eq:EIF}
  {\cal S} =  {\cal S}_{EH}[g_{\mu\nu}] + {\cal S}_{matter} [\psi_i , g_{\mu\nu}]
\ee
where the matter fields $\psi_i$ should be taken to mean every non-metric field, therefore including bosonic fields. 
\noindent
To envisage a theory which departs from GR, one can either change the matter sector or modify the gravitational one.
In the first case, then, one would violate the weak equivalence principle, i.e. the universality of free fall: this is one of the best tested of GR, at the point that it can be taken to represent the foundation of every (classical) theory of gravity~\cite{Will_LRR:2006}. 
To consider deviations from GR that are not already tightly constrained from experiments, one can choose to focus on modifying the Einstein-Hilbert action~(\ref{eq:SEH}). 
This does not mean having more metric fields, but more gravitational interactions, i.e. matter can couple to other fields, for example scalars, and these helicity-0 fields participate in setting up gravity together with the Einstein helicity-2 field.
Nevertheless, there is still one single metric field to which matter is minimally and universally coupled: of course, this metric field will not be given by $g_{\mu\nu}$ anymore but rather by $\tilde{g}_{\mu\nu} \equiv \tilde{g}_{\mu\nu} [g_{\mu\nu} ,\phi_a]$, with $\phi_a$ indicating one or more scalar fields.
%Choosing scalars as extra gravitational fields 
In full generality, one can represent the total set of gravitational fields (helicity-2 + helicity-0) with $\Phi$ and the matter variables with $\sigma$; the result is a compact reformulation of~\eq{eq:EIF}:
\be
\label{eq:PhiSi}
{\cal S} = {\cal S}_\Phi[\Phi] + {\cal S}_ {matter}[\s, \Phi]
\ee

\begin{figure}[t]
\begin{center}
\includegraphics[width=9cm]
{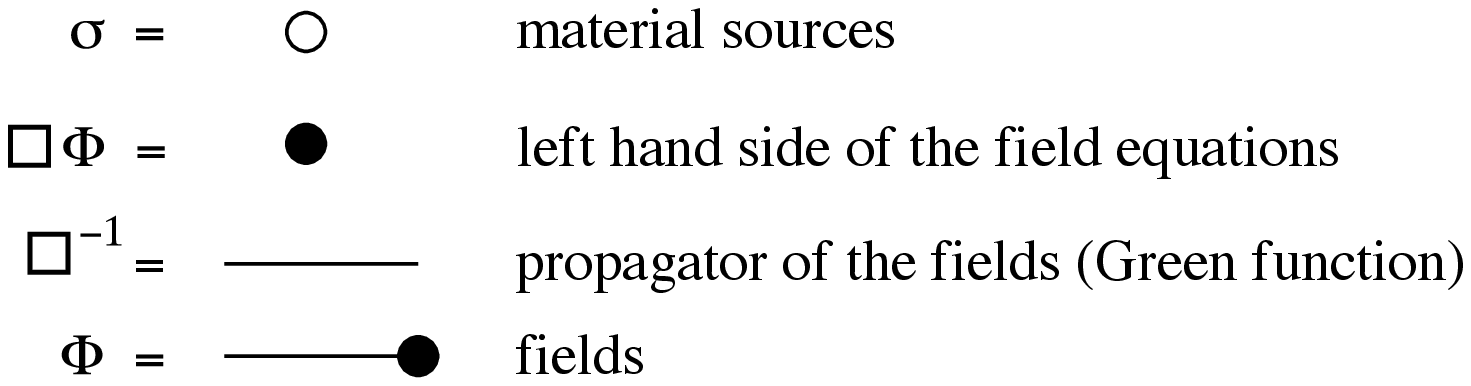} 
\caption{Diagrammatic representations of matter sources, fields and their propagators.
Figure taken from ref.~\cite{EF:09}. 
\label{fig:Dictionary} }
\end{center}
\end{figure}

\begin{figure}[t]
\begin{center}
\includegraphics[width=9cm]
{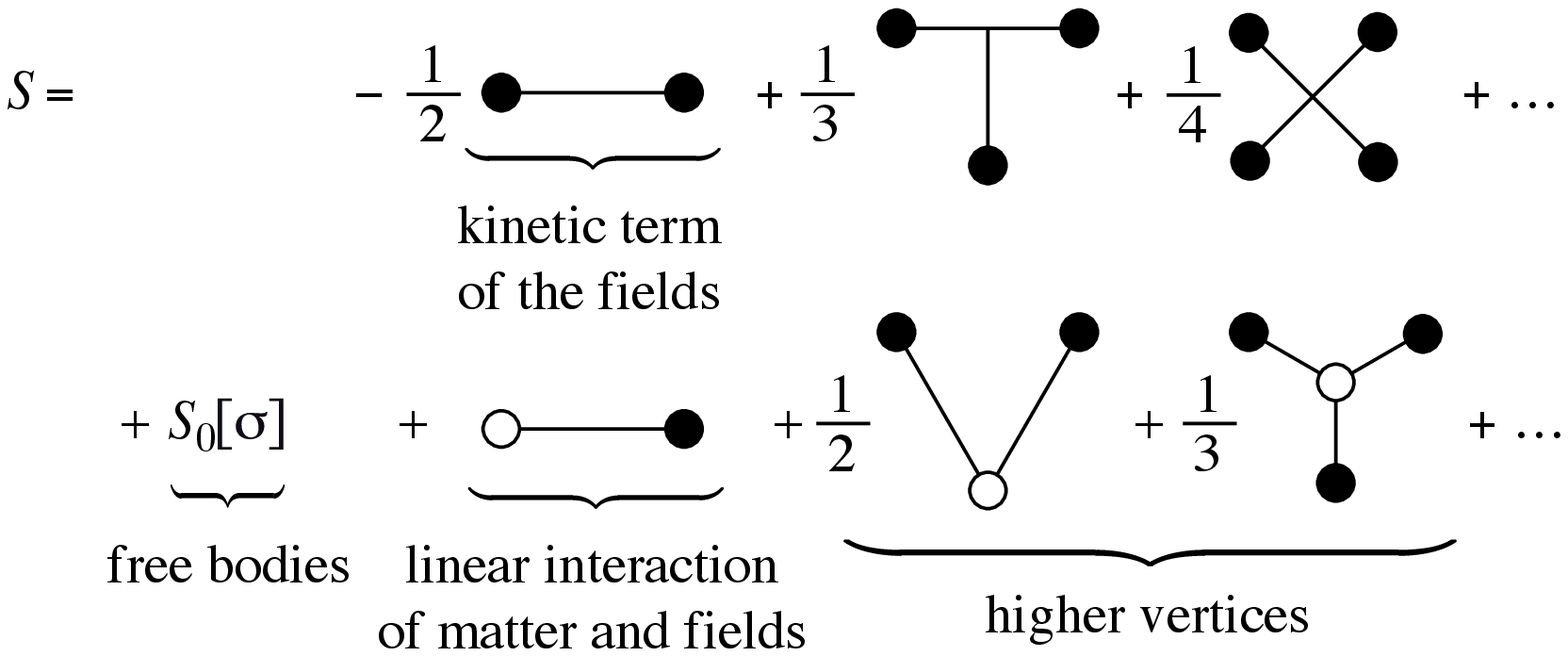} 
\caption{Diagrammatic representation of the full action of the theory, \eq{eq:PhiSi}, 
expanded in powers of the gravitational fields $\Phi$, which comprise helicity-2 and helicity-0 modes and are described by black blobs. 
The ﬁrst line corresponds to the ﬁeld action and the second one to the matter 
action (describing notably the matter-ﬁeld interaction). 
Figure taken from ref.~\cite{EF:09}. \label{fig:Fok_Exp} }
\end{center}
\end{figure}

The dynamics of the fields, as dictated by their equations of motion, can be conveniently represented through symbols like those of \fig{fig:Dictionary}:
by means of this "dictionary", one can expand the full gravity+matter action of~\eq{eq:PhiSi} in the diagrammatic fashion of \fig{fig:Fok_Exp}.
This is similar to what one does in particle physics, where the perturbative expansion of interaction amplitudes is expressed in the form of Feynman diagrams~\cite{Feyn_Diags}; we will extensively employ these diagrams starting from next chapter, where we present the EFT approach to binary systems. 

\begin{figure}[!t]
\begin{center}
\includegraphics[width=11cm]
{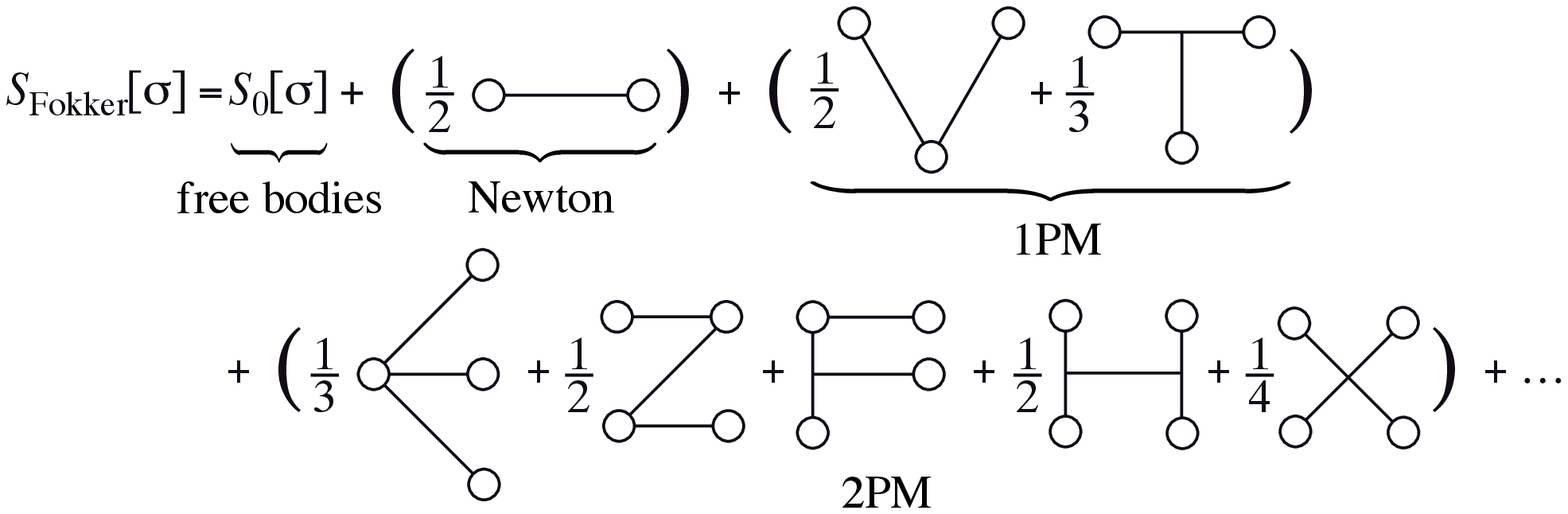} 
\caption{Diagrammatic representation of the Fokker action (4), which 
depends only on matter sources (white blobs).
The "dumbbell" diagram labelled “Newton” represents the Newtonian 
interaction $\propto G_N$, together with all velocity-dependent relativistic 
corrections. 
The 3-blob diagrams labelled “1PM” represent ﬁrst post-Minkowskian 
corrections, i.e. the ${\cal O}(G_N^2)$ PN terms as well as their full 
velocity dependence. 
The 4-blob diagrams labelled “2PM” represent second post-Minkowskian  
corrections ${\cal O}(G_N^3)$.
Figure taken from ref.~\cite{EF:09}. 
\label{fig:Fok_Matter} }
\end{center}
\end{figure}

In their studies of \sts, DEF aim at deriving the action that describes the motion of $N$ bodies subject to their mutual gravitational interaction.  
To this purpose, DEF integrate out the full set of gravitational fields (helicity-2 + helici\-ty-0) %at the level of the inter-body distance 
to obtain what they call a "Fokker" action, i.e. a functional which depends only on the matter variables:
\begin{equation} \label{eq:SFok}
{\cal S}_{\rm Fokker}[\sigma] \equiv
{\cal S}_\Phi\left[\bar\Phi[\sigma]\right]
+ {\cal S}_{\rm matter}\left[\sigma,\bar\Phi[\sigma]\right],
\end{equation} 
where $\bar\Phi[\sigma]$ denotes a solution of the field equation $\delta S/\delta\Phi = 0$ for given sources $\sigma$, with ${\cal S}$ given by \eq{eq:PhiSi}. 
In diagrammatic terms, this action is represented by \fig{fig:Fok_Matter}, where one has expanded the gravitational perturbations $\Phi$ up to the second post-Minkowskian (PM) order, i.e. including terms of ${\cal O}(G_N^2)$ with respect to Minkowski space-time. 
%~\footnote{Here $G$ denotes the effective coupling constant in the \st: we will see in the following how this is related to $G_N$.}.

%

In the EFT terminology of \chap{chap:EFT}, the Fokker action of DEF is said to be obtained by integrating out the modes that are responsible for binding the system, the so-called {\it potential modes}, as opposed to the {\it radiation modes}, which rather describe GWs.
%However, DEF build the PN limit of \sts\ in analogy with what is done in the case of GR so they do not operate the decomposition in potential and radiation modes, which we will see in the EFT framework. 
Another characteristic of the EFT approach is {\it power counting}, a generalized version of dimensional analysis: 
such a tool enables one to assess, ahead of any calculation, the order at which a given diagram will contribute to the perturbative series. 
For example, we will explicitly see in \chap{chap:EFT} how power counting explains why in calculating the $N$-body action it is justified to neglect diagrams containing loops of gravitons. %, while DEF do not say nor justify why these processes are not taken into account in their treatment. 
%Most notably, by means of power counting one does not have to pass through the "intermediate" post-Minkowskian expansion~\footnote{We revert for a while to the non-abbreviated versions of the PN and PM acronyms to better stress the difference between the post-Minkowskian and post-Newtonian expansions.}, which contains gravitational perturbations with their full dependence on the velocity. 
%This is probably the most powerful and appealing feature of the EFT treatment: that this framework is streamlined for the perturbative post-Newtonian expansion in the velocity of the bodies.

Coming back to \fig{fig:Fok_Matter}, the post-Minkowskian~\footnote{We revert for a while to the non-abbreviated versions of the PN and PM acronyms to better stress the difference between the post-Minkowskian and post-Newtonian expansions.} series needs to be further expanded in the small velocity parameter: 
this is because of the virial theorem, which makes terms of ${\cal O}(G_N)$ appear at the same time as terms of ${\cal O}(v^2)$ in the post-Newtonian expansion. 
Notably, to get Newton potential and its corrections in $v^2$, one has to take the diagrams labeled by "Newton" in the post-Minkowskian series of \fig{fig:Fok_Matter} and consider that the matter-gravity vertices depend on the velocity through the proper time of the particle (cfr. \eq{eq:prop_time}).  
%this dependence comes from the fact that the matter-gravity contains the proper time of the particle
The diagrams of \fig{fig:Fok_Matter} labeled by "1PM" are ${\cal O}(G_N^2)$ and contribute at 1PN through the zero-th order term of their velocity expansion. 
Summing up the relevant amplitudes gives a modified version of the Einstein-Infeld-Hoffmann (EIH) action~\cite{EIH}, where the usual contributions from the helicity-2 sector appear together with those of the helicity-0 terms. 
At 1PN, the deviations from the GR behavior produced by the scalars can be parametrized by two coefficients $\bppn$ and $\gppn$, where the index "PPN" is due to the fact that they belong to the general framework of the parametrized post-Newtonian (PPN) approach. 
Historically, these parameters were introduced by Eddington~\cite{Edd_book}, 
who was not considering a specific alternative theory with respect to GR but rather had a phenomenological attitude in mind.
This idea is at the heart of the PPN approach that we will describe in more detail in \Sec{sec:PPN}.

Labeling the massive bodies with capital letters $A$, $B$, \dots, the modified EIH action reads:
\begin{eqnarray} 
\label{eq:EIH_scal}
S_{\rm Fokker}&=&-\sum_A \int dt\, m_A c^2 \sqrt{1-{\bf v}_A^2/c^2} 
\\ \nn\\
&& + \frac{1}{2} \sum_{A\neq B} \int dt\, \frac{G_N\, m_A m_B}{r_{AB}} 
\Bigl[ 1+\frac{1}{2c^2} ({\bf v}_A^2+{\bf v}_B^2) -\frac{3}{2 c^2} ({\bf v}_A\cdot {\bf v}_B)
\nn\\ \nn\\
&&\hphantom{+ \frac{1}{2} \sum_{A\neq B} \int dt\, \frac{G_N\, m_A m_B}{r_{AB}}\Bigl[1}
-\frac{1}{2c^2}({\bf n}_{AB}\cdot{\bf v}_A) ({\bf n}_{AB}\cdot{\bf v}_B) 
+\frac{\gppn}{c^2} ({\bf v}_A -{\bf v}_B)^2 \Bigr]
\nn\\ \nn\\
&& -\frac{1}{2}\sum_{B\neq A\neq C} \int dt\, \frac{G_N^2 m_A m_B m_C}{r_{AB}
r_{AC}\, c^2} (2\bppn-1) \,+\, \mathcal{O}\left(\frac{1}{c^4}\right) \nn \,,
\end{eqnarray}
where:
\bi
\item $r_{AB}$ denotes the instantaneous distance between bodies $A$ and $B$;
\item ${\bf n}_{AB}$ is the unit 3-vector pointing from $B$ to $A$; 
\item ${\bf v}_A$ is the 3-velocity of body $A$; 
\item the sum over $B\neq A\neq C$ allows $B$ and $C$ to be the same body. 
\item the first line describes free bodies in special relativity and corresponds to what has been noted by $S_0[\sigma]$ in Figs.~\ref{fig:Fok_Exp} and \ref{fig:Fok_Matter}; 
\item the second line describes the 2-body interaction, i.e. the dumbbell diagram of Fig.~\ref{fig:Fok_Matter} labelled "Newton'': the lowest-order term of this diagram is the Newtonian potential, while the remaining terms are its first post-Newtonianian corrections of order $\mathcal{O}(v^2/c^2)$; 
\item the last line comes from the diagrams labelled "1PM'' in Fig.~\ref{fig:Fok_Matter}, at order zero in $v$ because of their gravity non-linearities.
\ei

\noindent
The theoretical content that the parameters $\bppn$ and $\gppn$ have in the context  of \sts\ can be conveniently addressed by means of a specific model. 
It turns out that the simplest possibility of a single scalar field is general enough to exhibit all the peculiarities of this type of theories. 
Let us then take for definiteness the action~\cite{Berg_ST:68,Nord_ST:70,Wag_ST:70}
\begin{equation}
\label{eq:actionST}
S = \frac {c^3}{16 \pi G_N}\int d^4x \sqrt{-g} \left(R -2g^{\mu\nu}\pa_\mu\varphi \pa_\nu\varphi\right) + S_{matter}\left[\psi_i ; 
\tilde g_{\mu\nu} \equiv e^{2a(\varphi)} g_{\mu\nu}\right] \,,
\end{equation}
where a potential $V(\varphi)$ has been discarded on the assumption that the field is massless and that it is not quickly varying on the scale of the $N$-body system. 
The field $\tilde g_{\mu\nu}$ is the physical metric, the one to which matter is universally coupled: it is the metric that defines the lengths and time intervals as measured by rods and clocks.  
As we anticipated, this metric depends on both the Einstein metric and the scalar field through the product of $g_{\mu\nu}$ and a function of the scalar $e^{2a(\varphi)}$. 
This function is what characterizes the coupling between matter and the scalar field. 
To enable a perturbative study, it is convenient to expand the coupling function around the background value $\varphi_0$ that the scalar field assumes far from any massive body: without loss of generality, this value can be assumed to vanish so that 
\begin{equation}
\label{eq:a}
a(\varphi) \simeq 
\alpha_0 \varphi + \frac12 \beta_0 \varphi^2 +\cdots
\end{equation}
where $\a_0$ defines the strength of the coupling between matter
and scalar excitations at linear level, while $\beta_0$ expresses a quadratic  coupling where a massive body sources two scalars.
The case $(\a_0 = 0, \b_0=0)$ represents GR, while the case $(\a_0 \neq 0, \b_0=0)$ corresponds to another type of mono-scalar-tensor theory, that of Brans-Dicke~\cite{Brans:1961sx,Will_TEGP}: in this theory the coupling function between matter and the scalar field is completely linear and characterized by a single parameter $\omega_{\rm BD}$ such that $2\omega_{\rm BD}+3 = 1/\alpha_0^2$. 

\begin{figure}[t]
\begin{center}
\includegraphics[scale=.75]{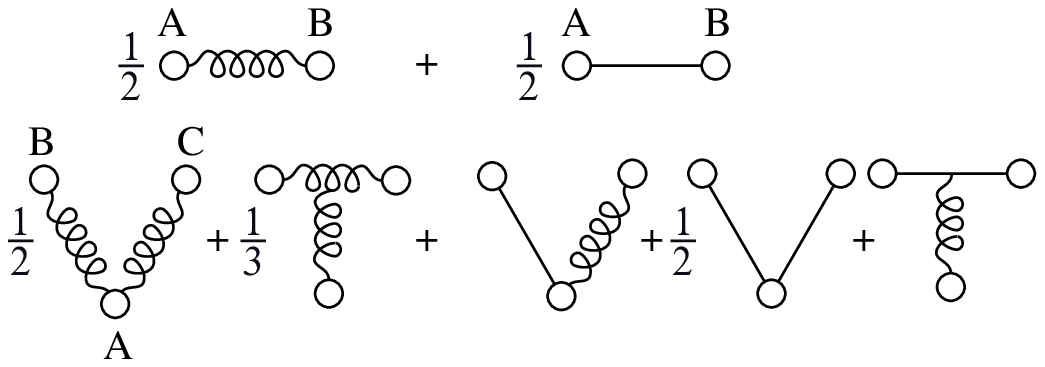}
\caption{The 1PN diagrams contributing to the modified EIH $N$-body action (\ref{eq:EIH_scal}) in presence of a mono-scalar-tensor theory of gravity. Gravitons are represented by curly lines, while scalars correspond to straight lines. Figure taken from ref.~\cite{EF:09}. }
\label{fig:EIH_scal}
\end{center}
\end{figure}

As we will now see, the pair of parameters $(\a_0 , \b_0)$ describes peculiar PN deviations of the class of mono-scalar-tensor theories under consideration with respect to GR. 
At Newtonian level, only $\a_0$ contributes through a scalar analog of Newtonian potential: however, this effect is unobservable because it just redefines the gravitational coupling as an effective constant: $G^{eff} = G_N (1+\alpha_0^2)$.
It is only at next-to-leading order that scalar effects are no longer degenerate with the GR predictions. 
This could have been anticipated from \eq{eq:EIH_scal}, where the PPN parameters are seen to modify only some terms at ${\cal O}$(1PN). 
%and not by means of overall factors. 
If one indicates gravitons by curly lines and scalars by straight lines, the 1PN order of the mono-scalar-tensor model is given by the diagrams of \fig{fig:EIH_scal}, where the left hand side column contains the usual GR terms and the velocity dependence of the vertices in the first line of diagrams has been left implicit (cfr. \fig{fig:EIH} in \chap{chap:EFT}).

The series of diagrams of \fig{fig:EIH_scal} has enabled DEF to calculate how 
$\a_0$ and $ \b_0$ enter the PPN parameters $\gppn$, $\bppn$ introduced in \eq{eq:EIH_scal}:
\begin{equation}
\label{eq:PPNscal}
\gppn = 1 - \frac{2 \alpha_0^2}{1+\alpha_0^2} \qquad,
\qquad \bppn = 1+\frac{1}{2}\,
\frac{\alpha_0\beta_0\alpha_0}{ {(1+\alpha_0^2)}^2 }\;.
\end{equation}
The factor $\alpha_0^2$ in $\gppn$ comes from the exchange of a scalar particle between two bodies: this is the case for the second diagram in the first line and for the third one in the second line; 
the factor $\alpha_0\beta_0\alpha_0$ in $\bppn$ comes from the purely scalar three-body process of the V-shaped diagram: here two bodies exchange one scalar with a third body that has a quadratic interaction weighted by $\b_0$.
Through \eq{eq:PPNscal}, these scalar effects are linked to the terms that are multiplied by the PPN parameters in the modified EIH action~(\ref{eq:EIH_scal}): therefore, experiments that probe the conservative dynamics of a system at 1PN can be used to bound $\alpha_0$ and $\beta_0$. 
\begin{figure*}[t]
\begin{center}
\includegraphics[width=0.6\textwidth,angle=0]
{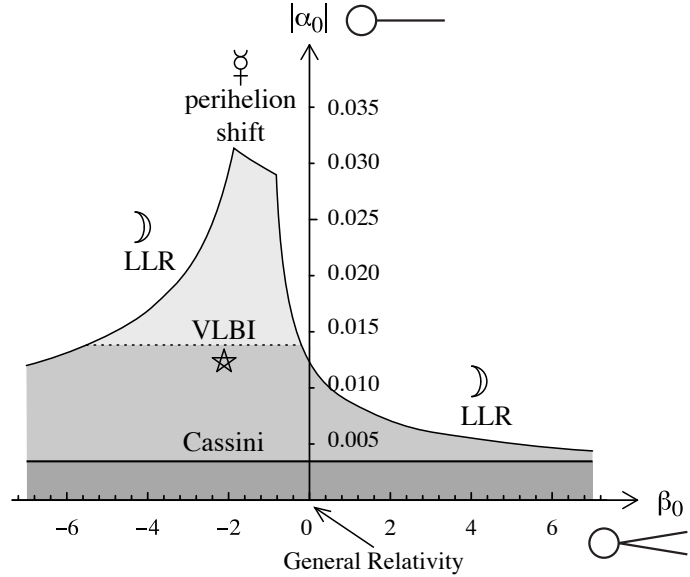}
\caption{Solar-System constraints on the parameters $(\a_0,\b_0)$ 
defining a mono-scalar-tensor theory of gravity. 
The dark grey region is allowed by all the tests, which  
are explained in the text.
Figure taken from ref.~\cite{EF:09}. } 
\label{fig:ST_Tests_SS}
\end{center}
\end{figure*}
In the case of the Solar System, these bounds are presented in \fig{fig:ST_Tests_SS}, where, for symmetry reasons, only the upper $\a_0$ plan is reported.
The various tests to which the figure refers are: 
\bd 
\item perihelion shift = measure of Mercury's perihelion advance~\cite{Shapiro_book}; 
\item LLR = Lunar Laser Ranging~\cite{LLR:04}, which monitors the Earth-Moon distance to check if there is a difference between the Earth's and the Moon's accelerations towards the Sun; this due to the so-called Nordtvedt effect~\cite{Nordt_eff}, which is absent in GR because of the strong equivalence principle (SEP): I will comment more on this effect in \Sec{sec:PPN};
\item VLBI = Very Long Baseline Interferometry~\cite{VLBI}, an experimental technique that measures the deflection of light in the curved spacetime of the Solar System;
\item Cassini = measure of the time-delay variation from Earth to the Cassini spacecraft when this last was near solar conjunction~\cite{Cassini}. 
\ed
\noindent 
As it is evident from the figure, the most constraining experiment for what concerns the linear coupling constant is Cassini~\cite{Cassini}, which implies that $\alpha_0$ must be smaller than $3\times10^{-3}$.
On the contrary, the quadratic coupling $\beta_0$ results basically unbound from Solar-System tests. 
The reason for this different behavior is due to the way $\alpha_0$ and $\beta_0$ enter the PPN parameters: $\bppn$ depends on the product between $\beta_0$ and $\alpha_0^2$, so that a small value of $\alpha_0$ as constrained by $\gppn$ prevents $\beta_0$ to be bound even in sign.

Every point in the diagram corresponds to a theory of gravity: 
GR, which can be regarded as the very peculiar theory with vanishing values of the scalar parameters, sits in the origin of the axes. 
The full vertical axis $\beta_0=0$ corresponds to Brans-Dicke theory: because $2\omega_{\rm BD}+3 = 1/\alpha_0^2$, a bound on $\alpha_0$ translates as $\omega_{\rm BD} \geq 4\times10^4$~\cite{Cassini}.
%Therefore, while Solar System tests are sufficient to tightly constrain $\omega_{\rm BD}$, other types of tests are necessary to bound $\beta_0$, 
%as it is clear from \fig{fig:ST_Tests_SS}. 

\begin{figure}[t]
\centerline{\epsfbox{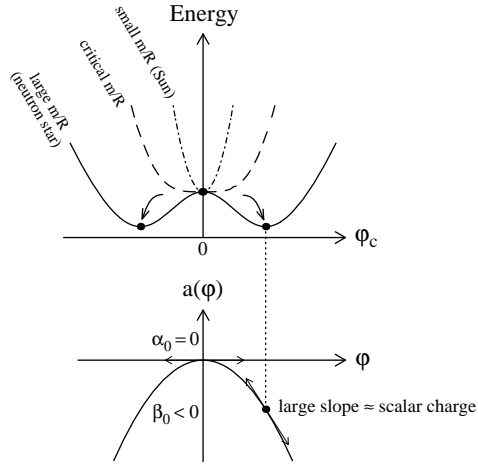}}
\caption{Heuristic argument to explain the phenomenon of spontaneous
scalarization. 
When $\beta_0 <0$ and the compactness $m\,G_N/Rc^2$ of a body is large enough, it is energetically favorable for the object to create a local scalar field different from the background value. 
The body becomes thus strongly coupled to the scalar field. 
Figure taken from ref.~\cite{EF:04}.
\label{fig:spont_scalrn}}
\end{figure}

The horizontal axis corresponds to theories that DEF call "perturbatively equivalent to GR", which are characterized as follows. 
From the structure of the PPN parameters \eq{eq:PPNscal} and of the diagrams of \fig{fig:EIH_scal}, one can foresee that the higher post-Newtonian orders of the \st\ $(\a_0,\b_0)$ will involve at least two factors $\alpha_0$, so that schematically 
\begin{equation}
\text{deviation from GR} = \alpha_0^2\times
\left[\lambda_0 + \lambda_1 \frac{m\,G_N}{Rc^2} + \lambda_2
\left(\frac{m\,G_N}{Rc^2}\right)^2+\cdots \right],
\label{eq:lambdas}
\end{equation}
where the $\lambda_i$'s are constants built from $\a_0$, $\b_0$ and the other parameters entering the coupling function $a(\varphi)$ at higher orders.
Because $\a_0^2$ is already constrained to be small, one could expect
the theory to be close to GR at any order. 
However, the scalar sector of the theory can exhibit non-perturbative effects~\cite{Dam_Far-Non_Pert_PRL} that make the square brackets of \eq{eq:lambdas} become large enough to compensate even a vanishingly small $\a_0^2$. 
These effects may occur in strong-field conditions and are analogous to the spontaneous magnetization of ferromagnets~\cite{Dam_Far-ST_Puls:PRD54}.
In fact, if the so-called compactness $m\,G_N/Rc^2$ of a body is greater than a critical value, an object like a neutron star can become strongly coupled to the scalar field and acquire a charge with respect to it.
A heuristic argument is enough to illustrate this peculiar phenomenon. 
As an example of a theory that is perturbatively equivalent to GR, one can take the limiting case of a model for which $\a_0$ vanishes strictly, so that  
the coupling function is a parabola $a(\varphi) = \frac{1}{2} \beta_0 \varphi^2$. 
The behavior of the scalar field can be described by the usual $1/r$ fall-off  outside the compact object and by the value $\varphi_c$ that the field assumes at the center of the body. 
The energy of the scalar is made up of two contributions, one that comes from its kinetic term and one from its coupling with matter (see \eq{eq:actionST} for both): as a rough estimate of the energy configuration, one can write
\begin{equation}
\label{eq:Ephic}
E(\varphi)\approx \int\left[\frac{1}{2}(\partial_i\varphi)^2
+\rho\, e^{\beta_0\varphi^2/2} \right]
\approx mc^2\left(\frac{\varphi_c^2/2}{G_N m/Rc^2} 
+ e^{\beta_0\varphi_c^2/2}\right) 
\equiv E(\varphi_c) \,,
\end{equation}
which, for negative values of $\b_0<0$, is the sum of a parabola and a Gaussian.
Fig.~\ref{fig:spont_scalrn} shows plots of the function $E(\varphi_c)$ for three different values of the ratio $m/R$ affecting the compactness: when this ratio is large enough, the function $E(\varphi_c)$ has the shape of a Mexican hat, which makes it energetically favorable for the star to create a nonvanishing scalar field $\varphi_c$, and thereby a nonvanishing scalar charge given by $a'(\varphi_c) =
\beta_0\varphi_c$. 

\begin{figure}[t]
\centerline{\epsfbox{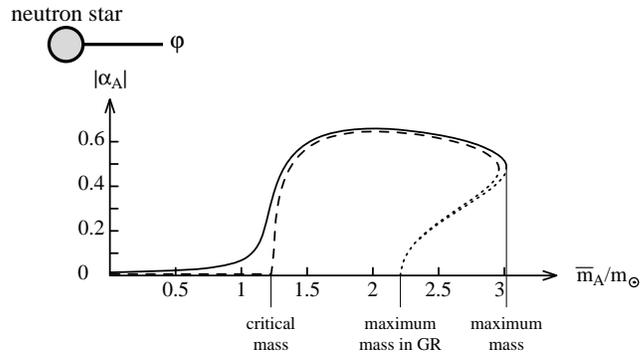}}
\caption{Scalar charge $\alpha_A$ versus baryonic mass $\overline m_A$,
for the model $a(\varphi) = -3\varphi^2$ (\textit{i.e.}, $\beta_0 =
-6$). The solid line corresponds to the maximum value of
$\alpha_0$ allowed by Solar-System tests and the dashed line to
$\alpha_0 = 0$. The dotted lines correspond to unstable configurations
of the star.
Figure taken from ref.~\cite{EF:04}.
\label{fig:alpha_A}}
\end{figure}

The heuristic argument has been confirmed by a full-fledged treatment of the physics involved in the phenomenon of spontaneous scalarization, which  requires taking into account the coupled differential equations of the metric and the scalar field, plus using a realistic equation of state to describe nuclear matter inside a neutron star~\cite{Dam_Far-Non_Pert_PRL,Dam_Far-ST_Puls:PRD54,Dam_Far_GW_Tests:98}. 
The main consequence is that the various PN-expanded predictions for the observables depend on the scalar parameters in much the same way they did in the weak regime: the only difference is a replacement of the parameters $(\a_0,\b_0)$ by their strong-field versions, which are body-dependent 
\begin{align}
\label{eq:AB}
&\a_A\equiv \pa\ln m_A/\pa\varphi_0 \\
&\beta_A\equiv\pa\,\a_A/\pa\varphi_0  \nn \;.
\end{align}
As a consequence, the effective gravitational constant between two bodies $A$ and $B$ reads 
\begin{align}
\label{eq:GAB}
G_{AB}^\text{eff} = G (1+\alpha_A\alpha_B) 
\end{align}
instead of $G^{eff} = G_N (1+\alpha_0^2)$: the fact that there is now a body-dependence makes this change observable when one is dealing with different objects. 
In a similar way, the strong-field analogues of the PPN parameters $\gppn$ and $\bppn$ (\ref{eq:PPNscal}) involve a factor of $\alpha_A$ for body~$A$ and a factor of $\alpha_B$ for body $B$, instead of an $\alpha_0$ for both bodies. %, and a factor of $\beta_A$ instead of $\beta_0$ .
Remarkably, the linear coupling is what plays the role of a scalar charge in the process of spontaneous scalarization: this takes place above a critical
mass, whose value decreases as "$-\beta_0$" grows.
On the other hand, if $\beta_0$ is positive, strong field effects do not occur because the energy function $E(\varphi)$ of \eq{eq:Ephic} does not develop the essential Mexican-hat configuration. 
The behavior of the scalar charge is plotted in \fig{fig:alpha_A} for the particular model $\beta_0 = -6$: it is seen that $\alpha_A$ can be as high as $0.6$, so that two factors of it can induce deviations from GR up to $36\%$. 

There is a major advantage in using systems made of compact objects like neutron stars as testing grounds for gravity with respect to the Solar System.
In the former situation, both the velocities and the gravitational fields at play are a factor of $10^3$ larger than our planetary playground: 
this gives access to higher orders in the PN expansion, notably to those which are odd in time and describe the dissipation of mechanical energy through the emission of gravitational radiation. 
%The resulting orbit is shrinking in radius so that the orbital period is changing in time.
Therefore, in the presence of binary pulsars, one is testing gravity is in its strong-field and radiative regime. 
We will discuss the astrophysical laboratories represented by two-body systems comprising neutron stars in more detail in \Sec{sec:PPK}, where we present the phenomenological framework employed to extract physical information from them, i.e. the parametrized post-Keplerian approach. 
For the time being, it is enough to discuss how the emission of gravitational radiation changes in a \st\ with respect to GR. 
This difference basically consists in the fact that in a \st\ the system can lose its mechanical energy through an extra "decaying channel" represented by the helicity-0 modes.
The resulting change in the orbital period is a different function of time than in GR.
The mechanical energy lost by the system at the orbital scale is received by an observer at infinity in the form of a flux of GWs; in the case of a \st, this flux is schematically given by
\begin{align}
\label{eq:flux_20}
\text{GW flux} = &\left\{\frac{\text{Quadrupole}}{c^5} +
\mathcal{O}\left(\frac{1}{c^7}\right)\right\}_\text{helicity-2}
\nonumber\\
+& \left\{\frac{\text{Monopole}}{c} +
\frac{\text{Dipole}}{c^3} + \frac{\text{Quadrupole}}{c^5} +
\mathcal{O}\left(\frac{1}{c^7}\right)\right\}_\text{helicity-0} \,
\end{align}
where the subscript "helicity-2" refers to the GR prediction, while the curly brackets labeled "helicity-0" contain the extra contributions predicted in \sts. 
Relying on the different powers of $1/c$, one would generically expect that the monopolar and dipolar contributions of helicity-0 are much larger that the usual helicity-2 quadrupole of GR~\footnote{For what concerns the helicity-0 quadrupole it can be neglected because it is roughly ${\cal O}(\a_0^2)$ with respect to the helicity-2 quadrupole of GR~\cite{Will:1994fb,Dam_Far_GW_Tests:98}.}. 
A closer inspection of the terms reveals that this is not always the case. 
In fact, the structure of the scalar monopole is of the type 
\begin{equation}
\label{eq:flux_mono}
\frac{\text{Monopole}}{c} =
\frac{G}{c}\left\{\frac{\partial(m_A\alpha_A)}{\partial t}
+\frac{\partial(m_B\alpha_B)}{\partial t} +
\mathcal{O}\left(\frac{1}{c^2}\right)\right\}^2,
\end{equation}
so that it reduces to $\mathcal{O}(1/c^5)$ if the stars are at equilibrium, $\pa_t(m_A\alpha_A) = 0$, which is the case for all binary pulsars detected so far. 
On the other hand, a collapsing star would make this monopolar 
term huge. 
For what concerns the scalar dipole it has the form
\begin{equation}
\label{eq:flux_dip}
\frac{\text{Dipole}}{c^3} = \frac{G}{3c^3}
\left(\frac{G^{eff}_{AB}m_Am_B}{r_{AB}^2}\right)^2 (\alpha_A-\alpha_B)^2 +
\mathcal{O}\left(\frac{1}{c^5}\right)\,,
\end{equation}
so that it can be efficiently excited only in presence of asymmetric systems, for which $\a_A \neq \a_B$; this is the case of a pulsar-white dwarf binary: the pulsar's scalar charge $\alpha_A$ may be of order unity, as we saw in \fig{fig:alpha_A}, whereas a white dwarf is a weakly self-gravitating body so its scalar charge is rather given by $\a_0$ which is known to be small.

\begin{figure*}[!t]
\begin{center}
\includegraphics[width=0.7\textwidth,angle=0]
{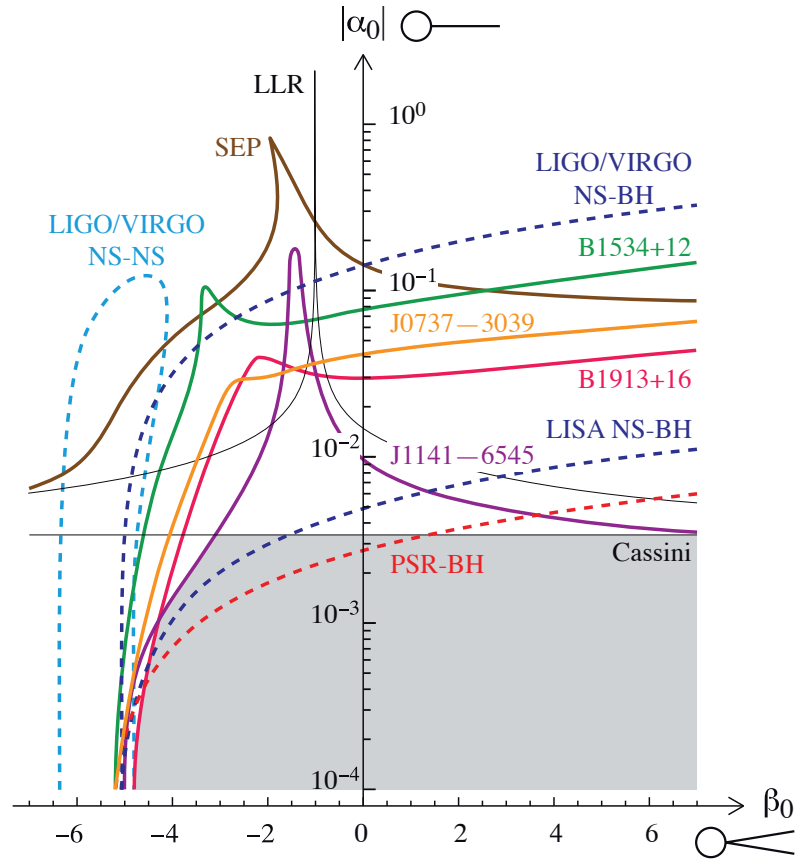}
\caption{Constraints on generic tensor-scalar theories imposed by tests of relativistic gravity in different dynamical regimes, notably: Solar-System experiments, classic binary-pulsar tests and future detections of inspiraling binaries with laser interferometers (see text for explanation of individual  acronyms and the significance of the various tests).
The hatched region is allowed by all the tests. 
%See text for comments on the significance of the various tests. 
Figure taken from ref.~\cite{EF:09}.} 
\label{fig:ST_Tests_Full}
\end{center}
\end{figure*}

Observations of binary pulsars and composite systems, like the pulsar-white dwarf binary, can be used to put upper bounds on the extra decaying channel represented by scalar waves. 
A similar analysis can be done also in the case of fiducial GW signals, like those awaited for detection at the LIGO~\cite{LIGO}, Virgo~\cite{Virgo} and LISA~\cite{LISA} interferometers.
A thorough investigation of the subject has been put forward by DEF and the most up-to-date constraints are reported in ref.~\cite{EF:09}, from which \fig{fig:ST_Tests_Full} is taken. 
This comprehensive graphical panel has been obtained translating the bounds on the strong-field parameters $\a_A,\b_A$ as constraints on their weak-field analogues $\a_0,\b_0$: these are indeed the parameters appearing in the action~\ref{eq:actionST} through the coupling function~(\ref{eq:a}). 
In such a way, one can compare the different probing power that various testing grounds have with respect to the \st\ characterized by $\a_0,\b_0$.
The allowed regions lie below the various curves, except for the case of the closed curve "LIGO-Virgo NS-NS" for which the region below the curve is excluded.
The grey region is consistent with all the tests: this region contains GR, which sits at $-\infty$ on the vertical axis because the scale of the plot is semi-logarithmic. 

From top to bottom, the new acronyms with respect to \fig{fig:ST_Tests_SS} are:
\bd 
\item SEP = Strong Equivalence Principle: this test refers to the bounds coming from a set of low-eccentricity pulsar-white dwarf binaries, which can best probe the effect being asymmetric systems;
\item B1534+12 refers to the homonymous pulsar \cite{B1534+12};
\item J0737-3039 is the double pulsar \cite{double_pulsar};
\item B1913+16 is Hulse-Taylor pulsar \cite{HT};
\item J1141-6545 is a pulsar-white dwarf binary \cite{J1141-6545};
\item PSR-BH is the curve that corresponds to a possible observation of a pulsar--black-hole binary: it is supposed that this discovery is not made by means of GW detection but rather through the analysis of electromagnetic signals.
\ed 
The equation of state used to describe neutron stars is the polytropic described in ref.~\cite{Dam_Far-Non_Pert_PRL} because it can represent also the behavior of more realistic models~\footnote{Private communication by Esposito-Farese.}.

From the figure it is interesting to remark that, for the purpose of constraining \sts, future GW detections seem not to be competitive with binary pulsar observations already at hand.
Moreover, it is very surprising to note that the curve "PSR-BH" is more constraining  than the one labeled "LISA NS-BH".
However, both these conclusions strongly depend on the type of alternative theory chosen and on the parametrization used for describing its PN deviations from GR.
I will comment on both these assumptions in what follows next, starting from the parametrization issue.

From \eq{eq:a} one can see that the pair $(\a_0,\b_0)$ parametrizes the first two orders of the coupling function between matter and the scalar field: as a consequence, the choice of the pair $(\a_0,\b_0)$ is adequate to confront \sts\ with GR only at the lowest radiative orders accessible with binary pulsars.  
In fact, these astrophysical laboratories have allowed one to probe the PN expansion only up to the ${\cal O}(1/c^5)$ terms in \eq{eq:flux_20}. 
On the other hand, the aim of GW astronomy is to probe gravity at least up to ${\cal O}(1/c^{12})$! 
In \Sec{sec:PPN_GW}, I will describe a phenomenological approach to  test the PN expansion up to this order; % of the most accurate waveforms presently available. 
later on, in Chapters \ref{chap:Group} and \ref{chap:Berti}, I will describe an attempt, which is similar in spirit but stems from a field-theoretical approach. 
By means of these extended PPN frameworks one can have a clear view on how deeply various tests are able to probe gravity in its strong-field and radiative regime:  
as it will follow from the presentation of these frameworks, GW observations have a unique potential of detecting and constraining high-order gravity non-linearities. 

Concerning the specific choice of alternative theory, it should be stressed that the scalar field studied by DEF is not "cosmological" in the sense that it does not modify the dynamics of the universe as described by GR. 
On the other hand, in the last few years, cosmological models have been investigated where a non-GR behavior of the universe on Hubble scales can be ascribed to some scalar field: 
for example, one can characterize the DGP cosmological model~\cite{DGP} in terms of a scalar field called the Galileon~\cite{Nico_Gali}. 
In DGP, the luminosity distance has a dependence on the redshift that is not the same of GR and affects the propagation of GWs on cosmological scales. 
The space-based \ifo\ LISA is being designed to detect coalescences up to redshift $z\simeq 20$~\cite{Sesana:2007sh}, so it will probe the Galileon \st\ in a way that is not possible with observations made with binary pulsars. 
Moreover, the frequency band of LISA will give access to sources in a 
mass range that is very different from that of binary pulsars: if one of the component object is a neutron star, the other one in the binary would be a $10^6 \msun$ black hole! 
This type of events are called extreme-mass-ratio inspirals and will allow one to map the space-time surrounding a black hole: in a such a way, it will be possible to test GR in a very peculiar way, through confirming the existence of black holes and  ruling out alternative types of very compact objects~\cite{Ryan:97}.
Another example of the limitation of binary pulsar tests in bounding alternative theories is offered by Chern-Simons modification of GR~\cite{Chern-Simons}: if this is the correct theory of gravity, we could only perceive a deviation from GR
by means of GW tests, % at 2PN order,
as we will see in \Sec{sec:PPN_GW}.
%Again, GW observations are a unique tool of testing gravity in ways which are not possible otherwise.  
%Therefore, the conclusions of DEF about how the various testing grounds can probe alternative theories are only valid for the type of \sts\ chosen and for lowest orders of the radiative regimes. 

There is another point to be stressed about the conclusions on \sts\ derived by DEF. 
The analysis summarized by \fig{fig:ST_Tests_Full} assumed that the constraints put by GW tests %of the radiative regime 
on the parameters $(\a_0,\b_0)$ refer to detection of single events. 
This partially explains~\footnote{The other reason has already been discussed before: it is the fact that one is limiting the confrontation to the lowest radiative orders given by \eq{eq:flux_20}.} why the curve "PSR-BH" is more constraining than the one labeled "LISA NS-BH", i.e. why detecting a neutron star-black hole system by pulsar timing could give more information than can be done by analyzing the same system through the emitted GWs with the most sensitive space based \ifo. 
This comparison does not take into account one of the advantages of entering the era of GW astronomy, i.e. the fact that, already with advanced ground based \ifos, it will be possible to observe many coalescences per year: for the case of a neutron star-black hole system, the number of detections per year is likely to be ten but could reasonably be as high as three hundreds~\cite{Adv_Rates:09} (see Table~\ref{table:det_rates} in \chap{chap:Berti}).   
This large statistics will bring crucial improvements in constraining parameters like $(\a_0,\b_0)$ for the following reasons. 
In fact, these parameters are peculiar of the theory of gravity and do not depend on the astrophysical system: therefore, they are extrinsic with respect to the emitting source.
As a consequence, if $N$ events are detected, each one can be interpreted as providing an independent measurement of an extrinsic parameter, for example $\b_0$: the total error on $\b_0$ would then be a fraction $1/\sqrt{N}$ of the error coming from a single observation. 
Detection of many systems has a further advantage in measuring extrinsic parameters: as we have seen for tests with binary pulsars, there might be systems that are better suited for measuring one parameter instead of another. 
For example, the scalar charge of a neutron star is better measurable if one can detect a dipolar flux from an asymmetric system like a pulsar-white dwarf binary (see \eq{eq:flux_dip}).
Moreover, the bound derived on a given parameter from one system might be used as a prior when testing a different parameter with another source: in this way, one should be able to disentangle the effects of various extrinsic parameters. %if they happen to be present in the GW template at the same PN order. 
A final note concerns the possibility to further increase the accuracy of GW experiments by cross-correlating the outputs of many \ifos, which enables one to dig signals that lie below the sensitivity of an individual instrument: this has proven crucial in obtaining the bound on the GW background of stochastic origin with the LIGO instruments~\cite{GW_bck-Nat} that I will discuss in \chap{chap:Ric}.

%\clearpage

%%%%%%%%%%%%%%%%%%%%%%%%%%%%%%%%%
\section[Probing the weak-field regime of gravity - \\
  The parametrized post-Newtonian formalism]
   {Probing the weak-field regime of gravity - \\ 
   The parametrized post-Newtonian formalism%
 \sectionmark{The parametrized post-Newtonian formalism}}
    \sectionmark{The parametrized post-Newtonian formalism}
 \label{sec:PPN}
   
    % where:    
% "[]" is for the table of contents
% the subsequent "{}" is for where the section actually is
% 1st "sectionmark" is for the first page containing the section
% 2nd "sectionmark" is for the other pages containing the section

%%%%%%%%%%%%%%%%%%%%%%%%%%%%%%%%%

In the previous section I have discussed \sts\ as an example of alternative theories of gravity. 
In the presentation we have seen that, at first PN order, the action governing the dynamics of \sts\ is similar to that of GR: \eq{eq:EIH_scal} shows that the structure underlying both theories is the same, the differences being encoded in parameters that multiply the PN expansion terms of ${\cal O}\left(v^2,m\,G_N/r\right)$. 
This similarity is not peculiar to \sts, it is a feature of the PN regime of metric theories of gravity; for this reason, in \eq{eq:EIH_scal} the parameters have been labeled with the index PPN, referring to a general framework that can be used to describe this regime, irrespective of the underlying theory. 
The use of parameters to describe the {\it first} PN limit of metric theories of gravity goes under the name of PPN approach. 
Among the proponents and major exponents of this framework is Clifford Will, who has treated the subject in two monographs on tests of gravity and confrontation among theories~\cite{Will_TEGP,Will_LRR:2006}. 
I refer to these reviews for extensive discussions and relevant citations of the PPN; in this section I give a concise account of only those features that overlap with \sts\ and that are more useful for the rest of the thesis.

The basic idea was formulated by Eddington~\cite{Edd_book} who wrote the Solar-System metric as the 1PN limit of \Sch metric in {\it isotropic} coordinates introducing some phenomenological parameters $\bppn$ and $\gppn$ in front of the different powers of the dimensionless ratio $m\,G_N/rc^2$:
\begin{subequations}
\begin{eqnarray}
\label{eq:g00}
g_{00}&=& -1 +2\frac{m\,G_N}{rc^2} - 2 \bppn \left(\frac{m\,G_N}{rc^2}\right)^2 + 
\mathcal{O}\left(\frac{1}{c^6}\right)\,, \\
\nn \\
\label{eq:gij}
g_{ij}&=&\delta_{ij}\left(1+2 \gppn\frac{m\,G_N}{rc^2}\right)
+ \mathcal{O}\left(\frac{1}{c^4}\right) \\
\nn \\
g_{0i}&=& 0 + \mathcal{O}\left(\frac{1}{c^5}\right)\,.
\end{eqnarray}
\end{subequations}
In this particular form of the metric the parameters $\bppn$ and $\gppn$ can be given a heuristic physical interpretation.
From \eq{eq:g00} the parameter $\bppn$ can be regarded as a measure of the amount of non-linearity that a given theory of gravity puts into the $g_{00}$ component of the metric. 
On the other hand the parameter $\gppn$ measures the amount of space-time curvature produced by a body of mass $m$ at a distance $r$ from it: in fact, to 1PN order, the spatial components of the Riemann tensor read~\cite{Will_TEGP} 
\be
R_{ijkl} = 3\gppn\,\frac{m\,G_N}{r^3} 
\left(n_j n_k \delta_{il} + n_i n_l \delta_{jk} 
- n_i n_k \delta_{jl} - n_j n_l \delta_{ik} 
- \frac23 \delta_{jk} \delta_{il} 
+ \frac23 \delta_{ik} \delta_{jl} \right) \,, \nn
\ee
where $n_i = x_i/r$.

The most general version of the PPN metric has a total of ten parameters: here I will only consider $\bppn$ and $\gppn$ because they are the only non-vanishing ones in \sts; within this choice, the metric reads~\cite{Will_LRR:2006}
\bees
\label{eq:gppn}
 g_{00} &=&
        -1 + 2U - 2 \bppn U^2 
        + (2 \gppn +2+ \alpha_3 ) \Phi_1 \phantom{\frac12} \\
        &&+ 2(3 \gppn - 2 \bppn + 1 ) \Phi_2
        + 2 \Phi_3 + 6 \gppn \Phi_4 + {\cal O}(\epsilon^3) \phantom{\frac12} \nn \\
        \nn \\
g_{0i} &=&
         - {1 \over 2 } (4 \gppn + 3 ) V_i
        - {1 \over 2} W_i + {\cal O}(\epsilon^{5/2}) \nn \\
        \nn \\
g_{ij} &=& 
(1 + 2 \gppn U ) \delta_{ij} + {\cal O}(\epsilon^2) \phantom{\frac12} \nn \;,
\ees
where $\eps$ is the PN expansion parameter of order $(v^2\,,m\,G_N/r)$ and capital letters indicate metric potentials such as: 
    \setlength{\arraycolsep}{0.18 em}
\begin{eqnarray}
U &=& 
	\int {{\rho' } \over {| {\bf x}-{\bf x}' |}} d^3x' \\
\Phi_1 &=&
        \int {{\rho' v'^2} \over {| {\bf x}-{\bf x}' |}} d^3x' \nn \\
%\Phi_2 &=&
%        \int {{\rho' U'} \over {| {\bf x}-{\bf x}' |}} d^3x' \\
%\Phi_3 &=&
%        \int {{\rho' \Pi'} \over {| {\bf x}-{\bf x}' |}} d^3x' \\
%\Phi_4&=&
%        \int {{p' } \over {| {\bf x}-{\bf x}' |}} d^3x' \\
V_i &=&
        \int {{\rho' v_i'} \over {| {\bf x}-{\bf x}' |}} d^3x' \nn \;,
%W_i&=&
%        \int {{\rho' [{\bf v}' \cdot ({\bf x}-{\bf x}')](x-x')_i} \over {| {\bf x}-{\bf x}' |^3}} d^3x'
\end{eqnarray} 
in which $\rho$ is the density of rest mass as measured in a local freely falling frame momentarily comoving with the gravitating matter and $v^i=(dx^i /dt)$ is the coordinate velocity of the matter.

The form (\ref{eq:gppn}) for the metric corresponds to the {\it standard} PPN gauge; this choice is only dictated by convenience and it relies on these assumptions:
\ben 
\item matter is modeled as a perfect-fluid; 
\item the spatial part of the metric is diagonal and isotropic;
\item the metric is expressed in a quasi-Cartesian coordinate system at rest with respect to the Universe rest frame.
\een 
With the explicit form (\ref{eq:gppn}) one can obtain the equations of motion in the PPN metric, which are given by a set of coupled equations for matter and non-gravitational field variables in terms of other matter and non-gravitational field variables. 
For example, the geodesics of light rays can be written as~\cite{Will_TEGP}
\bees
\label{eq:phot}
\frac{d^2\vx_p}{dt^2} &=& (1+\gppn) \left[ \nabla U - 2\hat n (\hat n \cdot \nabla U)\right] \\
\nn \\
\hat n \cdot \frac{d\vx_p}{dt} &=& -(1+\gppn) U \nn \;,
\ees
where $\vx_p$ is the deviation of the photon's path from uniform straight line motion. 
These equations enable one to compute the deflection of light and the time delay of an electromagnetic signal which are produced by a massive object; both these effects are characterized by the same factor $(1+\gppn)$ that appears in \eq{eq:phot}. 
The best constraint on the parameter $\gppn$ comes from the measure of the time-delay variation from Earth to the Cassini spacecraft when the latter was near solar conjunction~\cite{Cassini} and gives $\gppn \,\lsim \, 2.3 \times 10^{-4}$. 

\smallskip 

In the case of massive bodies with spherical symmetry, the equations of motion have a quasi-Newtonian form 
\be
{\bf a} = (m_{\rm p}/m) \nabla U 
\ee
where $m$ is the inertial mass of the body and $m_p$ its {\it passive} gravitational mass, i.e. the mass that determines the force on a body in a gravitational ﬁeld: this mass given by  
\be
\label{eq:}
m_{\rm p} = m- \eta_{\rm N} E_{\rm g} 
\ee
where $E_{\rm g}$ is the negative of the gravitational self-energy of the body ($E_{\rm g} >0$) and   
\be
\eta_{\rm N} = 4 \bppn - \gppn -3
\ee
is a parameter that measures the resulting violation of the massive-body equivalence principle, which is called the Nordtvedt effect.  
As we know from the discussion of \fig{fig:ST_Tests_SS}, this effect is absent in GR but present in scalar-tensor theories. 
%The existence of the Nordtvedt effect does not violate the weak equivalence principle since for laboratory-sized objects one has $E_{\rm g} /m \le 10^{-27}$, far below the sensitivity of current or future experiments. 
It refers to self-gravitating astronomical bodies, for which the ratio $E_{\rm g} /m$ may be significant: 
%$3.6 \times 10^{-6}$ for the Sun, $10^{-8}$ for Jupiter, 
$4.6 \times 10^{-10}$ for the Earth and $0.2 \times 10^{-10}$ for the Moon; therefore, if the Nordtvedt effect is present the Earth should fall toward the Sun with a slightly different acceleration than the Moon. 
%This perturbation leads to a polarization of the orbit of the Earth-Moon orbit that is directed toward the Sun: as seen from Earth, the apogee of the Moon will always be lying on the Earth-Sun line. 
%This polarization represents a perturbation in the Earth-Moon distance that 
This difference in accelerations affect the Earth-Moon distance that can be monitored with the lunar laser-ranging experiment (LLR): using a retroreflector left on the Moon by the most famous Apollo mission in August 1969, LLR has made regular measurements of the round-trip travel times of laser pulses sent from the Earth. 
%between a network of observatories and the lunar retroreflectors, 
The accuracies are at the level of 50\ ps, which means that one is sensitive to errors of 1\ cm in distance. %; soon, one may approach a factor of ten improvement i.e. 1\ mm! 
At this level of precision, it is important to make sure that one is really probing the {\it strong} version of equivalence principle, the one valid for massive self-gravitating bodies. 
In fact, it could be that the Earth-Moon system exhibits a violation of the {\it weak} equivalence principle due to the different chemical compositions of the bodies. 
To this end, the world-leader group E{\"o}t-Wash has tested laboratory-sized bodies whose chemical compositions mimic that of the Earth and Moon.  
In this way, one could constrain the difference in accelerations due to a violation of the {\it weak} equivalence principle to 1.4 parts in $10^{13}$~\cite{Eotwash:01}, which bounds the difference in accelerations due to a violation of the {\it strong} equivalence principle at the level of about 2 parts in $10^{13}$. 
For what concerns the Nordtvedt parameter, the resulting bound is $| \eta_{\rm N} | = (4.4 \pm 4.5) \times 10^{-4}$, which in turns constrains the parameter $\bppn$ to be smaller than $2.3 \times10^{-4}$~\cite{LLR:04}. 

\smallskip 
In Chapters~\ref{chap:Group} and \ref{chap:Berti} I will present a field-theoretical approach that allows a different interpretation of $\bppn$. 
To conclude this section, in \fig{fig:PPN_Tests_SS} I report the constraints on the PPN parameters $\gppn$ and $\bppn$ imposed by Cassini, LLR other Solar-System experiments.

\begin{figure*}[htbp]
\begin{center}
\includegraphics[width=0.7\textwidth,angle=0]
{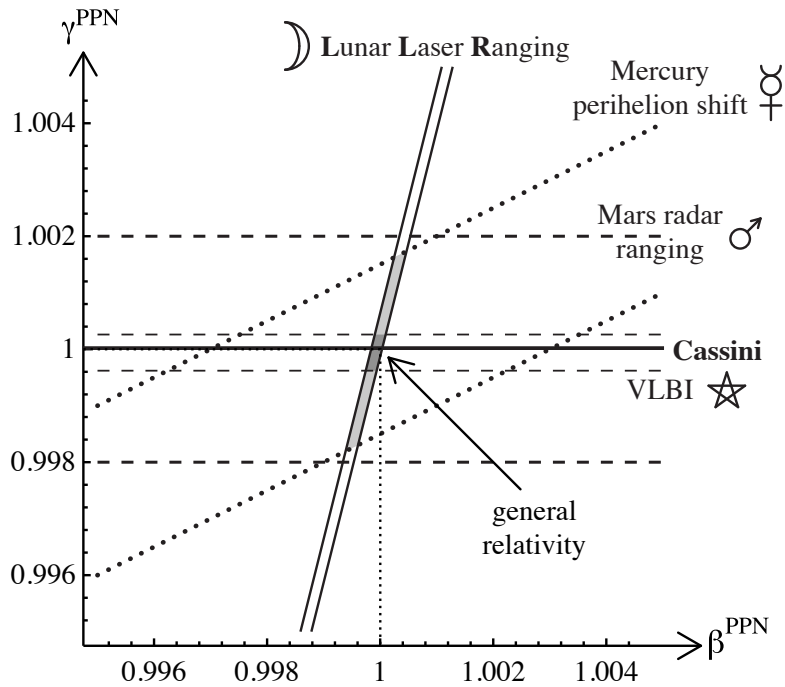}
\caption{Constraints on the PPN parameters $\gppn$ and 
$\bppn$ imposed by Solar-System experiments. 
"Mars radar ranging" refers to measures of radar echo delay between the Earth and Mars as a probe of the propagation of light in the curved space-time of the Solar System. 
For the other tests see the discussion following~\eq{eq:PPNscal}.
The region allowed by all the tests is the intersection of the almost vertical strip labeled "Lunar Laser Ranging" and the horizontal bold line labeled "Cassini". 
GR sits in the point $(\bppn=1,\gppn=1)$. 
Figure taken from ref.~\cite{EF:09}.} 
\label{fig:PPN_Tests_SS}
\end{center}
\end{figure*}

\clearpage

%%%%%%%%%%%%%%%%%%%%%%%%%%%%%%%%%
\section[Probing the strong-field/radiative regime of gravity - Part I  \\ 
 The parametrized post-Keplerian formalism]
 {Probing the strong-field/radiative regime of  \\ gravity - Part I \\
 The parametrized post-Keplerian formalism
 \sectionmark{The parametrized post-Keplerian formalism}}
 \sectionmark{The parametrized post-Keplerian formalism} 
\label{sec:PPK}

%where:    
% "[]" is for the table of contents
% 1st "{}" is for where the section actually is
% 1st "sectionmark" is for the first page containing the section
% 2nd "sectionmark" is for the other pages containing the section

%%%%%%%%%%%%%%%%%%%%%%%%%%%%%%%%%

In this section I describe a phenomenological approach to testing gravity first proposed by Damour in ref.~\cite{Dam-PPK} and then extended by Damour and Taylor in ref.~\cite{Dam_Tay:92}.
The presentation follows the one recently given by Damour at the SIGRAV  
lectures~\cite{Dam_Como}.

The equations of motion of $N$ compact bodies in GR can be derived from 
the following action 
\begin{equation}
\label{eq:action_start}
S =  \frac{c^3}{16\pi \, G}\int d^{D+1} \, x \ \sqrt g \ R(g) 
- \sum_{a=1}^N m_a \, c \int \sqrt{-g_{\mu\nu} (z_a^{\lambda}) 
\, dz_a^{\mu} \, dz_a^{\nu}} \, ,
\end{equation}
where the first piece is the usual Einstein-Hilbert action that accounts for the pure gravitational sector and the second term describes the coupling of a point-like source to gravity; the space dimension $D$ is kept as an arbitrary complex number until the end of the calculation when it is set to its physical value $D=3$: this serves to regularize the divergencies stemming from the point particle approximation.
The equations derived from this action can be solved applying the weak-field PM expansion that we discussed in \Sec{sec:Dam_Far}.
Using this method, Damour and collaborators have derived the equations of motion of two compact bodies at the 2.5PN order~\cite{DD81,Damour_CR,Dam-NH}: this level of accuracy is sufficient for describing the onset of dissipative contributions in binary pulsars' dynamics according to
\begin{eqnarray}
\label{eq:motion}
\frac{d^2 \, z_a^i}{dt^2} &=& A_{a0}^i (\bm{z}_a - \bm{z}_b) 
+ c^{-2} \, A_{a2}^i (\bm{z}_a - \bm{z}_b , \bm{v}_a , \bm{v}_b) \nn \\
\nn \\
&+ &c^{-4} \, A_{a4}^i (\bm{z}_a - \bm{z}_b , \bm{v}_a , \bm{v}_b , \bm{S}_a , 
\bm{S}_b) \nn \\
\nn \\
&+ &c^{-5} \, A_{a5}^i (\bm{z}_a - \bm{z}_b , \bm{v}_a - \bm{v}_b) 
+ {\mathcal O} (c^{-6}) \, 
\end{eqnarray}
where: 
\bd
\item $\bm{z}_a$ denotes the coordinate of the pulsar $a$, $\bm{v}_a$ its velocity and $\bm{S}_a$ its spin;
\item $A_{a0}^i = - m_b \,G_N (z_a^i - z_b^i) / \vert z_a - z_b \vert^3$ denotes 
the Newtonian acceleration, i.e. the 0PN order in the expansion; 
\item $A_{a2}^i$ is the term of order ${\cal O}(v/c)^2$, i.e. the 1PN correction;  
\item  $A_{a4}^i$ stands for the 2PN modification which comprises spin-orbit effects; 
\item $A_{a5}^i$ is the term that had a crucial role in the discovery of gravitational 
radiation; it is the first term in the expansion which is odd in time and describes the decrease of mechanical energy of the system by an amount which, on 
average, equals the energy lost in a flux of GWs at infinity. 
%The contribution $A_{a5}^i$ to the acceleration is derived in the so-called near zone of the system, i.e. within a distance smaller than the wavelength of the emitted GWs: in this zone, the 2.5PN term is a direct consequence of the fact that, in GR, the gravitational interaction between two bodies propagates at the speed of light~$c$. 
For this reason, the detection through radio pulses of the presence of $A_{a5}^i$ in binary pulsar dynamics constituted a direct proof of the existence of GWs.
\ed
In order to investigate the pulsar dynamics through experiments, one needs to link the equations of motion (\ref{eq:motion}) to observational effects through two steps: 
\ben
\item solve the equations of motion (\ref{eq:motion}) so to get 
the coordinate positions $\bm{z}_1$ and $\bm{z}_2$ as 
explicit functions of the coordinate time $t$; 
\item relate the coordinate motion $\bm{z}_a (t)$ to the 
pulsar observables, i.e. to the times of arrival of 
electromagnetic pulses on Earth.
\een
The first step has been accomplished in ref.~\cite{DD85}, where the relativistic two-body motion has been written in a simple quasi-Keplerian fashion. 
The second step %in relating the equations of motion to pulsar observations 
has been accomplished through the derivation of a relativistic timing formula~\cite{BT76,DD86}, a multi-parameter mathematical function that encodes many different physical phenomena: 
\bi
\item dispersion effects; 
\item travel time across the Solar System;
\item gravitational time delay due to the Sun and the planets; 
\item time dilation effects between the time measured 
on the Earth and the Solar-System-barycenter time; 
\item variations in the travel time between the binary pulsar and 
the Solar-System barycenter like those due to relative and proper motion; 
\item time delays happening within the binary system. 
\ei
For the purpose of discussing the relativistic timing formula, one can  
concentrate on the time delays that take place within the binary system. 
These are the "aberration" effect and the delays named after Einstein, Roemer and Shapiro; I will discuss them in what follows.

In describing pulsar timing, it is convenient to use a "multi-chart" approach which owes its name to the adoption of $N+1$ separate coordinate systems to account for the motion of $N$ bodies (see ref.~\cite{Dam_Como} for more details and full list of citations). 
The space-time inside and around each body $a$ is described by $N$ {\it local} coordinate charts denoted by $X_a^{\alpha}$, where $\alpha = 0,1,2,3$ and $a = 1,2,\ldots , N$. 
The spacetime outside the $N$ "tubes" that contain the bodies is described by 
one {\it global} coordinate chart denoted by~$x^{\mu}$ with $\mu = 0,1,2,3$.
The different coordinate systems are related among themselves through  expansions of the form
\begin{equation}
\label{eq:link}
x^{\mu} = z_a^{\mu} (T_a) + e_i^{\mu} (T_a) \, X_a^i + \frac{1}{2} \, f_{ij}^{\mu} (T_a) \, X_a^i \, X_a^j + \cdots \,,
\end{equation}
where $z_a^{\mu}$ denotes the global coordinates of the center of mass of the pulsar, $T_a$ the local (proper) time of the pulsar frame and where, %$e_i^{\mu}$ and $f_{ij}^{\mu}$ are functions of the proper time $T_a$, 
for instance, 
\begin{equation}
\label{eq3.8}
e_i^0 = \frac{v_i}{c} \left( 1+\frac{1}{2} \ \frac{\bm{v}^2}{c^2} + 3 \, \frac{m_b\,G_N}{c^2 \, r_{ab}} + \cdots \right) + \cdots \,.
\end{equation}
By means of the multi-chart decomposition, it is possible to combine the information contained in several expansions: in the global frame one can use a weak-field expansion %of the type 
\be 
\label{eq:gexp}
g_{\mu\nu} = h_{\mu\nu}^{(1)} + h_{\mu\nu}^{(2)} + \cdots \,,
\ee 
where the indices $(0),(1)$ in the exponents correspond to powers of Newton's constant; in the local frames one can use expansions of the type
\begin{equation}
\label{eq:Gexp}
G_{\alpha\beta} (X_a^{\gamma}) = G_{\alpha\beta}^{(0)} (X_a^{\gamma} ; m_a) + H_{\alpha\beta}^{(1)} (X_a^{\gamma} ; m_a , m_b) + \cdots \, ,
\end{equation}
where $G_{\alpha\beta}^{(0)} (X ; m_a)$ denotes the (possibly strong-field) metric generated by an isolated body of mass $m_a$.
The separate expansions (\ref{eq:gexp}) and (\ref{eq:Gexp}) are then matched in some overlapping domain of common validity of the type 
\be 
m_a\,G_N / c^2 \lesssim R_a \ll \vert \bm{x} - \bm{z}_a \vert \ll d \sim \vert \bm{x}_a - \bm{x}_b \vert \q \text{with} \; b \ne a. 
\ee 
With this setting in hand, it is possible to relate the time delays that happen within the binary pulsar to quantities which are observable in the Earth system.

In the rest frame attached to the pulsar $a$ $(X_a^0 = c \, T_a , X_a^i)$ the pulsar phenomenon can be modelled by the secularly changing rotation of a beam of radio waves:
\begin{equation}
\label{eq3.7}
\Phi_a = \int \Omega_a (T_a) \, d \, T_a \simeq \Omega_a \, T_a + \frac{1}{2} \, \dot\Omega_a \, T_a^2 + \cdots \, ,
\end{equation}
where $\Phi_a$ is the longitude around the spin axis. 
Rest frame quantities like $T_a$ and $\Phi_a$ have to be translated into the global coordinates $x^{\mu}$ that describe the dynamics of the binary system; 
this is done by using the link (\ref{eq:link}). 
Among other results, one finds that a radio beam emitted in the proper direction $N^i$ in the local frame appears to propagate, in the global frame, in the coordinate direction $n^i$ given by
\begin{equation}
\label{eq3.9}
n^i = N^i + \frac{v^i}{c} - N^i \, \frac{N^j \, v^j}{c} + {\mathcal O} \left( \frac{v^2}{c^2} \right) \,:
\end{equation}
this change in direction is called the aberration effect.

The Einstein time delay stems from the link between the pulsar proper time $T_a$ and the global coordinate time $t = x^0 / c = z_a^0 / c$ used in the orbital motion;  
to first PN order (${\cal O}(G_N,v^2)$), this time link is given by  
\begin{equation}
\label{eq:eins_delay}
T_a \simeq \int dt \left( 1 - \frac{2 \,m_b\,G_N}{c^2 \, r_{ab}} 
- \frac{\bm{v}_a^2}{c^2} \right)^{\frac{1}{2}} \simeq 
\int dt \left( 1 - \frac{m_b\,G_N}{c^2 \, r_{ab}} 
- \frac{1}{2} \ \frac{\bm{v}_a^2}{c^2} \right)
\end{equation}
i.e. by a sum of the special relativistic and general relativistic time 
dilation effects: the combined effect of this terms clearly deserves 
the name of Einstein time delay \cite{DD86}.

The Roemer and Shapiro time delays rather originate when 
one computes the global time needed to reach the Solar-System 
barycenter by a light beam emitted by the pulsar;  
%at the proper time $T$ (linked to $t_{\rm emission}$ by 
%(\ref{eq:eins_delay})) 
%in the initial global direction. 
this is achieved by imposing that this light beam follows a null geodesic according to:  
\begin{equation}
\label{eq3.12}
0 = ds^2 = g_{\mu\nu} (x^{\lambda}) \, dx^{\mu} \, dx^{\nu} \simeq 
- \left( 1-\frac{2U}{c^2} \right) c^2 \, dt^2 
+ \left( 1+\frac{2U}{c^2} \right) d\bm{x}^2 \,,
\end{equation}
where $U = m_a\,G_N / \vert \bm{x} - \bm{z}_a \vert + m_b\,G_N / \vert \bm{x} 
- \bm{z}_b \vert$ is the Newtonian potential within the binary system. 
Indicating by $t_e$ the time of emission and by $t_a$ the time of arrival, 
the interval between the two reads
\begin{equation}
\label{eq3.13}
t_a - t_e = \int_{t_e}^{t_a} dt 
\simeq \frac{1}{c} \int_{t_e}^{t_a} \vert d\bm{x} \vert 
+ \frac{2}{c^3}  \int_{t_e}^{t_a} \left( \frac{m_a\,G_N}{\vert \bm{x} - \bm{z}_a \vert} 
+ \frac{m_b\,G_N}{\vert \bm{x} - \bm{z}_b \vert} \right) \vert d\bm{x} \vert \, .
\end{equation}
The result of the first integral is $(1/c) \, \vert \bm{z}_{SSB} (t_a) - \bm{z}_a (t_e) \vert$, i.e. the light crossing time between the Solar-System barycenter (SSB) and the pulsar: the fact that the pulsar moves on an orbit $\bm{z}_a (t_e)$ gives rise to the so-called Roemer time delay.
The second integral rather accounts for the fact that the beam is 
propagating in the curved space-time generated by the companion: 
this effect is called the Shapiro time delay.

Once expressed as functions that relate the pulsar proper time to the 
Solar-System-barycenter time, the delays discussed so far 
will contribute to the GR timing formula as~\cite{DD86}
\begin{equation}
\label{eq:time_1}
t_{SSB} - t_0 = D^{-1} [T + \Delta_A (T) + \Delta_E (T) 
+ \Delta_R (T) + \Delta_S (T)] \,,
\end{equation}
where, here and henceforth, $T$ denotes the pulsar proper time, $D$ is an overall Doppler factor and the terms denoted by $\Delta_i$'s encode the contributions from the various physical effects: $A$ for the aberration,
$E$ for Einstein, $R$ for Roemer and $S$ for Shapiro. %of \eq{eq_polar_1}.
The structure of the timing formula (\ref{eq:time_1}) can be conveniently 
expressed in a condensed parametrized fashion 
\begin{equation}
\label{eq:time_2}
t_{SSB} - t_0 = F \, [T_N ; \{ p^K \} ; \{ p^{PK}\} ; \{ q^{PK}\}] \,,
\end{equation}
where: 
\bd
%\item $t_{\rm barycenter}$ denotes the solar-system barycentric 
%(infinite frequency) arrival time of a pulse, 
\item $T_N$ is the pulsar proper time corresponding to the 
$N^{\rm th}$ pulse, which is related to the integer $N$ by an equation 
of the form
\begin{equation}
\label{eq:T_N}
N_{pulse} = const + \nu_p \, T + \frac{1}{2} \, \dot\nu_p \, T^2 + \, \cdots  \q,
\end{equation}
in which $\nu_p$ and $\dot\nu_p$ are some spindown parameters;
\item $\{ p^K \} $ is the set of {\it Keplerian} parameters, like 
the orbital period $P_b$\,;
% $\{ p^K \} = \{ P_b , T_0 , e_0 , \omega_0 , x_0 \}$
\item $\{ p^{PK} \}$ is the set of {\it separately measurable post-Keplerian} 
parameters, like the derivative of the orbital period $\dot P_b$\,;
% $\{ p^{PK} = k , \gamma , \dot P_b , r, s, \delta_{\theta} , \dot e , \dot x \}$ 
\item $\{ q^{PK} \}$ is the set of {\it non separately 
measurable post-Keplerian} parameters, 
like the overall Doppler factor $D$\,; being non separately 
measurable, these parameters can be absorbed into changes of the 
other parameters and set to some arbitrary values.
% $\{ q^{PK} \} = \{ \delta_r , A, B, D \}$  
% set to some given fiducial values e.g. $\{ 0,0,0,1 \}$
\ed
The relativistic timing formula of \eq{eq:time_1} was derived by Damour 
and Deruelle~\cite{DD86} for GR but its validity is more 
general: remarkably, the mathematical form of its parametrized expression 
(\ref{eq:time_2}) was found to be applicable also to \sts. 
As we saw in \Sec{sec:Dam_Far}, these theories predict that binary pulsars observables differ markedly from their behavior in GR.  
Notwithstanding these differences, the parametrized form of the timing formula (\ref{eq:time_2}) holds in the case of \sts\ too, thus enabling one to treat binary pulsars observations by means of a single phenomenological approach: this is the so-called parametrized post-Keplerian (PPK) framework~\cite{Dam-PPK,Dam_Tay:92} and consists in the following steps.
By a least-square fit of the observed %sequence of pulsar 
arrival times %$t_N$ 
to the parametrized formula (\ref{eq:time_2}), one can 
extract from raw data the best fit values of both the Keplerian parameters 
%$\{ p^K \} = \{ P_b , T_0 , e_0 , \omega_0 , x_0 \}$, 
and a subset of post-Keplerian (PK) parameters: 
%$\{ p^{PK} \} = \{ k,\gamma, \dot P_b ,r,s,\delta_{\theta} , \dot e , \dot x \}$. 
as we will see below, three PK parameters could be measured for the Hulse-Taylor pulsar while, in the case of the double pulsar, one has obtained six PK parameters and a measure of the mass ratio. 
The parameter extraction is phenomenological in that it does not rely on the 
validity of a specific theory of gravity. 
However, each theory of gravity makes its peculiar predictions about 
how the PK parameters are related to the Keplerian ones and to the 
masses of the binary system.  
It is through these links that one can test the consistency of a theory of gravity with experiments. 
The masses are a priori unknown and have to be inferred from the 
measured values of the parameters. % in the following way. 
Because every PK parameter $p$ is a function of the binary masses $m_a$ and $ m_b$, it is possible to invert this function to obtain a relation of the type $m_a^p = f^p[m_b]$, i.e. a link between the values of $m_a$ and those of $m_b$ which are allowed by the parameter~$p$. 
\begin{figure}[!t]
\begin{center}
\includegraphics[width=0.65\textwidth]{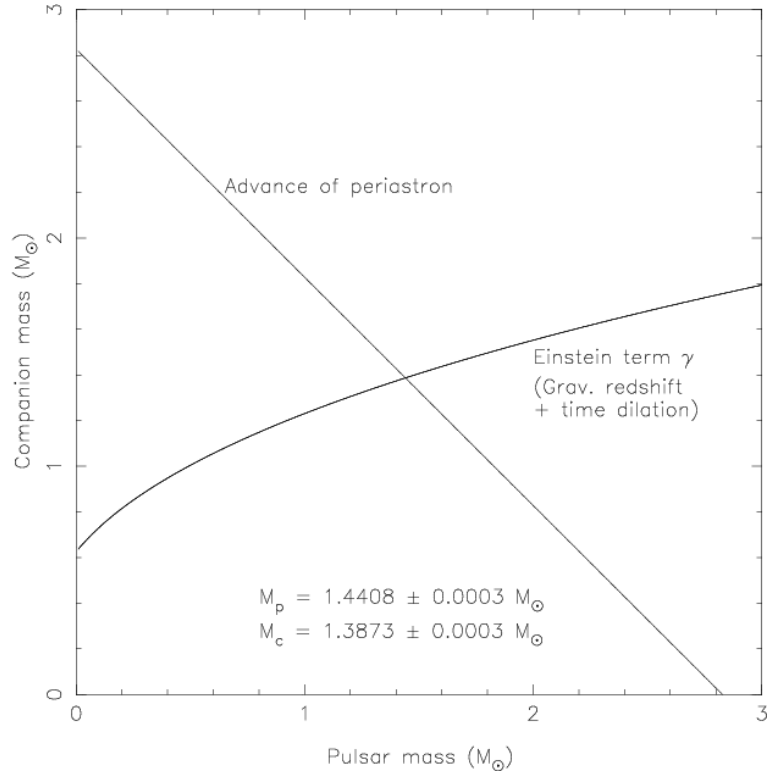}
\caption{Plot of the parameters $\gamma$ and $\dot \omega$ for the Hulse-Taylor pulsar as functions of the constituent masses. 
The errors on the parameters are smaller than the thickness of the respective curves. 
The meeting point of the two curves defines the measured values for the pulsar and its companion. 
Figure taken from ref.~\cite{Weis_Tay:03}. }
\label{fig:2PK}
\end{center}
\end{figure}
These links correspond to curves in the plane of the binary masses: the graphs of  two such curves overlap, within errors, in a common region, which gives the values of the binary masses together with their statistical errors. 
The situation is shown in \fig{fig:2PK}, where two PK parameters are plotted for the case of Hulse-Taylor binary pulsar . 
The parameter $\gamma$ measures the amplitude of the Einstein 
time delay, $\Delta_E$: it has the dimension of time and should not 
be confused with the dimensionless PPN parameter $\gppn$ probed by Solar-System experiments. 
The parameter $\dot \omega$ measures the periastron advance per orbit.
Both $\gamma$ and $\dot \omega$ are determined so precisely that the thickness of the respective curves is larger than the errors on the parameters; the inferred values for the masses are indeed very accurate~\cite{Weis_Tay:03}: $M_p=(1.4408\pm0.0003)\msun$ for the pulsar, $M_c= (1.3873\pm0.0003)\msun $ for the companion.

Once two curves (parameters) have been used to fix the values of the masses, a third one should consistently overlap in the same common region of the masses plane: the theoretical prediction for the third parameter can be regarded as a check of the consistency between the theory under examination and experiments. 
As a third parameter let us take the time derivative of the orbital period 
$\dot P_b$\,, which is a direct consequence of the term $A_5 \sim {\cal O}(v/c)^5$ in the equations of motion for a binary system~(\ref{eq:motion}): 
%As we have already discussed, this term is directly linked to the fact that the gravitational interaction between two strongly self-gravitating bodies propagates at the velocity of light;
therefore, any test involving the observable $\dot P_b$ will be a mixed type of test of the radiative/strong-field regime.
On the other hand, the two parameters $\dot \omega$ and $\g$ do not involve  radiative effects.  
%$k$ being a dimensionless parameter that measures the fractional periastron advance per orbit: 
%The parameter $\gamma$ measures the amplitude of the Einstein time delay, $\Delta_E$: it has the dimension of time and should not be confused with the dimensionless post-Newtonian Eddington parameter $\gppn$ probed by Solar-System experiments.
In the context of GR, the predictions for these parameters have been worked out in refs.~\cite{BT76, DD86,Dam_Q_contro}. 
Considering the fractional periastron advance per orbit
$k = \Delta\theta / 2\pi = \langle \dot\omega \rangle / n 
= \langle \dot\omega \rangle \, P_b / 2\pi$, where $n \equiv 2\pi/P_b$ 
indicates the orbital frequency, 
the GR predictions for $\kappa$, $\g$ and $\dot P_b$ read
\begin{eqnarray}
\label{eq:k}
k^{\rm GR} (m_a , m_b) &= &\frac{3}{1-e^2} \ \b_0^2 \, , \\
\nn \\
\label{eq:gamma_pk}
\gamma^{\rm GR} (m_a , m_b) &= &\frac{e}{n} \ X_b (1+X_b) \, \b_0^2 \, , \\
\nn \\
\label{eq:dotP_pk}
\dot P_b^{\rm GR} (m_a , m_b) &= &- \frac{192 \pi}{5} \ 
\frac{1 + \frac{73}{24} \, e^2 + \frac{37}{96} \, e^4}{(1-e^2)^{7/2}} 
\ X_a \, X_b \, \b_0 ^5 \, ,
\end{eqnarray}
where $e$ is the eccentricity of the orbit and
\begin{eqnarray}
\label{eq:M}
M &\equiv &m_a + m_b \\
\nn \\
\label{eq:Xa}
X_a &\equiv &m_a / M \, ; \quad X_b \equiv m_b/M \, ; 
\quad X_a + X_b \equiv 1 \\
\nn \\
\label{eq:gM}
\b_0 (M) &\equiv &\left( \frac{G_N Mn}{c^3} \right)^{1/3} \,. 
\end{eqnarray}

It is interesting to look at the scaling of $\b_0$ so as to compare the PN order of the various observables. 
By means of Kepler's law, $n^2=m/a^3$ where $a$ is the semi-major axis of the orbit; this enables to write $\b_0^2\simeq(M\,n)^{2/3}=m/a$, which, a part for a factor of $G_N$, is of order $\epsilon$\,. 
Therefore, $\kappa$ and $\g$ are seen to be ${\cal O}(\epsilon)$ corrections to the Keplerian motion, while $\dot P_b$ is ${\cal O}(\epsilon)^{5/2}$.

\begin{figure}[htbp]
\begin{center}
\includegraphics[width=1.0\textwidth]
{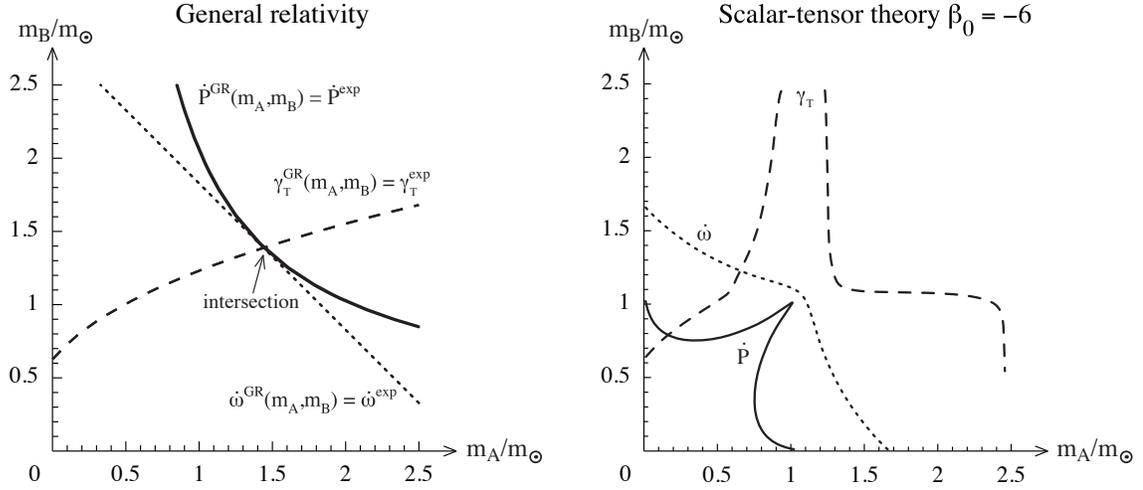}
\caption{A confrontation of the consistency tests realized by means of post-Keplerian parameters as measured in the Hulse-Taylor pulsar, with $m_A$ the pulsar's mass, $m_B$ the companion's. 
The panel on the left shows that GR passes the test with flying colors, the panel on the right indicates that a scalar-tensor theory with $\beta_0 = -6$ is ruled out.
The widths of the lines are larger than $1\sigma$ error bars. 
Figure taken from ref.~\cite{EF:09}. }
\label{fig:dotP_k_omega_GR_ST}
\end{center}
\end{figure}

In \sts\ the PK parameters $\kappa$, $\g$ and $\dot P_b$ will not look the same as \eqst{eq:k}{eq:dotP_pk}: first of all, they will depend on the coupling constants of the theory; secondly, they will have a different functional dependence on the masses and on Keplerian parameters like the eccentricity and the orbital frequency.  
The expressions for the PK parameters $\kappa$, $\g$ and $\dot P_b$ in \sts\  
% have been worked out in refs.~\cite{Dam_Far-ST_CQG,Dam_Far-ST_Puls:PRD54,Dam-PPK,Will_TEGP,Will:1989p140} 
are reported in ref.~\cite{Dam_Como} (to which we refer for the relevant citations) and are given by:
\begin{eqnarray}
\label{eq:kappa_st}
k^{ST} (m_A , m_B) &= &\frac{3}{1-e^2} \left( \frac{G^{eff}_{AB} (m_A +
m_B) \, n}{c^3} \right)^{2/3}  \\
\nn \\
&&\times \left[ \frac{1 - \frac{1}{3} \, \alpha_A \, \alpha_B}{1 + \alpha_A
\, \alpha_B} - \frac{X_A \, \beta_B \, \alpha_A^2 + X_B \, \beta_A \,
\alpha_B^2}{6 \, (1 + \alpha_A \, \alpha_B)^2} \right] \;, \nn \\ 
\nn \\
\label{eq:gamma_st}
\gamma^{ST} (m_A , m_B) &= &\frac{e}{n} \, \frac{X_B}{1 + \alpha_A
\, \alpha_B} \left( \frac{G^{eff}_{AB} (m_A + m_B) \, n}{c^3} \right)^{2/3} \\
\nn \\
&&\times[X_B (1 + \alpha_A \, \alpha_B) + 1 + k_A \, \alpha_B ] \;, \nn \\
\nn \\
\label{eq:dotP_st}
\dot P_{h2-quad}^{ST} (m_A , m_B) &= &- \frac{192 \pi}{5 (1
+ \alpha_A \, \alpha_B)} \, \frac{m_A \, m_B}{(m_A + m_B)^2} \\
\nn \\
&&\times \left( \frac{G^{eff}_{AB} (m_A + m_B) \, n}{c^3} \right)^{5/3} \frac{1 +
73 \, e^2 / 24 + 37 \, e^4 / 96}{(1-e^2)^{7/2}} \;, \nn
\end{eqnarray}
where $G^{eff}_{AB}$ is the effective gravitational constant of \eq{eq:GAB}, which comprises scalar effects, 
and where for the period decay we have indicated only the change in the quadrupole contribution of helicity-2.
Taking again Hulse-Taylor binary pulsar, a confrontation of the consistency tests realized by means of these PK parameters is reported in \fig{fig:dotP_k_omega_GR_ST}: the panel on the left shows that GR passes the test with flying colors, the panel on the right indicates that a scalar-tensor theory with $\beta_0 = -6$ is ruled out.

\begin{figure}[!t]
\begin{center}
\includegraphics[width=0.6\textwidth]{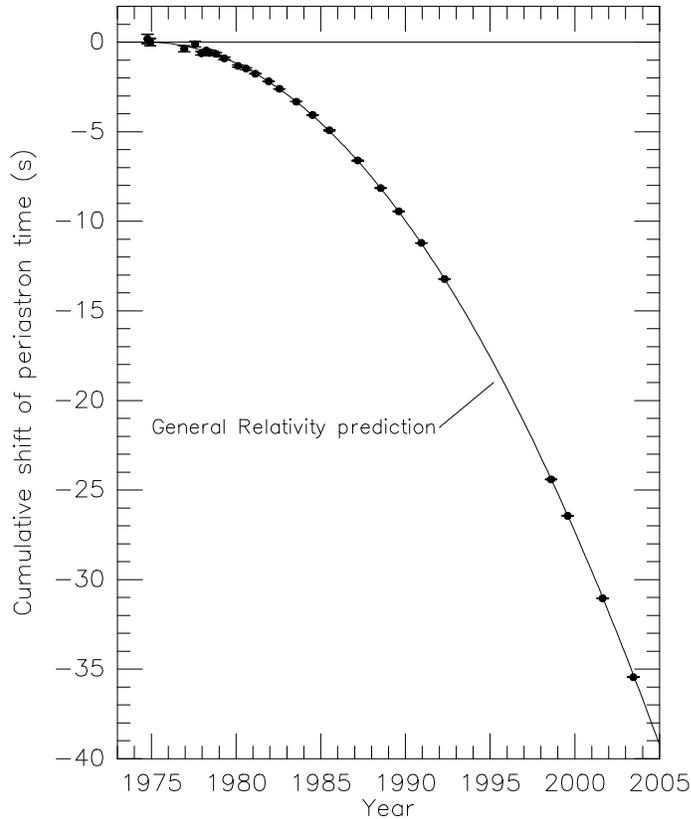}
\caption{Orbital decay of PSR B1913+16. The data points indicate 
the observed change in the epoch of periastron with date while the 
parabola illustrates the theoretically expected change in epoch for a 
system emitting gravitational radiation, according to GR.
Figure taken from ref.~\cite{Weisberg:2004hi}. }
\label{fig:hulse_taylor}
\end{center}
\end{figure}

Concerning Hulse-Taylor pulsar, it is interesting to report a test of GR realized by means of the periastron precession. 
In a way similar to the one used to determine the time of the $N$-th pulse emitted by the pulsar~(\ref{eq:T_N}), one can write a relation for the time of periastron passage $T_p$
\begin{equation}
\label{eq:T_peri}
N_{peri} = n \, T_p + \frac{1}{2} \, \dot n \, T_p^2 
\end{equation}
where higher derivatives of the orbital frequency $n$ have been discarded because of their smallness in Hulse-Taylor binary. 
The time of periastron passage $T_p$ will then differ from the ratio $N_{peri}/n$ as
\be
T_p - \frac{N_{peri}}{n} = -\frac{\dot n}{2 n} T_p^2 \,,
\ee
which is a parabola with coefficient $-\dot n/2n<0$; using the GR prediction for $\dot n$, this parabola lies on top of experimental data with the remarkable agreement shown in \fig{fig:hulse_taylor}.

An even more spectacular laboratory for testing GR through PK parameters is represented by the double pulsar J0737$-$3039, whose probing power for \sts\ has been reported in \fig{fig:ST_Tests_Full}.
The peculiarity of this system relies in the fact that both the component neutron stars are pulsating in the direction of the Earth, thus making it unique so far.
Besides being oriented favorably, this binary is so relativistic that one could measure more PK parameters than in the Hulse-Taylor case and in much less time. 
Notably, on top of the triple $(\g,\dot\omega,\dot P_b)$ it has been possible to determine three other parameters. 
The first two are the {\it shape} and {\it range} of the Shapiro time delay: with an obvious choice of symbols, they are indicated by $s$ and $r$. 
The third parameter is the rate of spin-precession $\Omega_{SO}$ caused by relativistic spin–orbit coupling. 
Moreover, the fact that both neutron stars are detectable as pulsars has enabled one to measure the orbit of both objects around the common center of mass: in other words, if $i$ is the inclination angle of the orbit with respect to the line of sight, the projected semi-axes $x_I=a_I \sin i$ is known for both objects. 
The ratio of the two semi-axis is identical to the inverse ratio of masses, which is theory independent. 
Therefore, by measuring the orbits of the two pulsars relative to the centre of mass, one obtains a precise measurement of the mass ratio parameter $R\equiv m_A/m_B$\,. 
The tests corresponding to all these parameters are reported in \fig{fig:Kramer_Test}, where the curves in the masses plane overlap in the region allowed by the geometrical constraint $\sin i < 1$. % coming from the individual mass functions.
As shown by the figure, GR passes all the tests corresponding to the measured parameters.
\begin{figure}[!t]
\begin{center}
\includegraphics[width=0.8\textwidth]{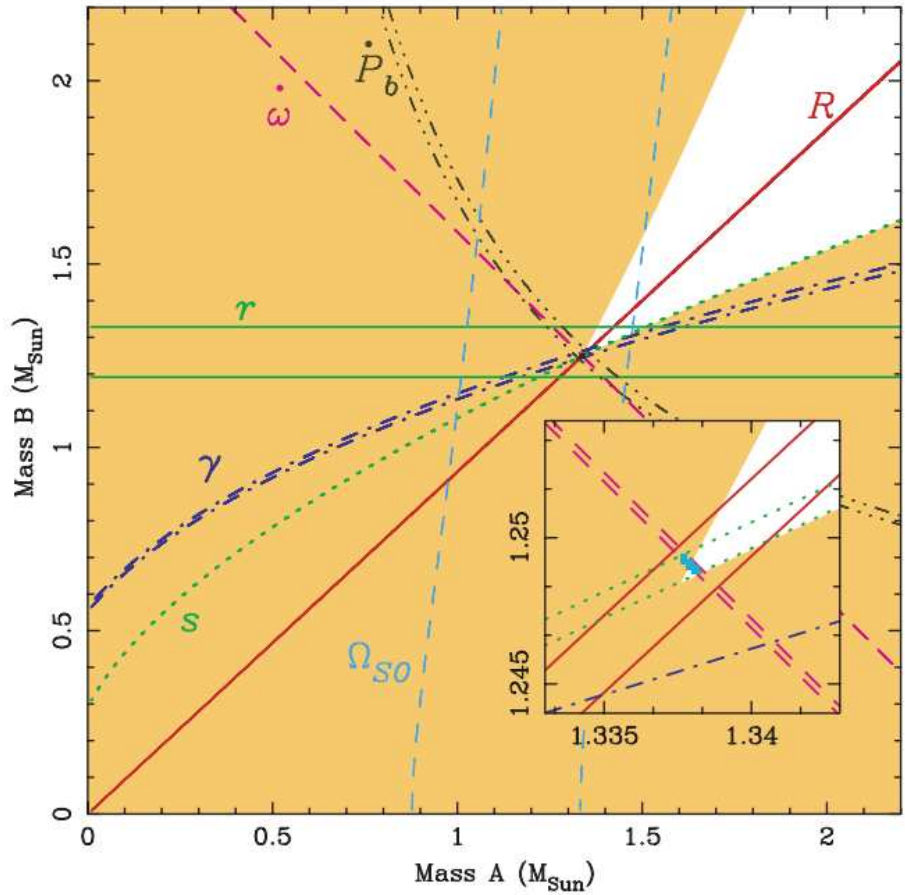}
\caption{Mass–mass diagram of the double pulsar system summarizing the measured PK 
parameters in combination with the derived mass ratio $R \equiv m_A/m_B = X_B/X_A$ (solid red line) and constraints given by the mass functions of the binary system. 
The latter are indicated by the coloured regions which mark areas in the diagram that are excluded by the Keplerian mass functions of the two pulsars and the condition that $\sin i < 1$. 
Further constraints are shown as pairs of lines enclosing permitted regions as predicted by GR. These are: 
the measurement of the advance of periastron~$\dot\omega$ (diagonal dashed line); 
the measurement of the gravitational redshift/time dilation parameter $\gamma$ (dot-dash line); 
the measurement of the Shapiro parameters $r$ (solid green line) and $s$ (dotted green line); 
the measurement of the orbital decay $\dot P_b$ (dot-dot-dot-dash line) 
and the rate of spin precession of B, $\Omega_{SO}$ (almost vertical 
dashed line). 
The inset is an enlarged view of the small square which encompasses the intersection of the tightest constraints. 
The permitted regions are those between the pairs of parallel lines and we see that an area exists which is compatible with all constraints, delineated by the solid blue region. 
Figure taken from ref.~\cite{Kramer:09}. }
\label{fig:Kramer_Test}
\end{center}
\end{figure}

\clearpage

%%%%%%%%%%%%%%%%%%%%%%%%%%%%%%%%%
\section[Probing the strong-field/radiative regime of gravity - Part II \\
Extending the parametrized post-Newtonian framework \\
to the gravitational-wave phasing formula]
 {Probing the strong-field/radiative regime of \\ gravity - Part II \\
Extending the parametrized post-Newtonian  \\ framework
to the gravitational-wave phasing \\ formula
 \sectionmark{Parametrizing the gravitational-wave phasing formula}}
 \sectionmark{Parametrizing the gravitational-wave phasing formula} 
\label{sec:PPN_GW}
 
%where:    
% "[]" is for the table of contents
% first "{}" for where the section actually is
% 1st "sectionmark" is for the first page containing the section
% 2nd "sectionmark" is for the other pages containing the section

%%%%%%%%%%%%%%%%%%%%%%%%%%%%%%%%%

In the PPN framework a phenomenological attitude is applied towards the first PN order in the expansion of the gravitational field: all metric theories share a common structure that is conveniently represented by means of parameters. 
The same philosophy can be applied to the mathematical expressions that describe the GW waveform.
In recent years, indeed, an extension of the PPN framework has been proposed where the coefficients to be bound by experiments are those of the phasing 
formula of a GW signal. 
This line of investigation has been pursued in a series of three papers from Arun, Iyer, Sathyaprakash and collaborators~\cite{Arun:2006yw,Arun:2006hn,Mishra:2010tp}.
I describe this framework in the present section, while in \chap{chap:Group} I will present an example of a study which is similar in spirit but has a field-theoretical interpretation.

Because \ifos\ are most sensitive to the phase evolution of a GW signal, 
one usually neglects the PN corrections in the amplitude 
with respect to those of the phase~\cite{Cutler:1992tc} and works 
in the so-called {\it restricted} post-Newtonian scheme.
This is one of the assumptions that can be made for both simplifying the 
treatment  and demonstrating the feasibility of the tests themselves. 
Other simplifications stem from neglecting the spin of the objects and the 
eccentricity of the orbits.

To obtain the phase, one can proceed as follows.
From \eq{eq:motion} we know that the equations of motion for a binary system contain terms that are odd in time and therefore describe dissipative effects in the dynamics; notably, the first term that corresponds to the loss of the system's mechanical energy is $A_5\sim(v/c)^5$: once averaged over time, this loss equals the leading order flux of GWs at infinity. 
As a consequence, the resulting change in the orbital phase of the binary system can be computed from the energy balance equation 
\bees
\label{eq:balance}
-\frac{dE(t-r/c)}{dt} = {\cal F}(t) \,,
\ees
where $E$ is mechanical energy of the system and $ \cal F$ the averaged flux of emitted gravitational radiation. 
The time arguments of the two quantities indicate that the energy carried by GWs at a time $t$ and a distance $r$ from the system are balanced by the system's loss at retarded time $t-r/c$.
Therefore, \eq{eq:balance} expresses the conservation of energy and a similar relation holds for the angular momentum. 
As we will see below, when one considers PN orders higher than $(v/c)^5$, non-linear phenomena affect the propagation of GWs in a very non-trivial way: for example, the back-scatter of radiation produces the so-called {\it tails} of the wave that have support also inside the light cone and therefore propagate with an effective speed smaller than $c$\,.
In this context it is not obvious {\it a priori} that the flux at time $t$ is exactly compensated by a loss at retarded time $t-r/c$. 
In the case of GR, \eq{eq:balance} has been checked to be valid up to relative 1.5PN order~\cite{Blan-1.5_bal}, while for alternative theories, like the \sts\ studied by DEF, the relation (\ref{eq:balance}) is usually assumed. 

The PN expressions for the mechanical energy and the GW flux are presently known, in the case of GR, up to order $v^7 $ or 3.5PN~\cite{Blanchet:2004ek} and are given by 
\bees
\label{eq:EF}
E &=& -\frac{1}{2}\nu \, v^2\sum_{k=0}^{3} E_k v^{2k} \nn \\ \nn \\ 
{\cal F} &=& \frac{32}{5}\nu^2 v^{10} \sum_{k=0}^{7} {\cal F}_k v^{k} 
- \frac{1712}{105}\ln (v) \, v^6
\ees
where $\nu=m_1 m_2/M^2$ is the symmetric mass ratio of the system and the explicit expression of the PN series coefficients $E_k$ and ${\cal F}_k$ as functions of $\nu$ can be found in ref.~\cite{Damour:2000zb}.
Expressing $E$ and $\cal F$ as functions of the velocity~$v$, the energy balance equation transforms into the following coupled ordinary differential equations 
\bees
\label{eq:diffeqs}
&&\frac{dv}{dt} = \frac{dE/dt}{dE/dv} = \frac {-\cal F}{E'(v)} \nn \\ \nn \\
&&\frac{d\phi}{dt} = \omega = \frac{v^3}{M}  
\ees
where $E'(v)\equiv dE/dv$ and $\phi$ is the orbital phase. 
Finally, the GR prediction for the phase $\varphi_{GW}$ of gravitational radiation is obtained, at leading order, by two times the orbital phase, $\varphi_{GW}(t) = 2\phi(t)$: for example, in the extreme-mass-ratio limit its expression reads
\begin{equation}
\label{eq:gwphase}
\varphi_{GW}(t) = \int_{t_0}^t d\tau\,\omega_{GW}(\tau) = {2\over G_N M_H}\int^{v_0} _{v(t)}d v' v'^3 {dE/dv'\over {\cal F}(v')}\,,
\end{equation}  
where $\omega_{GW}$ is the GW frequency and $M_H$ the mass of the heaviest object in the system.

From \eq{eq:diffeqs}, we can see that the terms of PN phasing formula are combinations of the PN coefficients of $E$ and $\cal F$ (\ref{eq:EF}): therefore, each PN term in the phase encodes non-linearities that are intrinsic to GR.  
One such peculiar non-linearity is the aforementioned phenomenon of GW {\it tails}~\cite{Blanchet:1987wq,Blanchet:1993ec,Blanchet:1997jj}, which arises starting from order 1.5PN in the phase.
This effect is due to the fact that waves which are sourced by the quadrupole moment of the system can scatter off the curved (Schwarzschild) background generated by the source: the scattering effectively delays these waves, which therefore propagate also inside the light cone (a Feynman diagram describing this effect is reported in \fig{fig:tail} of \chap{chap:EFT}).
Other viable theories of gravity have their own predictions for the PN coefficients of the phasing formula, in analogy with the alternative predictions for the PK parameters that I have discussed in~\Sec{sec:PPK}. 
Through the measure of GWs, an accurate scrutiny of the PN coefficients will be possible; as in the case of the PPN framework, this will serve a two-fold purpose: check the consistency of GR itself and provide bounds on the parameters characterizing alternative theories.
Therefore, it is relevant to consider how to extend the PPN framework to the strong-field/radiative regime probed by means of GWs. 
Such an extension has been put forward in refs.~\cite{Arun:2006yw,Arun:2006hn,Mishra:2010tp} and consists in regarding the PN coefficients of the GR phasing formula as parameters to be estimated.
To illustrate the procedure, we report the expressions for the waveforms that are valid in GR and then present the possible tests on its PN coefficients.
%This follows a series of three papers .

In the restricted PN approximation, the response of an interferometric
antenna to the incident radiation from a source at a luminosity distance
$D_L$ is
\begin{eqnarray}
h(t) &=& \frac{8{\cal M}}{5 D_L} \left [ \pi{\cal M}F(t) \right ]^{2/3} 
\cos \[\varphi_{GW}(t) \] \,,
\end{eqnarray}
where ${\cal M}=\nu^{3/5}M$ is the so-called {\it chirp mass} of the system  
and $F(t)$ is the instantaneous frequency of radiation given by 
%% {\bf introdurre dipendenza d M dal redshift}
$$F(t)\equiv \frac{1}{2\pi} \frac{d\varphi_{GW}(t)}{dt} \,.$$
For the purpose of testing the PN phasing coefficients it is useful to work with the Fourier transform of the signal 
$\tilde h(f) \equiv \int_{-\infty}^{\infty} h(t)\, \exp(2\pi i f t) dt$.  
Using the stationary phase approximation it has been shown~\cite{Thorne:1987af,Sathyaprakash:1991mt} that one can write the waveform in the frequency domain as
\bees
\tilde h(f) = {\cal A}\, f^{-7/6} \exp\left [i \Psi(f) + i \frac{\pi}{4}\right ],
\ees
with the Fourier amplitude ${\cal A}$ and phase $\Psi(f)$ defined by
\bees 
\label{eq:phase_spa}
{\cal A} & = & \frac{2}{5 D_L\pi^{2/3}} \sqrt{\frac{5}{24}} {\cal M}^{5/6}, \nn \\
\Psi(f) & = & 2\pi f t_c - \Phi_c + \sum_{j=0}^7 [ \psi_j + \psi_{jl}\, \ln f] f^{(j-5)/3} \,.
\ees
In the expression for the phase, $t_c$ and $\Phi_c$ are integration constants that represent respectively a fiducial epoch of merger and the phase of the signal at that epoch. 
Moreover, the $\psi$-coefficients of the PN expansion of the Fourier phase 
are given by 
\bees
\psi_j &=& \frac{3}{128\,\nu}(\pi M)^{(j-5)/3}\alpha_j \nn \\
\psi_{jl}&=&\frac{3}{128\,\nu}(\pi M)^{(j-5)/3}\alpha_{jl}
\label{eq:psi}
\ees
and the $\a$-coefficients are \cite{Arun:2004hn}
\bees
\label{eq:alpha}
\alpha_0 &=& 1 \q , \q \alpha_1 =0 \q , \q
\alpha_2 =\frac{3715}{756}+\frac{55}{9}\nu \q , \q
\alpha_3 =-16 \pi \q , \\ 
\nn \\
\alpha_4 &=& \frac{15293365}{508032}+\frac{27145}{504} \nu
+\frac{3085}{72} \nu^2 \q , \nn \\
\nn \\
\alpha_5 &=&\pi \left(\frac{38645}{756}-\frac{65}{9}\nu\right)
\left[1+\ln\left(6^{3/2}\pi M \right)\right] \q , \nn \\
\nn \\
\alpha_6 &=&\frac{11583231236531}{4694215680}
-\frac{640}{3}\pi^2-\frac{6848}{21} \gamma_E
+\left(-\frac{15737765635}{3048192}+ \frac{2255}{12}\pi ^2\right)\nu + \nn \\
\nn \\
&& \frac{76055}{1728}\nu^2-\frac{127825}{1296}\nu^3
-\frac{6848}{63}\ln\left(64\,\pi M \right) \q , \nn \\
\nn \\
\alpha_7 &=&\pi\left( \frac{77096675}{254016}+\frac{378515}{1512}
\nu -\frac{74045}{756}\nu ^2\right) \q , \nn \\
\nn \\
\alpha_{5l} &=& \pi\left(\frac{38645}{756}-\frac{65}{9}\nu\right) \q , \q
\alpha_{6l} = -\frac{6848}{63} \q , \q \alpha_{jl}=0 \q \text{for} \; j=0,1,2,3,4,7 \nn
\ees
where the constant appearing in the expression for $\alpha_6$ is Euler's constant, $\gamma_E \simeq 0.577 $.

There is a total of nine PN parameters: seven are the coefficients of 
${\cal O}(v^n)$ terms for $n=0,2,3,4,5,6,7$ and two are 
coefficients of ${\cal O}(v^n \ln v)$ terms for $n=5,6$. 
We stress the absence of $\psi_1$, the coefficient multiplying the term $v^1$: this feature is peculiar to GR as we will discuss also in \Sec{sec:beta3_half}.
%The parameters $\psi_{5l}$ and $\psi_{6l}$ multiply a logarithmic term in the frequency but, as suggested in \cite{Arun:2006yw}, this is a mild dependence in the bandwidth of \ifos. 
%Therefore, one can approximate the contributions of $\psi_{5l}$ and $\psi_{6l}$ by evaluating them, for example, for $f=f_{LSO}$, which is the frequency of the last stable orbit before the two objects merge.
On the other hand, the coefficient $\psi_5$ gives a contribution to the phase (\ref{eq:phase_spa}) that is constant as it multiplies $f^0$: for this reason, the coefficient $\psi_5$ cannot be used as a test parameter and only redefines the phase of coalescence.

The PN coefficients of~\eqs{eq:psi}{eq:alpha} have been quoted entirely to show explicitly that, in the absence of spin, GR predicts that they depend only on the masses of the system~\footnote{An obvious exception to this is represented by the coefficients which are constants, like $\a_3$.}, through the parameters $M$ and $\nu$. 
If the PN limit of GR is the correct description of the binary dynamics, 
the parameters of \eqs{eq:psi}{eq:alpha} must be consistent with each 
other within their respective error bars, much in the same way as the PK parameters inferred by binary pulsars observations (see \Sec{sec:PPK})~\footnote{\label{foot:EOB}
As suggested in ref.~\cite{Arun:2006yw}, there are also non-perturbative schemes that one could implement to check the consistency of GR using GWs: for example, one could test the full numerical relativity predictions. Another possibility is offered by Pad\'e-approximants~\cite{Damour:1997ub} and the effective one-body approach~\cite{Buonanno:1998gg,Buonanno:2000ef,Damour:2000we}; here, by means of re-summation techniques, one builds model waveforms that include both the inspiral and the merge. Implementation of these schemes to test gravity has not appeared in the literature yet.}. 
When talking about the set of the phasing parameters, we will collectively denote 
them by $\psi_j$'s, i.e. we will not use a different notation for the $\psi_{jl}$'s; 
similarly, the error bars will be denoted by $\Delta \psi_j$. %old a (look for b)

Notwithstanding the proven consistency of GR, the presence of the statistical errors leaves room for alternative theories. 
With respect to GR, the phasing coefficients of alternative theories have a different behavior: either their functional dependence on the constituents masses is not the same as \eqs{eq:psi}{eq:alpha}, or they depend on other types of parameters.
We already met an example of the latter category when we considered the \st\ of DEF in \Sec{sec:Dam_Far}: in fact, in \eqst{eq:flux_20}{eq:flux_dip} we saw that the parameter~$\a_A$, describing the linear coupling of a scalar field to matter in the strong-field regime, causes a change in the period decay of a binary pulsar at orders $(v/c)^1$ and $(v/c)^3$. 
In the language of the present section, these changes would bring two extra $\psi_j$'s of "negative" order with respect to the ${\cal O}(v/c)^5$ GR quadrupole: $\psi_{-2}$ for the monopolar term of ${\cal O}(v/c)^1$ and $\psi_{-1}$ for the dipolar term of ${\cal O}(v/c)^3$.
In ref.~\cite{Arun:2006yw} the focus was rather towards a more phenomenological alternative in which the graviton has a mass. 
This possibility had been previously entertained by Will~\cite{Will:1997bb}, who begins by writing the velocity a graviton would have in a local inertial frame 
\be
\label{eq:vma}
\frac{v_g^2}{c^2} = 1- \frac{m_g^2 c^4}{E^2} \,,% = 1- \left(\frac{m_g c^2}{hf}\right)^2 
\ee
where $m_g$ is the rest mass, $E$ the energy. 
This velocity difference could be tested by monitoring a source that emits both gravitons and photons: an example could be supernova located at a distance $D\simeq z c/H_0$, with $z<<1$ the redshift and $H_0$ the Hubble constant.   
With the appropriate detectors on Earth, we would measure a difference in the times of arrival of the two types of radiation 
\be
\label{eq:Delta_t}
\Delta t \equiv \Delta t_a - (1+z) \Delta t_e
\ee
where the index "$a$" refers to the arrival times, the index "$e$" refers to the emission times.
Such a time difference is related to the velocity difference by
\begin{equation}
\label{eq:vDis}
1- {v_g \over c}= 5 \times 10^{-17} 
\left ( {{200 {\rm Mpc}} \over D} \right ) 
\left ( {{\Delta t} \over {1 {\rm s}}} \right ) \,,
\end{equation}
where a typical value has been chosen for $D$.
However, for the purpose of detecting the presence of $m_g$, this technique is not the optimal one: in fact \eq{eq:Delta_t} contains the unknown $\Delta t_e$ which is model dependent. 
A better strategy is to make use of gravitational radiation only, as suggested by Will~\cite{Will:1997bb}. 
In terms of the frequency $f$, the energy reads $E=h f$, where $h$ is Planck's constant; \eq{eq:vma} can then be re-written as 
\bees
\label{eq:vlam}
\frac{v_g^2}{c^2} &=& 1- \frac{m_g^2 c^4}{E^2} = 1- \left(\frac{m_g c^2}{hf}\right)^2 \nn \\
&\simeq& 1- \frac12 \left(\frac{c}{\lambda_g f}\right)^2 \,,
\ees
where we have introduced the graviton Compton wavelength $\lambda_g$ and assumed that the frequency is such that $h f >> m_g c^2$.
Through \eq{eq:vlam} the velocity difference is related to the graviton frequency, which lends itself to be tested by detecting GW from compact binaries.  
In fact, during the observation of an inspiral, the frequency {\it chirps}: it enters the sensitivity band of an \ifo\ at its lower end, for example $\sim50$Hz in the case of ground based instruments where low-frequency seismic noise is relevant, and exits the band when the binary reaches the last stable orbit before merger, around $\sim500$Hz for first generation instruments. 
According to \eq{eq:vlam}, this frequency sweep will cause $v_g$ to increase towards $c$, which in turn modifies $\Delta t$ according to \eq{eq:vDis}, with a {\it shrinking} effect.
From the point of view of GW interferometry, this shrink will reduce the amount of time that a given number of cycles spends in the frequency band of the instrument. 
In order to detect this effect by means of \ifos, Will computes how the time change is reflected in the Fourier transform of the GW phase. 
Within a 1.5 PN accuracy, the result is that a finite graviton Compton wavelength introduces an extra term at 1PN order, i.e. it causes a shift~\cite{Will:1997bb}
\be
\label{eq:a2shift}
\a_2 \longrightarrow \a_2 - \frac{128\,\nu}{3}\, \frac{\pi^2\, D\, M }{\lambda_g^2 (1+z)}\,.
\ee

\smallskip %old b 
To study and constrain the effect of terms like \eq{eq:a2shift}, the strategy of refs.~\cite{Arun:2006yw,Arun:2006hn,Mishra:2010tp} is analogous to what is done for PK parameters: two $\psi_j$'s are used to plot two curves of the type $m_2^j (m_1,\, \psi_j\pm \Delta \psi_j)$ in order to fix the masses of the emitting binary (within errors); afterwards, to test the theory, one draws (at least) a third curve for a test parameter $\psi_T$ to see if it overlaps in a region of the mass-mass plane that is common with the first two parameters used. 
The difference with binary pulsars observation is that in the present case all the information about the emitting sources has to be extracted from the phase.
In GW astronomy in fact, signal detection and parameter extraction are realized through matched-filtering, a technique where the output time series of the \ifo\ is convoluted with a bank of theoretical waveforms called templates. 
Every template depends on specific values of the $\psi_j$'s so that, maximizing over the signal-to-noise ratio (SNR), one can identify the template and the $\psi_j$'s that matched the best: this theoretical waveform can then be considered as representative of the actual signal (within a certain confidence level).

A first possibility is to fit a GW signal through a template where all the $\psi_j$'s are treated as independent of one another and try to measure {\it each} of the PN coefficients~\cite{Arun:2006yw}. 
This type of tests requires SNRs as high as 1000, which could be the case of a supermassive black-hole binaries observed by the space interferometer LISA for one year before merger.
\begin{figure*}[t]
\begin{center}
\includegraphics[width=0.6\textwidth]{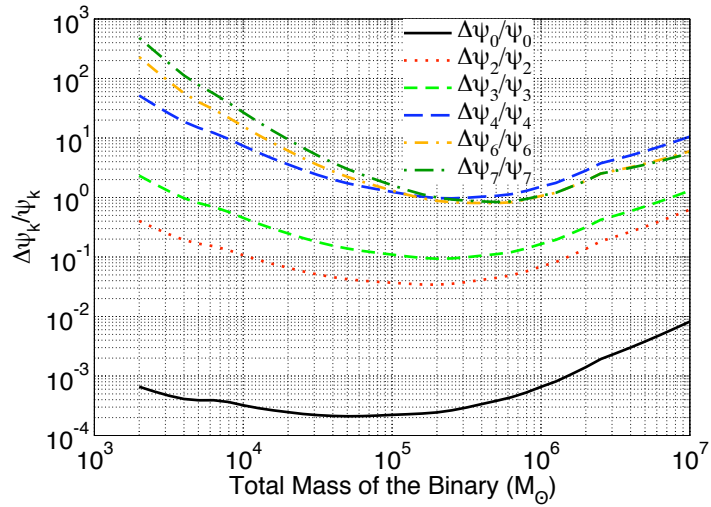}
\caption{Errors in the measurement of the PN coefficients $\psi_i$'s of the GW phasing formula (\ref{eq:phase_spa}) with LISA.
The coefficients $\psi_{5l}$ and $\psi_{6l}$ of \eq{eq:phase_spa} are here absorbed in $\psi_{5}$ and $\psi_{6}$, respectively. 
LISA is modeled as a single \ifo\ with no amplitude modulation (see text for explanation). 
The sources are optimally oriented supermassive black-hole binaries located at a luminosity distance of $D_L=3$ Gpc. 
Figure taken from ref.~\cite{Arun:2006yw}. }
\label{fig:LISA_err_full}
\end{center}
\end{figure*}
In order to study the feasibility of the tests, one can consider GW events in such a way as to minimize the errors in measuring the PN coefficients, which results from a compromise among many quantities: SNR, binary total mass and luminosity distance of the source from the experiment. 
Moreover, for the purpose of estimating phasing coefficients, one can model LISA as a single interferometer and neglect the orbital motion of its spacecrafts. 
These assumptions are justified by the fact that the angles describing the position of the source and the orientation of LISA are parameters that modify the amplitude of the signal and not its phase: for this reason, there is no correlation among the angles and the physical parameters contained in the phase.
Within this context, ref.~\cite{Arun:2006yw} obtains the errors reported in \fig{fig:LISA_err_full}, where the coefficients $\psi_{5l}$ and $\psi_{6l}$ are absorbed in $\psi_{5}$ and $\psi_{6}$, respectively.
Looking at the figure, one can see that the best candidate source for a test with LISA is a black hole binary with total mass of $M = 2\times 10^{5.1} \msun$: this is indeed the system chosen in ref.~\cite{Arun:2006yw}.
The resulting test is represented in \fig{fig:m1m2_psi_full} where the consistency among the parameters, comprising of their 1$\sigma$ error bands, is evident: the curves $m_2^j (m_1,\, \psi_j\pm \Delta \psi_j)$ have indeed a non-vanishing intersection in the point $M_1 = M_2 = 1\times 10^{5.1} \msun$. 

With one-year observations of coalescences in the mass range $(2 \times 10^4$\,--\,$2 \times 10^7) \msun$, ref.~\cite{Arun:2006yw} concluded that LISA tests could  distinguish massive gravity theories from GR at the 1PN level, provided the graviton Compton wavelength 
$\lambda_g$ is not larger than $(5.5 \times 10^{14}$\,--\,$3.8 \times 10^{15})$km, 
i.e. $(2 \times 10^{1}$\,--\,$1 \times 10^{2})$pc. 
It is interesting to convert this bound in a constrain on the graviton mass. 
In units $\hbar=c=1$ one has $1=\hbar c\simeq200$~MeV\,fm$^{-1}$, where fm=fermi=$10^{-15}$m: with a conversion factor of 1m=($2\times10^{-7}$ev)$^{-1}$, the highest bound on 
$\lambda_g$ gives $m_g\,\lsim\,10^{-25}$eV.

\begin{figure}[t]
\begin{center}
\includegraphics[width=0.6\textwidth]{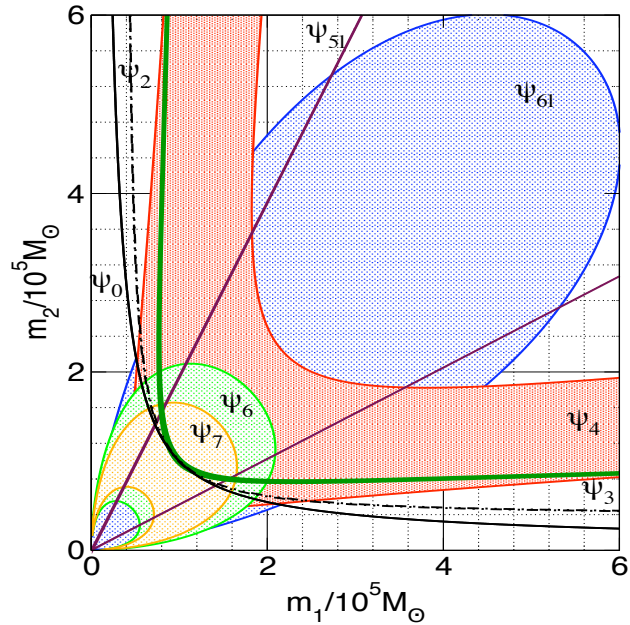}
\caption{
Plot showing the regions in the $M_1$-$M_2$ plane that correspond to 1$\sigma$ uncertainties in the PN parameters $\psi$'s. 
The source is a $2\times 10^{5.1} \msun$ black hole binary at luminosity distance of $D_L=1$Gpc: it is imagined to be observed for a year by LISA as described by a single interferometer with no amplitude modulation (see text for explanation).
Figure taken from ref.~\cite{Sathyaprakash:2009p197}. }
\label{fig:m1m2_psi_full}
\end{center}
\end{figure}

As we have seen, a test where one tries to measure the PN coefficients all at the same time is only possible for a very specific choice of the constituents masses and for very high SNR. 
In reality, the parameters are correlated among them: this covariance enhances the errors in parameter estimation and dilutes the accessible information.

A more effective test is the one proposed in ref.~\cite{Arun:2006hn}, where the consistency check is run on one coefficient at a time. 
The coefficients $\psi_0$ and $\psi_2$ play the role of a {\it base} with respect to which one expresses all the other parameters except the one on which the test is realized, which is taken as independent too.
The choice of $\psi_0$ and $\psi_2$ as a base is motivated by the fact that, being the least suppressed in the PN expansion, these parameters are the best measured; moreover they are the only ones which will not be modified by spin when one decides to introduce this degree of freedom in the treatment too.
Therefore, test parameter $\psi_T$ is chosen in the set $\psi_T=\psi_3, \psi_4, \psi_{5l}, \psi_6, \psi_{6l}, \psi_7$ and the GW signal is fitted with a template which depends on $\psi_0$, $\psi_2$ and $\psi_T$.

From \fig{fig:m1m2_psi_set_1} one can see that the choice of a restricted parameter set improves the accuracy in estimating the coefficients: all the regions spanned by the parameters in the $m_1$-$m_2$ plane are smaller than those of \fig{fig:m1m2_psi_full}, most notably for the higher PN order coefficients.
It is remarkable that, also in the present case, the coefficient $\psi_4$ is the worst determined: even the highest PN orders perform much better. 
This is mainly due to the structure of the phasing function (\ref{eq:phase_spa}), where a PN coefficient of order $k$ brings a correction $\propto f^{k/3}$ to the overall factor $f^{-5/3}$.
This is the reason why $\psi_5$ could be discarded from the very start of the study. Because of the phasing structure, terms which are close to $\psi_5$ in the expansion will be slowly varying with frequency. 
However, $\psi_6$ multiplies the term $f^{+1/3}$, while $\psi_4$ multiplies the term $f^{-1/3}$, which is a decreasing function of the frequency: therefore, the contribution of $\psi_4$ will be close to $f^0$ and then very much affected by the covariance with $\psi_5$. 
\begin{figure}[t!]
\begin{center}
\includegraphics[width=0.6\textwidth]{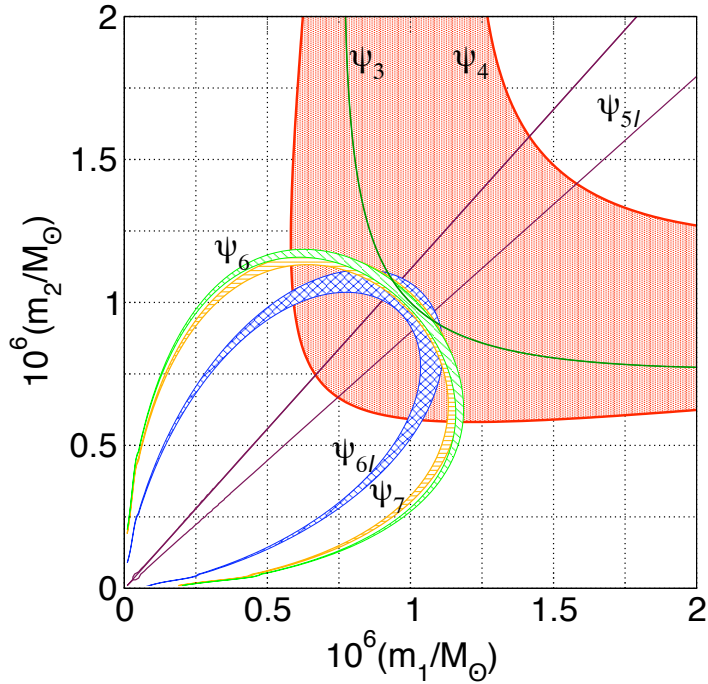}
\caption{
Plot showing the regions in the $m_1$-$m_2$ plane that correspond to 1$\sigma$ uncertainties in a test parameter chosen one at a time form the set $\psi_T=\psi_3, \psi_4, \psi_{5l}, \psi_6, \psi_{6l}, \psi_7$. 
The source is a supermassive black hole binary with masses $(10^6,\,10^6) \msun$, located at a redshift of $z=1$; it is supposed that the source is observed for a year by a single \ifo\ of LISA type (see text for explanation). 
Note that the 1$\sigma$ uncertainty in $\psi_3$ is smaller than the thickness of the line. 
Comparison with \fig{fig:m1m2_psi_full} shows that using a single test parameter at a time enhances the estimation precision. 
%The test is realized on one parameter at a time.
Figure taken from ref.~\cite{Sathyaprakash:2009p197}. }
\label{fig:m1m2_psi_set_1}
\end{center}
\end{figure}

The results presented so far refer to tests conducted with a possible GW signal detected by the space \ifo\ LISA.
Ref.~\cite{Arun:2006hn} considered also tests realized using ground-based \ifos, of second and third generation, like Advanced LIGO~\cite{ALIGO} and the Einstein Telescope (ET)~\cite{ET}. 
%~\footnote{At the time of ref.~\cite{Arun:2006hn} the possible third generation \ifo\, had been tentatively named EGO, after the European Gravitational Observatory consortium. 
%Since then, the project has entered a design study and nowadays the name for the future detector is Einstein Telescope. 
%We then implement the new nomenclature to refer to EGO too.}
The main conclusion is that already a second generation detector can provide tests which are sufficiently accurate to start probing GR through GWs.
The existence of such a potential has been confirmed by a very recent study~\cite{Mishra:2010tp}, which extended the investigation to the use of so-called {\it full} waveforms (FWF) as opposed to the restricted ones (RWF).
% :depending on the systems investigated and on the \ifo, the use of amplitude corrections allows one to measure most parameters with accuracies better than 10\%.
Investigating the mass range $(11-110)\msun$ in case of Advanced LIGO, it is found that one could measure $\psi_3$ with a fractional accuracy better than 6\% and $\psi_{5l}$ with a fractional accuracy better than 23\%, corresponding to an improvement varying from 3 to 100 with respect to the use of RWF.
Most importantly, ref.~\cite{Mishra:2010tp} analyzes in detail the efficiency of the test realized with the reduced set of coefficients as compared to the full set. 
In the latter case, if, for example, the 1.5PN coefficient did not overlap with the regions allowed by the other parameters, one could directly ascribe this to a failure of GR at 1.5PN order. 
This correspondence could not be so evident when using the reduced set of coefficients. 
Let us stick with $\psi_3$ as the testing parameter and suppose, for example, that the correct theory is Chern-Simons gravity~\cite{Chern-Simons}, which starts deviating from GR at 2PN order. 
The template used for fitting the signal will have three independent parameters: $\psi_0, \psi_2$ and $\psi_3$, meaning that $\psi_0$ and $\psi_2$ 
are parametrizing {\it all}  the coefficients of order higher than $k=2$, except for $\psi_3$. 
Employing such a parametrization in presence of the deviation implies that one is describing non-GR effects by means of GR terms: this can affect the estimation of the test parameter $\psi_3$, even if the 1.5 PN order is not the one from which the deviation actually starts.
With these caveats, it is possible to look at the consistency checks of \fig{fig:m1m2_22Msun}, which refers to the third generation \ifo\ ET: from the various plots one can see that each test parameter investigated consistently overlaps with the basis pair~$(\psi_0,\psi_2)$.
\begin{figure*}[t]
\begin{center}
\includegraphics[width=1\textwidth,angle=0]
{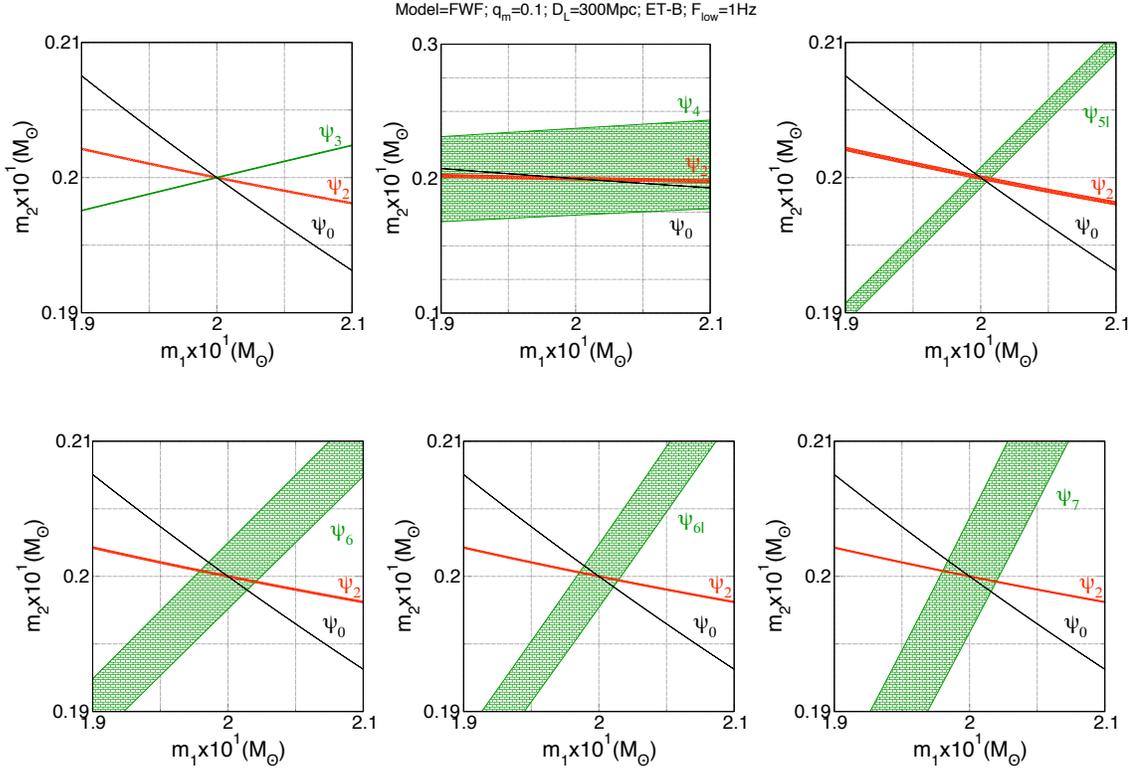}
\caption{
Plots showing the regions in the $m_1$-$m_2$ plane that correspond to 1$\sigma$ uncertainties in the basis parameters $\psi_0$, $\psi_2$ and in the test parameter chosen one at a time form the set $\psi_T=\psi_3, \psi_4, \psi_{5l}, \psi_{6}, \psi_{6l}, \psi_7$. 
The source is a black-hole binary with masses $(2,20)\msun$, located at a luminosity distance of $D_L=300$ Mpc; it is supposed that the source is observed by the Einstein Telescope (ET), a third-generation ground based \ifo. 
The basis parameters $\psi_0$ and $\psi_2$ are those from which one can measure the masses of the two black holes. 
Each test parameter, together with its error, spans a given region in the $m_1$-$m_2$-plane: if GR is the correct theory of gravity, then the three parameters $\psi_0$, $\psi_2$ and $\psi_T$ should have a non-empty intersection in the $m_1$-$m_2$ plane. 
A smaller region leads to a stronger test.
Notice that all panels have the same scaling except the top middle panel in which Y axis has been scaled by a factor of~10. 
With respect to \fig{fig:m1m2_psi_set_1}, the waveforms used for the tests presented here comprise amplitude corrections. 
Figure taken from ref.~\cite{Mishra:2010tp}. } 
\label{fig:m1m2_22Msun}
\end{center}
\end{figure*}

\smallskip

Before ending this section, it should be mentioned that very recently, another phenomenological framework has been proposed to discuss deviations of the GW phasing formula from the GR prediction. 
In ref.~\cite{Yunes_PPE1}, Yunes and Pretorius have put forward what they call the {\it parametrized post-Einstein} (PPE) approach, where they parallel the PPN spirit: while in the latter one parametrizes the metric in a model-indipendent way, in the former one parametrizes the full waveform, i.e. both the phase and the amplitude.  
This is achieved using the existing knowledge about various alternative theories of gravity such as Brans-Dicke~\cite{Brans:1961sx}, massive graviton effects~\cite{Will:1997bb} and Chern-Simons theory~\cite{Alexander:2009p224} to write down a \textit{generic} parametrized waveform in the Fourier domain. 
Apart from extending the use of parameters to the amplitude too, another difference with refs.~\cite{Arun:2006yw,Arun:2006hn,Mishra:2010tp} is that the PPE also considered the contributions from the merger and ringdown phases of the binary evolution beyond the inspiral. 
In both approaches spin degrees of freedom have not been included for the moment.

It is in this active domain of research that the parametrized framework presented in this thesis sets itself; in Chapters~\ref{chap:Group} and \ref{chap:Berti} I will discuss a field-theoretical extension of the PPN that stems from the EFT approach to inspiraling binaries: as we will see,  this is another convenient attitude to build parametrized waveforms for GW astronomy.

%% file: Chap_EFT/Chap_EFT.tex
%%%%%%%%%%%%%%%%%%%%%%%%%%%%%%%%%%
\chapter{The effective field theory approach to inspiraling compact binaries} 
\label{chap:EFT}
%%%%%%%%%%%%%%%%%%%%%%%%%%%%%%%%%%

%%%%%%%%%%%%%%%%%%%%%%%%%%%%%%%%%%%%
\section{An invitation}
%%%%%%%%%%%%%%%%%%%%%%%%%%%%%%%%%%%%

During the inspiral phase of a binary system the constituents are 
separated enough for their dynamics to be non-relativistic, i.e. the 
leading order interaction between the two bodies is Newtonian gravity 
and corrections to it can be organized in the PN approximation of GR.

Being perturbative in nature, the PN expansion can also be interpreted 
from a field theory standpoint and reformulated in terms of Feynman 
diagrams.  
As I have discussed in \Sec{sec:Dam_Far}, a first application of these instruments to the two-body problem in classical gravity dates back to 1960~\cite{BP};  more recently ref.~\cite{Damour:1992we} by DEF has started a series of works in which the PN limit of scalar-tensor theories of gravity is conveniently studied by means of diagrammatics.
In the last few years the use of these instruments in classical gravity has revived thanks to the introduction of an EFT approach~\cite{NRGR_paper}. 
In this context the non-relativistic dynamics of the two-body system is studied in a way analogous to bound states in QED and QCD~\cite{Caswell:1985ui}: for such a reason the EFT framework of ref.~\cite{NRGR_paper} has been dubbed non relativistic General Relativity (NRGR). 

The use of an EFT framework in the two-body problem of GR is 
motivated by the fact that the binary dynamics is a typical situation 
in which the disparity of physical scales allows one to separate the relevant 
degrees of freedom in subsets with decoupled dynamics.
For example, because the extension of the radiating source is much 
smaller than the wavelength of gravitational perturbations, the modes 
corresponding to the internal dynamics of the compact object can be 
{\it integrated out} and the source is {\it effectively} described as a 
point particle.

At high enough order the point particle approximation breaks down 
and causes  divergences to appear. 
However this happens in any ﬁeld theory coupled to point sources and only 
means that the theory needs to be supplemented by a more complete 
(fundamental) model. 
Quoting ref.~\cite{NRGR_paper}, the NRGR attitude towards the 
use of the point particle description and the related problem of 
divergences can be summarized as follows: 
"Rather than trying to resolve the point particle singularities by using 
a specific model of the short distance physics, in an EFT framework 
we systematically parametrize our ignorance of this structure by 
including in the effective point particle Lagrangian the most general 
set of operators consistent with the symmetries of the well understood, 
long wavelength physics. 
In our case, this long wavelength physics is GR and 
the symmetry that constrains the dynamics is just general coordinate 
invariance.".
This parametrization is a {\it model-independent} description of the 
compact source in terms of a point particle.
In fact, the EFT is "tuned" to reproduce the low energy regime of the full theory by means of the so-called {\it matching}, a procedure which consists of two stages. 
In the first, one calculates a convenient observable in the full theory and expands the result in the low energy limit. 
This form of the result is then compared with the prediction of the EFT which depends on the coefficients of the effective operators: by equating the two results one can then fix these coefficients.

\begin{figure}[t]
\centering 
\includegraphics[width=0.9\textwidth]{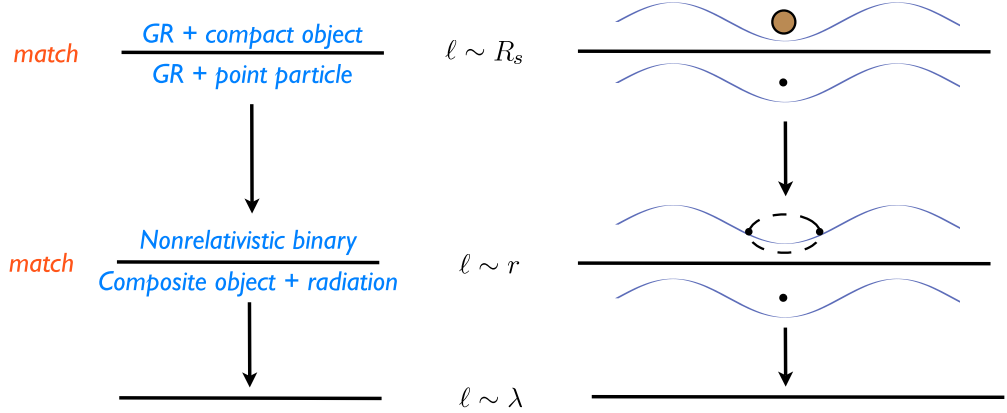}
\caption{Graphical representation of the step-like construction of 
a tower of EFT theories for the two-body problem in GR. 
From top to bottom, the characteristic values of the length scale ${\ell}$ 
are: the size of the compact object $R_S$, the orbital radius $r$ and 
the radiation wavelength $\lambda$.
Figure courtesy of Chad Galley.}
\label{fig:eft_tower}
%\end{center}
\end{figure}

Making use of the action for gravity as dictated by GR and the sum of two point particle actions is an adequate description of the two-body system as far as the objects are isolated, 
i.e. until the focus is on length scales smaller than the orbital radius. 
At this scale another EFT has to be built to describe 
the coupling of radiation with a composite, two-body source.
This step-like approach is conveniently represented by 
Fig.~\ref{fig:eft_tower} 
which graphically describes the construction of a tower of EFT theories. 
It should be stressed that these theories are built one at a time and 
one for each of the characteristic values of the length scale ${\ell}$, 
which, with reference to the figure, are: 
the size of the compact object $R_S$, the orbital radius $r$ and 
the radiation wavelength $\lambda$.
It is this separation of the intricacies of the binary problem that makes NRGR an ideal framework to study gravitational radiation: 
NRGR is effective in the sense that it is optimized to streamline the perturbative calculations of the PN regime.
Moreover, NRGR makes a systematic use of Feynman diagrams, which can be dealt with by automatic calculation on a computer. 
By means of these vantage points, it has been possible to obtain previously un-computed spin contributions to the source multipoles at next-to-leading order~\cite{Porto:2010zg}: 
this provides the last missing ingredient required to determine the phase evolution to 3PN accuracy including all spin effects.

Besides being a convenient tool for calculations, Feynman diagrams can be used to recast the non-linearities of GR in a field-theoretical language: by means of such a reformulation, it is possible to have a different point of view on the physical information that can be extracted from experiments of relativistic gravity. 
This is the standpoint which is taken in the present thesis and it is exposed in
Chapters \ref{chap:Group} and \ref{chap:Berti}: there I show how an EFT approach allows one to investigate the extent to which experiments can constrain the values of the non-Abelian vertices of GR.
This represents a specific field-theoretical interpretation of observations but has a general relevance because the analysis is model-indipendent. 
In this chapter I set the stage for these investigations by giving an introduction to NRGR based on refs.~\cite{NRGR_paper, NRGR_lectures}; along the treatment, I find it convenient to reproduce some formulae of ref.~\cite{NRGR_lectures} as they stand, notably those where Feynman diagrams are drawn inside equations thanks to the {\it feyn} package for \LaTeX.
With respect to the field-theoretical investigations of DEF that I have sketched in \Sec{sec:Dam_Far}, here I content myself with GR and I do not take extra fields into account: in such a way, I can keep the discussion self-contained and describe the original spirit of the approach.
An example of the application of the EFT to the case of one extra scalar 
field is the subject of \chap{chap:Ric}, where I report a study of the 
renormalization of the energy-momentum tensor for point-like and string-like sources.

%%%%%%%%%%%%%%%%%%%%%%%%%%%%%%%%%%
\section{Effective field theories} 
\label{sec:efts}
%%%%%%%%%%%%%%%%%%%%%%%%%%%%%%%%%%

\begin{figure}[t]
\begin{center}
\includegraphics[width=9cm]{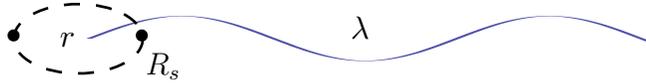} 
\caption{A schematic description of an inspiraling compact binary where the relevant length scales are shown. The disparity of these scales makes EFT techniques very well suited to describe binary systems. 
Figure courtesy of Chad Galley.}
\label{fig:bin_scales}
\end{center}
\end{figure}

The EFT treatment of binary systems pioneered in ref.~\cite{NRGR_paper} is motivated by the presence of three length scales in the problem (see \fig{fig:bin_scales}):
\bees
\label{eq:scales}
&R_S& = \q \mbox{size of compact objects}, \nn \\
&r& = \q\mbox{orbital radius}, \nn \\
&\lambda &= \q\mbox{wavelength of emitted radiation} 
\ees
which turn out to be interconnected by means of the relative velocity~$v$. 
For a gravitationally bound system, in fact, one can use the virial theorem 
to obtain the estimate  
\be
\label{eq:virial}
{2 G_N m \over r} = {R_S\over r} \sim v^2 
\ee
where we have introduced the \Sch radius $R_S \equiv 2 G_N m$; 
moreover, because the binary's motion is circular 
with frequency $\omega_o$, one can write 
\be
v = \omega_o r \sim {r\over\lambda}
\ee
in which use has been made of the fact that the GW frequency $\omega_{GW}$ is proportional to the orbital one, $\omega_{GW} \simeq \omega_o$.

Because of the link (\ref{eq:virial}) between the velocity of the sources and their gravitatio\-nal-field strengths, 
the length scales in (\ref{eq:scales}) are indeed connected one another through the PN expansion parameter~$v$: this shows that high PN orders encode physics from different scales at the same time.
Through this link it is also evident that the scales respect the following 
hierarchy: $R_S << r << \lambda$. 
This type of problems can be conveniently dealt with by means of 
an EFT where the physics which is relevant at a definite scale is 
treated separately and "once for all".

In particle physics, EFT schemes are typically used to study the impact of some ultra-violet (UV) scale $\Lambda$ on the physics at low energy characterized by the infrared (IR) scale~$\theta$, with $\theta<<\Lambda$. 
For the sake of concreteness, let us consider a theory described by the action functional ${\cal S}[\phi, \Phi]$, with $\phi$ collectively denoting light (or massless) degrees of freedom and $\Phi$ standing for some heavy fields with masses $M_\Phi\sim \Lambda$\,.
If the focus of the investigation is an energy regime $E\sim\theta$, it is not necessary to resort to the full theory ${\cal S}[\phi, \Phi]$: because the fields $\Phi$ are too heavy to be excited at energies $E\sim\theta$, only the fields $\phi$ will have a non-trivial dynamics; therefore, it is convenient to solve for the fields~$\Phi$ and plug the solution back in the action (cfr. the Fokker action in~\Sec{sec:Dam_Far}).
Formally, this is achieved by {\it integrating out} the heavy modes through the path integral 
\be
\label{eq:path}
e^{i S_{\mathit{eff}}[\phi]} = \int D\Phi(x) \, e^{i S[\phi,\Phi]} 
\ee 
and results in an {\it effective} action of the type
\be
\label{eq:local}
S_{\mathit{eff}}[\phi]= \sum_i c_i  \int  d^4 x \,{\cal O}_i(x) \,,
\ee
where the (local) operators ${\cal O}_i(x)$ are, in principle, infinitely many and the coefficients of the expansion $c_i$, called Wilson coefficients, are energy-dependent.
This effective action does not represent a "new" theory but rather approximates the full one in some regime; as a consequence the operators ${\cal O}_i(x)$ must respect the original symmetries.
For practical purposes, only a finite number of such operators contribute to \eq{eq:local}. 
The effective action is supposed to reproduce the predictions of the full theory for the low energy dynamics up to some corrections, i.e. \eq{eq:local} is a perturbative series in a small parameter $\eps=\theta/\Lambda$.
This is related to the fact that, because every operator ${\cal O}_i$ has some mass dimensions $\Delta_i$, the value of its Wilson coefficient at an energy scale $\mu=\Lambda$ can be shown to be~\footnote{Remember we are using units in which $c=\hbar=1$ so the scalings are naturally expressed in terms of mass dimensions.}
\be
\label{eq:cmu}
c_i(\mu=\Lambda)= {\a_i\over \Lambda^{\Delta_i-4}} \;,
\ee  
where $\a_i\sim{\cal O}(1)$: therefore, the higher the mass dimension $\Delta_i$ is with respect to 4, the lower the contribution of the corresponding operator is to the effective action; operators with $\Delta_i \leq 4$ have their coefficients {\it renormalized}, i.e. redefined in order to take the effect of the cutoff into account.
As a consequence, the impact of the high-energy physics on the IR dynamics is very simple: all the UV dependence appears directly in the coefficients of the effective action, which makes it possible to determine how a low energy observable depends on~$\Lambda$. 
More specifically, this {\it decoupling} between the two scales allows one to determine which operators are needed at a given order in the perturbative expansion by using a generalized version of dimensional analysis called {\it power counting}. 
Let us see how this works in practice. 
Because $\phi$ accounts for the low energy dynamics of the theory, its four-momentum is of the order of the IR scale, i.e. $k^\mu\sim \theta$: the same scaling will characterize derivatives $\pa_\mu\sim \theta$, while for coordinates one has $x^\mu\sim \theta ^{-1}$. 
With these rules, one can power-count every term in the action~(\ref{eq:local}), notably the scalings of different operators ${\cal O}_i$ with respect to each other. 
The leading order term in the perturbative series~(\ref{eq:local}) is the kinetic term, which scales as 
\be
\label{eq:KinTerm}
\int d^4 x (\pa_\mu \phi)^2 \sim \( \frac{\theta}{\Lambda} \)^0 \,,
\ee
where the overall mass dimension is zero because the action is a scalar under Lorentz transformations.
Eq.~(\ref{eq:KinTerm}) tells us that the scaling of the field itself is $\phi\sim\theta$.
Therefore, an interaction term like $c_8(\pa_\mu \phi \, \pa^\mu \phi)^2$ will be power-counted as $(\theta/\Lambda)^8$ and, when integrated to form an action, it will give a contribution which is suppressed with respect to the leading order term~(\ref{eq:KinTerm}) by $(\theta/\Lambda)^4$.
Power-counting terms in the action enables one to determine which operators are needed to compute physical observables at a given perturbative order.

\begin{figure}[t]
\begin{center}
\includegraphics[width=9cm]{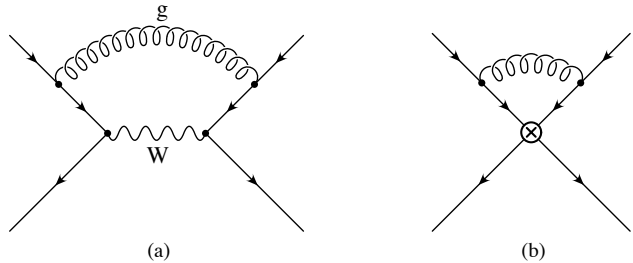} 
\caption{QCD correction to a typical four-Fermi process in the Standard 
Model, calculated in (a) the full theory with propagating $W$ bosons 
and (b) the effective Fermi theory of weak interactions. 
Graph (b) reproduces graph (a) up to corrections suppressed by powers 
of $E^2/m^2_W\ll 1$. 
Figure taken from ref.~\cite{NRGR_lectures}. \label{fig:Fermi} }
\end{center}
\end{figure}

Building an effective action is convenient in two typical situations, depending on whether the full theory is known or not. 
An example of the first case is represented by the electroweak interactions: if one is interested in energies $E\ll m_{W,Z}$, it makes sense to integrate out the $W,Z$ bosons because they are too heavy to have a dynamics. 
The result is Fermi theory of weak interactions, which makes it simpler to calculate amplitudes like the one depicted in~\fig{fig:Fermi}, where the weak process receives a QCD correction~\cite{Buras:1998raa}. 
Even when the full theory is not completely known the decoupling allows one to treat the low energy dynamics. 
The structure of the effective action is dictated by the symmetries that survive at low energies: by writing down the most general set of operators consistent with these  symmetries, one is accounting for the UV physics in a completely \emph{model independent} way.    
For instance, the Standard Model of particle physics is believed to be an EFT below a scale $\Lambda \simeq 1\,\mbox{TeV}$. 
The indirect manifestation of this scale has already been constrained 
with the LEP experiment at CERN and with the SLD experiment at 
SLAC~\cite{Han:2004az}, while the direct probe of the TeV scale 
is awaited from the LHC at CERN.

\begin{figure}[t]
	\centering 
	\subfigure[]{
		\includegraphics[scale=.50]{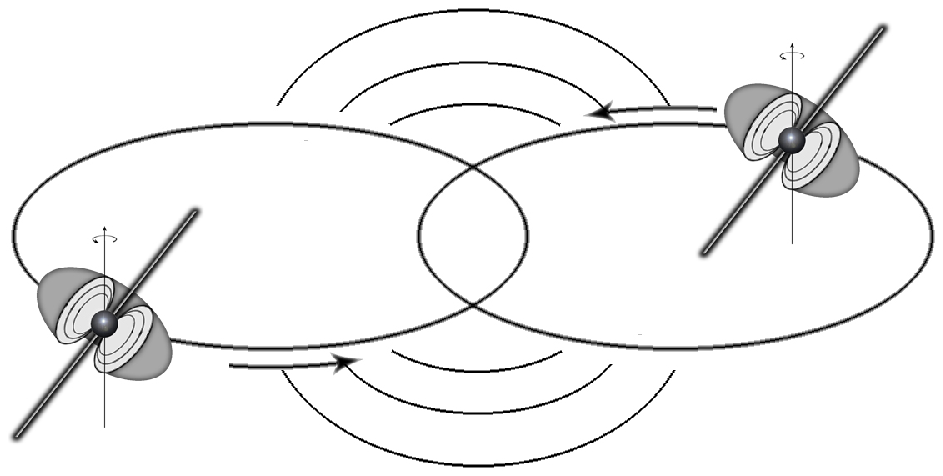}
%		\label{fig:} 
} 
	\subfigure[]{
		\includegraphics[scale=.45]{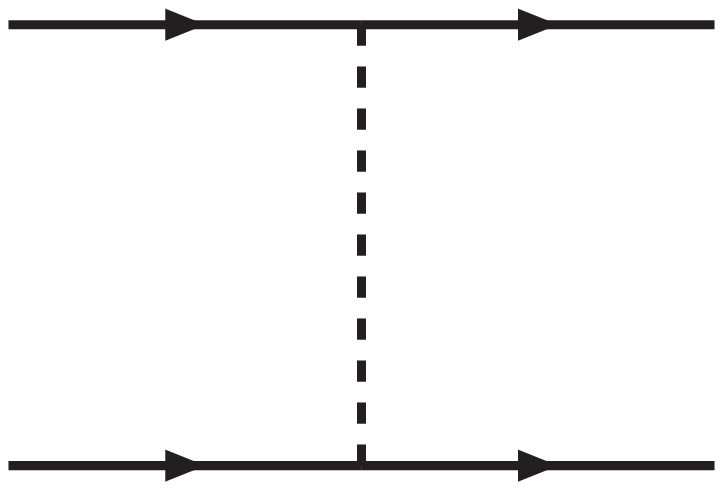}
%		\label{fig:} 
} 
	\caption[Optional caption for list of figures]{In analogy with \fig{fig:Fermi}, 
	this picture is a very schematic 	representation of how non-relativistic 
	compact binaries are treated in an effective field theory, where  
	the perturbative PN expansion is expressed in terms of Feynman diagrams. 
	Panel (a) represents the orbital motion of a binary pulsar and the 
	gravitational radiation emitted by the system, panel~(b) is the 
	Feynman diagram which accounts for the lowest order interaction between 
	the binary constituents, i.e. the Newtonian potential.}
	\label{fig:pert_insp}
\end{figure}

A final case where an EFT approach is useful is represented by a theory that is known but strongly interacting. 
The example that mainly concerns the present thesis is GR: 
this can be regarded as an effective theory of quantum gravity with predictive power below the strong coup\-ling scale $\mpl\sim 10^{19}\,$GeV. 
The EFT interpretation of GR is reviewed in detail in refs.~\cite{Donoghue:1995cz,Burgess:2003jk,Burgess:2007pt} and allows to address some issues that emerge within a quantum field theory reformulation of Einstein's theory.
For example, in an EFT framework, the fact that GR is non-renormalizable does not constitute a problem: the UV divergences that start arising at first quantum order are dealt with by renormalizing the parameters of higher derivative terms in the action; perturbatively, only a finite number of terms is required for each order.
The specific EFT of ref.~\cite{NRGR_paper} is designed to study the classical dynamics of binary systems so that, as we will see, NRGR does not need to care about quantum corrections. 
From the following sections, it will be clear that even in a {\it classical} context, it is useful to resort to {\it quantum} field theory tools like Feynman diagrams: in analogy with \fig{fig:Fermi}, the NRGR description of non-relativistic compact binaries can be schematically described as in \fig{fig:pert_insp}. 

The considerations spelled out so far are not exhaustive of EFT features but they suffice to move to the discussion of the two-body problem: we will come back to the EFT ingredients directly when they are at play.

%

%%%%%%%%%%%%%%%%%%%%%%%%%%%%%%%%%%
\section{Non-Relativistic General Relativity} 
\label{sec:NRGR}
%%%%%%%%%%%%%%%%%%%%%%%%%%%%%%%%%%

In the dynamics of a binary system there are three separated length scales (\ref{eq:scales}), so the EFT treatment will make use of the path integral (\ref{eq:path}) in two stages: the first for integrating out the physics pertaining the internal structure of the compact object, the second for that of the orbital dynamics. 
This is what we will see explicitly in the following, starting 
with the scale represented by the \Sch radius $R_S$.

% % % % % % % % % % % % % % % % % % % % % % % % % % 
\subsection{The effective theory for isolated compact objects} 
\label{sec:eft_iso}
% % % % % % % % % % % % % % % % % % % % % % % % % %

The first length scale that has to be integrated out is the size of the compact object.
According to the EFT philosophy, one can build an effective action by identifying  the degrees of freedom which are relevant at the scale of interest: including the most general operators consistent with the symmetries provides a model-independent treatment of a system.
To describe GW observables, the relevant degrees of freedom are the long wavelength gravitational perturbations and the world-lines of effective point particles.
In a first approximation one can neglect the spins of the objects: as we saw in \eq{eq:motion}, spin effects in the equations of motion do not show up until 2PN order.

\begin{figure*}[t]
\begin{center}
\def\size{7cm}
\includegraphics[width=7cm]{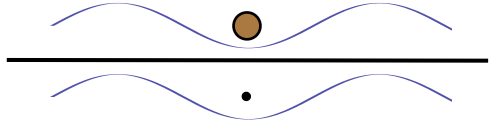} 
\caption{A schematic representation of what the EFT for isolated compact objects is about: the extended astrophysical object can be effectively regarded as a point-particle probe of the long-wavelength gravitational perturbations.
Figure courtesy of Chad Galley.} 
\label{fig:eft_iso} 
\end{center} 
\end{figure*}

The symmetries that govern the dynamics of these degrees of freedom are
\begin{enumerate}
\item general coordinate invariance, $x^\mu\rightarrow x^{\bar\mu}(x)$;
\item worldline reparametrization invariance (RPI), $\s\rightarrow {\bar\s}(\s)$;
\item invariance under $SO(3)$ transformations, which guarantees that the compact object is perfectly spherical and that it has no permanent moments relative to its own rest frame.
\end{enumerate}
Such EFT is the appropriate one for describing non-rotating (Schwarzschild) 
BHs interacting with external gravitational fields.   
The effective action consistent with these criteria is then
\begin{equation}
\label{eq:actions}
S_{eff}[x^\mu, g_{\mu\nu}] = S_{EH}[g]+S_{pp}[x,g],
\end{equation}
where the action for gravity is the usual Einstein-Hilbert action~\footnote{With respect to \Sec{sec:Dam_Far}, here we adopt the "mostly minus" convention for the metric $\eta_{\mu\nu}$=\text{diag}(+,--,--,--).} 
\begin{equation}
\label{eq:EH}
S_{EH}=- 2 \mpl^2 \int d^4 x \sqrt{g} R(x),
\end{equation}
with $\mpl^{-2}=32\pi G_N$ and $R(x)$ the Ricci scalar, 
and the point particle term is the usual minimal coupling 
\begin{equation}
\label{eq:pp}
S_{pp}= -m\int d\tau \equiv -m\int [g_{\mu\nu}dx^\mu dx^\nu]^{1/2}\,,
\end{equation}  
where $m$ is the particle mass. 
Both actions can be supplemented with terms containing higher order powers of the curvature.
For the gravitational action (\ref{eq:EH}) these terms are suppressed by powers of $M_{Pl}$, the scale where quantum gravity is expected: these contributions  are completely negligible in the classical regime of binary dynamics. 
Concerning the point particle action (\ref{eq:pp}) the situation is different. 
Higher order operators that contain the Ricci tensor $R_{\mu\nu}$ turn out to be redundant because they can be removed by a field redefinition (see the appendix of ref.~\cite{NRGR_lectures} for details). 
A second possibility is to have higher powers of the Riemann tensor $R_{\mu\nu\alpha\beta}$, notably through its components of electric and magnetic type parity
$E_{\mu\nu}$ and $B_{\mu\nu}$, respectively, which are given by
\begin{eqnarray}
\label{eq:EB}
E_{\mu\nu} &=&  R_{\mu\alpha\nu\beta} {\dot x}^\alpha {\dot x}^\beta,\\
B_{\mu\nu} &=& \eps_{\mu\alpha\beta\rho} {R^{\alpha\beta}}_{\nu \sigma} 
{\dot x}^\rho {\dot x}^\sigma \nn \,.
\end{eqnarray}
Therefore, one can re-write the point particle action (\ref{eq:pp}) as 
\begin{equation}
\label{eq:finsize}
S_{pp}=-m\int d\tau + c_E\int d\tau E_{\mu\nu} E^{\mu\nu} + c_B\int d\tau B_{\mu\nu} B^{\mu\nu}+\cdots \;:
\end{equation}
The operators constructed from $E_{\mu\nu}$ and $ B_{\mu\nu}$ are the first in an infinite series of terms that systematically encode finite-size effects, which are due to the physical extension of the source, in our case a black hole.  
One way to see this is to compute the matching coefficients $c_{E,B}$: the explicit matching calculation~\cite{NRGR_paper,NRGR_lectures} shows that both $c_E$ and $c_B$ scale as $\mpl^2 R_S^5$, so they approach zero quite fast for a vanishing $R_S$.
Therefore, the operators built from $E_{\mu\nu}$ and $ B_{\mu\nu}$ do not contribute to the binary dynamics of \Sch black holes until 5PN: this gives a proof of the effacement principle of GR in the EFT context.
Another way of seeing how the non-minimal terms are associated to finite-size effects is to look at the equations of motion that derive from \eq{eq:finsize}; ahead of the calculation, one can see that the particle will not move on a geodesic of the background gravitational field: in fact, this is what one gets by extremizing the minimal action~(\ref{eq:pp}).
Indeed, geodesic deviation implies stretching by tidal forces, which occurs when one considers the motion of {\it extended} objects in a gravitational field.   
As a consequence, one can rely on the minimal point particle action~(\ref{eq:pp}) up to the very high 5PN order for what concerns the dynamics of two non-rotating black holes. 
Schematically, this is described by \fig{fig:eft_iso} which points at the first EFT to be constructed in the tower of theories of \fig{fig:eft_tower}.

To study the non-relativistic dynamics of a binary system, one can expand both the gravitational action (\ref{eq:EH}) and the point particle action (\ref{eq:pp})
by defining the gra\-vitational field as a small perturbation over flat space-time 
in the following way~\footnote{This choice is not the only one possible, cfr. \eq{met_nr} of \chap{chap:Ric}.}   
\begin{equation}
\label{eq:h}
g_{\mu\nu} \equiv \eta_{\mu\nu} + {h_{\mu\nu}\over M_{Pl}} \, .
\end{equation}
The perturbative expansion can be very conveniently dealt with by means of Feynman diagrams, i.e. by associating drawing signs to the various fields and interaction vertices. 
We have already met an example of a symbolic representation dictionary in \Sec{sec:Dam_Far}, where the rules for diagrammatic computations in the context of \sts\ have been reported in \fig{fig:Fok_Exp}.
In the case of Einstein gravity, the Feynman rules for perturbative calculations can be found in the classic lectures by Veltman~\cite{Veltman} and the more recent review by Donoghue~\cite{Donoghue:1995cz}. 
It is custom to assign curly or wavy lines to the gauge bosons of a theory, like photons and gravitons, and continuous lines to fermions in the matter sector. 
For NRGR one can stick to these conventions. 
However, as we will see, the use of continuous lines for matter sources should not be taken as meaning that these are quantum degrees of freedom. 
 
Using the expansion (\ref{eq:h}) for the purely gravitational sector~(\ref{eq:EH}) one has 
\begin{eqnarray} 
\label{eh_fd}
-2 \mpl^2 \int d^4 x \sqrt{g} R(x) &\rightarrow& \int d^4 x \left[(\partial h)^2 
+ {h (\partial h)^2\over \mpl} + {h^2 (\partial h)^2\over \mpl}
+\cdots\right] \\ \nn
&=& \qq
(\Diagram{g})^{-1}\,\,\,+ 
\Diagram{gd\\
& g\\
gu}\,\,\, +
\Diagram{gd & gu\\
gu & gd}\,\,\, +
\cdots \,,
\end{eqnarray}
where gravitational self-interactions are represented by Feynman vertices containing many graviton lines. 
The first term in the expansion (\ref{eh_fd}) represents free propagation of a graviton from a space-time point to another; in a compact fashion, this can be written as 
\begin{equation}
\label{eq:hhprop}
\langle h_{\mu\nu}(x) h_{\alpha\beta}(y) \rangle = 
D_F(x-y) P_{\mu\nu;\alpha\beta} \;,
\end{equation}
where the Feynman propagator $D_F(x-y)$ is given by
\begin{equation}
\label{eq:Fprop}
D_F(x-y)=\int {d^d k\over (2\pi)^d} {i\over k^2+i\epsilon} e^{-i k\cdot (x-y)} 
\end{equation}
and the tensorial structure reads
\begin{equation}
\label{eq:Pprop}
P_{\mu\nu;\alpha\beta}={1\over 2}\left[\eta_{\alpha\mu}\eta_{\beta\nu}+\eta_{\alpha\nu}\eta_{\beta\mu}-{\eta}_{\mu\nu}\eta_{\alpha\beta}\right]\,.
\end{equation}

For what concerns the interaction of the gravitational field with the point particle, plugging the definition (\ref{eq:h}) in \eq{eq:pp} gives
\begin{align}
\label{eq:pp_exp}
&-m \int \[\(\eta_{\mu\nu} + \frac{h_{\mu\nu}}{\mpl}\) dx^\mu dx^\nu\]^{1/2} \;=\;
-m\int d{\bar\tau}\sqrt{1+{h_{\mu\nu} {\dot x}^\mu {\dot x}^\nu\over \mpl}} \nn \\
& = -m\int d{\bar\tau} 
-{m\over 2 \mpl} \int d{\bar\tau} h_{\mu\nu} {\dot x}^\mu {\dot x}^\nu
- {m\over 8 \mpl^2}\int d{\bar\tau}(h_{\mu\nu} {\dot x}^\mu {\dot x}^\nu)^2 
+\cdots,
\end{align}
where $d{\bar\tau}^2=\eta_{\mu\nu} dx^\mu dx^\nu$ and ${\dot x}^\mu\equiv dx^\mu/d{\bar\tau}$.  
In order to draw the Feynman diagrams corresponding to these vertices it is important to remark that the matter sources to which gravity couples are macroscopic objects. 
As a consequence, the binary constituents are non-relativistic particles endowed with typical three-momentum of the order ${\bf p}\sim m v$, with $v$ the orbital velocity; on the other hand, gravitons will rather have three-momentum ${\bf k}\sim\hbar$\,, where $\hbar$ is the Planck constant and we use $c=1$.
When a compact object emits a single graviton, momentum is effectively \emph{not} conserved and the non-relativistic particle recoils of a fractional amount roughly given by 
$$
\frac{|\delta{\bf p}|}{|{\bf p}|}\simeq \frac{|\vk|}{|{\bf p}|}
\simeq \frac\hbar{L}\,,
$$  
where $L\sim mvr$ is the angular momentum of the system: it is clear that for macroscopic systems such quantity is utterly negligible.
To summarize, in NRGR one describes the massive constituents of binary systems as static, background sources of gravitons. 
In this context, the Feynman diagrams corresponding to the vertices of \eq{eq:pp_exp} are given by 
\begin{equation}
\label{pp_fd}
\Diagram{fvA \\
%g\\
fvA} \q +  \q % \,\,\,\,
\Diagram{fvA \\
g\\
fvA} \,\,\, +  \,\,\,\,
\Diagram {fvA\\
fvA}
\Diagram{gu\\
gd} \,\,\, + \,\,\,\, \cdots \,,
\end{equation}
where: 
\bd
\item the continuous lines represent matter sources and they are straight to indicate that there is no recoil in the emission of gravitons; 
\item the arrows on the continuous lines describe the flow of time of the particle world-line;
\item each curly line represents a factor $h_{\mu\nu}\,\dot x^\mu\dot x^\nu$ (see \eq{eq:pp_exp}). 
\ed
\begin{figure}[b]
\centering
\def\size{4cm}
\hbox{\vbox{\hbox to \size {\hfil \includegraphics[width=2.5cm]
{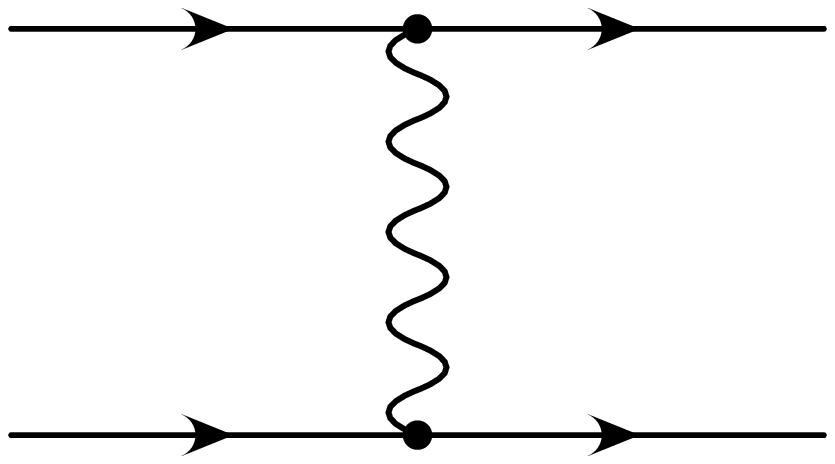} \hfil }\hbox to \size {\hfil(a)\hfil}}
\vbox{\hbox to \size {\hfil \includegraphics[width=2.5cm]
{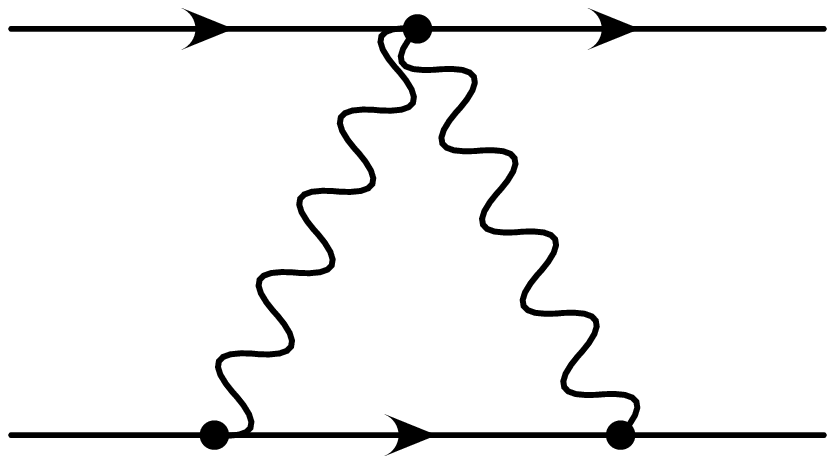} \hfil}\hbox to \size {\hfil(b)\hfil}}
\vbox{\hbox to \size {\hfil \includegraphics[width=2.5cm]
{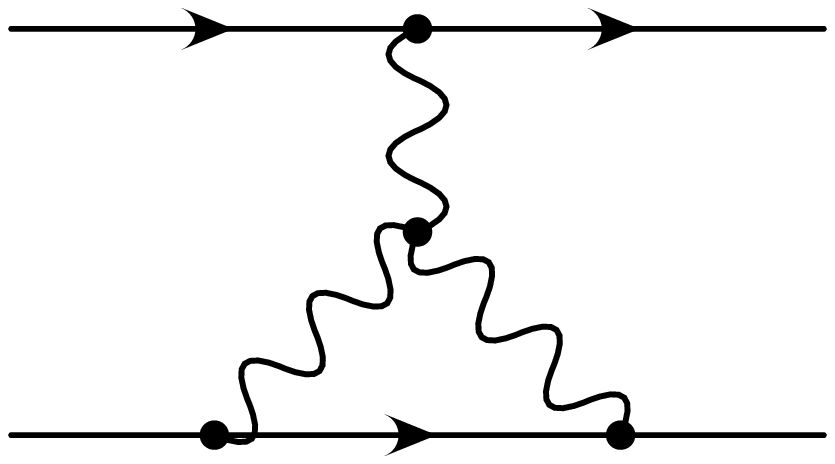} \hfil}\hbox to \size {\hfil(c)\hfil}}}
	\caption{The first few diagrams contributing to $S_{\mathit eff}(x_a)$ 
	in Lorentz covariant perturbation theory.
	Figure taken from ref.~\cite{NRGR_lectures}. 
  \label{fig:relexp} }
\end{figure}
Composing these matter-gravity vertices with those of the purely gravitational sector (\ref{eh_fd}), one ends up with diagrams of the type of \fig{fig:relexp} that describe the interaction between two sources.   
By calculating the corresponding amplitudes, one is integrating out the 
gravitational perturbations $h_{\mu\nu}$ of \eq{eq:h}. 
The result is an effective action where the only degrees of freedom which are present are the particles' world-lines: this corresponds to the functional $S_{\mathit eff}(x_a)$ defined via a path integral of the type of \eq{eq:path}
\begin{equation}
\label{eq:PI_h}
\exp\left[i S_{\mathit eff}(x_a)\right] = \int D h_{\mu\nu}(x) \exp\left[ i S_{EH}(h) + iS_{pp}(h,x_a)\right],
\end{equation}
where the particle world-lines $x_a^\mu(\tau)$ that source the gravitational perturbations $h_{\mu\nu}$ are held fixed.   
Once the effective action (\ref{eq:PI_h}) has been obtained, it is possible to calculate physical observables. 
Notably, taking the real part of \eq{eq:PI_h} generates the coupled equations of motion for the two-body system and consequently the mechanical binding energy. 
On the other hand, the imaginary part of the effective action measures the total number of gravitons emitted by a ﬁxed two-particle conﬁguration $\{x^\mu_a\}$ over an arbitarily large time $T\rightarrow\infty$ 
\begin{equation}
\label{eq:Im}
{1\over T} \,\mbox{Im} S_{eff}[x_a] =  {1\over 2} \int dE d\Omega {d^2\Gamma\over dE d\Omega} \, ,
\end{equation}
where $d\Gamma$ is the differential rate for graviton emission from the binary system. 
%. 
Although the graviton number is not a well defined observable classically, the classical power spectrum can be obtained by integrating $dP = E d\Gamma$ and reads 
\begin{equation}
P= \int dE d\Omega \, E {d^2\Gamma\over d\Omega dE}.
\end{equation}
As discussed in \Sec{sec:PPN_GW}, once we have $E$ and $P$, we can derive the GW phase $\varphi_{GW}$ seen by the detector; for example, in the mass ratio limit the phase reads
\begin{equation}
\label{eq:phase}
\Delta \varphi_{GW}(t) = \int_{t_0}^t d\tau\,\omega_{GW}(\tau) = {2\over G_N M}\int^{v_0} _{v(t)}d v' v'^3 {dE/dv'\over P(v')}\,,
\end{equation}  
where $\omega_{GW}$ is the GW frequency, $M$ the mass of the heavier body in the system and where, with respect to \eq{eq:gwphase}, we wrote the GW flux as $P$ instead of ${\cal F}$.

%%%%%%%%%%%%%%%%%%%%%%%%%%%%%%%%%%
\subsubsection{An example of amplitude computation}
\label{sec:amp_ex}
%%%%%%%%%%%%%%%%%%%%%%%%%%%%%%%%%%

To see how the computation of scattering processes like those of Fig.~\ref{fig:relexp} proceeds, here I will take as an example the case of the one-graviton exchange diagram in panel~(a).
The gravity-matter interaction is given by the linear term in \eq{eq:pp_exp}: for the $a$-th source this reads
\be
\label{eq:pp_lin}
-{m_a\over 2 \mpl} \int d{\bar\tau_a} h_{\mu\nu}(x_a) {\dot x_a}^\mu {\dot x_a}^\nu \,, \nn 
\ee
where I have specified the space-time dependence of the field $h_{\mu\nu}$ and where I remind that the four-velocity of the particle satisfies ${\dot x_a}^\mu {\dot x_a}^\nu\,\eta_{\mu\nu}=1$\,. 
Including an analogous term for particle $b$ gives a factor of $h_{\a\b}(x_b){\dot x_b}^\a{\dot x_b}^\b$ and therefore the propagator of the gravitational field (\ref{eq:hhprop}):
\be
\label{eq:hhprop_2}
\langle h_{\mu\nu}(x_a) h_{\alpha\beta}(x_b) \rangle = 
D_F(x_a-x_b) P_{\mu\nu;\alpha\beta} \,.
\ee
The tensorial part of \eq{eq:hhprop_2} brings the following contraction between the four-velocities of the particles:
\begin{align}
\label{eq:contract}
{\dot x_a}^\mu {\dot x_a}^\nu\,P_{\mu\nu;\alpha\beta}\,{\dot x_b}^\a{\dot x_b}^\b 
&={1\over 2}{\dot x_a}^\mu {\dot x_a}^\nu \[\eta_{\alpha\mu}\eta_{\beta\nu} 
+\eta_{\alpha\nu}\eta_{\beta\mu}-{\eta}_{\mu\nu}\eta_{\alpha\beta}\] {\dot x_b}^\a{\dot x_b}^\b \\ \nn
&={1\over 2} \[ ({\dot x_a} \cdot {\dot x_b}) + ({\dot x_a} \cdot {\dot x_b}) - ({\dot x_a} \cdot {\dot x_a}) ({\dot x_b} \cdot {\dot x_b}) \] \\ \nn
&={1\over 2} \[ 2 ({\dot x_a} \cdot {\dot x_b})^2 - {\dot x_a}^2 {\dot x_b}^2 \] \\ 
&={1\over 2} \[ 2 ({\dot x_a} \cdot {\dot x_b})^2 - 1 \] \nn
\end{align}
Combining all the terms, the amplitude for the one-graviton exchange diagram in Fig.~\ref{fig:relexp}(a) gives the following contribution to the effective action 
\begin{equation}
\label{eq:1grav_ex}
i{\cal S}_{eff} = \sum_{a\neq b} {m_a m_b\over 8 \mpl^2} \int d\tau_a d\tau_b \left[1-2 ({\dot x}_a\cdot {\dot x}_b)^2\right] D_F(x_a-x_b) \,,
\end{equation}   
where the numerical factor of $1/8$ comes from the two matter-gravity vertices together with the propagator, while the sign difference with respect to \eq{eq:contract} is due to the fact that every Feynman rule is defined with a factor of $i$.
As it stands, \eq{eq:1grav_ex} cannot be assigned a unique power of $v$ to assess the perturbative order at which it contributes to the effective action and, then, to GW observables. 
This is because the perturbative series of \fig{fig:relexp} is fully covariant and, as such, is not optimal to perform calculations in the limit of small three-velocity.   
As we will see below, for a term like the one-graviton exchange it is easy to do the expansion for small velocity; on the other hand, in a diagram like the one in Fig.~\ref{fig:relexp}(c), it would be more cumbersome to keep track of all the necessary terms at a given order in $v$. 
Since in the end one is interested in computing GW observables up to a fixed order in the velocity, it is crucial to develop a set of rules that assign a unique 
power of $v$ to each diagram in the theory.
Such a set of rules constitutes the \emph{power counting}, a characteristic ingredient of every EFT as we discussed in \Sec{sec:efts}.   
For NRGR this is presented in the following sections.

%%%%%%%%%%%%%%%%%%%%%%%%%%%%%%%%%
\subsection[Power counting and mode decomposition \\ for gravitational perturbations]
    {Power counting and mode decomposition for gravitational perturbations %
  \sectionmark{Power counting and mode decomposition in NRGR} } 
    \sectionmark{Power counting and mode decomposition in NRGR}
\label{sec:pchH}

% where:    
% "[]" is for the table of contents
% the subsequent "{}" is for where the section actually is
% 1st "sectionmark" is for the first page containing the section
% 2nd "sectionmark" is for the other pages containing the section

%%%%%%%%%%%%%%%%%%%%%%%%%%%%%%%%%

To establish a power counting scheme, it is very useful to recognize that one can decompose the gravitational perturbations as follows: on one hand there is a background whose range of variability is the wavelength of gravitational radiation~$\lambda$; on the other, there is a set of perturbations that have support only on scales below the orbital radius~$r$. 
In other words, one can distinguish between {\it hard} modes and {\it soft} modes. 
A similar decomposition of gravitational perturbations is adopted to distinguish GWs from a background field and to define the energy-momentum tensor carried by GWs~\cite{Maggiorebook}~\footnote{A complete treatment of the so-called {\it short-wave} approximation can be found in Secs.~1.4 and 1.5 of ref.~\cite{Maggiorebook}, while the "Further reading" at the end of Chap.~1 contains the relevant citations.}.
Because gravitational radiation is characterized by the scale $\lambda$ which is large compared to $r$, one can take $\lambda$ as the IR scale of the full theory 
and $r$ as the UV one. 
This is reminiscent of the general discussion on EFTs that I presented in \Sec{sec:efts}. 
Because $\lambda$ and $r$ are two separate scales, the physics characterizing each one is decoupled from the other: in these situations an EFT approach is very convenient. 
On these grounds, one can think of GW observables as regarding the IR sector of the theory, while the gravitational dynamics below the orbital radius is a short distance phenomenon and corresponds to the UV region. 
The high energy modes are responsible for binding the binary system through forces of Newtonian type: for this reason, the UV modes are referred to as {\it potential} modes. 
Because retardation effects are negligible on the orbital scale, Newtonian dynamics is {\it instantaneous}: from a field-theoretical perspective, potential modes are then {\it off-shell} and can only correspond to internal lines in a Fenyman diagram. 
The soft modes which have support on the IR scale $\lambda$ correspond to GWs and are called radiation modes: according to the particle physics parlance of EFT, physical radiation is described by {\it on-shell} modes that are represented by external lines in a Feynman diagram. 

The separation of scales made explicit by the EFT framework is physically equivalent to the matched asymptotic expansions adopted in the standard PN approach, which have been introduced in refs.~\cite{Burke:71,Blanchet:1986p332}.
Indeed, the consistency of the scale factorization proposed in NRGR has been recently verified up to order 3PN in the GW flux~\cite{Goldberger:2009qd}.

To make contact with the standard PN treatment, one can observe that potential modes are those obeying a Poisson equation, so they correspond to quasi-static gravitational fields whose profile is set by the matter sources. 
On the other hand, radiation modes are the gravitational perturbations that satisfy a wave equation, i.e. they represent GWs.

In order to illustrate more vividly both the meaning and the consequences of the mode decomposition, now I will go back to the one-graviton exchange amplitude (\ref{eq:1grav_ex}) and use it to compute the effective action in the non-relativistic limit: with this in hand, I will calculate gravitational observables to lowest order along the ways discussed after \eq{eq:PI_h}.
Let us start the evaluation of \eq{eq:1grav_ex} from the following term  
\begin{equation}
\label{eq:tba}
\int d\tau_a d\tau_b D_F(x_a-x_b) = \int d\tau_a d\tau_b \int{d^4 k\over (2\pi)^4} {i\over k^2+i\epsilon} e^{-ik\cdot (x_a-x_b)} \,;
\end{equation}
this integral can be split up into contributions from the following two regions of momenta: 
\bd

\item[$\bullet$] 
Space-like momenta of the form
\begin{equation}
\label{eq:pot}
k^\mu \rightarrow (k^0\sim {v\over r}, {\bf k}\sim {1\over r}) \, .
\end{equation}
In units $c=1$, the typical three-velocity $v$ is much smaller than one, therefore $k^2=k^\mu k_\mu\neq 0$ and these modes can never be on-shell; as a consequence, they never contribute to the emitted power through the imaginary part of the effective action: indeed, \eq{eq:tba} can have an imaginary when the $i\epsilon$ term in the propagator becomes important, i.e. when $k^2=0$. 
On the other hand, these modes will determine the equations of motion through the real part of the effective action. 
We recognize these modes as being the potential gravitons. 

\item[$\bullet$] 
Null momenta whose scaling is given by 
\begin{equation}
\label{eq:rad}
k^\mu \rightarrow (k^0\sim {v\over r},{\bf k}\sim {v\over r}) \, .
\end{equation}
These modes satisfy $k^2\simeq 0$, so they can give rise to the imaginary part of the effective action and, therefore, to the radiation that propagates out to the detector. 
It is straightforward to identify these modes as the radiation modes.
\ed

\noindent 
Since only potential modes contribute to the real part of the effective action, we may calculate the non-relativistic limit of this quantity by expanding the propagator in \eq{eq:tba} as
\begin{equation}
{1\over k^2_0 - {\bf k}^2} =-{1\over {\bf k}^2}\left[1 + {k^2_0\over {\bf k}^2}+\cdots\right] =-{1\over {\bf k}^2}\left[1 + {\cal O}(v^2)\right] \,;
\end{equation}
the leading order term ${\bf k}^{-2}$ gives the $1/r$ dependence of Newtonian potential once Fourier-transformed according to  
\begin{equation}
\label{eq:Fourier_newton}
\int {d^4 k\over (2\pi)^4} e^{-i k\cdot x} {1\over {\bf k}^2} = {1\over 4\pi |{\bf x}|} \delta(x^0) \, .
\end{equation}

\begin{figure*}[t]
\centering
\includegraphics[width=0.6\textwidth,angle=0]{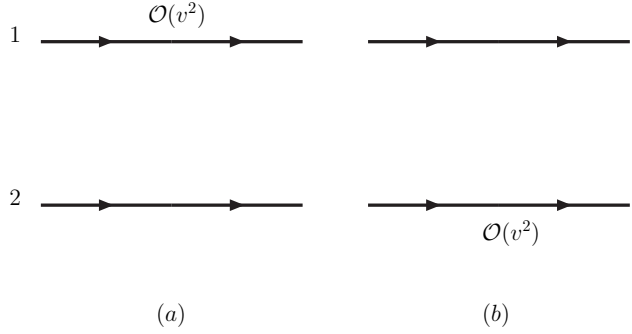}
\caption{Feynman diagrams describing free propagation of matter sources, i.e. the kinetic energy of the particles at ${\cal O}(v^2)$. 
Because of the virial theorem, these diagrams are of the same order as the Newtonian potential, which is ${\cal O}(G_N)$.}
\label{fig:free}
\end{figure*}

Let me stress that we are computing the effective action with observables in mind: on one hand we will take the real part of the action to obtain the energy through the equations of motion, on the other we will take the imaginary part as an intermediate step towards the power emitted in GWs. 
To compute the equations of motion we have to consider another process with respect to the one-graviton exchange.
This is because of the virial theorem, by which terms ${\cal O}(G_N)$ contribute at the same order as those ${\cal O}(v^2)$.
As a consequence, the Newtonian contribution to the one-graviton exchange~(\ref{eq:1grav_ex}) coming from \eq{eq:Fourier_newton} is of the same order as the kinetic energy of the particles, which is represented by the diagrams in \fig{fig:free}. 
To evaluate these diagrams, one has to expand the point particle action in the velocity, rather than in the gravitational field as we did in \eq{eq:pp_exp}; to first non trivial order in $v$, this expansion gives 
\begin{equation} 
\label{eq:prop_time}
 d\tau_a = 
 \[ \eta_{\mu\nu} dx^\mu dx^\nu \]^{1/2} \;=\;  \[ ( dx^0)^2 -{\bf v}^2_a \]^{1/2}
 \simeq dx^0\left[1 -{1\over 2}{\bf v}^2_a\right]\,.
\end{equation} 

With the expansions of the propagator and of the proper time, we can write the non-relativistic limit of the real part of the action as
\begin{equation}
\label{eq:ReSeff}
\mbox{Re} S_{eff}[x_a] = {1\over 2}\sum_a \int dx^0 m_a {\bf v}^2_a 
-\sum_{a,b}\int dx^0 {G_N m_a m_b\over |{\bf x}_a-{\bf x}_b|} + \cdots,
\end{equation}
where we have used $G_N\equiv (32\pi\mpl^2)^{-1}$.
As it could have been awaited, this is just the action for classical non-relativistic particles interacting  through a Newtonian potential.  
Note that the second term contains divergent self-energy contributions whenever $a=b$ in the sum. 
Diagrammatically, they are described by a potential graviton which is emitted and reabsorbed by the same source. 
These divergences can be absorbed by renormalizing the masses of the particles.   
Formally, they can just be set to zero, by evaluating the momentum integral in dimensional regularization using the formula
\begin{equation}
\int {d^D {\bf k}\over (2\pi)^D} e^{-i{\bf k}\cdot {\bf x}} {1\over {({\bf k}^2)}^\alpha} = {1\over (4\pi)^{D/2}} {\Gamma(D/2-\alpha)\over \Gamma(\alpha)} \left({{\bf x}^2\over 4}\right)^{\alpha-D/2},
\end{equation} 
and taking the limit ${\bf x}\rightarrow 0$ before setting $D=3$.

To calculate the imaginary part of the effective action one can use
\begin{equation}
\mbox{Im} {1\over k^2 + i\epsilon} = - i\pi\delta(k^2) 
= - i\pi \frac{1}{2|{\bf k}|} \delta(k_0=|{\bf k}|)\,,
\end{equation}
which guarantees that only on-shell particles contribute to the radiated power. 
Then, the imaginary part of \eq{eq:1grav_ex} gives 
\begin{equation}
\label{eq:ImSeff}
\mbox{Im} S_{eff}(x_a) 
= {1\over 16 \mpl^2}\int {d^3 {\bf k}\over (2\pi)^3} {1\over 2|{\bf k}|} 
\left|\sum_a m_a \int d\tau_a e^{-i k\cdot x_a}\right|_{k^0=|{\bf k}|}^2\,.
\end{equation} 
This results in a differential power given by
\begin{equation}
\label{eq:dpow}
{d P\over d\Omega d{|\bf k|}} = {1\over T} {G_N\over 4\pi^2} {\bf k}^2\left|\sum_a m_a \int d\tau_a e^{-i k\cdot x_a}\right|_{k^0=|{\bf k}|}^2 \,,
\end{equation}
where 
\be
\int d\tau_a e^{-i k\cdot x_a} \simeq \int dt_a e^{i k_0 t} e^{-i{\bf k}\cdot {\bf x}} 
= \delta{(k_0)} e^{-i{\bf k}\cdot {\bf x}} \nn
\ee
so that ${\bf k}^2=k_0^2=0$ and one obtains that the system does not source radiation through its mass monopole. 
The expression (\ref{eq:dpow}) coincides with what one would find both from the classical calculation of the energy-momentum tensor at an asymptotically large distance from the source and from the tree-level amplitude for single graviton emission by the binary source. 
The latter alternative will be used later on in \Sec{sec:eft_bin}.

\smallskip

Coming back to the mode decomposition of the gravitational field, we split it up in potential modes $H_{\mu\nu}(x)$ propagating on the slowly-varying background field ${\bar h}_{\mu\nu}(x)$, which represents the long-wavelength radiation: 
\begin{equation}
\label{eq:Hh}
h_{\mu\nu}(x)= H_{\mu\nu}(x) + {\bar h}_{\mu\nu}(x).
\end{equation}
For what concerns Feynman diagrams, we will draw the radiation modes ${\bar h}_{\mu\nu}$ with wavy or curly lines and the potential modes $H_{\mu\nu}$ with dashed lines.
As a consequence of the scaling of their momenta \eqs{eq:pot}{eq:rad} we have the following power counting for the two sets of modes 
\begin{align}
\ds 
\pa_0 H_{\mu\nu} \sim {v\over r} H_{\mu\nu} \; & , \;
\pa_i H_{\mu\nu}\sim {1\over r} H_{\mu\nu} \\ 
\nn \\
\pa_0 {\bar h_{\mu\nu}} \sim {v\over r} {\bar h}_{\mu\nu} \; & , \;
\pa_i {\bar h_{\mu\nu}} \sim {v\over r} {\bar h}_{\mu\nu} \; .
\end{align}
In order to treat all derivatives acting on fields on the same footing and assign them a common scaling $\pa_\mu\sim v/r$, it is convenient to re-write $H_{\mu\nu}$ in terms of its Fourier transform
\begin{equation}
\label{eq:FT}
H_{\mu\nu}(x)=\int {d^3{\bf k}\over(2\pi)^3} 
e^{i{\bf k}\cdot {\bf x}} H_{{\bf k}\mu\nu}(x_0)\,.
\end{equation}
With this redefiniton, spatial derivatives on $H_{\mu\nu}$ are replaced by factors of ${\bf k}$ multiplying $H_{{\bf k}\mu\nu}(x_0)$ and the only derivative left is $\pa_0 H_{{\bf k}\mu\nu}(x_0) \sim (v/r) H_{{\bf k}\mu\nu}(x_0)$.

Having split the gravitational perturbations $h$ into two sets of modes, the path integral of \eq{eq:PI_h} now needs to be worked out in two stages: one over $H$, another over $\bar h$.  
This would look like one is doubling the calculations to be done; rather on the contrary, this is one of the advantages of the EFT approach, as we defined it in the introduction: the EFT "allows one to separate the relevant degrees of freedom in subsets with decoupled dynamics".
Dividing the calculation of the path integral (\ref{eq:PI_h}) in two stages is the aim of the EFT approach to binary systems: in fact, this choice eases the treatment of the intricacies which are inherent to the problem. 

In a first stage, one performs the path integral over the hard modes $H_{{\bf k}\mu\nu}$ 
\begin{eqnarray}
\label{eq:PI_H}
e^{i S_{NR}({\bar h},x_a)} =\int D H_{{\bf k} \mu\nu}(x^0) e^{i S_{EH}({\bar h}+H) + S_{pp}({\bar h}+H,x_a)}
\end{eqnarray}
where ${\bar h}_{\mu\nu}$ is treated as a non-dynamical background. 
This is a formal way of doing the matching to the long-distance EFT containing ${\bar h}_{\mu\nu}$ and the particle world-lines: in practice, it will be obtained through the calculation of Feynman diagrams that contain potential modes as internal lines. 
This action, which is derived in \eq{eq:Lv2}, constitutes the theory valid at the orbital radius:  because of the matching, it contains explicitly the short distance scale $r$ in the coefficients of operators. 

%%%%%%%%%%%%%%%%%%%%%%%%%%%%%%%%%%%
\begin{figure}
\begin{center}
\includegraphics[width=0.4\textwidth]{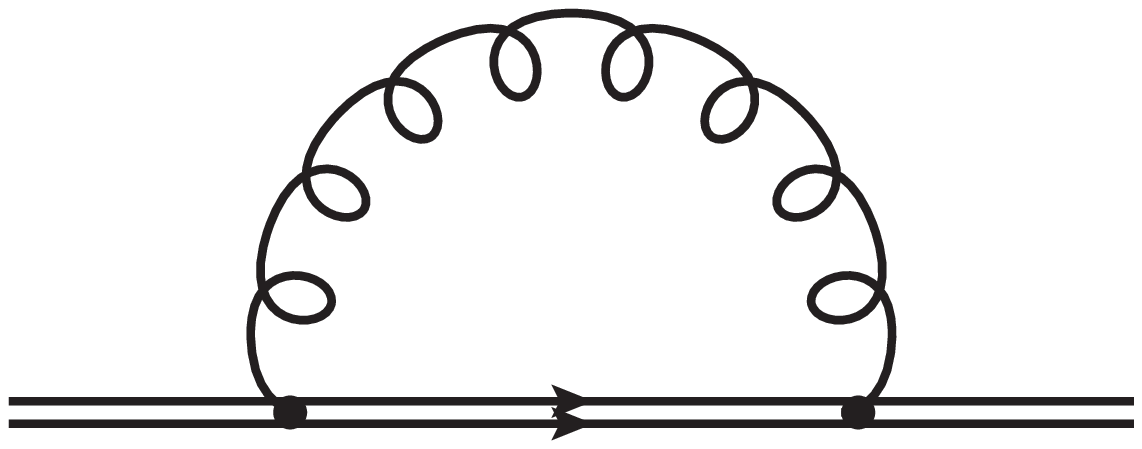}
\caption{The self-energy diagram whose imaginary part gives the 
radiated power (see text for discussion).}
\label{fig:self} 
\end{center}
\end{figure}
%%%%%%%%%%%%%%%%%%%%%%%%%%%%%%%%%%%

In order to compute observables, the EFT prescription requires obtaining an effective action that depends only on the particle world-lines. 
Therefore, given the action $S_{NR}({\bar h},x)$, the second stage in the NRGR approach is to perform the path integral over the radiation modes ${\bar h}_{\mu\nu}$
\begin{equation} 
\label{eq:PI_hbar}
\exp[iS_{\mathit eff}(x_a)] = \int D{\bar h}_{\mu\nu}(x) e^{i S_{NR}({\bar h},x)}\,.
\end{equation}
Rather than solving this path integral, to obtain the action $S_{\mathit eff}(x_a)$ one can resort to computing the diagrams that have no external graviton lines like the one in \fig{fig:self} where the double line describes the binary system as a composite source; the orbital scale is un-resolved because one has integrated it out by solving for the potential modes.  
Feynman diagrams are easier to calculate in this theory than in the full theory, i.e. the one in which gravitational perturbations are not split into potential and radiation modes: in fact, momentum integrals are characterized by a single scale $v/r$.
One can use these diagrams to compute the GW observables of \eqs{eq:ReSeff}{eq:ImSeff}: 
for example, from the imaginary part of the diagram in \fig{fig:self} follows the leading order quadrupole radiation formula
\begin{equation}
\label{eq:ImP_1}
P= {G_N\over 5}\langle \stackrel{\ldots}{Q}_{ij} \stackrel{\ldots}{Q}_{ij} \rangle \,,
\end{equation}
where the brackets denote a time average, dots stand for time derivatives and $Q_{ij}$ is the quadrupole moment of the system that sets the strength of the interaction with radiation gravitons in the vertices. 
The time evolution of $Q_{ij}(t)$ can be obtained by solving the equations of motion for the world-lines ${\bf x}_a(t)$, which follow from the real part of $S_{NR}({\bar h},x)$. 

Before moving to calculating $S_{NR}({\bar h}, x_a)$, it should be noted that  \eq{eq:PI_H} does not contain, as it stands, the gauge fixing terms necessary to make sense of the path integral; 
this is because the explicit form of such terms is not needed in the subsequent discussion.
Nevertheless, it is worth stressing that the most convenient way to do the gauge fixing is the so-called background field method~\cite{Abbott-bck}, where the gauge is chosen in such a way as to preserve the invariance under diffeomorphisms that transform the background metric ${\bar h}_{\mu\nu}$. 
If such a gauge fixing scheme is chosen, the action for ${\bar h}_{\mu\nu}$ is guaranteed to be gauge invariant. 
This requirement places strong restrictions on the form of the EFT that describes radiation.

%%%%%%%%%%%%%%%%%%%%%%%%%%%%%%%%%
\subsection[The conservative dynamics at first post-Newtonian order - \\
The Einstein-Ilbert-Hoffmann Lagrangian in the effective field theory approach]
    {The conservative dynamics at first post-Newtonian order - 
The Einstein-Ilbert-Hoffmann Lagrangian in the effective field theory approach %
  \sectionmark{Effective field theory derivation of the EIH Lagrangian} } 
    \sectionmark{Effective field theory derivation of the EIH Lagrangian}
\label{sec:EIH}

% where:    
% "[]" is for the table of contents
% the subsequent "{}" is for where the section actually is
% 1st "sectionmark" is for the first page containing the section
% 2nd "sectionmark" is for the other pages containing the section

%%%%%%%%%%%%%%%%%%%%%%%%%%%%%%%%%

With the aim of calculating $S_{NR}({\bar h}, x_a)$ as a sum of Feynman diagrams, we now proceed to deriving the power counting rules for the interaction terms in \eqs{eh_fd}{eq:pp_exp} involving only potential modes, i.e. radiation will not be considered for the moment and we postpone its treatment to next section.

%%%%%%%%%%%%%%%%%%%%%%%%%%%%%%%%%%%%
\subsubsection{Power counting and scalings for diagrams with only potential modes}
%%%%%%%%%%%%%%%%%%%%%%%%%%%%%%%%%%%
   
\begin{figure*}[t]
\centering
\includegraphics[width=0.3\textwidth,angle=0]{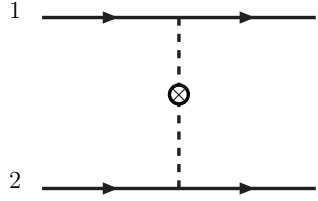}
\caption{The Feynman diagram stemming from the sub-leading contributions to the propagator: in NRGR, this ${\cal O}(v^2)$ correction corresponds to the  insertion of an operator, denoted by $\otimes$.}
\label{fig:full_prop}
\end{figure*}

The scaling of $H_{\bf k\mu\nu}$ in terms of kinematic variables can be obtained by looking at its propagator, i.e. at terms in the gravitational action that 
are quadratic in $H_{\bf k\mu\nu}$:  
\begin{align}
\label{eq:H2full}
S_{H^2} = -{1\over 2} \int dt \frac{d^3{\bf k}}{(2\pi)^3}
& \left[ {\bf k}^2 H_{{\bf k}\mu\nu} H_{-{\bf k}}^{\mu\nu} 
-{{\bf k}^2\over 2} H_{\bf k}  H_{-{\bf k}}\right. \nn \\
 & \left. -\partial_0 H_{{\bf k}\mu\nu} \partial_0 H_{-{\bf k}}^{\mu\nu} 
+ {1\over 2} \partial_0 H_{\bf k} \partial_0 H_{-\bf k}\right] \,,
\end{align}
where $H_{\bf k}=H^\mu_{\mu \bf k}$. 
Because potential modes have $\pa_0\sim v\,\pa_i$, the terms in the second line of this equation are suppressed relative to the first line by $v^2$ and can be treated as perturbative corrections.
Therefore, to leading order, the propagator for $H_{{\bf k}\mu\nu}$ reads 
\begin{equation}
\label{eq:Hprop}
\langle T H_{\bf k\mu\nu}(x^0) H_{\bf q\alpha\beta}(0)\rangle = -{i\over{\bf k}^2} (2\pi)^3 \delta^3({\bf k}+{\bf q})\delta(x^0) P_{\mu\nu;\alpha\beta} \,,
\end{equation}
where $T$ stands for {\it time-ordering} and the tensor structure is given by \eq{eq:Pprop}. 
Because of the factor ${\bf k}^{-2}$, \eq{eq:Hprop} is essentially the Fourier transform of the Newton potential (cfr. with \eq{eq:Fourier_newton}),
which will be then expressed by a one-graviton-exchange like \fig{fig:relexp}(a).
Since the tensor structure (\ref{eq:Pprop}) contains factors of $\eta_{\mu\nu}$, one has  $P_{\mu\nu;\alpha\beta}\sim{\cal O}(1)$ and the scaling of $H_{\bf k\mu\nu}$ is determined by the other terms in \eq{eq:Hprop}
\begin{equation}
H^2_{\bf k\mu\nu}\sim \left({1\over r}\right)^{-2} \times \left({1\over r}\right)^{-3}\times \left({r\over v}\right)^{-1} = r^4 v,
\end{equation}
so that 
\begin{equation}
H_{\bf k\mu\nu}\sim r^2 \sqrt{v} \, , 
\end{equation}
where $\delta^3({\bf k})$ has been power-counted as $({\bf k})^{-3}$.
In terms of $H_{\bf k\mu\nu}$ the linear interaction between gravity and the matter source of \eq{eq:pp_exp} reads 
\begin{align}
\label{eq:pp_H}
-{m_a\over 2 \mpl} \int d{\bar\tau} h_{\mu\nu}\,{\dot x}^\mu {\dot x}^\nu  
&= -{m\over 2 \mpl}\int dx^0 {d^3 {\bf k}\over (2\pi)^3} e^{i{\bf k}\cdot {\bf x}(x^0)} H_{\bf k\mu\nu} \, {\dot x^\mu} {\dot x^\nu} \\ \nn
&= -{m\over 2 \mpl}\int dx^0 {d^3 {\bf k}\over (2\pi)^3} e^{i{\bf k}\cdot {\bf x}(x^0)} 
\[ H_{{\bf k} 0 0} + 2 H_{{\bf k} 0 i}\, v^i + H_{{\bf k} ij}\, v^i v^j \] \,.
\end{align}
Therefore, to lowest order in velocity, the matter-gravity interaction vertex scales as 
\begin{eqnarray}
\label{eq:00vert}
\nonumber
-{m\over 2 \mpl}\int dx^0 {d^3 {\bf k}\over (2\pi)^3} e^{i{\bf k}\cdot {\bf x}(x^0)} H_{{\bf k}00} 
&\sim& {m\over \mpl} \times \left({r\over v}\right) \times \left({1\over r}\right)^3 \times \left (r^2 v^{1/2}\right) \nn \\ 
\nn \\ 
&=& v^{-1/2} {m\over \mpl} \,.
\end{eqnarray}
To complete this expression we need the scaling of $m/\mpl$: this is fixed by means of the virial theorem,
\begin{equation}
v^2\sim {G_N m\over r}\Rightarrow {m^2\over \mpl^2} \sim m v^2 r  = L v,
\end{equation}
where we have used $G_N\sim 1/\mpl^2$ and we have introduced the orbital angular momentum $L=m vr$.   
The interaction in \eq{eq:00vert} then scales as  
\begin{equation}
\label{eq:00scal}
\Diagram{fvA\\
h\\
fvA} =  
-{m\over 2 \mpl}\int dx^0 {d^3 {\bf k}\over (2\pi)^3} e^{i{\bf k}\cdot {\bf x}(x^0)} H_{{\bf k}00}\sim L^{1/2}\,;
\end{equation}
the exchange diagram with two insertions of this term corresponds to the Newton potential between the point particles, its scaling would then be as one power of the angular momentum
\begin{eqnarray}
\Diagram{fvA & fvA\\
h\\
fvA & fvA} &\sim& L \,.
\end{eqnarray}
This simple result has an important implication: it shows that, once translated in NRGR, the perturbative series of PN corrections will be of the type:
\be
L(1+v^2+v^4\cdots)\,,
\ee
where I have used only even powers of the velocity because the conservative dynamics I am dealing with now is time symmetric.
Having assessed the scaling of Newtonian potential, it is straightforward to see that a correction of order 1PN comes from the sub-leading terms in the quadratic action for $H_{\vk\mu\nu}$~(\ref{eq:H2full}). 
In fact these perturbations give a propagator which reads
\be 
\label{eq:fullprop}
\langle H_{\vk\mu\nu} (x^0)H_{{\bf q}\alpha\beta}(0)\rangle_{\otimes} = 
-(2\pi)^3\delta^3(\vk + {\vq}){i\pa_t^2\over \vk^4}\delta(x^0)
P_{\mu\nu;\alpha\beta} \,,
\ee
where the subscript $\otimes$ is to distinguish this propagator from the one in \eq{eq:Hprop} and where
\be
{\pa_t^2\over \vk^4} \sim {v^2\pa_i^2\over \vk^4} 
\sim {v^2\vk^2\over \vk^4} \sim {v^2\over \vk^2} \,,
\ee 
which is indeed suppressed by $v^2$ with respect to \eq{eq:Hprop}.
The diagram for the corresponding one-graviton exchange is reported in \fig{fig:full_prop}, where the correction to the propagator is described by an operator insertion denoted by the symbol $\otimes$.

The propagator is just the first term in the gravitational action: in \eq{eh_fd} we wrote its analytical and diagrammatic expansions as
\begin{eqnarray} 
\label{eh_fd_2}
-2 \mpl^2 \int d^4 x \sqrt{g} R(x) &\rightarrow& \int d^4 x \left[(\partial h)^2 
+ {h (\partial h)^2\over \mpl} + {h^2 (\partial h)^2\over \mpl}
+\cdots\right]  \\
\nonumber
&=& \qq
(\Diagram{g})^{-1}\,\,\,+ 
\Diagram{gd\\
& g\\
gu}\,\,\, +
\Diagram{gd & gu\\
gu & gd}\,\,\, +
\cdots \,,
\end{eqnarray}
where it is understood that each term in the series is suppressed with respect to the previous one: now, we can substantiate this statement by looking at the scalings of potential graviton self-interactions. 
In terms of $H_{{\bf k}\mu\nu}$, the cubic term of the gravitational action (\ref{eh_fd_2}) has the structure
\begin{equation}
\label{eq:H3}
S_{H^3} \sim {1\over \mpl} \int dx^0 (2\pi)^3\delta^3\left(\sum_n {\bf k}_n\right) {\bf k}^2 \prod_{n=1}^3 {d^3 {\bf k}_n\over (2\pi)^3}  H_{{\bf k}_n},
\end{equation}
where we have used 
\be 
\int d^3 {\bf x} \,exp\, \left\{ i \left(\sum_n {\bf k}_n \right) \right\}=\delta^3\left(\sum_n {\bf k}_n\right)\,.
\ee 
The novelty of \eq{eq:H3} with respect to the scalings worked out so far is in the factor~${\bf k}^2$, which is due to the fact that the Ricci scalar has two derivatives acting on the metric. 
Therefore, the full scaling of the interaction term (\ref{eq:H3}) is 
\begin{eqnarray}
\label{eq:3gscal}
\nonumber
\Diagram{hd\\ & h \\ hu}
&\sim& {1\over \mpl}\times\left({r\over v}\right)\times \left({1\over r}\right)^{-3}\times\left({1\over r}\right)^2\times\left[\left({1\over r^3}\right)\times \left(r^2\sqrt{v}\right)\right]^3\\
&=&  {v^2\over\sqrt{L}}\,,
\end{eqnarray}
which gives the following scaling of the corresponding exchange between two particles  
\begin{equation}
\label{eq:3gpot}
\Diagram{fvA \\fvA}\Diagram{hd\\ & h \\ hu}\Diagram{fvA\\ fvA}\sim \left(\sqrt{L}\right)^2\times {v^2\over \sqrt{L}}\times \sqrt{L} = L v^2,
\end{equation}
i.e. a term in the two-body potential that is suppressed by $v^2$ relative to the leading order Newtonian exchange diagram.   

\begin{figure*}[t]
\centering
\includegraphics[width=0.2\textwidth,angle=0]{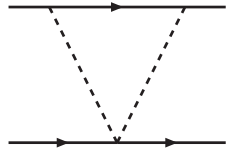}
\caption{The so-called seagull diagram, which owes its name to the two-graviton vertex sourced by matter. 
In NRGR the seagull diagram represents one of the corrections to the Newtonian potential of ${\cal O}({m\,G_N}/r)$ with respect to it.}
\label{fig:seagull}
\end{figure*}

The three-graviton potential (\ref{eq:3gpot}) is not the only term which is suppressed by $v^2$ with respect to Newtonian potential because of gravity 
non-linearities~\footnote{Remember that the PN expansion parameter is 
$\eps\sim(m\,G_N/r) \sim v^2$.}: there is indeed another diagram which is shown in \fig{fig:seagull} and comes from the quadratic term $(h\,\dot x\,\dot x)^2$ in the matter-gravity action~(\ref{eq:pp_exp}). 
To lowest order in the velocity, this term reads $H_{00} H_{00}$ (cfr. \eq{eq:pp_H}), so the quadratic interaction term and its scaling read 
\begin{eqnarray}
\label{eq:seagull}
S_{pp}&=&{m\over 8 \mpl^2}\int dx^0\int_{{\bf k},{\bf q}} e^{i({\bf q}+{\bf k})\cdot {\bf x}(x^0)} H_{{\bf k}00}(x^0) H_{{\bf q}00}(x^0) \nn \\
\nn \\
&\sim& {L\, v \over m} \times \left({r\over v}\right) \times \left({1\over r^3}\right)^2 \times \left (r^4 v\right) \nn \\ 
\nn \\
&=& {L \over m} \times \left({v\over r}\right) = v^{2} \,.
\end{eqnarray}

Working at order $\eps \sim(m\,G_N/r) \sim v^2$, one should also consider terms which are suppressed with respect to Newtonian potential because of velocity corrections, as it is the case for matter-gravity vertices. 
In \eq{eq:pp_H} we have written down the terms coming from the velocity expansion of $H_{{\bf k}\mu\nu}\,{\dot x^\mu}{\dot x^\nu}$, the lowest order one being $H_{{\bf k}00}$ which scales as $L^{1/2}$\,, as determined in \eq{eq:00scal}. 
Analogously, one can see that 
\bees
\label{eq:vH}
H_{{\bf k} 0 i}\,v^i \phantom{v^j}&\sim& L^{1/2}v^1 \\
H_{{\bf k} i j}\,v^i v^j &\sim& L^{1/2}v^{2} \,.
\ees 
Another vertex scaling as $L^{1/2}v^2$ comes from expanding the proper time of the $a$-th particle as in \eq{eq:prop_time} that we report here for convenience
\begin{equation} 
\label{eq:prop_time_2}
 d\tau_a = dx^0\sqrt{1-{\bf v}^2_a}\simeq dx^0\left[1 -{1\over 2}{\bf v}^2_a\right]\,;
\end{equation} 
taking the $v^2$ suppression from here, the polarization that couples to a matter source can only be $H_{{\bf k}00}$ (cfr. \eq{eq:pp_H}). 
The resulting vertex completes the point particle action (\ref{eq:pp_H}) to ${\cal O}(v^2)$, to this PN level the linear coupling of a potential mode to a world-line reads
\begin{align}
\label{eq:pp_H_v2} 
{\cal S}_{pp}^{(v^2)} =& -{m_a\over 2 \mpl} \int d{\bar\tau} h_{\mu\nu}\,{\dot x}^\mu {\dot x}^\nu  
e^{i{\bf k}\cdot {\bf x}(x^0)} H_{\bf k\mu\nu} \, {\dot x^\mu} {\dot x^\nu} \\ \nn
=& -{m\over 2 \mpl}\int dx^0 {d^3 {\bf k}\over (2\pi)^3} e^{i{\bf k}\cdot {\bf x}(x^0)} 
\[ H_{{\bf k} 0 0} + 2 H_{{\bf k} 0 i}\, v^i + H_{{\bf k} ij}\, v^i v^j + H_{{\bf k} 00}\, v^2 \] \,.
\end{align}

\begin{figure*}[t]
\centering
\includegraphics[width=0.50\textwidth,angle=0]
{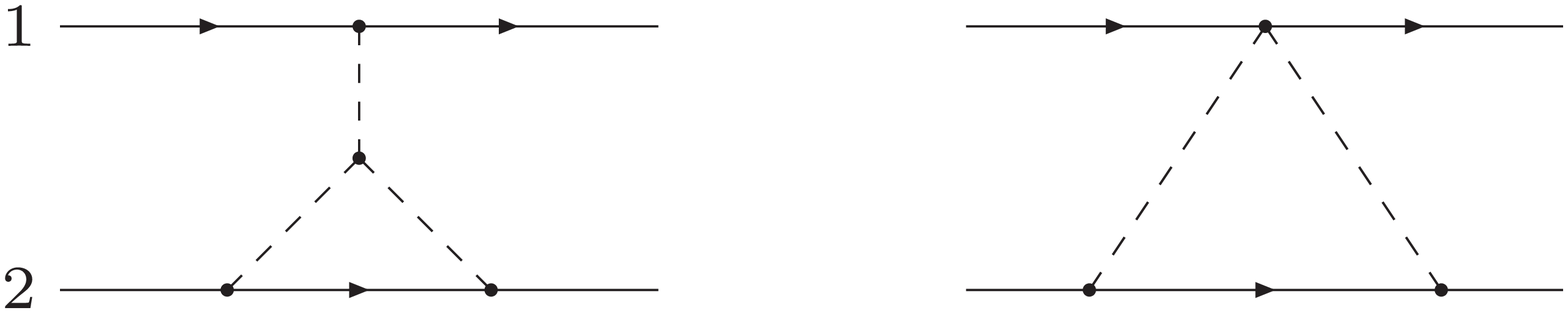}
\vskip 1.0cm
\includegraphics[width=0.80\textwidth,angle=0]
{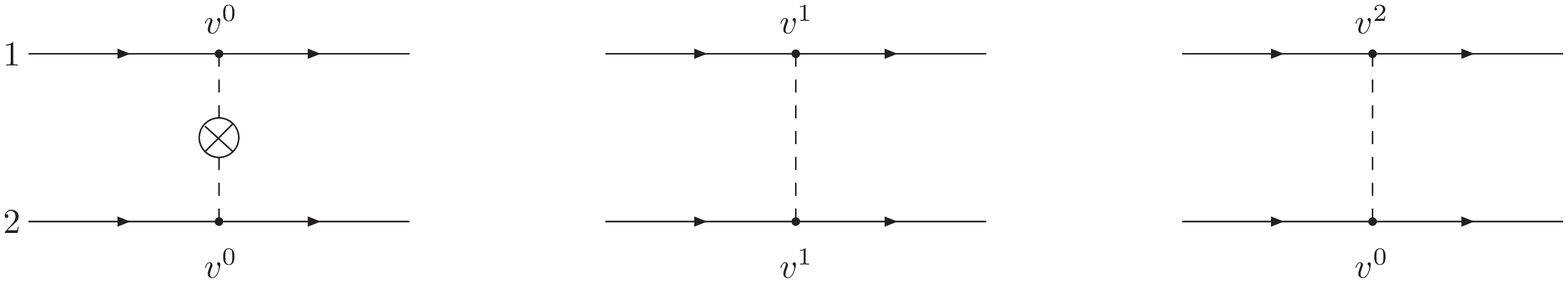}
\caption{The Feynman diagrams that, in NRGR, represent the next-to-leading 
order contributions to the conservative dynamics, i.e. corrections to the 
Newtonian potential which are ${\cal O}({m\,G_N}/r)$, upper row, and 
${\cal O}(v^2)$, lower row, with respect to it (cfr. \fig{fig:EIH_scal} in \Sec{sec:Dam_Far}). 
Diagrams taken from ref.~\cite{NRGR_paper}. }
\label{fig:EIH}
\end{figure*}

If we combine the various diagrams and interaction vertices we have treated so far, we can build five diagrams for two-body potentials that are suppressed by $v^2$ with respect to the Newtonian one.  
These diagrams are presented in \fig{fig:EIH}: as shown in ref.~\cite{NRGR_paper}, the sum of the corresponding amplitudes reproduces the famous Einstein-Infeld-Hoffman Lagrangian~\cite{EIH}, i.e.  the 1PN corrections to the two-body non-relativistic motion 
\begin{eqnarray}
\label{eq:EIH}
\nonumber
\ds L_{v^2} &=& \({1\over 8}\sum_a m_a {\bf v}^4_a \)
- {G^2_N m_1 m_2 (m_1+m_2)\over 2 |{\bf x}_{12}|^2} \\
& &+ {G_N m_1 m_2\over |{\bf x}_{12}|}\left[3({\bf v}^2_1 
+ {\bf v}^2_2) - 7 {\bf v}_1\cdot {\bf v}_2 {} 
-{({\bf v}_1\cdot {\bf x}_{12})({\bf v}_2\cdot {\bf x}_{12})\over |{\bf x}_{12}|}\right] \, ,
\end{eqnarray}
where ${\bf x}_{12}={\bf x}_1-{\bf x}_2$ is the orbital radius and: 
\bd
\item the first term is the leading order relativistic correction to the particles kinetic energies: it is reported in round brackets because it does not come from the diagrams of \fig{fig:EIH}; rather, it comes from diagrams of the type of \fig{fig:free}, with the $v$-expansion pushed up to order $v^4$;
\item the second term comes from the gravity self-interactions of \eqs{eq:3gpot}{eq:seagull}; 
\item the terms in square brackets arise from diagrams where the one-graviton exchange is accompanied by velocity-suppressed interactions; these are: the $v^2$ correction of the propagator (\ref{eq:fullprop}) and the velocity-dependent vertices of  \eq{eq:vH}. 
\ed

This constitutes a first step toward the derivation of the path integral over potential modes (\ref{eq:PI_H}).
The EIH Lagrangian (\ref{eq:EIH}) contains the gravitational forces between the non-relativistic particles which arise from diagrams with no external factors of the radiation ﬁeld. 
The functional integral in \eq{eq:PI_H} also generates couplings of matter to radiation from diagrams with one or more external radiation gravitons which will be the subject of next section.

\begin{figure}[t]
  \begin{center} 
    \includegraphics[width=.6\linewidth]{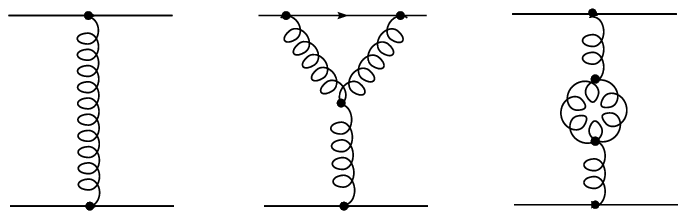}\\
    \hspace{0.5cm}{(a) Newton\hspace{2.3cm} (b) 1PN\hspace{2.0cm}(c) Quantum loop}
    \caption{Contributions to the gravitational scattering amplitude of two 
    massive objects. In this picture curly lines indicate gravitational perturbations 
    over flat space-time and do not specifically refer to radiation gravitons of the 
    mode decomposition~(\ref{eq:Hh}).
    From left to right, the diagrams represent respectively the leading Newtonian 
    approximation, a \emph{classical} contribution to the 1PN order and a 
    negligible \emph{quantum} loop process.}
    \label{fig:class_quant}
  \end{center}
\end{figure}

Before moving to the description of radiation in NRGR, let us take a deeper look at scalings.
Even if to deal with GWs we only need a classical field theory, it is interesting to see what the scalings in powers of $\hbar$ would be. 
Let us then assess the scaling of a process where two world-lines exchange gravitons through a quantum loop: by comparing this process with the two-body potentials worked out in \Sec{sec:eft_iso}, here we will justify why we did not need to keep this kind of diagrams into account~\footnote{This argument is also discussed in \chap{chap:Ric} because contained in the publication of mine that is reported there. However, for a more complete treatment of NRGR it nicely fits this chapter, too.}. 
The confrontation among diagrams is reported in \fig{fig:class_quant}, where, by means of curly lines, we show Newtonian interaction together with the three-graviton potential and the quantum loop one. 
With reference to \fig{fig:class_quant} the following scaling laws can be associated to the different contributions:
\renewcommand{\arraystretch}{1.4}
$$
\bea{rcl}
(a)&\sim& \ds\pt{\frac m{M_{Pl}}}^2\paq{dt\,d^3\vk}^2
\paq{\delta(t)\delta^{(3)}(\vk)\vk^{-2}}\sim L\,,\\
\nn \\
(b)&\sim& \ds\pt{\frac m{M_{Pl}}}^3\paq{dt \, d^3\vk}^3
\paq{\delta(t)\delta^{(3)}(\vk)\vk^{-2}}^3 
\paq{\frac{\vk^2}{M_{Pl}}dt\,\delta^{(3)}(\vk)\pt{d^3\vk}^3} \sim Lv^2\,,\\ 
\nn \\
(c)&\sim& \ds\pt{\frac m{M_{Pl}}}^2\paq{dt\,d^3\vk}^2
\paq{\delta(t)\delta^{(3)}(\vk)\vk^{-2}}^4
\paq{\frac{\vk^2}{M_{Pl}}dt\, \delta^{(3)}(\vk)\pt{d^3\vk}^3}^2\sim v^4\,.
\eea
$$
\renewcommand{\arraystretch}{1}

\noindent where the terms in the first square brackets come from the matter-gravity coupling, those in the second come from the graviton propagators and finally the terms in the third square brackets come from the three-graviton vertices.

To restore factors of Planck's constant $\hbar$, one can apply a rule that relates the number $\mathcal I$ of internal graviton lines (graviton propagators) to the 
number $\mathcal V$ of vertices and the number $\mathcal L$ of graviton loops
\be
\label{eq:vert_rule}
\mathcal L =\mathcal I- \mathcal V+1 \,;
\ee
then, taking into account that each internal line brings a power of $\hbar$ and each interaction vertex a $\hbar^{-1}$ from the interaction Lagrangian, the total scaling for diagrams where the only external lines are massive particles is 
$\hbar^{{\cal L}-1}$. 
According to this rule the third diagram of \fig{fig:class_quant} involves one more power of $\hbar$ than the first two.
The diagram with a graviton loop is then suppressed with 
respect to the Newtonian contribution, apart from some powers of $v$, by a 
factor $\hbar/L\ll 1$, whereas the second diagram in \fig{fig:class_quant} is a 
1PN contribution which does not involve any power of $\hbar$. 
Equivalently one can notice that there is no kinetic term in the Lagrangian for the matter source since the massive object is not a propagating degree of freedom, so the 1PN diagram is not a loop one.
As I will discuss in \Sec{sec:point_part}, these scaling arguments remain unchanged when one adds other fields/particles, like a scalar, and/or when one introduces another mass scale~\cite{Porto:2007pw}, provided
that the virial relation (\ref{eq:virial}) correctly accounts for the leading 
interaction.

%%%%%%%%%%%%%%%%%%%%%%%%%%%%%%%%%
\subsection[The effective theory for binary systems - \\
Incorporating radiation in the treatment]
    {The effective theory for binary systems - 
Incorporating radiation in the treatment %
  \sectionmark{The effective theory for radiation} } 
    \sectionmark{The effective theory for radiation}
\label{sec:eft_bin}

% where:    
% "[]" is for the table of contents
% the subsequent "{}" is for where the section actually is
% 1st "sectionmark" is for the first page containing the section
% 2nd "sectionmark" is for the other pages containing the section

%%%%%%%%%%%%%%%%%%%%%%%%%%%%%%%%%

In the previous section we have started the calculation of the path integral over potential modes (\ref{eq:PI_H}): the aim is to get to a theory which is valid at scales larger then the orbital radius and only contains radiation gravitons.
The first step consisted in integrating out potential modes from the two-body potentials. 
To complete the theory below the orbital radius we now have to add radiation modes and treat their couplings with both potential modes and matter. 
These couplings are indeed generated by the functional integral in \eq{eq:PI_H} and it is only by integrating potential modes out of them too that we will obtain a theory sensitive to radiation only. 

\begin{figure*}[t]
\centering
\includegraphics[width=0.45\textwidth,angle=0]{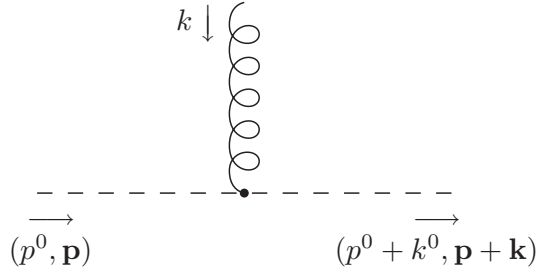}
\caption{A three-graviton vertex where a radiation graviton (curly line) interacts with two potential gravitons (dashed lines). In order to have a definite power counting, the radiation graviton should not impart any momentum to the potential graviton. This exemplifies the necessity of multipole-expanding the radiation modes as done for QCD in ref.~\cite{QCD_mult} (see text for discussion).
Figure taken from ref. \cite{NRGR_lectures}.}
\label{fig:mult}
\end{figure*}

There is no big conceptual issue in incorporating radiation, as it can be done following much of the same steps we described in the previous section. 
For example, one can start again from the propagator to assess the scaling of radiation gravitons $\bar h$: in a suitable gauge~\cite{NRGR_paper}, this is given by 
\begin{equation}
\label{eq:radprop}
\langle T {\bar h}_{\mu\nu}(x) {\bar h}_{\alpha\beta}(0)\rangle = \int {d^4 k\over (2\pi)^4} {i\over k^2+i\epsilon} e^{-ik\cdot x} P_{\mu\nu;\alpha\beta} \,,
\end{equation}
where the tensorial structure $P_{\mu\nu;\alpha\beta}$ is the same as in \eq{eq:Pprop}.
Because $\bar h$ represents radiation, its momentum will scale as the inverse of the wavelength, so $k^\mu\sim v/r$; \eq{eq:radprop} then implies that radiation modes should scale as
\begin{equation}
{\bar h}_{\mu\nu}\sim {v\over r}\,.
\end{equation}
This rule allows one to power count terms in the action containing the radiation field; however, this is not enough in order to obtain an EFT for radiation that has manifest velocity power counting. 
Let us see why through the following example. 
Consider the interaction vertex depicted in \fig{fig:mult}, which involves both  potential and radiation modes. 
Following the direction of the momentum $p^\mu$, after the interaction the potential mode has a three-momentum ${\bf p}+{\bf k}$ so that its propagator will scale as
\begin{eqnarray}
\label{eq:pk}
\mbox{Fig.~\ref{fig:mult}}\sim{1\over ({\bf p}+{\bf k})^2} = {1\over {\bf p}^2}\left[1 -2 {\bf p}\cdot {\bf k}+\cdots \right]\,;
\end{eqnarray}
because the individual momenta scale as $|{\bf p}|\sim 1/r$ and $|{\bf k}|\sim v/r$, this propagator contains an infinite number of powers of $v$ and cannot be properly power-counted in the velocity.   
To ensure that this does not happen, it is necessary to arrange that radiation gravitons do not impart momentum to the potential modes, as first pointed out in the context of non-relativistic gauge theories in ref.~\cite{QCD_mult}.
This is achieved by multipole-expanding the radiation field at the level of the action 
by means of
\begin{align}
\label{eq:mult}
{\bar h}_{\mu\nu}({\bf x},x^0) =  {\bar h}_{\mu\nu}({\bf X},x^0) & + 
\delta {\bf x}^i \pa_i {\bar h}_{\mu\nu}({\bf X},x^0)  \\
& + {1\over 2} \delta {\bf x}^i \delta {\bf x}^j \pa_i \pa_j {\bar h}_{\mu\nu}({\bf X},x^0) + \cdots \nn
\end{align} 
where ${\bf X}$ is an arbitrary point, which in the case of binary systems can be very conveniently chosen as the center of mass defined by ${\bf X}_{cm}=\sum_a m_a{\bf x}_a/\sum_a m_a$.  
The reason why \eq{eq:mult} avoids any transfer of momentum from radiation to potential modes, thus preserving the velocity power counting, is the following.
The expansion~(\ref{eq:mult}) is a redefinition of the radiation field, which now is only a function of time because the space argument in $\bar h({\bf X},x^0)$ is a fixed point in coordinate space (that can be taken as the origin of the coordinate system). 
This means that the couplings of the radiation field to either potential modes or point particles will not be Fourier expanded for what concerns $\bar h({\bf X},x^0)$ (cfr. \eq{eq:hHH}): 
in this way, terms like (\ref{eq:pk}) will not appear; rather, the factors $\delta {\bf x}^i \pa_i \sim v$ in \eq{eq:mult} will bring the necessary ingredients to form the multipoles of the system.

\begin{figure*}[t]
\centering
\includegraphics[width=0.60\textwidth,angle=0]{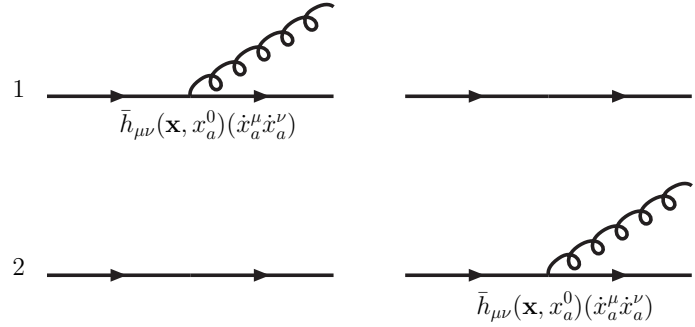}
\caption{Feynman diagrams describing the coupling~(\ref{eq:matt_rad}) of a radiation graviton to the particle word-lines.}
\label{fig:multi_pp}
\end{figure*}

Let us start by treating the coupling of the radiation field to the particle word-lines:
\begin{align}
\label{eq:matt_rad}
&-\sum_a{m_a \over 2 \mpl} \int d\tau_a^0 
{\bar h}_{\mu\nu}({\bf x},x_a^0) (\dot x_a^{\mu}\dot x_a^{\nu}) \simeq  \\ 
&-\sum_a{m_a \over 2 \mpl} \int dx_a^0 \left\{ 
{\bar h}_{00}({\bf x},x_a^0) + 2 {\bar h}_{0s}({\bf x},x_a^0)v_a^s 
+ {\bar h}_{rs}({\bf x},x_a^0) v_a^r v_a^s + \frac 12 v_a^2{\bar h}_{00}({\bf x},x_a^0) 
%+ {\cal O}(v^3)
\right\} \nn \,, 
\end{align} 
where we have performed the velocity expansion up to ${\cal O}(v^2)$ like we did for potential modes in \eq{eq:pp_H_v2} . 
In diagrammatic terms the action (\ref{eq:matt_rad}) is described by \fig{fig:multi_pp}. 
According to \eq{eq:mult}, every term in this action has to be multipole-expanded: 
we display these expansions explicitly in what follows comprising terms up to order $v^2$. 

For the polarization ${\bar h}_{00}$, we have 
\begin{align}
\label{eq:rad_00}
-\sum_a{m_a \over 2 \mpl} \int dx_a^0 & (1+\frac 12 v^2) 
{\bar h}_{00}({\bf x},x_a^0) \simeq \nn\\
-\sum_a{m_a \over 2 \mpl} \int dx_a^0 &\left\{ {\bar h}_{00}(0,x_a^0) 
\phantom{\frac12} \right. \nn\\
& + \delta {\bf x_a}^i \pa_i {\bar h}_{00}(0,x_a^0) \nn\\
& + \left. {1\over 2} \delta {\bf x_a}^i \delta {\bf x_a}^j \pa_i \pa_j {\bar h}_{00}(0,x_a^0) 
+ \frac 12 v_a^2 {\bar h}_{00}(0,x_a^0) \right\} \,,
\end{align} 
where we have separated the various terms in lines according to their scaling in the velocity and where we have made explicit the choice ${\bf X}={\bf X}_{cm}=0$ for the origin of the coordinate system.  
The leading order of \eq{eq:rad_00} is the coupling of ${\bar h}_{00}$ to the mass monopole of the system $M=\sum_a m_a$. 
In \eq{eq:dpow} we saw that the mass monopole does not lead to the emission of radiation, so
this coupling does not contribute to the power: 
indeed, ${\bar h}_{00}$ is not a physical degree of freedom in GR; moreover, one can show that the mass monopole is a conserved quantity, so it cannot be a source of radiation.
The second term on the right hand side of \eq{eq:rad_00} is of order $v$ with respect to the leading order because of the term $\delta {\bf x_a}^i \pa_i$; it can be transformed into 
\be
\label{eq:MX}
M {\bf X}_{cm}^i \pa_i {\bar h}_{00}(0,x_a^0)
\ee
using the definition of the center of mass coordinate to write $\sum_a m_a {\bf x}_a^i=M{\bf X}_{cm}^i$; because we could choose the origin of the system to be in the center of mass, we have ${\bf X}_{cm}=0$, so that this coupling of radiation to matter is zero.
The terms in the last line of \eq{eq:rad_00} are of order~$v^2$ and cannot be made to vanish.

For the polarization ${\bar h}_{0i}$, we have 
\begin{align}
\label{eq:rad_0i}
-\sum_a{m_a \over 2 \mpl} \int dx_a^0 & 2 {\bar h}_{0s}({\bf x},x_a^0)v_a^s \simeq \nn\\
-\sum_a{m_a \over 2 \mpl} \int dx_a^0 & 2\left\{ {\bar h}_{0s}(0,x_a^0)v_a^s \phantom{\frac12} \right. \nn\\
& \left. \phantom{\frac12} + \delta {\bf x_a}^i \pa_i {\bar h}_{0s}(0,x_a^0)v_a^s \right\} \,.
\end{align} 
Here the first term is of order $v$ and can be re-written as  
\be
2 {\bar h}_{0s}(0,x^0) {\bf P} _{cm}^s = 0
\ee
where we have defined ${\bf P}_{cm} =\sum_a m_a v_a$ as the total linear momentum of the system and used the fact that it is zero in the center of mass frame. 
This term would have contributed to radiation at order $v$ as the second term of \eq{eq:rad_00}: the fact that both these couplings vanish corresponds to the absence of dipole radiation in GR. 
For what concerns the second term on the right hand side of \eq{eq:rad_0i}, it is of order~$v^2$ and cannot be made to vanish.

Finally, concerning the term proportional to the polarization ${\bar h}_{rs}$ in \eq{eq:matt_rad}, we have 
\begin{align}
\label{eq:rad_ij}
-\sum_a{m_a \over 2 \mpl} \int dx_a^0 & {\bar h}_{rs}({\bf x},x_a^0)v_a^r v_a^s \simeq \nn\\
-\sum_a{m_a \over 2 \mpl} \int dx_a^0 & \left\{ {\bar h}_{rs}(0,x_a^0)v_a^r v_a^s \phantom{\frac12} \right\} \,,
\end{align} 
which is of order $v^2$ and does not vanish.

\begin{figure*}[t]
\centering
\includegraphics[width=0.70\textwidth,angle=0]{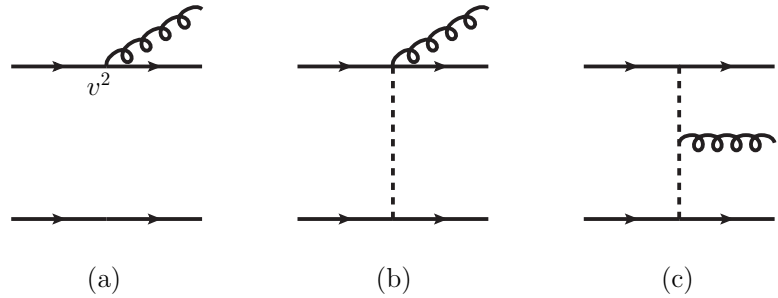}
\caption{Feynman diagrams representing the radiation sector at first non trivial order in GR. 
See text for discussion.  
Figure taken from ref.~\cite{NRGR_paper}.}
\label{fig:Lrad}
\end{figure*}

Collecting the non-vanishing terms of order~$v^2$ from \eqss{eq:rad_00}{eq:rad_0i}{eq:rad_ij}, 
we have the ${\cal O}(v^2)$ Lagrangian expressing the coupling of radiation to point particles \begin{align}
\label{eq:rad_multi}
-\sum_a{m_a \over 2 \mpl} \int dx_a^0 & \left\{ 
 \frac 12 v_a^2 {\bar h}_{00}(0,x_a^0) 
+ {1\over 2} \delta {\bf x_a}^i \delta {\bf x_a}^j \pa_i \pa_j {\bar h}_{00}(0,x_a^0) \right. \nn \\
& \left. \phantom{\frac12} + \delta {\bf x_a}^i \pa_i {\bar h}_{0s}(0,x_a^0) v_a^s
+ {\bar h}_{rs}(0,x_a^0) v_a^r v_a^s \right\} \,.
\end{align} 
As remarked in ref.~\cite{NRGR_paper}, this Lagrangian has two problems: it is not gauge invariant under infinitesimal coordinate transformations and seems to predict that the un-physical modes ${\bar h}_{00}$ and ${\bar h}_{0s}$ can be sourced by the system. 
These two problems are related. 
The solution comes from considering the couplings of radiation to potential modes. 
These contributions are represented by the second and third diagram of \fig{fig:Lrad}, which encodes the complete radiative Lagrangian at order $v^2$. 
Indeed, by means of power counting, one can check that 
\begin{eqnarray}
\Diagram{gd\\ hu} \Diagram{fvA\\ fvA}\sim v^{5/2}\;, 
& \Diagram{gd & hu \\ hu }\sim {v^{5/2}\over \sqrt{L}}\;,
\end{eqnarray}
so that every graph in \fig{fig:Lrad} scales as $\sqrt{L} v^{5/2}$. 
Because the monopole contribution scales as $\sqrt{L} v^{1/2}$, the diagrams in \fig{fig:Lrad} are suppressed by $v^2$: in fact, as we will see below, the third diagram gives rise to the coupling of radiation with the system's quadrupole, which is the term of order $v^2$ in the multipole expansion.

Let us now go through the diagrams of \fig{fig:Lrad} more in detail.
The first graph corresponds to \eq{eq:rad_multi} and it has been discussed above.

The second graph comes from a two-graviton coupling of the point particle similar to the one that gives rise to the seagull diagram~(\ref{eq:seagull}). 
As in that case, the lowest order contribution to a vertex $(h\,\dot x\,\dot x)^2$ corresponds to taking the polarization $00$ for both modes: the only difference is that here we have a potential and a radiation graviton, so that only the first is integrated over. 
The explicit result reads~\cite{NRGR_paper}: 
\begin{equation}
\label{eq:rad_pot}
\mbox{Fig.~\ref{fig:Lrad}(b)} = 
-{i\over 2\mpl}\int dx^0 {G_N m_1 m_2\over |{\bf x}_1 -{\bf x}_2|} {\bar h}_{00}\,,
\end{equation}
which we will comment about when we sum all the diagrams to obtain the final expression.

For the diagram in Fig.~\ref{fig:Lrad}(c), we need the vertex ${\bar h} H^2$, whose expression reads~\cite{NRGR_paper} 
\begin{eqnarray}
\label{eq:hHH}
{\cal L}_{{\bar h} H^2} &=&  {\bf k}^2 {\bar h}_{00} 
\left[
-{1\over 4} H_{\bf k}^{\mu\nu} {H_{-{\bf k}}}_{\mu\nu} 
+ {1\over 8} H_{\bf k}  H_{-{\bf k}} + {H_{\bf k}}^{\mu 0} {H_{-{\bf k}}}_{\mu 0} 
- {1\over 2} {H_{\bf k}}_{00} H_{\bf -k}\right] \\ \nonumber 
\\ \nonumber
& & + {\bf k}^2 {\bar h}_{0i}\left[2  H^{00}_{\bf k} {H_{-\bf k}}_{0i} 
- {H_{\bf k }}_{0i} H_{-\bf k} 
\right] 
\\ \nonumber 
\\ \nonumber
& & + {\bar h}_{ij} 
\left[ 
{1\over 2} {\bf k}_i {\bf k}_j  H_{\bf k}^{\mu\nu} {H_{-{\bf k}}}_{\mu\nu} 
+ {\bf k}^2 {H_{\bf k}}_{i \mu} {H_{-{\bf k}}}_j^\mu 
- {1\over 2}{\bf k}^2  {H_{\bf k}}_{ij} H_{\bf -k} 
- {1\over 4} {\bf k}_i {\bf k}_j  H_{\bf k}  H_{-{\bf k}} \right. \\ \nonumber 
\\ \nonumber
& & \left. \phantom{\qq}
-\delta_{ij}{\bf k}^2 \left(-{1\over 4} H_{\bf k}^{\mu\nu} {H_{-{\bf k}}}_{\mu\nu} 
+ {1\over 8} H_{\bf k}  H_{-{\bf k}} 
\right) \right],
\end{eqnarray}
where the integral over momentum ${\bf k}$ has been suppressed and where the dependence $({\bf X}_{cm},x^0)$ of ${\bar h}_{\mu\nu}$ has been left implicit. 
Given this term, it is easy to show that 
\smallskip
\begin{eqnarray}
\nonumber
\mbox{Fig.~\ref{fig:Lrad}(c)} = {i m_1 m_2\over 8 m^3_{Pl}}\int dx^0\int_{{\bf k}} e^{-i{\bf k}\cdot ({\bf x}_1-{\bf x}_2)} {1\over {\bf k}^4} \left[{3\over 2} {\bf k}^2 {\bar h}_{00} +{1\over 2}{\bf k}^2  {\bar h}_{ii} - {\bf k}_i {\bf k}_j {\bar h}_{ij}\right].
\end{eqnarray}  
\smallskip
To solve the last of the momentum integrals, one needs a generalized version of the Fourier transform~(\ref{eq:Fourier_newton}) 
\begin{equation}
\int_{\bf k} e^{-i{\bf k}\cdot {\bf x}} {{\bf k}_i {\bf k}_j \over {\bf k}^4}  = {1\over 8\pi |{\bf x}|} \left[\delta_{ij} - {{\bf x}_i {\bf x}_j\over |{\bf x}|^2}\right]\,;
\end{equation}
using as well as the equations of motion at leading order, the final result is 
\begin{eqnarray}
\label{eq:rad_quad}
\nonumber
\mbox {Fig.~\ref{fig:Lrad}(c)} = {i\over m_{Pl}} \int dx^0 
\left[{3 G_N m_1 m_2\over 2 |{\bf x}_1 -{\bf x}_2|} {\bar h}_{00} 
-{1\over 2}\sum_a m_a {{\bf x}_a}_i {\ddot{\bf x}_a}{}_j  {\bar h}_{ij}\right].
\end{eqnarray}

Adding together the results of \eqs{eq:rad_multi}{eq:rad_quad} to what one obtains by supplementing \eq{eq:rad_pot} with its mirror image under exchange of the particle label, one finds the complete Lagrangian at order $\sqrt{L} v^{5/2}$
\begin{eqnarray}
\label{eq:Lv2}
\nonumber
L_{v^2}[{\bar h}] &=& 
- {1\over 2 \mpl} {\bar h}_{00}\left[{1\over 2}\sum_a m_a {\bf v}^2_a 
- {G_N m_1 m_2\over |{\bf x}_1-{\bf x}_2|}\right] 
- {1\over 2 \mpl} \epsilon_{ijk} {\bf L}_k \partial_j {\bar h}_{0i}\\
 & & +{1\over 2 \mpl}\sum_a m_a {\bf x}^i_a {\bf x}^j_a R_{0i0j}\,,
\end{eqnarray}
where:
\bd
\item The first term is the coupling of ${\bar h}_{00}$ to the Newtonian energy of the two-particle system; this term can be regarded as a correction to the mass monopole of the source given by the kinetic and gravitational energy
\begin{equation}
\sum_a m_a\rightarrow \sum_a m_a \left(1+{1\over 2}{\bf v}^2_a\right) 
- {G_N m_1 m_2\over |{\bf x}_1-{\bf x}_2|}\,.
\end{equation}
\item The second term is a coupling of the graviton ${\bar h}_{0i}$ to the total mechanical angular momentum of the system, ${\bf L}=\sum_a {\bf x}_a\times m_a {\bf v}_a$. 
Both the mass monopole and the angular momentum are conserved at this order in the velocity expansion: therefore, ${\bar h}_{00}$ and ${\bar h}_{0i}$ do not represent physical contributions to radiation. 
\item The last term is the coupling of the source moment $\sum_a m_a {\bf x}^i_a {\bf x}^j_a$ to the (linearized) Riemann tensor of the radiation field, which reads
\begin{equation}
R_{0i0j} = \frac12 \left( \pa_0^2 {\bar h}_{ij} + \pa_i \pa_j {\bar h}_{00} - 
\pa_0 \pa_i {\bar h}_{0j} - \pa_0 \pa_j {\bar h}_{0i} \right)  \,;
\end{equation}
one can show that $R_{00}= R_{0i0i}=0$ for on-shell graviton matrix elements, so that radiation only couples to the traceless quadrupole moment of the source
\begin{equation}
Q^{ij} = \sum_a m_a \left({\bf x}^i_a {\bf x}^j_a
-{1\over 3}{\bf x}^2_a\delta_{ij}\right) \,.
\end{equation}
\ed

\begin{figure*}[t]
\centering
\includegraphics[width=0.50\textwidth,angle=0]{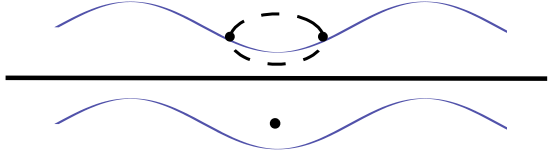}
\caption{A schematic representation of what the EFT for binary systems is about: at length scales larger than the orbital radius, and smaller than the typical wavelength of radiation, the two-body system can be regarded as a point particle with suitable gravitational interactions. Figure courtesy of Chad Galley.} 
\label{fig:bin_eft}
\end{figure*}

\noindent 
The derivation of the radiation Lagrangian is an interesting example of the systematic approach of EFT. 
In fact, because of power counting, the radiation Lagrangian at ${\cal O}(v^2)$ cannot just contain terms from the multipole expansion of the point particles couplings to radiation, i.e. it cannot be given just by \eq{eq:rad_multi}; at the same order in the velocity one finds contributions from diagrams with graviton self-interactions. 
This is consistent with the strong equivalence principle, according to which gravity couples in the same way to all sources of energy-momentum, including the gravitational field generated by the source.

\begin{figure*}[t]
\centering
\includegraphics[width=0.25\textwidth,angle=0]{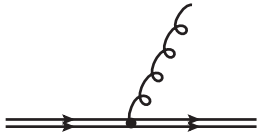}
\caption{Feynman diagram representing the one-graviton emission amplitude from the binary system. At length scales larger of the order of the orbital radius, and smaller than the typical wavelength of radiation, the two-body system can be regarded as a point particle with suitable gravitational interactions~(see \eq{eq:Lv2}).}
\label{fig:rad_vert}
\end{figure*}

\smallskip
What \eq{eq:Lv2} represents is the matching of the short-distance theory valid below the orbital scale $r$ to the long-wavelength theory valid at scales $\ell \simeq r$\,. 
This matching constitutes the second step in the construction of NRGR as a tower of effective theories that we discussed in the introduction of this chapter and is described in \fig{fig:bin_eft}: as it is evident from this figure, the orbital separation between the constituent objects is not {\it resolved} at this level and the binary system itself is now described as an effective point particle. 
As anticipated, the scale $r$ is only present in the Wilson coefficients of the operators of the theory~(\ref{eq:Lv2}). 
The Feynman diagrams corresponding to this theory are of the type of \fig{fig:rad_vert}, where we have used a double-line notation to indicate that this is a vertex in the EFT above the scale $\ell \simeq r$. 
The strength of the interaction vertex is one of the couplings of \eq{eq:Lv2}. 

Once we have calculated the action $S_{NR}[{\bar h},(x_a)]$, the EFT prescription requires that, in order to compute observables, we integrate out the radiation modes and obtain a theory where the only degrees of freedom are the particle world-lines. 
As discussed at the end of \Sec{sec:pchH}, this action can be obtained by summing the Feynman diagrams that have no external graviton lines like the one of \fig{fig:self} that we report here for convenience.  
\begin{figure}
\begin{center}
\includegraphics[width=0.4\textwidth]{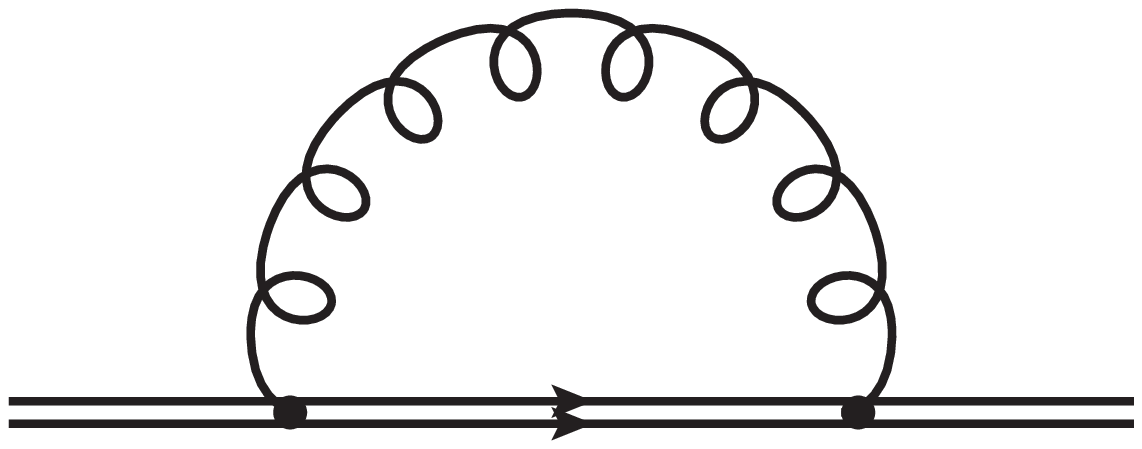}
\caption{The self-energy diagram whose imaginary part gives the 
radiated power.} 
\label{fig:self_bis} 
\end{center}
\end{figure}
Now we know that the interaction vertices in this diagram are those of \eq{eq:Lv2}, among which only the last one is a physical radiative coupling: taking the imaginary part of a diagram with two such couplings in the vertices, one can then derive the leading order quadrupole radiation formula. 
An alternative to taking the imaginary of diagrams like the one in \fig{fig:self_bis} is to compute  the square of the one-graviton emission amplitudes, whose corresponding Feynman diagram is the one in \fig{fig:rad_vert}. 
This is indeed what has been found in \eq{eq:dpow} by direct calculation and replaces {\it de facto} the explicit computation of the effective action $S_{\mathit eff}(x_a)$.

\smallskip

To end this section I would like to emphasize that, for the purpose of characterizing a GW signal, it is not enough to consider the diagrams of \fig{fig:Lrad}, i.e. the dissipative dynamics of the system. 
In fact, future GW detections at \ifos\ will probe the non-linear dynamics of GR in both its conservative and its radiative regime through monitoring the phase of the waveform: as I have already shown more than once by now~\footnote{See the discussions about \eq{eq:phase} in \Sec{sec:eft_iso} and \eq{eq:gwphase} in \Sec{sec:PPN_GW}.}, a typical expression for the phase is the one corresponding to an extreme-mass-ratio 
\begin{equation}
\label{eq:phase_emri}
\Delta \varphi_{GW}(t) = \int_{t_0}^t d\tau\,\omega_{GW}(\tau) = {2\over G_N M}\int^{v_0} _{v(t)}d v' v'^3 {dE/dv'\over P(v')}\,,
\end{equation}  
where it is evident that the phase receives contributions from both the energy function $E$ and the power emitted $P$. 
As a consequence, tests of gravity which involve the orbital decay, but are not limited to it, should take both regimes into account. 
This is indeed the case of the parametrized post-Keplerian test $(\gamma, \dot \omega\,, \dot P)$, that I have discussed in \Sec{sec:PPK}, where the orbital decay is measured together with parameters that do not belong the dissipative dynamics.
Another example is represented by the phenomenological tests that I have investigated personally. 
These studies have been inspired by the NRGR approach and are discussed in the following two chapters.

%%%%%%%%%%%%%%%%%%%%%%%%%%%%%%%%%%
\subsection{Further examples of diagrams corresponding to known effects}
\label{sec:examples}
%%%%%%%%%%%%%%%%%%%%%%%%%%%%%%%%%%

In order to further elucidate the rationale behind the decomposition of gravitational perturbations in high frequency modes and a slowly varying background (\ref{eq:Hh}), it is instructive to present here some Feynman diagrams that correspond to physical effects pertaining the different regimes of conservative and radiative dynamics.
An interesting case is the one related with spin phenomena. 
The inclusion of spin degrees of freedom in NRGR has been addressed in refs.~\cite{Porto-Spin_Intro,Porto-Spin_EIH}: here, the conservative dynamics is studied at leading order and the corresponding spin-orbit and spin-spin potentials are reproduced. 
Spin effects are incorporated by adding the world-line degrees of freedom $\Lambda_a^J(\lambda)$, where $\lambda$ is the affine parameter of the world-line and $\Lambda_a^J$ is the boost that transforms the locally flat frame, labelled by $a$, to the co-rotating frame labelled by $J$. 
These frames are defined by {\it vierbeins}, which are given by $e^\mu_a$ and $e^\mu_I= e^\mu_a \Lambda^a_I$ for the locally flat and co-rotating frame respectively and verify $e^a_{\mu}e^b_{\nu} g^{\mu\nu} = \eta^{ab}$. 
Then one can introduce the generalized angular velocity given by $\Omega^{\mu\nu}=e^{\mu J}(De^\nu_J/d\lambda)$. 
Finally, the spin is defined as the tensor $S_{\mu\nu}$ that is the conjugate momentum to $\Omega_{\mu \nu}$ so that the form of the world-line action reads~\cite{Porto-Spin_EIH}
\begin{equation} 
\label{eq:action_spin}
S=-\sum_i \left(\int p^\mu_i u^i_{\mu}d\lambda_i + \int
\frac{1}{2}S_i^{IJ}\Omega^i_{IJ} d\lambda_i\right),
\end{equation}
where the sum extends over the binary constituents and $S^{IJ} \equiv S^{\mu\nu}e^I_\mu e^J_\nu$.  
From this action one can derive the Mathisson-Papapetrou equations of motion for spinning test particles~\cite{Mathisson,Papapetrou}. 
The spin-gravity coupling in (\ref{eq:action_spin}) can be rewritten by introducing the Ricci rotation coefficients $\omega_\mu^{ab} = e^b_\nu D_\mu e^{a\nu}$~\cite{Porto-Spin_Intro}
\begin{equation} 
\label{eq:spin_grav}
S_{spin-grav} =  -\frac{1}{2} \int S_{Lab}\omega^{ab}_\mu u^\mu d\lambda\,,
\end{equation}
with $S_L^{ab} \equiv S^{\mu\nu}e_{\mu}^a e_{\nu}^b$, the spin in the locally flat frame. 
Expanding (\ref{eq:spin_grav}) in the weak gravity limit one obtains the Feynman rules for NRGR with spin~\cite{Porto-Spin_Intro,Porto-Spin_EIH}; for example, at lowest order, the spin  coupling of a potential graviton to a matter source reads 
\be
\label{eq:Lspin}
L_{SPIN}^{(0)} = \frac{1}{2\mpl}S_L^{ik}\pa_k H_{i0}\;.
\ee
By making use of similar rules for higher order terms, one can derive the conservative dynamics for a binary system in the case of spinning bodies. 
At leading order the spin-orbit and spin-spin interactions are represented by the Feynman diagrams of \fig{fig:S1S2_cons}: as one can see from the figure, with respect to Newtonian potential, the spin-orbit interaction is ${\cal O}(v^3)$ while the spin-spin is ${\cal O}(v^4)$, 
i.e. these interactions constitute higher PN corrections than the EIH Lagrangian of \fig{fig:EIH}: this explains why we did not take spin into account so far.
From \Sec{sec:PPK} we know that it has recently become possible to measure and constrain the general relativistic spin-orbit interaction by pulsar timing, even if with a slightly worse accuracy than other post-Keplerian parameters (cfr. \fig{fig:Kramer_Test}). 

As discussed at the end of the previous section, future GW detections at \ifos\ will probe the non-linear dynamics of GR in both its conservative and its radiative regimes through monitoring the phase of the waveform. 
Concerning the dissipative dynamics, the presence of spin affects the radiative multipole moments, like the mass quadrupole $I_{ij}$ which equals $Q_{ij}$ at leading order. 
At next-to-leading order, $I_{ij}$ receives contributions from the spin of both objects: these terms have been recently calculated in ref.~\cite{Porto:2010zg} and the corresponding diagrams are represented in \fig{fig:S1S2_rad}.
The confrontation of \fig{fig:S1S2_cons} with \fig{fig:S1S2_rad} constitutes another manifestation of the different roles that potential and radiation modes have in NRGR.

\begin{figure}[t]
 \centering 
  \subfigure[]{
   \includegraphics[width=6cm]{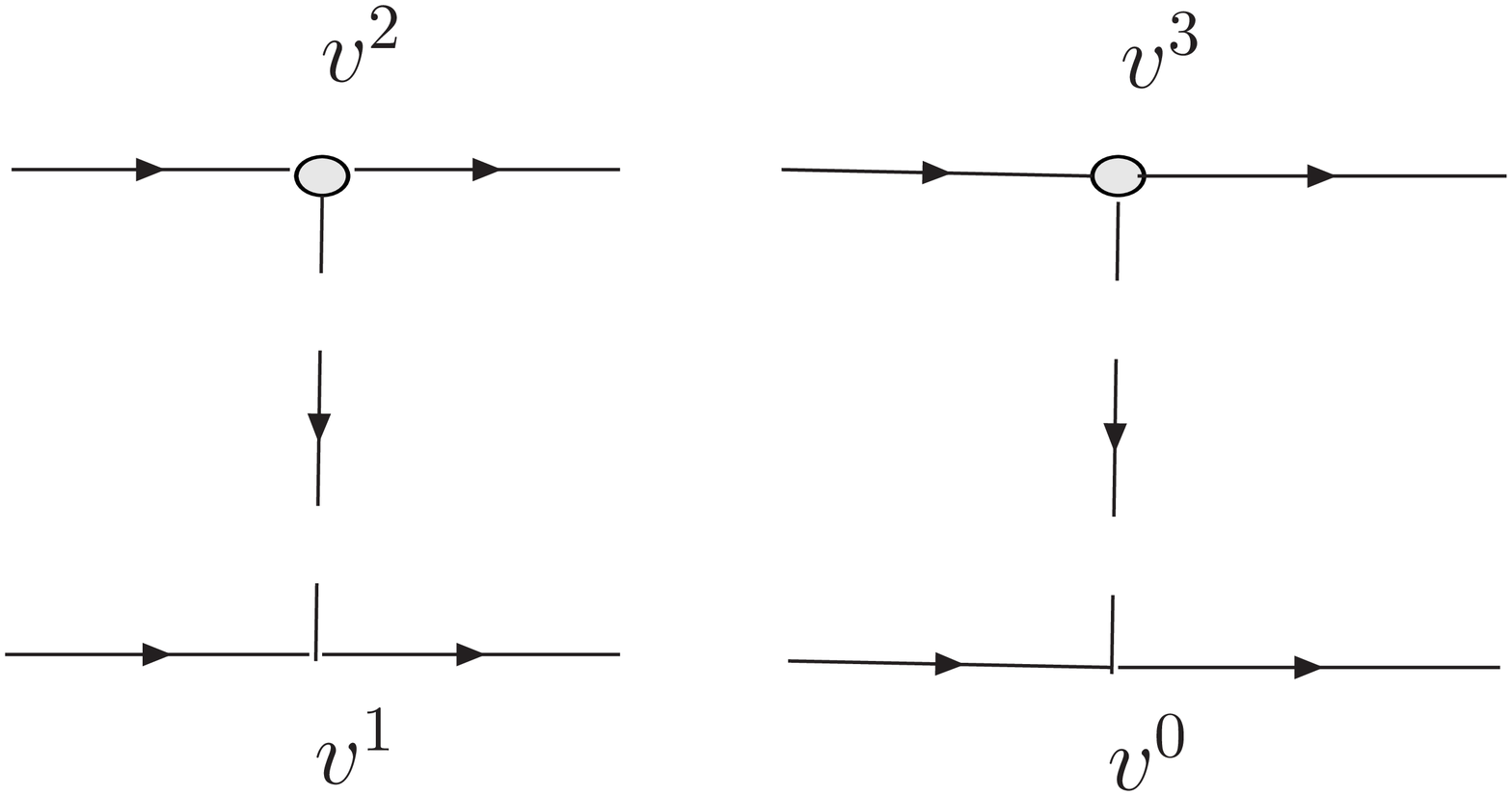} } 
%		\label{fig:} 
     \subfigure[]{
      \includegraphics[width=3.1cm]{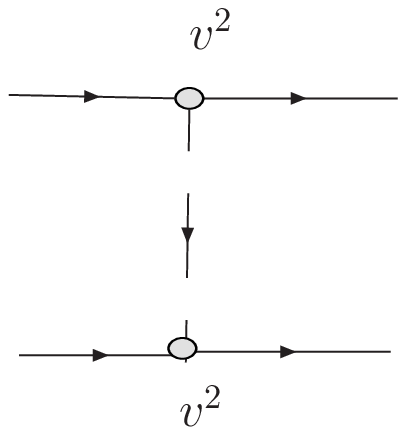} }
       \caption{
       Panel (a): Feynman diagrams representing the spin-orbit interaction at leading order. 
       Panel (b): Feynman diagram representing the spin-spin interaction at leading order. 
       A grey blob represents an insertion of the spin operator in the point particle action; 
       the factors of $v^n$ correspond to the suppression with respect to the leading order 
       coupling of a potential mode to a world-line (see \eq{eq:00vert}).
Figures taken from ref.~\cite{Porto-Spin_Intro}. }
\label{fig:S1S2_cons}
\end{figure}
\begin{figure}[t]
    \centering
    \includegraphics[width=11cm]{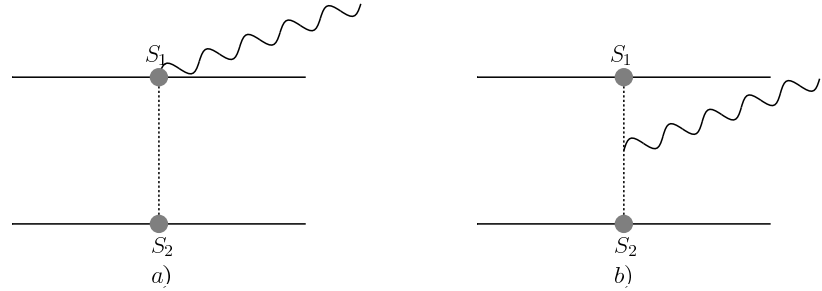}
\caption{NRGR diagrams describing some of the spin contributions to the quadrupole moment at next-to-leading order. 
The spin-graviton vertex of particle $i$ is still represented with a grey blob in these diagrams it is labeled by $S_i$. 
Figure taken from ref.~\cite{Porto:2010zg}.} 
\label{fig:S1S2_rad}
\end{figure}

Another interesting example of the complementarity of $H$ and ${\bar h}$ is provided by phenomena which are still to be investigated in NRGR: the tail of the radiation, discussed in ref.~\cite{Blanchet:1987wq}, and the non-linear memory effect, predicted in refs.~\cite{Christo-Mem,Blanchet:1992p41}. 
These are peculiar non-linearities of GR and their detection would then be an invaluable confirmation of the validity of the theory. 
The possibility of testing GR through detection of the tail effect has been considered in ref.~\cite{Blan_Sathya-Tail} while the importance of the non-linear memory has been recently investigated in ref.~\cite{Fava-Mem}.

The tails of GWs result from the non-linear interaction between the quadrupole radiation generated by an isolated system with total mass--energy $M$ and the static monopole field associated with $M$. 
Their contributions to the field at large distances from the system include a particular effect of modulation of the phase in the Fourier domain, which has $M$ as a prefactor and depends on the frequency $\omega_{GW}$ as "$\omega_{GW} \ln \omega_{GW}$". 

On the other hand, the non-linear memory effect is a slowly-growing, non-oscilla\-tory contribution to the GW amplitude. 
It originates from the fact that, in a non-linear theory like GR, GWs can themselves constitute the source of GWs. 
In an ideal interferometer a GW with memory causes a permanent displacement of the test masses that persists after the wave has passed. 
Remarkably, the non-linear memory affects the signal {\it amplitude} starting at leading order, i.e. the same as the Newtonian quadrupole; however, due to the {\it non-oscillatory} behavior, detecting the memory at \ifos\ is cumbersome.

In both the tail and memory phenomena, gravity non-linearities affect the propagation of the emitted waves. 
Therefore, in NRGR both effects will be represented by diagrams where the orbital scale has been integrated out and the outermost field line is a radiation graviton $\bar h$ resulting from a non-linear interaction with other gravitons. 
The situation is described in \fig{fig:tail_mem} where the arrows on the double lines indicate the flow of time for the composite binary source.
In the case of the non-linear memory, panel (b), the outgoing graviton results from the interaction between a wave emitted at time $t_1$ and a wave emitted at time $t_2>t_1$, so the three-graviton vertex involves only gravitons of radiative type. 
For what concerns the tail instead, panel (a), the GW produced at a time $t_2$ back-scatters off the pre-existent background gravitational field~\footnote{This long-range static field is due to the two-body system so it is described by a potential mode that is not the same as the one integrated out to take care of the orbital dynamics.}: for this reason, one of the gravitons in the three-vertex is indicated with a dashed line.

\begin{figure}[htbp]
	\centering 
	\subfigure[]{
		\includegraphics[scale=.50]{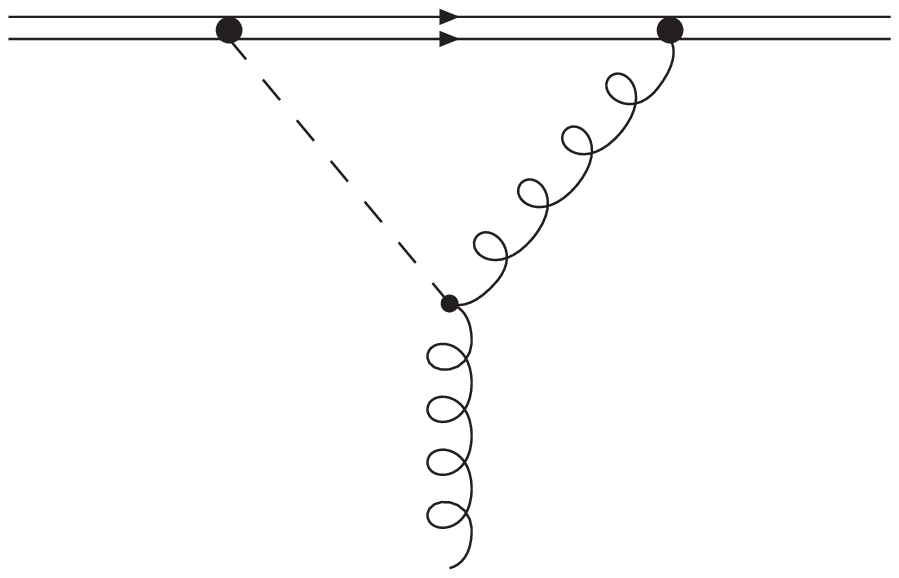}
		\label{fig:tail} } 
	\subfigure[]{
		\includegraphics[scale=.55]{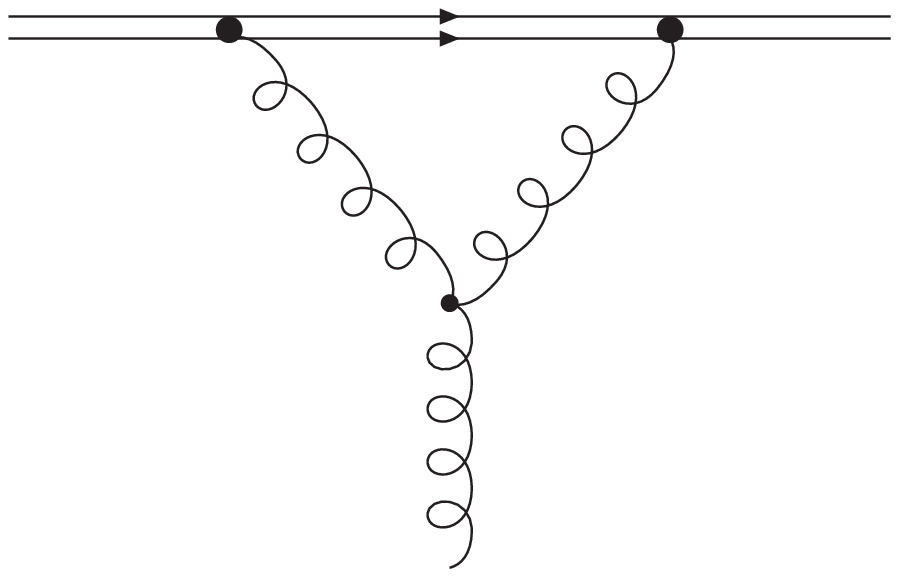}
		\label{fig:memory} } 
	\caption[Optional caption for list of figures]
	{NRGR description of the non-linearities affecting the vacuum propagation 
	of a GW: panel (a) reports the tail of the radiation, panel (b) the non-linear 
	memory. 
	Figures stemming from a discussion with Ira Rothstein at Carnegie Mellon University.}
	\label{fig:tail_mem}
\end{figure}
%

%%%%%%%%%%%%%%%%%%%%%%%%%%%%%%%%%%
\section{Final remarks} \label{sec:eft_remarks}
%%%%%%%%%%%%%%%%%%%%%%%%%%%%%%%%%%

NRGR is adequate for describing the emission of gravitational radiation from binaries of comparable masses. 
In this case the perturbative expansion is a series in the weak-field/slow-motion parameter $\eps$ of the usual PN approximation.
Comparable-mass inspirals are the relevant binary sources in the frequency band of first-genera\-tion gravitational antennas; in the future, the space-based interferometer LISA will also be sensitive to binaries which are very asymmetric in the component masses, the so-called extreme-mass-ratio inspirals (EMRIs). 
A representative example of these systems is constituted by a stellar-mass BH inspiraling a super-massive BH so that their mass ratio $\nu=m_1 m_2 / (m_1 + m_2)^2$ is typically of the order of $10^{-6}$ and naturally lends itself to be used as a perturbative expansion parameter different from the PN $\eps \sim v^2 \sim m\,G_N/r$. 
The light object can be thought of as a test particle moving in the gravitational background produced by the super-massive BH so the standard picture in this context is BH perturbation theory. 
Analogously, the EFT approach tailored to describe EMRIs will take advantage of the smallness of $\nu$, which is not the case for NRGR. 
I do not present this EFT here since I did not make use of it for the investigations presented in this thesis.
The interested reader can find more details about it in the works of refs.~\cite{Galley:2008ih,Galley:2009p69} by Chad Galley who took up its studying in recent years for accurately modeling EMRIs in view of LISA.

To end this chapter, it should be mentioned that the NRGR treatment of the dynamics of binary systems has some analogies with the approaches that were traditionally used before its introduction, notably the one from Blanchet, Damour and collaborators (see the review~\cite{Blan-LRR:06} for a complete treatment and full list of citations). 
In this approach space-time is divided in zones and the matching is done 
between two different expansions which are possible for the gravitational 
field according to the zone.
With reference to Fig.~\ref{fig:dam_blan}, the procedure can be sketched 
as follows. 
The zone comprising the binary system and its neighbors until a 
length scale $\ell = R_1$ is the so-called \emph{near zone}: outside 
this zone the metric admits the weak-field post-Minkowskian expansion 
which is controlled by Newton's constant~$G_N$. 
The slow-motion PN expansion is rather possible below 
a length scale $\ell = R_2$. 
This scale is of the order of the radiation wavelength, so it is bigger 
than $R_1$: there exists a region where both approximations are 
valid which makes it possible to match the two expansions.

\begin{figure}[t]
\begin{center}
\includegraphics[width=0.8\textwidth]{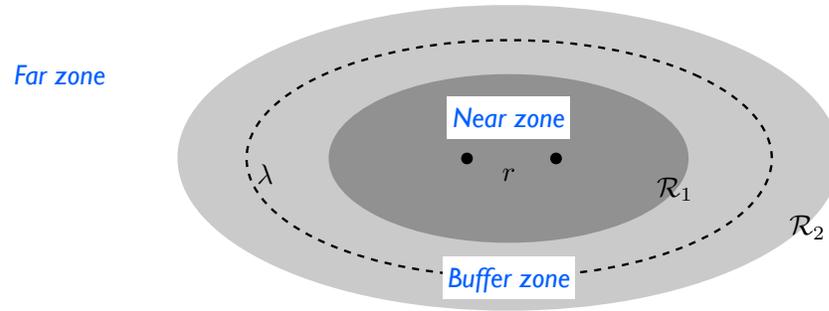}
\caption{A pictorial description of the approach from Blanchet and Damour 
(see the review~\cite{Blan-LRR:06} for a complete treatment and full list of citations).
Figure courtesy of Chad Galley.}
\label{fig:dam_blan}
\end{center}
\end{figure}

For what concerns the differences between the NRGR framework and the standard approach one could remark what follows.
First of all, NRGR operates at the level of the action, which is a scalar under Lorentz transformations; on the contrary, the standard approach works with the equations of motion and has to worry about vector and rank-two tensor indices. 
The standard approach then matches metric components, while NRGR matches actions. 
The former keeps higher order terms until the end when they drop out; the latter can rather take advantage of an explicit power counting that, ahead of any calculation, enables one to select the interaction terms needed at the desired perturbative order.

%% file: Ric_Paper/Ric_Paper.tex
%%%%%%%%%%%%%%%%%%%%%%%%%%%%%%%%%
\chapter[Classical energy-momentum tensor renormalization \\ via 
effective field theory methods]
    {Classical energy-momentum tensor renormalization via 
effective field theory methods %
  \chaptermark{Classical energy-momentum tensor renormalization} } 
    \chaptermark{Classical energy-momentum tensor renormalization}
\label{chap:Ric}

% where:    
% "[]" is for the table of contents
% the subsequent "{}" is for where the chapter actually is
% 1st "chaptermark" is for the first page containing the chapter
% 2nd "chaptermark" is for the other pages containing the chapter

%%%%%%%%%%%%%%%%%%%%%%%%%%%%%%%%%

In this chapter I present the work I have conducted with Riccardo Sturani, who has been part of my group at the University of Geneva during my PhD studies.
In this work~\cite{UC:08} we applied the EFT of ref.~\cite{NRGR_paper} to study point-like and string-like sources in the context of scalar-tensor theories of gravity. 
Within this framework we computed the classical renormalization of the energy-momentum tensor at next-to-leading order: for what concerns helicity-2 gravitons, this means 
up to ﬁrst PN order, whereas regarding helicity-0 modes this amounts to the (non-derivative) trilinear terms in the interaction. 
In such a way, we could write down the corrections to the standard (Newtonian) gravitational potential and to the extra-scalar potential for both types of sources we investigated.
In the case of point particles, we obtained the usual results concerning the extra scalar~$\psi$, which we endowed with a mass~$m_\psi$: at momenta $q$ higher than the mass, the effective potential mediated by $\psi$ has a logarithmic profile, rather than the $1/r^2$ behavior typical of 1PN terms in Einstein gravity; at low momenta, $q\ll m_\psi$, the Yukawa suppression takes place as usual.
In the case of one-dimensional objects, we calculated the renormalization of the tension for cosmic strings. 
Before us, this issue had been investigated in ref.~\cite{Dabholkar:1989jt}, by Dabholkar and Harvey (DH), and in ref.~\cite{Buonanno:1998kx}, by Buonanno and Damour (BD).  
These two works conducted different analysis, leading to apparently conflicting results for the string-tension renormalization. 
Even if this discrepancy had already been explained in ref.~\cite{Buonanno:1998kx}, we found it interesting to re-analyize the whole subject in the context of NRGR. 
Now, to enter the subject in more detail, I will partially follow the introduction given in BD. 

The importance of cosmic strings is due to the fact that the existence of these one-dimensional topological defects is a generic feature of grand uniﬁed theories and of cosmological models motivated by string theory. %: in this last context, they are also referred to as super-strings. 
It is believed that cosmic strings form abundantly during phase transitions in the early universe and that they represent relevant sources of GWs in the form of a stochastic background~\cite{Vilenkin:81,GW_bck-Maggiore}. 
In this context, a significant scientific result of first generation ground-based \ifos\ has been accomplished in 2009: in fact, the LIGO-Virgo collaboration could put a direct upper bound on GWs of stochastic origin~\cite{GW_bck-Nat} in the same frequency region of integral constraints previously obtained from Big Bang nucleosynthesis~\cite{GW_bck-Maggiore,GW_bck-BBN} and from the cosmic microwave background (CMB)~\cite{GW_bck_CMB}. 
The situation is described in \fig{fig:GW_bck} taken from ref.~\cite{GW_bck-Nat}: we refer to this work for complete discussion and full list of references. 
%Deriving the effective action of a cosmic string and studying its parameters is relevant to GW observations. 
%What is important for the present discussion is that theoretical modeling of cosmic strings is relevant before direct detection of a GW signal from cosmic strings, constraining the amplitude of a stochastic background can teach us something about the dynamics of cosmic strings and the range of values of parameters like the string tension~\cite{Steer-GW_strings}.
%This is the aim of theoretical studies like those of DH, BD and the one I have conducted.
%As an introduction to my work, I summarize here the investigation by BD. 

Loops of oscillating cosmic strings can produce other types of radiation than the gravitational one, according to the fields they couple to: for example, one can consider a dilaton scalar $\Phi$ and an axion field $B_{\mu\nu}$ at the same time as the gravitational field~$g_{\mu\nu}$. 
All these fields back-react on the string source, thus leading to a self-interaction of the string with itself. 
In the limit of an infinitely thin string, the self-interaction is divergent, much in the same way as the self-interaction of a point-like electron with its own electromagnetic field. 
In this last case, Dirac noticed that the divergence could be cured by renormalizing the mass, i.e. by assuming that the bare mass of the electron depends on a cutoff radius $\delta$~\cite{Dirac:1938nz}: 
\be
\label{eq:m_delta}
m(\delta) = m_R - \frac{e^2}{2\delta} \,,
\ee
where $m(\delta)$ is the UV divergent bare mass of the electron, $m_R$ its renormalized mass and $e$ its electric charge.
Notably, this cutoff dependence is consistent with the assumption that $m(\delta)$ represents the total mass-energy of the particle plus that of the electromagnetic field contained in a sphere of radius $\delta$; in other words:
\be 
\label{eq:m_diff}
m(\delta_2) - m(\delta_1) = +\int_{\delta_1}^{\delta_2} d^3x \, T^{00}_{field}(x)
\ee
where $T^{00}_{field}(x) = E(r)^2/(8\pi)=e^2/(8\pi r^4)$, with $r$ the radial coordinate.
If one generalizes this argument to the case of a string in four dimensions, the linearly-divergent electron mass translates as a logarithmically-divergent string tension~$\mu$
\be
\label{eq:mu_delta}
\mu (\delta) = \mu_R + C \log \(\frac{\Delta_R}{\delta}\) \,,
\ee
where $C$ is a renormalization coefficient that we will discuss next and $\Delta_R$ is an arbitrary length which has to be introduced because of the logarithmic dependence of $\mu$ on $\delta$.
The coefficient $C$ receives contributions form each fundamental field with which the string interacts: if one considers the case of the gravitational field $g_{\mu\nu}$, the dilaton scalar $\Phi$ and the axion field $B_{\mu\nu}$, $C$ is given by
\be
\label{eq:Cs}
C\equiv C_{g} + C_{\Phi} + C_B \,.
\ee
In the spirit of Dirac's argument, the individual renormalization coefficients $C_i$'s would have to be determined by the energy of the corresponding field. 
This is the method used by DH and us: it gives the correct vanishing value for $C$ in the case of fundamental strings, for which a peculiar link exists among the couplings of the various fields to the string (see the discussion below \eq{eq:az_str}). 
On the other hand, the individual values of the $C_i$'s obtained by DH and us do not match the ones calculated by means of other approaches in the literature, notably that of BD.
This is due to the fact that BD calculated the renormalization of the string tension by means of a different quantity, as explained by BD and also by us in ref.~\cite{UC:08}. 
While DH and we computed the energy-momentum tensor for a string, BD derived its effective action. 
In the latter case, by taking the coincidence limit in 4-D coordinate space, BD found divergent  contributions to the total interaction energy which are localized on the source. 
These terms are not obtained within the approach based on the energy-momentum computation. 
However, the effect of the divergent pieces is to contribute to the renormalization of the {\it bare} unobservable tension of the string: therefore, on physical grounds, the results of the two approaches are equivalent. 

In the EFT terminology that we used, this difference can be grasped more clearly once expressed in terms of Feynman diagrams. 
DH and we calculated a quantity that corresponds to \fig{str_1PN}, where there is an external line corresponding to the gravitational field; on the contrary, BD obtained the amplitude of the self-interaction process depicted in \fig{damour}.
A more complete discussion is contained in our work that I report next. 
In this publication a different notation is adopted with respect to the rest of the thesis for what concerns fields in Feynman diagrams: here, curly lines are used to represent the scalar degree of freedom of the gravitational field (see \eq{met_nr} for the metric parametrization); whereas  
a double line here describes a string source, instead of an unresolved binary system as in Chapters \ref{chap:EFT} and \ref{chap:Group}.

%As pointed out by BD, the reason for this discrepancy is due to the fact that, in the case of the gravitational and dilaton scalar fields, the total interaction energy cannot be unambiguously localized only in the field itself: there are indeed contributions which are localized on the sources. 
%This is not the case for the axion field because its coupling to the string does not depend on the metric $g_{\mu \nu}$: therefore, the $B$ field does not contribute to the total energy-momentum tensor $T^{\mu\nu}\propto \delta S/\delta g_{\mu \nu}$, where $S$ is the total action for a string.

%For these reasons, BD set out to calculate the effective action of a cosmic string up to first order in interaction for the three aforementioned fields.

%
\begin{figure}
\begin{center}
\includegraphics[width=4in]{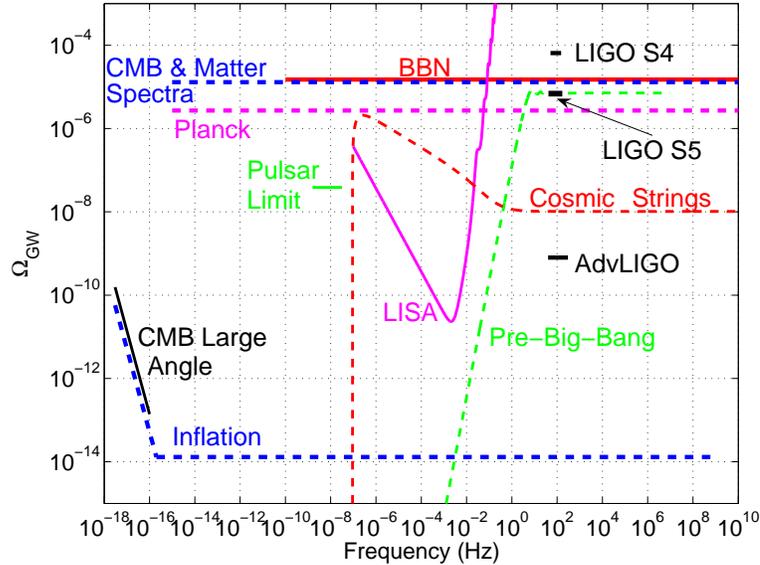}
\caption{
Comparison of different measurements and models of stochastic backgrounds of GW.  
The upper bound derived in ref.~\cite{GW_bck-Nat} is the one pointed at by the arrow and refers to LIGO's~\cite{LIGO} fifth scientific run S5: at 95\% confidence level, it is given by $\Omega_0 < 6.9 \times 10^{-6}$ and applies in the frequency band 41.5-169.25 Hz. 
This bound should be compared with the previous LIGO S4 result and with the projected sensitivity of Advanced LIGO~\cite{ALIGO}. 
%Note that the corresponding S5 95\% upper bound on the total gravitational-wave energy density in this band, assuming frequency independent spectrum, is $9.7 \times 10^{-6}$. 
The bound due to BBN \cite{GW_bck-Maggiore,GW_bck-BBN} is an indirect integral bound: it applies 
%to $\Omega_{\rm BBN} = \int \Omega_{\rm GW} (f) d(\ln f)$ (and not to the density $\Omega_{\rm GW} (f)$) 
over the frequency band denoted by the corresponding horizontal line. 
A similar integral bound (over the range $10^{-15}-10^{10}$ Hz) can be placed using CMB and matter power spectra~\cite{GW_bck_CMB}. 
Projected sensitivities of the satellite-based CMB experiment Planck~\cite{Planck} and the space-based GW \ifo\ LISA~\cite{LISA} are also shown.
The pulsar bound \cite{GW_bck_pulsar} is based on the fluctuations in the pulse arrival times of millisecond pulsars and applies at frequencies around $10^{-8}$ Hz. 
Measurements of the CMB at large angular scales constrain the possible redshift of CMB photons due to the stochastic backgrounds of GW: this limits the amplitude of the backgrounds at largest wavelengths (smallest frequencies)~\cite{GW_bck-BBN}. 
Examples of inflationary, cosmic strings and pre-Big-Bang models are also shown: however, the amplitude and the spectral shape can vary significantly as a function of model parameters.  
Figure taken from ref.~\cite{GW_bck-Nat}: refer to it for complete discussion and full list of references.}
\label{fig:GW_bck}
\end{center}
\end{figure}

\newpage

\begin{center}
{\large General Relativity and Gravitation {\bf 42}, 2491-2509, (2010)}
\par\end{center}{\large \par}

\begin{center}
\textbf{\Large Classical energy-momentum tensor renormalization via 
effective field theory methods}
\par\end{center}{\Large \par}

\begin{center}
{\large Umberto Cannella and Riccardo Sturani}
\par\end{center}{\large \par}

%\maketitle
\vspace{1.2cm}

%newcommands******************************* + defs Maggiore
%\newcommand{\be}{\begin{eqnarray}}
%\newcommand{\ee}{\end{eqnarray}}
%\newcommand{\bdm}{\begin{displaymath}}
%\newcommand{\edm}{\end{displaymath}}
%\newcommand{\ds}{\displaystyle}
%\newcommand{\ba}{\begin{array}} % -> \bea    % OK
%\newcommand{\ea}{\end{array}}  % -> \eea       % OK
%\newcommand{\pt}[1]{\left(#1\right)}  %  parentesi tonda
%\newcommand{\paq}[1]{\left[#1\right]} %  parentesi quadra
%\newcommand{\dpa}{\partial} % -> \pa  % OK
%\newcommand{\az}{\mathcal{S}}         % OK
%\newcommand{\K}{{\bf k}} % -> \vk      % OK
%\newcommand{\Q}{{\bf q}} % -> \vq     % OK
%******************************************

%%%%%%%%%%%%%%%%%%%%%%%%%%%%%%%%%%%%%%%
\subsubsection{abstract}
%%%%%%%%%%%%%%%%%%%%%%%%%%%%%%%%%%%%%%%

We apply the Effective Field Theory approach to General Relativity, introduced 
by Goldberger and Rothstein, to study point-like and string-like sources in the
context of scalar-tensor theories of gravity.
Within this framework we compute the classical energy-momentum tensor 
renormalization to 
first Post-Newtonian order or, in the case of extra scalar fields, up to the 
(non-derivative) trilinear interaction terms: this allows to write down the 
corrections to the standard (Newtonian) gravitational potential and to the 
extra-scalar potential. 
In the case of one-dimensional extended sources we give an alternative 
derivation of the renormalization of the string tension enabling a re-analysis 
of the discrepancy between the results obtained by Dabholkar and Harvey in one 
paper and by Buonanno and Damour in another, already discussed in the latter.

\vspace{0.8cm}

\noindent
{\small DOI: 10.1007/s10714-010-0998-0 \hfill 
 PACS numbers: 04.20.-q,04.50.Kd,11.10.-z}

\vspace{1.0cm}

%%%%%%%%%%%%%%%%%%%%%%%%%%%%%%%%%%%%
\section{Introduction} \label{intro}
%%%%%%%%%%%%%%%%%%%%%%%%%%%%%%%%%%%%

We consider in this work the \emph{classical} renormalization of the 
energy-momen\-tum tensor (EMT) of fundamental particles and strings 
due to their interaction with long range fundamental fields, including 
standard gravity.
The gravitational self-energy of a massive body for instance, arises 
because of gravitons' self-interactions, it can be described as an 
effective renormalization of the massive body EMT
and it is fully classical having its analog in Newtonian physics.
Such self-interactions, even if they involve point-like particles, are not 
divergent when gravity is present, as on general grounds General 
Relativity imposes a lower limit on the size of massive objects: 
their Schwartzchild radii.

In the case of one-dimensional extended objects like strings, no 
horizon analog is present and no fundamental lower limit can be 
imposed on their size: classical contributions to the EMT
due to self-interactions of gravity can (and do indeed) diverge in this case. 
Letting the source size shrink to zero and keeping fixed other physical
parameters like mass and charge (and eventually neglecting gravity), 
usually one encounters infinities, or equivalently,  
physical quantities depending critically on the source size.
Dirac~\cite{Dirac:1938nz} emphasized that the cutoff dependence of 
the energy of the electromagnetic field sourced by an electron can be 
absorbed by an analog dependence of the bare electron mass, 
to provide a finite, physically observable invariant mass.
However the usual way to calculate mass renormalization is by considering
the virtual process of emission and reabsorption of a massless field, like for
mass renormalization of the electron in standard electrodynamics, rather than
a renormalization of the EMT, i.e. of the particle coupling to gravity, as we are
going to do here.
The above mentioned virtual processes are usually considered in the context 
of quantum field theory, but they show their effects also classically, 
when heavy, non-dynamical, non-propagating sources
are considered, as we will show.

In order to compute these quantities we make use of the formalism 
introduced in \cite{NRGR_paper,NRGR_lectures}, which is an 
effective field theory (EFT) method borrowed from particle physics, where 
it originated from studying non-relativistic bound state problems in the 
context of quantum electro- and cromo-dynamics \cite{Caswell:1985ui,
Luke:1999kz}; 
for this reason, it has been coined Non Relativistic General Relativity 
(NRGR) (see also \cite{Dam_Far_2PN_Diags:PRD53} for 
the first application of field theory techniques to gravity problems).
Here we apply NRGR in the framework of scalar-tensor theories of 
gravity for computing next-to-leading order corrections to the EMT renormalization, which in turn define, via the usual Einstein equations, 
the profile of the graviton generated by the sources.

An example of such a renormalization has been worked out in 
\cite{BjerrumBohr:2002ks} for point particles
in the GR case and by \cite{Dabholkar:1989jt,Buonanno:1998kx} 
for string-like sources coupled to an extra scalar, the dilaton, and an 
anti-symmetric tensor, the axion. See also \cite{Lund:1976ze,
Battye:1994qa,Quashnock:1990wv} for the string sources interacting 
with axionic and gravitational fields.
We find particularly worth of interest the different analysis performed in 
\cite{Dabholkar:1989jt,Buonanno:1998kx}, leading to apparently 
conflicting results for the string-tension renormalization.
The explanation of the discrepancy is actually given already in 
\cite{Buonanno:1998kx}, but here we re-analyize it with the fresh 
insight available thanks to NRGR.

The plan of the paper is as follows.
In sec.~\ref{sec:eft_string} we summarize the basic ingredients of NRGR and 
set the notation for the case at study.
In sec.~\ref{sec:point_part} we apply EFT methods to a model where a scalar
and the standard graviton field mediate long range interactions, to compute 
the effective EMT of a massive body. 
In sec.~\ref{sec:string} we present the analogous computation for a 
one-dimensional-extended object in four dimensions.
%In presence of codimension 2 objects (like strings in a 
%3+1-dimensional spacetime) 
%there is a subtlety about the conservation of the EMT that we 
%analyze in the appendix.
Finally we draw our conclusions in sec.~\ref{conclusion}.

%%%%%%%%%%%%%%%%%%%%%%%%%%%%%%%%%
\section{Effective field theory} \label{sec:eft_string}
%%%%%%%%%%%%%%%%%%%%%%%%%%%%%%%%%

We start by describing the basis of NRGR: in doing so we closely follow 
the thorough  presentation given in \cite{NRGR_paper}, to which 
we refer for more details, with the exception of the metric signature, 
as we adopt the "mostly plus'' convention: 
$\eta_{\mu\nu}\equiv(-,+,+,+)$.

In order to be able to exploit the manifest velocity-power counting, which is 
at the heart of PN expansion, we must first identify the relevant physical 
scales at stake. 
If, for simplicity, we restrict to binary systems of equal mass 
objects it is enough to introduce one mass scale $m$ and two 
parameters of the relative motion, namely the separation $r$ 
and the velocity $v$. It turns out that, up to the very last 
stages of the inspiral, the evolution of the system can be 
modelled to sufficiently high accuracy by non-relativistic dynamics, 
i.e. the leading order potential between the two bodies is 
the Newtonian one. The virial theorem then allows to relate 
the three afore-mentioned quantities according to 
\be
\label{virial}
v^2\sim \frac{G_Nm}{r}
\ee
(where $G_N$ is the ordinary gravitational constant) and tells 
that an expansion in the (square of the) typical three-velocity of 
the binary is at the same time an expansion in the strength of 
the gravitational field.

The compact objects being macroscopic, they can 
be considered fully non-relati\-vistic ($v<<c$) so that from a field 
theoretical point of view, and with scaling arguments in mind,
the binary constituents are non-relativistic particles endowed with 
typical four-momentum of the order $p_\mu\sim (E\sim mv^2, 
{\bf p}\sim m {\bf v})$ (boldface characters are used to denote 
3-vectors). Concerning the motion of the bodies subject to mutual 
gravitational potential, it is convenient to consider only the
\emph{potential} gra\-vitons, i.e. those responsible for binding the 
system as they mediate instantaneous interactions: their 
characteristic four-momentum 
$k_\mu$ will thus be of the order 
$$
\label{pot_g}
k_\mu\sim (k^0\sim \frac v{r} , \vk\sim\frac 1{r})
$$
so that these modes are always off-shell ($k_\mu k^\mu\neq 0$).\\
When a compact object emits a single graviton, 
momentum is effectively \emph{not} conserved 
and the non-relativistic particle recoils of a fractional amount roughly 
given by 
$$
\frac{|\delta{\bf p}|}{|{\bf p}|}\simeq \frac{|\vk|}{|{\bf p}|}
\simeq \frac\hbar{L}\,,
$$  
where $L\sim mvr$ is the angular momentum of the system:
it is clear that for macroscopic systems such quantity is 
negligibly small.
To summarize, an EFT approach describes massive compact objects 
in binary systems as non-dynamical, background sources 
of point-like type: quantitatively this corresponds to having 
particle world-lines interacting with gravitons. 
The action we consider is then given by 
\be
\az = \az_{EH}+\az_{pp}\,,
\ee
where the first term is the usual Einstein-Hilbert action
\be
\label{az_EH}
\az_{EH}= 2 M^2_{Pl}\int d^4x\sqrt{-g}\ R(g)\,,
\ee
with the Planck mass defined (non canonically) as 
$M^{-2}_{Pl}\equiv 32\pi G_N\simeq 1.2\times 10^{18}$GeV,
and the second term is the point particle action
\be
\az_{pp}=-m\int d\tau = -m\int \sqrt{-g_{\mu\nu} dx^\mu dx^\nu}\,,
\ee
in which $g_{\mu\nu}$ is the metric field that we write as
$g_{\mu\nu}\equiv \eta_{\mu\nu}+ h_{\mu\nu}$\,.
To make the graviton kinetic term invertible, 
one should also include a gauge fixing term like
\be
\az_{gf}=-M^2_{Pl}\int d^4x\ \Gamma_\mu\Gamma^\mu\,,
\ee 
with $\Gamma_\mu\equiv \pa^\nu h_{\mu\nu}-1/2\,\pa_\mu h^\nu_\nu$\,.

We now parametrize the metric following \cite{Kol:2007bc}, instead  
of \cite{NRGR_paper}, as
\be
\label{met_nr}
g_{\mu\nu}=\pt{
\bea{cc}
-e^{2\varphi} & -e^{2\varphi}a_j \\
-e^{2\varphi}a_i &\quad e^{-2\varphi}\gamma_{ij}-e^{-2\varphi}a_ia_j\\
\eea
}\,,
\ee
where  $\mu,\nu=0,..,3$ and $i,j=1,2,3$.
We define $\gamma^{ij}$ as the inverse matrix of $\gamma_{ij}$, 
so that $\gamma^{ij}\equiv\pt{\gamma^{-1}}_{ij}$ and 
$a^i\equiv \gamma^{ij}a_j$\,.
It is also useful to introduce 
$\varsigma_{ij}\equiv \gamma_{ij}-\delta_{ij}$ 
(so that $\varsigma^{ij}=\varsigma_{ij}$ to first order) 
and $\varsigma\equiv\varsigma_{ij}\delta^{ij}$.
Then, to quadratic order, the following action for non-canonically 
normalized fields is obtained
\begin{align}
\label{az_2}
\left.\az_{EH}\right|_{\rm quadratic} + \az_{gf} 
= -\frac{M_{Pl}^2}2 & \int dt\, d^3{\bf x} \\ 
\times & \paq{\pa_\mu \varsigma_{ij}\pa^\mu \varsigma_{ij}
- \frac 12\pa_\mu\varsigma\pa^\mu\varsigma
+ 8\pa_\mu\varphi\pa^\mu\varphi - 2\pa_\mu a_i\pa^\mu a_i} \nn \,.
\end{align}

The non-relativistic parametrization of the metric 
(\ref{met_nr}) allows to write down all the terms 
that do not involve time derivatives in a simple way
\bees
\label{az_stat}
\az_{EH}|_{static}=
2M_{Pl}^2\int dt\,d^3{\bf x}\sqrt{-\gamma}\paq{R(\gamma)
-2 \pa_i\varphi\pa_j\varphi \, \gamma^{ij} +
\frac 14e^{4\varphi} F_{ij}F_{kl}\gamma^{ik}\gamma^{jl}}\,,
\ees
where $F_{ij}\equiv \pa_ia_j-\pa_ja_i$ is the usual field strength 
tensor.\\
The canonically normalized fields $\sigma_{ij},\phi,A_i$ can be 
defined as
\bees
\bea{ccl}
\sigma_{ij} &\equiv & M_{Pl} \, \varsigma_{ij}\,,\\
\phi &\equiv & 2\sqrt 2M_{Pl} \, \varphi\,,\\
A_i &\equiv & \sqrt 2 M_{Pl} \, a_i\,. 
\eea
\ees
The only interaction term we will need, as it will be explained, is the cubic 
one $\sigma\phi^2$ given by
\bees
\left.\az_{EH}\right|_{\sigma\phi^2}&=&\ds\frac 1{2M_{Pl}}\int dt\,d^3{\bf x}
\paq{\pa_i\phi\pa_j\phi\pt{\delta_{ik}\delta_{jl}-
\frac 12\delta_{ij}\delta_{kl}}\sigma_{kl}}\,.
\ees
The world-line coupling to the graviton thus reads 
\renewcommand{\arraystretch}{1.4}
\bees
\label{matter_grav}
\az_{pp} &=& -m\ds \int d\tau = -m \int \sqrt{g_{\mu\nu}d x^\mu x^\nu} \\ 
&=& \ds -m\int dt\ e^{\phi/(2\sqrt 2 M_{Pl})}
\sqrt{\pt{1-\frac{A_i}{\sqrt 2M_{Pl}}v^i}^2
-e^{-\sqrt 2\phi/M_{Pl}}\gamma_{ij}v^iv^j} \nn \\
&\simeq& \ds -m\int dt\ e^{\phi/(2\sqrt 2 M_{Pl})}\pt{1-\frac 12 v^2 
+ \frac{\phi}{2\sqrt 2 M_{Pl}} -\frac{A_i}{\sqrt 2M_{Pl}}v^i+\ldots} \nn \,.
\ees
\renewcommand{\arraystretch}{1}

\noindent The propagators we use are given by the following non-relativistic 
expressions, as we are treating the time derivatives in the kinetic terms as 
perturbative contributions,
\renewcommand{\arraystretch}{1.4}
\be
\bea{ccl}
\ds\rnode{s1}\sigma_{ij}(t,\vk)\rnode{s2}\sigma_{kl}(t',\vk')&=&\ds
(2\pi)^3\delta(t-t')\delta^{(3)}(\vk-\vk')\frac i{\vk^2}P_{ij,kl}
\ncbar[nodesep=2pt,angle=-90,armA=4pt,armB=4pt]{s1}{s2}\\
\ds\rnode{a1}A_i(t,\vk)\rnode{a2}A_j(t',\vk')&=&
\ds(2\pi)^3\delta(t-t')\delta^{(3)}(\vk-\vk')\frac i{\vk^2}\delta_{ij}
\ncbar[nodesep=2pt,angle=-90,armA=4pt,armB=4pt]{a1}{a2}\\
\ds\rnode{f1}\phi (t,\vk)\rnode{f2}\phi (t',\vk')&=&
\ds (2\pi)^3\delta(t-t')\delta^{(3)}(\vk-\vk')\frac i{\vk^2}
\ncbar[nodesep=2pt,angle=-90,armA=4pt,armB=4pt]{f1}{f2}\\\\
\eea
\ee
\renewcommand{\arraystretch}{1}

\noindent where 
\be
\label{prop_ten}
P_{ij,kl}\equiv\frac 12\pt{\delta_{ik}\delta_{jl}+\delta_{il}\delta_{jk}-
2 \delta_{ij}\delta_{kl}}\,.
\ee

\begin{figure}[t!]
  \begin{center} 
    \includegraphics[width=.7\linewidth]{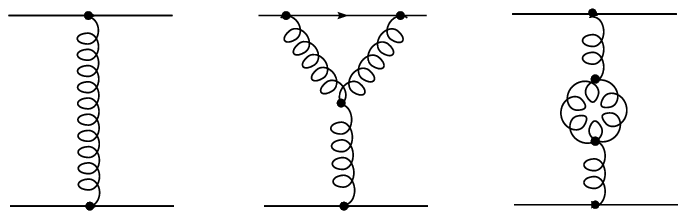}\\
    \hspace{0.5cm}{(a) Newton\hspace{2.3cm} (b) 1PN\hspace{2.0cm} (c) Quantum loop}
    \caption{Contributions to the scattering amplitude of two massive objects. 
      From left to right the diagrams represent respectively the leading Newtonian 
      approximation, a \emph{classical} contribution to the 1PN order and a 
      negligible \emph{quantum} 1-loop diagram.}
    \label{class_quant}
  \end{center}
\end{figure}

As far as we are only concerned in scaling we can set $k\sim 1/r$, $t\sim r/v$ 
and, by virtue of the virial theorem (\ref{virial}), $m/M_{Pl}\sim \sqrt{Lv}$. 
We can then immediately estimate what are the scalings of the contributions to 
the scattering amplitude of two massive objects: each of the three diagrams 
reported in fig.~\ref{class_quant}, for instance, contributes to such process.
By assigning a factor $[\frac m{M_{Pl}}dt \, d^3\vk]$ to a graviton-worldline
coupling not involving velocity, a factor $[\delta(t)\delta^{(3)}(\vk)\vk^{-2}]$ 
for each propagator, and a factor 
$[\frac{\vk^2}{M_{Pl}} dt\,\delta^{(3)}(\vk)\pt{d^3\vk}^3]$ for a 
three-graviton vertex, the following scaling laws can be associated to the 
different contributions of fig.~\ref{class_quant}:
\renewcommand{\arraystretch}{1.4}
$$
\bea{rcl}
(a)&\sim& \ds\pt{\frac m{M_{Pl}}}^2\paq{dt\,d^3\vk}^2
\paq{\delta(t)\delta^{(3)}(\vk)\vk^{-2}}\sim L\,,\\
(b)&\sim& \ds\pt{\frac m{M_{Pl}}}^3\paq{dt \, d^3\vk}^3
\paq{\delta(t)\delta^{(3)}(\vk)\vk^{-2}}^3 
\paq{\frac{\vk^2}{M_{Pl}}dt\,\delta^{(3)}(\vk)\pt{d^3\vk}^3} \sim Lv^2\,,\\ 
(c)&\sim& \ds\pt{\frac m{M_{Pl}}}^2\paq{dt\,d^3\vk}^2
\paq{\delta(t)\delta^{(3)}(\vk)\vk^{-2}}^4
\paq{\frac{\vk^2}{M_{Pl}}dt\, \delta^{(3)}(\vk)\pt{d^3\vk}^3}^2\sim v^4\,.
\eea
$$
\renewcommand{\arraystretch}{1}

\noindent Even if we are actually dealing with a classical field theory, 
it is interesting to give a look at the scalings in powers of $\hbar$.
To restore $\hbar$'s one can apply the usual rule that relates the 
number $\mathcal I$ of internal graviton lines (graviton propagators) to the 
number~$\mathcal V$ of vertices and the number $\mathcal L$ of graviton loops
\be
\mathcal L =\mathcal I- \mathcal V+1 \,;
\ee
then, taking into account that each internal line brings a power of $\hbar$ and each 
interaction vertex a $\hbar^{-1}$ from the interaction Lagrangian, the total scaling 
for diagrams where the only external lines are massive particles is 
$\hbar^{{\cal L}-1}$. 
According to this rule the third diagram of fig.~\ref{class_quant} involves one more 
power of $\hbar$ than the first two.
The diagram with a graviton loop is then suppressed with 
respect to the Newtonian contribution, apart from some powers of $v$, by a 
factor $\hbar/L\ll 1$, whereas the second diagram in fig.~\ref{class_quant} is a 
1PN contribution which does not involve any power of $\hbar$. 
Equivalently one can notice that 
since the massive object is not propagating (there is no kinetic 
term in the Lagrangian for such a source), the 1PN diagram is not a 
loop one.
These scaling arguments remain unchanged when other particles are added, 
like a scalar field, and/or another mass scale is introduced 
\cite{Porto:2007pw}, as we will discuss in sec.~\ref{sec:point_part}, provided
that the virial relation (\ref{virial}) correctly accounts for the leading 
interaction.

%%%%%%%%%%%%%%%%%%%%%%%%%%%%%%%%%%
\section[Effective energy-momentum tensor in scalar gravity: Part I \\
the point particle case]{Effective energy-momentum tensor in \\ scalar gravity: 
the point particle case
\sectionmark{Effective EMT in scalar gravity: 
the point particle case}}
 \sectionmark{Effective EMT in scalar gravity: 
the point particle case}
\label{sec:point_part}
%%%%%%%%%%%%%%%%%%%%%%%%%%%%%%%%%%

The usual way to obtain an effective action $\Gamma$ out of a fundamental 
action $\az_{fund}$ is by integrating out the degrees of freedom we do not 
want to propagate to infinity according to the formal rule 
\be
e^{i\Gamma}\equiv \int{\mathcal D}\Phi\,e^{i\az_{fund}}\,,
\ee
where $\Phi$ denotes the generic field to integrate out.
  
In practice this non-perturbative integration is replaced by a perturbative computation, performed with the aid of Feynman diagrams like those of fig.~\ref{class_quant} which shows some contributions to the effective action of two 
particles interacting gravitationally. 
At lowest order (Newtonian interaction) the diagram in fig.~\ref{class_quant}(a) represents the term responsible for the Newtonian $1/r$ potential between two massive objects. 
Stripping away one of the two external lines in this diagram an amplitude for 
the coupling of a single particle to a graviton is obtained: this amplitude is
linear in the external graviton wave-function and defines the \emph{effective} EMT of the particle.
Thus, at Newtonian level the two diagrams in fig.~\ref{emt_newt} give the 
following contributions to the effective action
\begin{align} 
\label{g0p}
\ds \Gamma^{(0)}=\Gamma_\phi^{(0)}+\Gamma_\sigma^{(0)}&=\ds
\frac 1{2\sqrt 2 M_{Pl}}\int\phi(x)\paq{T_{00}(x)+T_{ij}(x)\delta_{ij}}\,
d^4x \nn \\
&\ds =\frac{m}{2\sqrt 2M_{Pl}}\int\phi(t,{\bf x_p}(t))\,dt
\end{align} 
where ${\bf x_p}$ is the three-vector of the position of the source particle 
and use has been made of the Newtonian value of the EMT defined as 
usual as 
\be
\label{def_emt}
T_{\mu\nu}(x)\equiv \frac{-2}{\sqrt{-g}} \left.
\frac{\delta \az}{\delta g^{\mu\nu}(x)}\right|_{g_{\mu\nu}=\eta_{\mu\nu}}\,.
\ee
Note that the contribution from $\Gamma^{(0)}_\sigma$ is vanishing as 
the $\sigma_{ij}$ 
part of the metric field does not couple directly to a static massive source for 
which $T_{ij}(x)=0$, $T_{00}(x)=m\delta^{(3)}({\bf x}-{\bf x_p})$.

\begin{figure}[t]
\centering
    \includegraphics[width=.7\linewidth]{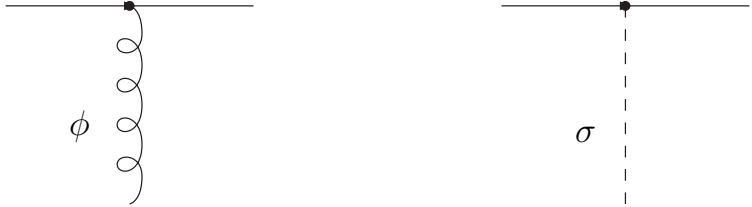}
      \caption{Feynman diagrams describing the gravitational contributions to 
	the effective energy-momentum tensor of a particle at Newtonian level 
	according to the parametrization (\ref{met_nr}) used for the metric.}
  \label{emt_newt}
  \end{figure}

The second diagram in fig.~\ref{class_quant} is a representative  
contribution of the 1PN corrections to the Newtonian potential between two 
particles. Stripping away again one of the two external particle lines the diagram 
showed in fig.~\ref{pp_1PN} is obtained, whose contribution to the 
effective action at next-to-leading order is
\renewcommand{\arraystretch}{1.8}
\begin{align} %\bea{rl}
\label{si_1PN} 
\ds \Gamma^{(I)}_\sigma =&\ds \frac 1{M_{Pl}}
\int d^4x\,\sigma_{ij}(x){T^{ij}}^{(I)}(x) \nn \\
=&\ds \frac 1{M_{Pl}}
\int dt\,\frac{d^3\vq}{\pt{2\pi}^3} \sigma_{ij}(t,-\vq){T^{ij}}^{(I)}(t,\vq)
e^{i\vq\cdot {\bf x}_p} \nn \\
=&\ds\frac{m^2}{8M^3_{Pl}}\int dt \frac{d^3\vq}{\pt{2\pi}^3}
\frac{d^3\vk}{\pt{2\pi}^3}\frac{k^ik^j-k^iq^j}{\vk^2\pt{\vk-\vq}^2}
\pt{\delta^l_i\delta^m_j-\frac{\delta_{ij}\delta^{lm}}2}\sigma_{lm}(t,-\vq)
e^{i\vq\cdot {\bf x}_p} \nn \\
=&\ds \frac{m^2}{2^{10} M^3_{Pl}}\int dt\,\frac{d^3\vq}{(2\pi)^3}
\sigma_{ij}(t,-\vq)\pt{-\delta^{ij}q + \frac{q^iq^j}{q}} 
e^{i\vq\cdot {\bf x}_p}\,,
\end{align} %\eea
\renewcommand{\arraystretch}{1}

\noindent where $q\equiv \sqrt{{\bf q}\cdot{\bf q}}$ and we have used eqs.~(\ref{kikj}).
The analogous quantity for $\phi$ vanishes as there is no $\phi^3$ vertex, see eq.~(\ref{az_stat}).
Incidentally, we note that the EMT obtained from eq.~(\ref{si_1PN}) is transverse, 
consistently with the request that the effective EMT has to be conserved order by order (see \cite{Sundrum:2003yt} for an interesting discussion of scalar gravity at interacting level).

\begin{figure}[t]
  \begin{center}
    \includegraphics[width=.25\linewidth]{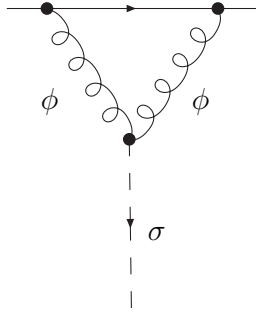}\\
    \caption{Feynman diagram describing the gravitational contribution to the 
      effective energy-momentum tensor of a particle at first post-Newtonian order 
      according to the parametrization used for the metric (\ref{met_nr}).} 
    \label{pp_1PN}
  \end{center}
\end{figure} 

Another check of the correctness of our result can be obtained by reconstructing 
the metric out of this effective EMT. The linearized equations of motion for gravity give
\be
\phi(t,\vk)=-\frac{1}{\vk^2}
\left.\frac{\delta \Gamma_\phi}{\delta\phi(t,\vk)}\right|_{\phi=0=\sigma_{ij}}
\ee
which, using the first of eqs.~(\ref{int_k}), allows to compute the metric component $\varphi$ according to
\be
\varphi(x) \equiv \frac{\phi(x)}{2\sqrt{2}M_{Pl}}=-\frac m{8 M_{Pl}^2}
\int\frac{d^3\vk}{\pt{2\pi}^3} 
\frac {e^{i\vk\cdot\pt{{\bf x}-{\bf x_p}}}}{\vk^2}=-\frac{G_Nm}r\,,
\ee
where $G_N$ has been reinstated in the final result and 
$r\equiv |{\bf x}-{\bf x_p}|$.
Analogously, for $\varsigma_{ij}$ one has
\be
\varsigma_{ij}(t,\vk)=-\frac 1{\vk^2}\frac 1{M_{Pl}}P_{ij;kl}
\left.\frac{\delta S}{\delta \sigma^{kl}(t,\vk)}\right|_{\phi=0=\varsigma_{kl}}
\ee
which, again using eqs.~(\ref{int_k}), leads to
\begin{align}
\ds \varsigma_{ij}(t,x) =&\ds P_{ij;kl}\int\frac{d^3\vk}{\pt{2\pi}^3} 
\frac{m^2}{2^{10}M_{Pl}^4}\pt{\delta_{kl}k-\frac{k_k k_l}{k}}
\frac 1{\vk^2} e^{-i\vk\cdot\pt{{\bf x}-{\bf x_p}}} \nn \\
=&\ds -\frac{\pt{G_Nm}^2}{r^2}\pt{\delta_{ij}-\frac{x_ix_j}{r^2}}\,.
\end{align}
Given the metric parametrization (\ref{met_nr}) we obtain 
\renewcommand{\arraystretch}{1.4}
\be
\bea{rcl}
\ds g_{00}&=&\ds -1+\frac{2G_Nm}{r}-2\frac{\pt{G_Nm}^2}{r^2} \\
\ds g_{0i}&=&0 \\
\ds g_{ij}&=&\ds \pt{1+\frac{2G_Nm}r+\frac{\pt{G_Nm}^2}{r^2}}\delta_{ij}+
\frac{\pt{G_Nm}^2}{r^2}\frac{x^ix^j}{r^2}
\eea
\ee
\renewcommand{\arraystretch}{1}

\noindent which is the Schwarzschild metric to 1PN order in the harmonic gauge,
see \cite{BjerrumBohr:2002ks}.

Let us now consider an extra degree of freedom with respect to ordinary 
gravity, that is a massive scalar field $\psi$ whose action is given by
\be
\az_{\psi}= -\frac 12\int d^4x\sqrt{-g}  \paq{g^{\mu\nu}
\pa_\mu\psi\,\pa_\nu\psi +m_\psi^2\psi^2+ \lambda \psi^3} \;,
\ee 
where a cubic self-interaction has been allowed. The interaction with the 
gravitational field $\sigma_{ij}$, embodied by the trilinear term $\psi\psi\sigma$, 
can be derived from the kinetic term, namely
\be
\label{pps}
\left.\az_\psi\right|_{\psi\psi\sigma} = \frac 1{2M_{Pl}}\int dt\,d^3{\bf x}\, 
\pa_i\psi \pa_j\psi \pt{\sigma^{ij}-\frac 12\delta^{ij}\sigma}\,.
\ee
There are no trilinear terms such as $\phi\psi\psi$ 
or $\phi\phi\psi$ because of the specific metric parametrization we chose
(\ref{met_nr}).
The field $\psi$ is assumed to couple to matter in a metric type
in analogy with (\ref{matter_grav}):
$$
\az'_{pp}=-me^{\alpha \psi/\pt{2\sqrt 2M_{Pl}}}\int d\tau \quad ,
$$
for some dimensionless parameter $\alpha$. Therefore the tree-level coupling 
of $\psi$ to matter at lowest order is very similar to the diagram on the left of 
fig.~\ref{emt_newt}:
\be
\Gamma^{(0)}_\psi=\frac{\alpha \,m}{2\sqrt 2M_{Pl}}\int\psi(t,{\bf x_p}(t))
\, dt\,.
\ee 

\begin{figure}[t]
  \begin{center}
    \includegraphics[width=.25\linewidth]{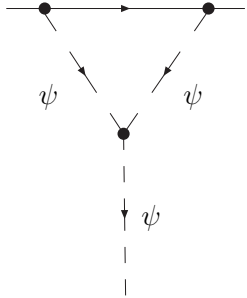}
    \caption{Feynman diagram representing the self-interaction contribution of  
      the massive scalar field $\psi$ to the energy-momentum tensor of a particle 
      at next-to-leading order.}
    \label{3psi}
  \end{center}
\end{figure}

At next-to-leading order we have two possible contributions. The first comes from 
a diagram like that of fig.~\ref{pp_1PN} where the two $\phi$'s  are replaced with 
two $\psi$'s: the amplitude is almost the same as eq.~(\ref{si_1PN}), apart from 
an extra factor $\alpha^2$. 
The second contribution comes from the cubic $\psi$ self-interaction, depicted in the 
diagram of fig.~\ref{3psi}:
\be
\Gamma^{(I)}_\psi=\frac{\lambda m^2\alpha^2}{64\pi M^2_{Pl}}
\int dt\int\frac{d^3\vq}{(2\pi)^3}e^{i{{\bf q}\cdot{\bf x_p}}}
\psi(t,\vq) \frac{1}{q}\arctan{\left(\frac{q}{2m_\psi}\right)}\,.
\ee
Note that at high momentum transfer ($q\gg m_\psi$) the integrand goes as $q^{-1}$, 
whereas in the gravity case (\ref{si_1PN}) we had $T_{ij}^{\sigma}(q)\propto q$:
this difference leads to an effective potential due to the $\psi$~mediation which has a logarithmic profile, rather than the $1/r^2$ behavior typical of 1PN terms in 
Einstein gravity derived in \cite{Porto:2007pw}; at low momenta ($q\ll m_\psi$) 
the Yukawa suppression takes place as usual.

\vspace{0.5cm}

%%%%%%%%%%%%%%%%%%%%%%%%%%%%%%%%%%
\section[Effective energy-momentum tensor in scalar gravity: Part II \\
the string case]{Effective energy-momentum tensor in \\ scalar gravity: 
the string case
\sectionmark{Effective EMT in scalar gravity: the string case}}
 \sectionmark{Effective EMT in scalar gravity: the string case}
\label{sec:string}
%%%%%%%%%%%%%%%%%%%%%%%%%%%%%%%%%%

In the case of a one-dimensional extended source we consider the 
Nambu-Goto string with action $\az_s$ given by
\be
\label{eq:az_str}
\az_s=\mu \int_\Sigma\sqrt{-\gamma}\ e^{\alpha\Phi/\pt{\sqrt 2M_{Pl}}}\,
d\tau d\sigma-
\frac{\beta\mu}{2\sqrt 2M_{Pl}}\int_\Sigma \pa_\alpha x^\mu\pa_\beta x^\nu 
\eps^{\alpha\beta} B_{\mu\nu}\,d\tau d\sigma\,,
\ee
where $\gamma\equiv {\rm det} \gamma_{\alpha\beta}$, with 
$\gamma_{\alpha\beta}\equiv \pa_\alpha x^\mu\pa_\beta x^\nu g_{\mu\nu}$, 
$x^\mu$ are coordinates in the 4-dimensio\-nal space, $\sigma$ 
and $\tau$ are the coordinates on the world-sheet $\Sigma$ 
spanned by the string in its temporal evolution.
Such an action describes a fundamental string interacting with gravity 
via a string tension $\mu$, with a scalar field $\Phi$ through a coupling 
$\alpha\mu/(\sqrt 2M_{Pl})$ and with the antisymmetric tensor $B_{\mu\nu}$ 
through the coupling $\beta\mu/(2\sqrt 2M_{Pl})$. 
In this notation a supersymmetric string corresponds to $\alpha=\beta=1$.\\ 
The convention for indeces is the following:
$\alpha,\beta$ denote the two directions parallel to the world-sheet while 
$\mu,\nu,\ldots$ are generic 4-dimensional indeces, then Latin 
letters $i,j,\ldots$ denote 3-space indeces and we will use $a,b$ or $c$ 
to denote the (two) spatial dimensions orthogonal to the string.\\
The action $\az_f$ determining the dynamics of the fields is
\be
\label{az_bulk}
\az_f=\int d^4x \sqrt{-g} \paq{2M_{Pl}^2R -\frac 12\pt{\pa\Phi}^2-
\frac 1{12}e^{-\sqrt 2\alpha\Phi/M_{Pl}}H_{\mu\nu\rho}H^{\mu\nu\rho}}\,,
\ee
where $H_{\mu\nu\rho}\equiv\pa_\mu B_{\nu\rho}+\pa_\rho B_{\mu\nu}+
\pa_\nu B_{\rho\mu}$. The only new propagator we will need with respect to 
the point-particle study is
\be
\label{prop_b}
\rnode{b1}B_{\mu\nu}(t,{\bf k})\rnode{b2}B_{\rho\sigma}(t',{\bf k'})=\frac 12
\pt{\eta_{\mu\rho}\eta_{\nu\sigma}-\eta_{\mu\sigma}\eta_{\nu\rho}}
\pt{2\pi}^3\delta(t-t')\delta^{(3)}(\vk-\vk')\frac i{\vk^2}\,.
\ncbar[nodesep=2pt,angle=-90,armA=4pt,armB=4pt]{b1}{b2}
\ee
Analogously to diagrams in fig.~\ref{emt_newt}, the effective action 
for the linear coupling to the string source of the fields 
$\phi$, $\sigma_{ij}$, $\Phi$ and $B_{\mu\nu}$ is 
$\Gamma^{(0)}=\Gamma^{(0)}_{\phi}+\Gamma^{(0)}_{\sigma_{ij}}+\Gamma^{(0)}_{\Phi}+
\Gamma^{(0)}_{B_{\mu\nu}}$ with 
\renewcommand{\arraystretch}{2.8}
\bees \label{gammazero}
\bea{rcl}
\Gamma^{(0)}_\phi &=&\ds\int\phi\pt{T_{00}+T_{ij}\delta^{ij}}\,d^4x=0\,,\\
\Gamma^{(0)}_{\sigma_{ij}} &=&\ds\int\varsigma_{ij}T^{ij}\,d^4x=
-\frac\mu{M_{Pl}} \int_\Sigma \sigma_{11}(x(\tau,\sigma))\,d\tau d\sigma\,,\\
\Gamma^{(0)}_\Phi &=&\ds\frac{\alpha}{2\sqrt 2M_{Pl}}
\int\Phi\pt{T_{00}-T_{ij}\delta^{ij}}\,d^4x=
\frac{\alpha\mu}{\sqrt 2M_{Pl}}\int_\Sigma \Phi(x(\tau,\sigma))\,d\tau d\sigma\,,\\
\Gamma^{(0)}_{B_{\mu\nu}} &=&\ds\frac{\beta\mu}{2\sqrt 2M_{Pl}}
\int_\Sigma\partial_\alpha x^\mu\partial_\beta x^\nu\eps^{\alpha\beta}
B_{\mu\nu}\,d\tau d\sigma = \frac{\beta\mu}{\sqrt 2M_{Pl}} 
\int_\Sigma B_{01}(x(\tau,\sigma))\,d\tau d\sigma\,,
\eea
\ees
\renewcommand{\arraystretch}{1} 

\noindent where use has been made of the explicit parametrization of 
a static string: 
$x^0=\tau$, $x^1=\sigma$, and of the definition (\ref{def_emt}) for the 
string EMT $T^s_{\mu\nu}$ giving
\be
\label{emt_s}
T_{\mu\nu}^s={\rm diag}(\mu,-\mu,0,0)\delta^{(2)}(x^a)\,.
\ee

Following the same reasoning as in sec.~\ref{sec:point_part}, 
the contributions to the renormalization of the effective EMT due 
to the dilaton and the antisymmetric tensor interaction can be 
computed, see fig.~\ref{str_1PN}. 
We thus restrict to those trilinear interaction terms involving a graviton 
field, either a $\phi$ or a $\sigma$, as an external line 
(in a completely analogous way the renormalization of the $\Phi$ and 
$B_{\mu\nu}$ coupling could be computed).
We then have:
\begin{align}
\label{s_s3}
\ds\az_{3} = \frac 1{2M_{Pl}}\ds\int dt\,d^3{\bf x} 
& \left\{ \frac 12\paq{\partial_i\Phi\partial_j\Phi
\pt{\delta^{il}\delta^{jm}-\frac 12\delta^{ij}\delta^{lm}}
\sigma_{lm}} \right. \\
\ds & \left. + \frac 12 \paq{\partial_iB_{01}\partial_jB_{01}
\pt{T_B^{iljm} } \sigma_{lm}} %-\frac{\sqrt 2\alpha\Phi}6 H^2
\right\} \nn \,,
\end{align}
where we have defined the tensor 
$$
T_B^{iljm} \equiv \pt{\delta^{il}\delta^{jm}+\delta^{ij}\delta^{l1}\delta^{m1}
-\frac 12\delta^{ij}\delta^{lm}
+\delta^{il}\delta^{j1}\delta^{m1}+\delta^{jm}\delta^{i1}\delta^{l1}}
$$ 
and we have specified the antisymmetric tensor polarization indices to "$01$" ,
as this is the only polarization involved in this interaction, and omitted rewriting 
the terms coming from the pure gravity sector, i.e. $\sigma^3$ and 
$\phi^2\sigma$, because they read the same as in (\ref{az_stat}).
\begin{figure}[t]
  \begin{center} 
    \includegraphics[width=.95\linewidth]{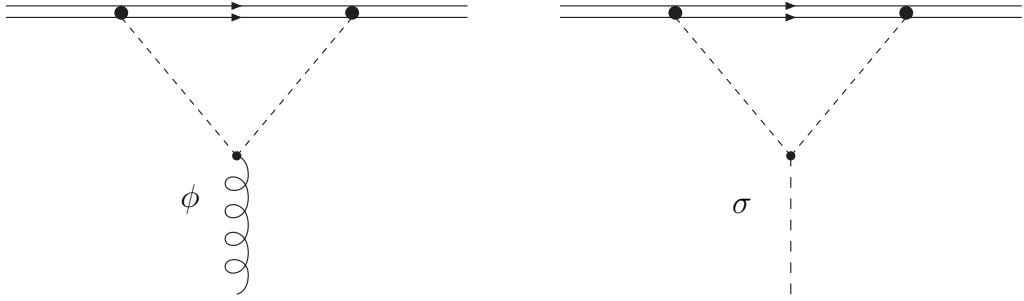}
    \caption{Diagrams reproducing the coupling to $\phi$ (curly line) and to 
     $\sigma_{ij}$ (long-dashed), or the effective energy-momentum tensor, of 
      a string at next to lowest order in interaction. The diagram on the left
      vanishes (see discussion in the text).}
    \label{str_1PN}
  \end{center}    
\end{figure}

The diagram on the left in fig.~\ref{str_1PN} is actually vanishing because 
no $\phi$ can attach directly to the string and no trilinear 
term with only one $\phi$ is present in the action (\ref{az_bulk}), as it can be 
seen from (\ref{az_stat}) or (\ref{s_s3}): this implies that the relation 
$T_{00}=-T_{ij}\delta^{ij}$ holds also at next-to-leading order.
We are thus left with the diagram on the right in fig.~\ref{str_1PN}, 
where the particles propagating in the internal dashed lines can be 
either two dilatons or two antisimmetric tensors or two gravitons of 
the type $\sigma_{ij}$.
The contribution to $\Gamma^{(I)}_{\sigma_{ij}}$ from the diagram 
involving two dilatons is 
\renewcommand{\arraystretch}{1.6}
\begin{align}
\label{str_1pn}
\ds \Gamma_{\sigma\Phi\Phi}^{(I)} =
-\ds \frac{\mu^2\alpha^2}{8M_{Pl}^3}
\int d\tau\frac{d^2q}{\pt{2\pi}^2} e^{-iq_ax_s^a} & 
\pt{\delta^{ai}\delta^{bj} -\frac 12\delta^{ab}\delta^{ij}} \sigma_{ij}(\tau,q)
\int \frac{d^2k}{\pt{2\pi}^2} \frac{k_ak_b-k_aq_b}{k^2\pt{k-q}^2} \nn\\
= -\ds\frac{4G_N\mu^2\alpha^2}{M_{Pl}} \int d\tau \frac{d^2q}{\pt{2\pi}^2}
e^{-iq_a x^a_s} & \pt{C\delta_{ab}-\frac{q_aq_b}{q^2}} 
\pt{\delta^{ai}\delta^{bj} \frac 12\delta^{ab}\delta^{ij}} \sigma_{ij}(\tau,q) \nn\\
= \ds \frac{4G_N\mu^2\alpha^2}{M_{Pl}} \int d\tau\frac{d^2q}{\pt{2\pi}^2}
    e^{-iq_a x^a_s} & 
    \left[ \pt{ -\frac 12\delta^{ab} +\frac{q^aq^b}{q^2} }
    \sigma_{ab}(\tau,q) \right. \nn\\ 
& \ds \left. \phantom{\frac 12 \delta^{ab}}
+ \pt{ C-\frac 12 } \sigma_{11}(\tau,q) \right] \,,
\end{align}
\renewcommand{\arraystretch}{1}

\noindent with $C$ a divergent quantity, coming from the last integration in 
the first line, whose value can be read from eq.~(\ref{kikj_2})
\be
\label{div_const}
C=\lim_{\eps\to 0}\ -\frac 1\eps
\paq{1+\frac \eps 2\pt{\gamma-2+\log\paq{q^2/(4\pi)}+o(\eps)}}\,; 
\ee
here dimensional regularization has been used, as this entry of the effective
EMT is expected to be (logarithmically) UV divergent, see e.g. 
\cite{Dabholkar:1989jt,Buonanno:1998kx}. 
Note that the divergent constant only enters the $T_{11}$ component of the 
effective EMT.

For the $B_{\mu\nu}$ interaction a simiflar result is obtained 
\begin{align}
\Gamma_{\sigma BB}^{(I)} = \frac{4G_N\mu^2\beta^2}{M_{Pl}}
\int d\tau\frac{d^2q}{\pt{2\pi}^2} e^{-iq_ax^a_s} 
& \left[ \left( \frac 12\delta^{ab} -\frac{q^a q^b}{q^2} \right)  
\sigma_{ab}(\tau,q) \right. \nn \\
& \phantom{\frac 12\delta^{ab} }
+ \left. \left( C-\frac 12 \right)  \sigma_{11}(\tau,q) \right] \,.
\end{align}
 
The contribution to the 1PN effective action due to purely gravitational 
process, i.e. by the diagram on the right of fig. \ref{str_1PN} with three 
$\sigma$'s, can be computed by making use of the three graviton 
point function:
\be
\langle\sigma_{11}(k_1)\sigma_{11}(k_2)\sigma_{ij}(q)\rangle=
-\frac 12\delta^{(3)}(k_1+k_2+q) q^2\delta_{i1}\delta_{j1}\,,
\ee
which has been obtained thanks to the Feyncalc tools \cite{Mertig:1990an} 
for Mathematica; the result is
\begin{align}
\label{gamma3}
\Gamma^{(I)}_{\sigma\sigma\sigma} =& \ds -\frac{\mu^2}{4M^3_{Pl}}
\int d\tau\frac{d^2q}{\pt{2\pi}^2}e^{-iq_a x^a_s}
\delta^{a1}\delta^{b1}\sigma_{ab}(\tau,q)
\int \frac{d^2k}{\pt{2\pi}^2}\frac{q^2}{k^2(k-q)^2} \nn \\
=&\ds \frac{8G_N\mu^2}{M_{Pl}} 
\int d\tau\frac{d^2q}{\pt{2\pi}^2}e^{-iq_a x^a_s}D\,\sigma_{11}(\tau,q) 
\end{align}
where $D$ is a divergent constant, again entering the $T_{11}$ 
component only, given by 
\be
\label{nuovoD}
D=\lim_{\epsilon\to 0} \frac 1\epsilon\paq{1+\frac \epsilon 2
\pt{\gamma +\ln \paq{q^2/\pt{4\pi}}}}\,.
\ee
 
The conserved effective EMT is thus given by the sum of the three 
contributions just calculated and reads
\bees
\label{eemt_s}
T_{ij}^{(I)}(q) &=& 4 G_N\mu^2 \\
&\times &
\left( 
\bea{cc}
\left( 2-\alpha^2-\beta^2 \right) D +\alpha^2+\beta^2 & 0 \\
0 & \ds \left( \alpha^2-\beta^2 \right)
\left( -\frac{\delta_{ab}}{2} +\frac{q_aq_b}{q^2} \right) 
\eea
\right) \nn
\ees
together with $T_{00}=-T_{11}$ and $T_{0i}=0$\,.
The coordinate space counterpart of (\ref{eemt_s}) is reported in the 
Appendix.

We note that in the directions orthogonal to the string the EMT is still 
vanishing for $\alpha^2=\beta^2$, thus preserving the no-force
condition valid for supersymmetric strings of the same type (charge).
The divergent part of the entry $T_{11}$ is also vanishing in the 
supersymmetric case due to a cancellation among the different terms: 
therefore, the \emph{superstring tension}, given by $T_{11}$, does not 
receive divergent contribution.
This confirms the result of Dabholkar and Harvey \cite{Dabholkar:1989jt} 
obtained through the analysis of the EMT's 
%{\bf TENSORS (FOR B AND PHI) AND} pseudo-tensor 
on the (linearized) GR solution around a string.

\begin{figure}[t]
  \begin{center}
    \includegraphics[width=0.3\linewidth]{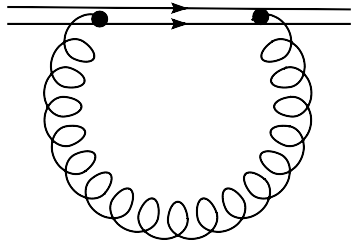}\\
    \caption{Feynman diagram representing the string tension renormalization 
    as computed in \cite{Buonanno:1998kx}. 
    The internal wavy line stands for all possible fields interacting with the string: 
    dilaton, antisymmetric tensor and graviton of type $\sigma$.}
    \label{damour}
  \end{center}
\end{figure}

In \cite{Buonanno:1998kx} Buonanno and Damour also found a 
non-renormalization, but via a different cancellation. 
The authors of \cite{Buonanno:1998kx} analyzed a 
physical quantity which is described by a diagram of the 
type depicted in fig.~\ref{damour}, where it is understood that each of the 
fields interacting with the string can propagate in the internal line.
We now take a closer look at the different contributions to this process.
Letting a $\sigma_{ij}$ propagate in the wavy line of fig.~\ref{damour} 
yields a vanishing 
result given that the amplitude for such a process has the following behavior 
\be 
\label{null}
fig.~\mbox{\ref{damour}}|_{\sigma_{ij}}\propto 
T^{ij}\rnode{g1}\sigma_{ij}\rnode{g2}\sigma_{kl}T^{kl}\propto\mu^2P_{11;11}=
0\,,
\ncbar[nodesep=2pt,angle=-90,armA=4pt,armB=4pt]{g1}{g2}
\ee
as it can be explicitly checked from eq.~(\ref{prop_ten}). 
This diagram 
vanishes for the same reason why two straight, static,
parallel strings do not exert a force on each other: the amplitude for one 
graviton exchange between two such strings is proportional to the 
same vanishing quantity 
$P_{11;11}$. 
The dilaton contribution to the amplitude of fig.~\ref{damour} is
\be
\label{ren}
fig.~\mbox{\ref{damour}}|_{\Phi}=\frac{\alpha^2\mu^2}{2M_{Pl}^2}\int \frac {d^2k}{k^2}\,,
\ee
whereas to find the effect of the antisymmetric tensor it is enough to 
replace "$\alpha^2$" with "$-\beta^2$" in eq.~(\ref{ren}), as can be checked 
using (\ref{prop_b}) and (\ref{gammazero}).
These three amplitudes, condensed in the representation of fig.~\ref{damour}, 
have a close correspondence with what is found in \cite{Buonanno:1998kx} and 
show that the contributions to the superstring renormalization are different when 
calculated by looking at the self-energy as in \cite{Buonanno:1998kx} 
other than through the (effective) EMT as in \cite{Dabholkar:1989jt} and in 
the present work; 
nonetheless, the non-renormalization property of superstrings is preserved 
in both approaches.

The source of the discrepancy is explained in \cite{Buonanno:1998kx}
where it is observed that the difference in the two ways of computing 
the renormalization of the string tension amounts to a (divergent) 
source-localized term, "as the interaction-energy cannot be unambiguously 
localized only in the field, there are also interaction-energy contributions 
which are localized in the sources" which are missed in one approach 
but accounted for in the other. 
Moreover, the contribution of the antisymmetric tensor to the string tension 
renormalization turns out to be the same with the two methods because this 
coupling to the string is metric-independent, so it does not contribute to the total 
EMT given by $T^{\mu\nu}\equiv 2 g^{-1/2} \delta S / \delta g_{\mu\nu}$.
Of course the physical result cannot depend on the details of the calculation 
method: indeed the source-localized contribution just renormalizes the 
\emph{bare} tension of the string and does not give physical effects.
As observed in \cite{Buonanno:1998kx}, this constrasts Dirac's argument 
\cite{Dirac:1938nz} about the connection bewteen the renormalization of a point 
charge and its divergent field self-energy.

%We then confirmed that the renormalization of the string tension and the
%(divergent) source-localized contribution to the field self-energy 
%cannot be matched.

Therefore, we support the explanation of the discrepancy given by Buonanno 
and Damour \cite{Buonanno:1998kx} and provide a computation of the 
renormalization of the EMT with a completely different technique than in Dabholkar and 
Harvey \cite{Dabholkar:1989jt}, confirming their result.

Following the track of the EFT methodes we employed, one could also compute the 
renormalization of the couplings of $\Phi$ and 
$B_{\mu\nu}$. 
For the dilaton coupling the relevant diagrams are two, both of the type 
fig.~\ref{str_1PN}, with a $\Phi$ as outer wavy line and either two $B_{\mu\nu}$'s
or a $\Phi$ and a $\sigma_{ij}$ as dashed inner lines.
For the antisymmetric tensor case, the external $B_{\mu\nu}$ can be attached to 
either a $\sigma_{ij}$ and a $B_{\mu\nu}$ or to a $\Phi$ and a $B_{\mu\nu}$. 
All the above mentioned trilinear vertices have the same dependence on external 
momentum as the gravity case. 

One final remark is needed about result (\ref{eemt_s}). A tensor $T_{ab}(x)$ 
is conserved if $T_{ab}^{\phantom{ab},b}(x)=0$ which, in Fourier space, 
translates naively to
\be
\label{naive_conserv}
\pa^aT_{ab}(q)\stackrel ?=-i q^a T_{ab}(q)\qquad {\rm NO!}
\ee
Clearly, with an EMT of the form (\ref{eemt_s}), for $\alpha^2\neq\beta^2$ 
the right hand side of eq.~(\ref{naive_conserv}) does not vanish. This happens 
because $T_{ab}(q)$ is not square integrable, thus it is not ensured that the 
derivative operation and the Fourier transform commute with each other, 
and indeed they do not in this case, see Appendix for details.

\section{Conclusions}
\label{conclusion}

We have studied point-like and one-dimensional-extended sources in the 
context of scalar-tensor gravity and we have computed the effects of fields 
self-interactions to the renorma\-lization of the effective 
energy-momentum tensor.

The calculations have been performed within the framework provided by 
the effective field theory methods applied to gravity \cite{NRGR_paper,
NRGR_lectures}, exploiting the powerful tool of a systematic expansion 
in terms of Feynman diagrams.

The classical ``dressing'' of the sources by long range interactions has the 
effect of smearing the source, consistently with coordinate covariance, 
and implies energy-momentum tensor conservation.
We obtained perturbative solutions valid to first post-Newtonian order or, in 
the case of extra scalar fields, up to the trilinear interaction terms.

In the case of a string source we reviewed the renormalization of both its 
effective energy-momentum tensor and its tension, which has
been subject of investigation with apparently conflicting results in the past 
\cite{Dabholkar:1989jt,Buonanno:1998kx}.
We exposed the fully satisfactory explanation of the discre\-pancy given by 
Buonanno and Damour \cite{Buonanno:1998kx} and confirmed that the 
renormalization of the energy-momentum tensor and the renormalization 
of the string tension differ by source-localized contributions.

\subsubsection*{Acknowledgements}

This work is supported by the Fonds National Suisse. 
The work of R.~S. is supported by the Boninchi foundation.
The authors wish to thank M.~Maggiore for discussions and support, 
W.~Goldberger for his prompt and helpful correspondence and R.~Porto 
for introducing them to Feyn Calc.
R.~S. wishes to thank A. Nicolis for useful discussions and the Aspen Center 
for Physics for organizing a very stimulating workshop on gravitational wave 
astronomy.

\section{Appendix}

To second order the metric (\ref{met_nr}) can be rewritten as
\be
g_{\mu\nu}=\pt{
\bea{cc}
-1 -2\varphi -2\varphi^2 & a_j+2\varphi a_j \\
a_i +2\varphi a_i & \quad\delta_{ij} -2\varphi(\delta_{ij}+\varsigma_{ij})
+2\varphi^2\delta_{ij}+\varsigma_{ij}-a_ia_j\\
\eea
}\,,
\ee
where $\gamma_{ij}\equiv \delta_{ij}+\varsigma_{ij}$ (exact).
It is also useful to have the form of the inverse metric  
\be
g^{\mu\nu}=\pt{
\bea{cc}
-e^{-2\varphi}\pt{1-e^{4\varphi}\gamma^{ij}a_ia_j} & e^{2\varphi}a^j\\
e^{2\varphi}a^i & e^{2\phi}\gamma^{ij}\\
\eea
}\,.
\ee
To second order one has
\be
g^{\mu\nu}\simeq\pt{
\bea{cc}
-1+2\varphi-2\varphi^2+\delta_{ij}a_ia_j & a_j+2\varphi a_j-\varsigma_{jk}a_k\\
a_i+2\varphi a_i -\varsigma_{ik}a_k & 
\delta_{ij}+2\varphi\delta_{ij}-\varsigma_{ij}+
2\varphi\pt{\varphi\delta_{ij}-\varsigma_{ij}}\\
\eea
}\,.
\ee

The relevant integrals for computing Feynman diagrams like the one 
represented in fig.~\ref{pp_1PN} (see for instance \cite{Itzykson:1980rh} 
and \cite{BjerrumBohr:2002ks}) are
\renewcommand{\arraystretch}{1.4}
\be
\label{kikj}
\bea{rcl}
\ds\int \frac{d^3\vk}{\pt{2\pi}^3}\frac{k^ik^j}{k^2(\vk+\vq)^2} &=&\ds
\frac 1{64}\pt{-\delta^{ij}q+3\frac{q^iq^j}q}\,,\\
\ds\int \frac{d^3\vk}{\pt{2\pi}^3}\frac{k^i}{k^2(\vk+\vq)^2}&=&
\ds\frac{q^i}{16q}\,.
\eea
\ee
The integral relevant for fig.~\ref{3psi} is
\be
\int \frac{d^3\vk}{\pt{2\pi}^3}\frac 1{(k^2+M^2)\paq{(\vk+\vq)^2+M^2}}=
\frac {1}{4\pi q}\arctan\pt{\frac q{2M}}\,,
\ee
and to reconstruct the metric out of the effective EMT we used
\be
\label{int_k}
\bea{lcl}
\ds\int \frac{d^3\vq}{\pt{2\pi}^3}\,e^{i\vq\cdot{\bf x}}\,\frac 1{q^2}&=&
\ds\frac 1{4\pi |{\bf x}|}\,,\\
\ds\int \frac{d^3\vq}{\pt{2\pi}^3}\,e^{i\vq\cdot{\bf x}}\,\frac 1{q}&=&\ds
\frac 1{2\pi^2 |{\bf x}|^2}\,,\\
\ds\int \frac{d^3\vq}{\pt{2\pi}^3}\,e^{i\vq\cdot{\bf x}}\,\frac {q^iq^j}{q^3}&=&
\ds\frac 1{2\pi^2 |{\bf x}|^2}\pt{\delta_{ij}-2\frac{x^ix^j}{|{\bf x}|^2}}\,.
\eea
\ee

The relevant integrals for computing Feynman diagrams like the one 
represented in fig.~\ref{str_1PN} are (see again \cite{Itzykson:1980rh})
\begin{align}
\label{kikj_2}
\ds\int \frac{d^{2+\eps}k}{\pt{2\pi}^{2+\eps}}\frac{k^ik^j}{k^2(k+q)^2}=
&\ds\frac {\pt{q^2}^{\eps/2}}{\pt{4\pi}^{1+\frac \eps 2}}\left[
\frac 12\delta^{ij}\Gamma\pt{-\frac \eps 2}\int_0^1 
\paq{x(1-x)}^{\frac \eps 2}dx+\right.\\
&\ds\qquad\qquad\left.\,\frac{q^iq^j}{q^2}\Gamma\pt{1-\frac\eps 2}
\int_0^1x^2\paq{x(1-x)}^{\frac\eps 2-1}dx\right] \nn \,,\\
\ds\int \frac{d^{2+\eps}k}{\pt{2\pi}^{2+\eps}}\frac{k^i}{k^2(k+q)^2} =
&\ds\frac{q^i}{\pt{q^2}^{1-\eps/2}\pt{4\pi}^{1+\frac\eps 2}}
\Gamma\pt{1-\frac\eps 2}\int_0^1 x\paq{x\pt{1-x}}^{\frac \eps 2-1}dx\,.
\end{align}

Other useful formulas to anti-Fourier transform the string effective EMT at 
next-to-leading order, are
\bees
\int \frac{d^2q}{2\pi^2} \log(q) e^{iqx} &=&-\frac 1{x^2}\\
\int_{q_\eps} \frac{d^2q}{\pt{2\pi}^2}\frac 1{q^2}e^{iqx}&=&
-\frac 1{2\pi}\log(xq_\eps)+\frac{\ln 2-\gamma}{2\pi}+\frac{r^2q_\eps^2}{16\pi}
+o\paq{\pt{xq_\eps}^3}\,,
\ees
where a disk of radius $q_\eps$ around the origin has been cut out of the 
integral. Moreover  
\be
\int_0^{2\pi}e^{ix\cos\theta}d\theta=2\pi J_0(x)\,,
\ee
where $J_0$ is the Bessel function of zero-th order.
To derive the metric out of the string effective EMT the following integral
\be
\int_{q_\eps} \frac{d^2q}{\pt{2\pi}^2}\frac 1{q^4}e^{iqx}=
\frac{x^2}{2\pi}\paq{\frac 1{2q_\eps^2r^2}+\frac 18\log\pt{q_\eps^2r^2}+
\frac{\gamma-\ln 2-1}4+o(xq_\eps)}
\ee
is helpful.

The effective EMT (\ref{eemt_s}) in coordinate space is
\bees
\label{str_eemtI}
T_{ij}^{(I)}(x) &=& -\frac 4\pi G_N\mu^2 \\
&\times &
\left( 
\bea{cc}
\left( \alpha^2-\beta^2 \right) \left( C' \delta^{(2)} (x^a)+1/r^2 \right) & 0 \\
0 & \ds \frac{\alpha^2+\beta^2}{r^2}
\left( -\frac 12\delta_{ab} +\frac{x_a x_b}{r^2} \right) 
\eea
\right) \nn
\ees
where $r$ denotes the distance to the string in the transverse 
two-dimensional space. 
Here $C'$ denotes the $q$-independent part of the quantity 
defined in text in (\ref{div_const}).

To explicitly check conservation in the Fourier space of the string 
effective EMT (\ref{eemt_s}), let us write down the conservation 
of the EMT in $q$-space, keeping only the components transverse 
to the string world-sheet:
\be
\label{c_emt}
\pa^aT_{ab}(q)=\int d^2x\paq{\pa^a T_{ab}(x)}e^{iqx}=
\int d^2x\paq{\pa^a\pt{T_{ab}e^{iqx}}-iq^aT_{ab}(q)
e^{iqx}}\,,
\ee
which has an extra piece with respect to (\ref{naive_conserv}).
Let us restrict for simplicity to the total derivative term and let us fix 
the index $b=2$.
To make sense of the integral we have to integrate over a region $\Omega$ 
obtained by cutting out of the plane the the two regions $r<r_\eps$ and $r>R$, 
and we will finally (but after taking the other limits first) let 
$r_\eps\to 0$ and $R\to\infty$.\\
By changing coordinates from $y,z$ to $\rho,\theta$ according to 
$y=r\cos\theta$, $z=r\sin\theta$ and using the Green-Gauss theorem
one obtains
\be
\bea{rl}
\ds -\frac{\pi}{4G_N\mu^2\pt{\alpha^2+\beta^2}}
\int_\Omega d^2x &\pa^a\paq{T_{a2}(x)e^{iq_ax^a}}=
\ds \int_{\pa\Omega}\frac 1{2r}\cos\theta e^{iqr\cos\theta} d\theta\\
=&\ds\frac 1{2R}\int_0^{2\pi}\cos\theta e^{iqR\cos\theta}d\theta -
\frac 1{2r_\eps}\int_0^{2\pi} \cos\theta e^{iqr_\eps\cos\theta}d\theta\,.
\eea
\ee
The first integral is clearly vanishing in the limit $R\to\infty$. Expanding 
the exponential in the second integral, taking the limit $r_\eps\to 0$ and
finally plugging this result into (\ref{c_emt}), one has
$$
\pa^aT_{ab}(q)\propto \frac{iq^a}2-
iq^a\pt{-\frac{\delta_{ab}}2+\frac{q_aq_b}{q^2}}=0\,,
$$ 
qed.

%% file: Group_Paper/Group_Paper.tex
%%%%%%%%%%%%%%%%%%%%%%%%%
\chapter{Extracting the three- and four-graviton vertices
from \\ binary pulsars and coalescing binaries} 
\chaptermark{Extracting the three- and four-graviton vertices 
from experiments}
\label{chap:Group}
%%%%%%%%%%%%%%%%%%%%%%%%%

In \Sec{sec:Dam_Far} I have discussed the confrontation of \sts\ with GR by means of tests of relativistic gravity. 
This confrontation followed the studies of DEF and is summarized in \fig{fig:ST_Tests_Full}, where future GW detections seem not to be competitive with binary pulsar observations already at hand.
Surprisingly, even the detection of a neutron star-black hole system with the most sensitive space-based \ifo\ LISA is found to be less constraining than the electromagnetic observation of the same system performed by means of pulsar timing.
In the course of presenting these conclusions by DEF I have pointed out how they strongly  depend on the assumptions made.
Notably, DEF parametrized a certain class of \sts\ by considering the couplings of matter to the scalar field only up to second order in perturbation. 
Within this choice, DEF confronted different tests of gravity in the radiative regime only to the lowest orders probed by binary pulsars, i.e. up to terms of ${\cal O}(v/c)^5$ in the dynamics. 
On the contrary, GW detections will probe gravity up to ${\cal O}(v/c)^{12}$!
Therefore, there is the necessity to envisage testing frameworks for gravity in the dynamical regime probed by GWs. 
In \Sec{sec:PPN_GW}, I have described one such test where, extending the PPN approach to the radiative regime, one applies a phenomenological attitude towards the phasing coefficients of the GW formula. 
Another example of extended PPN tests is the one I present in the present chapter.
This is a study that I have conducted with my supervisor and the rest of his group at the University of Geneva: inspired by the NRGR method of ref.~\cite{NRGR_paper}, that I have  described in \chap{chap:EFT}, we constrained GR non-linearities from a field-theoretical perspective~\cite{UC:09}. 
Here I report our publication, while in the following chapter I discuss more details in the context of a follow-up of our work.

%\vspace{1cm}
\newpage

\begin{center}
{\large PHYSICAL REVIEW D {\bf 80}, 124035 (2009)}
\par\end{center}{\large \par}

\begin{center}
\textbf{\Large Extracting the three- and four-graviton vertices
from binary pulsars and coalescing binaries}
\par\end{center}{\Large \par}

\begin{center}
{\large Umberto Cannella, Stefano Foffa, 
Michele Maggiore, Hillary Sanctuary \\ and 
Riccardo Sturani}
\par\end{center}{\large \par}

%\maketitle
\vspace{1.2cm}

%%%%%%%%%%%%%%%%%%%%%%%%%%%%%%%
\subsubsection{abstract}
%%%%%%%%%%%%%%%%%%%%%%%%%%%%%%%

Using a formulation of the post-Newtonian  expansion in terms of Feynman 
graphs, we discuss how various tests of General Relativity (GR) can be 
translated into measurement of the three- and four-graviton vertices.
In problems involving only the conservative dynamics of a system, a
deviation of the three-graviton vertex from the GR prediction is
equivalent, to lowest order, to the introduction of the parameter $\bppn$
in the parametrized post-Newtonian formalism, and its strongest bound 
comes from lunar laser ranging, which measures it at the 0.02\% level. 
Deviation of the three-graviton vertex from the GR
prediction, however, also affects the radiative sector of the theory. We
show that  the timing of the  Hulse-Taylor binary
pulsar provides a bound on the deviation of the three-graviton vertex 
from the GR prediction at the 0.1\% level.
For coalescing binaries at  interferometers we find that,  
because of degeneracies with other parameters in the template 
such as mass and spin, the effects of modified
three- and four-graviton vertices is just to induce an error in the
determination of these parameters and, at least in the restricted PN
approximation,  it is not possible to use coalescing binaries for
constraining  deviations of the vertices
from the GR prediction.

\vspace{0.8cm}

\noindent
{\small \textbf{DOI}: 10.1103/PhysRevD.80.124035 \hfill 
\textbf{PACS} numbers: 04.30.--w, 04.80.Cc, 04.80.Nn}

\vspace{1.0cm}

%%%%%%%%%%%%%%%%%%%%%%%%%%%%%%%%%
\section{Introduction}
%%%%%%%%%%%%%%%%%%%%%%%%%%%%%%%%%

Binary pulsars, such as the Hulse-Taylor \cite{HT} and 
the double pulsar~\cite{double_pulsar}, are
wonderful laboratories for testing  General
Relativity (GR).  
They have given the first experimental confirmation of the existence 
of gravitational radiation
\cite{Taylor:1982zz,Weisberg:2004hi}, provide stringent tests of GR
and allow for comparison with alternative theories of gravity, such 
as scalar-tensor theories~\cite{Dam_Tay:91,Dam_Tay:92,
Taylor:1993zz,Kramer:2006nb, Will:1994fb,Dam_Far_GW_Tests:98,
Dam_Far-ST_CQG,Dam_Far-Non_Pert_PRL,
Dam_Far_2PN_Diags:PRD53,Dam_Far-ST_Puls:PRD54} 
(see \cite{Stairs:2003eg,Will_LRR:2006,Maggiorebook} for reviews). 
Another very sensitive probe of the non-linearities of GR is given
by the  gravitational wave (GW) emission during the last stages of the
coalescence of compact binary systems made of black holes
and/or neutron stars, which is one of the
most promising signals for GW interferometers such as LIGO~\cite{LIGO} and Virgo~\cite{Virgo}, especially in their advanced stage, and for the space interferometer LISA~\cite{LISA}.
Various investigations have been devoted to the possibility of using
the observation of coalescing binaries at GW interferometers to probe 
non-linear aspects of GR \cite{Will:1994fb,Dam_Far_GW_Tests:98,
Blan_Sathya-Tail,Blanchet:1994ez,Will:1997bb,Berti:2004bd,
Berti:2005qd,Arun:2006yw,Arun:2006hn}.

Compact binary systems probe both the radiative sector of the theory, 
through the emission of gravitational radiation, and the
non-linearities intrinsic to GR which are already present in the
conservative part of the Lagrangian.
In a field-theoretical language, these non-linearities can be traced
to the non-Abelian vertices of the theory, such as the three-
and four-graviton vertices. It is therefore natural to ask
whether from binary pulsars or from future observations of coalescing
binaries at interferometers one can extract a measurement of these
vertices, much in the same spirit in which the triple
and quartic gauge boson couplings have been measured at LEP2 and at the 
Tevatron~\cite{deRujula:1978p144,Ellison:1998uy,Abbott:1998dc,Schael:2004tq}.

In this paper we tackle this question.  The organization of the paper
is as follows. In Section~\ref{sec:tag} we discuss how to ``tag''
the contribution of the three- and four-graviton vertices to various
observables in a consistent and gauge-invariant
manner, and we compare it with other approaches, such as the
parametrized post-Newtonian (PPN) formalism~\cite{Will_LRR:2006}.
In particular, we
find that the introduction of a modified three-graviton vertex
corresponds -- in the conservative sector of the theory and at first 
Post-Newtonian order (1PN) -- to the
introduction of a value for the PPN parameter $\bppn$ different from the
value $\bppn=1$ of GR. However, a modified three-graviton vertex also
affects the radiative sector of the theory, which is not the case for
the PPN parameter $\bppn$. We also discuss  subtle issues related to
the possible breaking of gauge invariance which takes place when one modifies
the vertices of the theory.
In Section~\ref{sec:comp} we present our computations with
modified vertices, and in Section~\ref{sec:exp} we compare these computations
with experimental results obtained from the timing of
binary pulsars and 
with what can be expected from the detection of gravitational waves
(GWs) at ground-based interferometers or with
the space interferometer LISA. 
Section~\ref{sec:group_concls} contains our
conclusions.

%%%%%%%%%%%%%%%%%%%%%%%%%%%%%%%%%%%%%%%
\section{Tagging the three- and four-graviton vertices}\label{sec:tag}
%%%%%%%%%%%%%%%%%%%%%%%%%%%%%%%%%%%%%%%

Our aim is to quantify how well the
non-linearities of GR can be tested by various existing or planned 
experiments/observations. Historically, there have been several approaches
to this problem and, 
basically, one can identify two complementary strategies. The
first is to develop a purely phenomenological approach in which
deviations from GR are expressed in terms of a number of parameters,
without inquiring at first whether such a
deformation of GR can emerge from a fundamental theory.
An example of such an approach is the
parametrized post-Newtonian (PPN) formalism. In its simpler version,
it consists of writing the 1PN metric generated by a source, treated
as a perfect fluid with density $\rho(\vx)$ and velocity field
$v(\vx)$, 
in the form
\bees
g_{00}&=&-1+2U-2\bppn U^2\, ,\label{g00}\\
g_{0i}&=&-\frac{1}{2}(4\gppn+3)V_i\, ,\\
g_{ij}&=&(1+2\gppn U)\d_{ij}\, ,\label{gij}
\ees
where
\bees
U(\vx )&=&\int d^3x'\, \frac{\rho(x')}{|\vx-\vx'|}\, ,\\
V_i(\vx )&=&\int d^3x'\, \frac{\rho(x')v_i(x')}{|\vx-\vx'|}\, ,
\ees
and the standard PPN gauge has been used~\cite{Will_LRR:2006,Will_TEGP}
(we use units $c=1$).
General Relativity corresponds to $\bppn=1$ and $\gppn =1$. More
phenomenological parameters can be introduced  by working at higher PN
orders, see \cite{Will_LRR:2006}.
One then investigates how deviations of $\bppn$ and $\gppn$ from
their GR values affect various experiments.
Writing $\bppn=1+\bar{\b}$ and $\gppn=1+\bar{\g}$, the best current
limits (at 68\% c.l.) are
\be
\bar{\g}=(2.1\pm 2.3)\times 10^{-5}\, 
\ee
from
the Doppler tracking of the Cassini spacecraft, and
\be\label{Cassini44}
4\bar{\beta}-\bar{\g}=(4.4\pm 4.5)\times 10^{-4}\, 
\ee
from lunar laser ranging. 
This bound comes from the Nordtvedt effect, i.e. from the fact that, 
in a theory with $\bar{\beta}$ and $\bar{\g}$ generic, the weak 
equivalence principle is violated and the Earth and the Moon can 
fall toward the Sun with different accelerations, which depend  on their 
gravitational self-energy. 
The effect is studied by monitoring the Earth-Moon distance with 
lunar laser ranging.
The perihelion shift of Mercury gives instead the bound
$|\bar{\b}|<3\times 10^{-3}$~\cite{Will_LRR:2006,Amsler:2008zzb}.

To get these bounds, we do not need to know the fundamental
theory that gives rise to values of $\bppn$ and $\gppn$ 
that differ from their
GR values. However, it is of course interesting to see that consistent  
field theories exist
that give rise to values of $\bppn$ and $\gppn$
different from one. For instance,
a Brans-Dicke theory with parameter $\omega_{\rm BD}$ gives
$\bppn=1$ and $\gppn =(1+\omega_{\rm BD})/(2+\omega_{\rm BD})$, with the GR
value $\gppn=1$ recovered for $\omega_{\rm BD}\ra\infty$, while  more
general tensor-scalar theories can produce both $\gppn\neq 1$ and 
$\bppn\neq 1$. 
However, in the PPN approach, one can also explore other possibilities, 
such as PPN parameters that correspond to preferred-frame effects or
to violation of the conservation of total momentum, which are not
necessarily well-motivated in terms of  current field-theoretical
ideas on possible extensions or UV completions of GR.
It is also important to observe 
that the parameters $\bppn$ and $\gppn$ are gauge-invariant,
and therefore observables, because they have been defined with respect
to a specific gauge, namely the standard PPN 
gauge in which the metric takes the 
form~(\ref{g00})--(\ref{gij}).

A second, complementary, approach to the problem is to study a specific
class of field-theoretical extensions of GR. A typical well-motivated
example is provided by multiscalar-tensor theories. 
These have been studied
in detail and compared with experimental tests of relativistic gravity
in refs.~\cite{Dam_Tay:92,Will:1994fb,Dam_Far_GW_Tests:98,
Dam_Far-ST_CQG,Dam_Far-Non_Pert_PRL,
Dam_Far_2PN_Diags:PRD53,
Dam_Far-ST_Puls:PRD54}. 
This approach has the advantage that one is testing a specific 
and well-defined fundamental theory. 
On the other hand, an experimental  bound on the parameters of a 
given scalar-tensor theory, such as for instance the bound 
$\omega_{\rm BD}>40000$ on the parameter $\omega_{\rm BD}$ 
of Brans-Dicke theory obtained from the tracking of the Cassini 
spacecraft~\cite{Cassini} is, strictly speaking, only a 
statement about that particular extension of GR and not about GR itself.

In this paper we quantify how well GR performs with respect
to experiments of relativistic gravity by studying how much these
experiments constrain the values of the non-Abelian vertices of the
theory, in particular the three-graviton vertex and the four-graviton
vertex. We proceed as follows. 
After choosing a gauge (the De~Donder gauge, corresponding 
to harmonic coordinates) we multiply the 
three-graviton vertex by a factor $(1+\b_3)$ and
the four-graviton vertex by a factor $(1+\b_4)$, with constants
$\b_3$ and $\b_4$. For $\b_3=\b_4=0$ we recover GR. 
Observe that, since $\b_3$ and $\b_4$ are defined with respect to 
a given gauge choice, they are gauge-invariant
by definition. This is in fact the same logic used to define in a
gauge-invariant manner the PPN parameters $\bppn$ and $\gppn$.

We then use a Feynman diagram approach to compute the 
modifications induced by $\b_3$
and $\b_4$ on various
observables in classical GR. Diagrammatic approaches and 
field-theoretical methods have been in use in classical GR for 
a long time, see e.g.~\cite{Dam_Far_2PN_Diags:PRD53,Damour:2001bu,Blanchet:2003gy}. 
We make use of the effective field theory formulation proposed 
in~\cite{NRGR_paper}, which provides a clean 
and systematic separation of the effects that depend 
on (model-dependent) short-distance physics from 
long-wavelength gravitational dynamics, and in the non-relativistic 
limit (after performing a multipole expansion)
has manifest power counting in the typical velocity 
$v$ of the source.

In the next section we see how these deformed vertices give
additional terms in the PN effective Lagrangian. 
In particular we find that, in the conservative sector of the theory, 
the introduction of $\b_3$ is phenomenologically equivalent, 
at 1PN level, to the introduction of a non-trivial value of $\bppn$ 
given by $\bppn=1+\b_3$. 
However, $\b_3$ also affects the radiative sector of the theory, 
i.e. the Lagrangian describing the interaction between the
matter fields and the gravitons radiated at infinity.

Before entering into the technical aspects, however, let us 
clarify the meaning of the introduction of $\b_3$ and $\b_4$. 
In ordinary GR, with $\beta_{3,4}=0$, coordinate transformation
invariance ensures that the negative norm states decouple. 
After gauge fixing (in the De~Donder gauge for instance), 
the kinetic terms for all of the ten components of the metric are 
invertible, but four of them have the wrong sign, 
i.e. they give rise to negative norm states. 
In the De Donder gauge the six positive-norm states that
diagonalize the kinetic term are 
\be
\tilde{h}_{ij}\equiv h_{ij}+\frac
12\delta_{ij}(h_{00}-\delta^{lm}h_{lm})\, ,
\ee
while the four "wrong-sign'' components are given by the spatial
vector $h_{0i}$ and by the scalar
\be
\tilde{h}_{N}\equiv h_{00}-\delta^{lm}h_{lm}\, .
\ee
In standard GR the existence of these negative-norm states 
do not create difficulties because they are coupled to four integrals 
of motion (energy, and the three components of angular momentum), 
so they cannot be produced. 
In contrast, the remaining six "healthy'' components couple to the 
source multipole moments. 
After complete gauge fixing one finds that among the six positive-norm 
states, four obey Poisson-like equations, so they do not radiate
(even though they are non-radiative physical 
degrees of freedom), while the remaining two 
are the radiative degrees of freedom representing 
GW's \cite{Flanagan:2005yc}.

Allowing $\beta_3\neq 0$ has the effect that the negative 
norm state $\bar h_{N}$ now couples, already at lowest order, 
to a non conserved quantity, namely to a combination of the 
Newtonian kinetic and potential energy of the binary system. 
This means that, in general, a modification of GR in which we just
change the strength of the three-graviton vertex cannot be taken as a
fundamental field theory, neither at the quantum level, nor even at
the classical level, since the negative-norm state  contributes to
the classical radiated power (a related concern is that, for
$\beta_3\neq 0$, the energy--momentum tensor is in general 
not conserved).
A consistent classical and quantum field theory could in principle 
emerge from a simultaneous modification of all the vertices of the 
theory, such as the three-, four- and higher-order graviton vertices, 
together with a related modification of the graviton-matter couplings. 
As a trivial example, an overall rescaling of the gauge coupling in 
a Yang-Mills theory, or of Newton's constant in GR, results in a 
combined modification of all the vertices, but obviously introduces 
no pathology. 
Anyway, our approach to the problem is purely phenomenological. 
We introduce $\b_3$ and $\b_4$ simply as ``tags'' that allow us 
to track the contribution of the three- and four-graviton vertices 
throughout the computations. 
As long as $|\b_3|\ll 1$ and $|\b_4|\ll 1$, the corrections that they 
induce to the radiated power are small compared to the standard 
GR result, so the total radiated power is given by the GR result 
plus a small correction, and in particular the total radiation emitted 
is positive. 
At this phenomenological level the introduction of modified vertices 
is therefore acceptable, and provides a simple and, most importantly, 
gauge invariant manner of quantifying how well different observations 
constrain the non-linear sector of GR, in a way which is intrinsic to 
GR itself, without reference to any other specific field theory.

In this sense, our approach is  close in spirit to the phenomenological 
PPN approach, and can be seen as an extension of it where the 
radiative sector of the theory is  also modified.
Another approach which is related to ours is the one proposed
in ref.~\cite{Arun:2006yw,Arun:2006hn}. 
They consider the phase of the GW emitted during the coalescence 
of compact binaries, which up to 3.5PN has the form
\be\label{Psipsik}
\Psi(f)=2\pi f t_c-\Phi_c+
\sum_{k=0}^7[\psi_k+\psi_{kl}\ln f] f^{(k-5)/3}\, ,
\ee
where $f$ is the GW frequency, and $t_c$ and $\Phi_c$ are the 
time and the phase at merger.
The seven non-zero coefficients $\psi_k$ with $k=0,2,3,\ldots 7$ 
and the two non-zero coefficients $\psi_{kl}$ with $k=5,6$ are 
known from the PN expansion, in terms of the two masses 
$m_1$ and $m_2$. 
In ref.~\cite{Arun:2006yw,Arun:2006hn} they study how the 
template is affected if these coefficients are allowed to vary,
so that two of them, the 0PN coefficient $\psi_0$ and
the 1PN coefficient $\psi_2$, are used to fix the masses 
$m_1$ and $m_2$ of the two stars, while varying any of the 
remaining coefficients with $k\geq 3$ provides a test of GR.
In the case of coalescing binaries at
interferometers, our introduction  of $\b_3$  and $\b_4$
is a particular case of a more general
analysis in which one treats  the quantities $\psi_k$ and $\psi_{kl}$
as free parameters, but it has a sharper field-theoretical meaning
since $\b_3$ and $\b_4$ measure 
the deviation of the three- and four-graviton vertices from
the GR prediction. For the same reason, we are also able to compare
the effect of $\b_3$ on the waveform of coalescing binaries  with its effect
on binary pulsar timing and on solar system experiments, while in the
phenomenological approach in which the parameters  $\psi_k$ and
$\psi_{kl}$ of the GW phase are taken as free parameters, a
modification of the waveform of coalescing binaries cannot be related
to a modification of the binary pulsar timing formula.

Another  issue is whether a modification of the vertices 
of this form (typically with $\b_3$, $\b_4$, etc. not independent, but
related to each other by some
consistency conditions) could
emerge from a plausible and consistent extension of GR. Actually,
a typical UV completion of GR
at an energy scale $\Lambda$ 
will rather generate corrections to the vertices that are
suppressed by inverse powers of $\Lambda$, so it would give rise to an energy
dependent $\beta_3$, e.g. $\beta_3=E^2/\Lambda^2$, which furthermore, at the
energy scales that we are considering and for any sensible choice of
$\Lambda$, would be utterly negligible. Still, let us remark that
this kind of behavior is not
a theorem. It assumes  the
UV-IR decoupling typical of effective field theories, and one can
exhibit counterexamples. For instance,
in non-commutative Yang-Mills theories there is a UV-IR mixing, such that 
low-energy processes receive contributions from loops where
very massive particles are running, and these contributions are independent 
of the mass of these particles~\cite{Minwalla:1999px}. Anyway, again  our
aim here 
is not to test any given consistent extension of GR, but rather provide 
a simple and phenomenologically
consistent way of quantifying how well various experiments can test
the non-linearities of GR, and  quantify  how the results of 
different experiments compare among themselves. 

It is also interesting to observe that, even when  $\b_3$ and $\b_4$
are non-zero, the graviton remains massless at the classical level, since 
 $\b_3$ and $\b_4$ affect interaction terms, but not the kinetic term. 
 The breaking of diffeomorphism invariance induced by  $\b_3$ 
and $\b_4$ could in principle generate a graviton mass at the one-loop level. 
However, even if we are using the language of quantum field theory, 
in the end we are only interested in the classical theory, since quantum 
loops are suppressed by powers of $\hbar/L$, where $L$ is the angular 
momentum of the system, so they are completely negligible for a 
macroscopic system~\footnote{At the quantum level, if a mass is 
generated, it will be power divergent with the cutoff, and will have 
to be fine tuned order by order in perturbation theory.
However, the fact that quantum divergences
have to be subtracted order by order is a generic problem of the
standard quantum extension of GR, independently of  $\beta_3$. 
In any case, again, our approach is purely phenomenological, and 
the introduction of $\beta_3$ is simply a tool for tracking a specific 
contribution to the computation.}.

%%%%%%%%%%%%%%%%%%%%%%%%%%%%%%%%%%%%%%%
\section{Effective Lagrangian from a modified three-gra\-viton vertex}\label{sec:comp}
%%%%%%%%%%%%%%%%%%%%%%%%%%%%%%%%%%%%%%%

To perform our computations we use the effective field theory 
formulation proposed in ref.~\cite{NRGR_paper}.
Computations of the conservative dynamics at 2PN level have been performed 
using this effective field theory technique 
\cite{Chu:2008xm,Gilmore:2008gq}
and the results are in agreement with the classic 2PN results 
of refs.~\cite{DD81,Damour_CR}, while the full 3PN result for non-spinning
particles to date has only been obtained with the standard
PN formalism using dimensional
regularization ~\cite{Damour:2001bu,Blanchet:2003gy}
(see ref.~\cite{Blan-LRR:06}, or
Chapter~5 of  ref.~\cite{Maggiorebook},
for a pedagogical introduction to the PN
expansion and for a more complete list of references).
Spin-spin contributions at 3PN order in the conservative
two-body dynamics  have recently been  
computed both with the effective field theory
techniques~\cite{Porto:2008tb,Levi:2008p25} and with the ADM 
Hamiltonian formalism~\cite{Steinhoff:2008ji} (see also
\cite{Galley:2008ih,Galley:2009px} for other applications of the EFT technique 
related to gravitational radiation).

In the formalism of ref.~\cite{NRGR_paper},
after integrating out length-scales
shorter than the size of the compact objects, the action 
becomes
\be
S=S_{\rm EH}+S_{\rm pp}\, , 
\ee
where $S_{\rm EH}$ is the
Einstein-Hilbert action and
\be
S_{\rm pp}=-\sum_{a}m_a\int d\tau_a\, 
\ee 
is the point particle action.
Here $a=1,2$ labels the two bodies in the binary system and
$d\tau_a=\sqrt{\gmn(x_a) dx_a^{\mu}dx_b^{\nu}}$.
One then observes that, in the binary problem, the gravitons appearing
in a Feynman diagram can be divided into two classes: the forces
between the two bodies with relative distance $r$ and relative speed
$v$ are mediated by
gravitons whose
momentum $k^{\mu}$ scales as $(k^0\sim v/r,|{\bf k}|\sim 1/r)$. These
are called ``potential gravitons'' and are off-shell, so they can
only appear in internal lines. The gravitons radiated to infinity
rather have
$(k^0\sim v/r,|{\bf k}|\sim v/r)$.
One then writes $\gmn=\emn+\hmn$ and 
separates $\hmn$ into two parts,
$\hmn(x)=\bhmn(x)+H_{\mu\nu}(x)$
with $\bhmn(x)$ describing the radiation gravitons and
$H_{\mu\nu}(x)$ the potential gravitons.  
One fixes the  de Donder gauge and,
expanding the action in powers of
$\bhmn$ and $H_{\bf k\mu\nu}(x_0)$, one can
read off the propagators and the vertices, 
and write down the Feynman rules of the theory. Then, using
standard methods from quantum field theory, one can
construct an effective Lagrangian that, used at tree level, reproduces
the amplitudes computed with the Feynman graphs. For a classical
system, whose angular momentum $L\gg\hbar$, only tree graphs
contribute, and reproduce the classical Lagrangian that is usually 
derived from GR using the PN expansion. 
For instance,
the conservative dynamics of the two-body problem, at 
1PN level, is given by the 
Einstein--Infeld--Hoffmann Lagrangian, 
\bees\label{2L2PN}
L_{\rm EIH}&=&\frac{1}{8}m_1 v_1^4 + \frac{1}{8}m_2 v_2^4
+\frac{G_Nm_1m_2}{2r}\, \left[ 3 (v_1^2+v_2^2) \phantom{\frac12} \right. \nn\\
&&\hspace*{-4mm}\left. -7{\bf v}_1\bdot{\bf  v}_2
-(\hatr\bdot{\bf v}_1)(\hatr\bdot{\bf v}_2) -\frac{G_N(m_1+m_2)}{r}\right].
\ees
In the language of the effective field theory of ref.~\cite{NRGR_paper}, 
this result is obtained from Feynman diagrams involving the exchange 
of potential gravitons. 
In particular, the terms linear in $G_N$ in \eq{2L2PN} are obtained from 
a single exchange of potential gravitons between two matter lines, 
see Fig.~4 of ref.~\cite{NRGR_paper}, while the term proportional 
to $G_N^2$ is obtained from the sum of the two graphs shown in 
Fig.~\ref{fig1}.
In the derivation of the conservative 1PN Lagrangian, 
the three-graviton vertex only enters through the graph 
in Fig.~\ref{fig1}a. Multiplying this vertex
by a factor $(1+\b_3)$ and repeating the same computation as in
ref.~\cite{NRGR_paper} we get
the additional contribution to the conservative part of the Lagrangian
\begin{eqnarray}
\label{lb1}
\D{\cal L}_{\rm cons}=-\beta_3\frac{G_N^2 m_1 m_2 (m_1+m_2)}{r^2}\, .
\end{eqnarray}

%%%%%%%%%%%%%%%%%%%%%%%%%%%%%%%%%%%%%%%
\begin{figure}
\begin{center}
\includegraphics[width=0.6\textwidth]{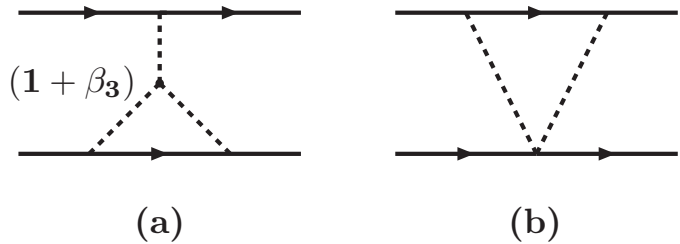}
\caption{The diagrams that give the terms proportional to $G_N^2$ in the 1PN
Lagrangian. Dashed lines denote potential gravitons, solid lines the
point-like sources.} \label{fig1}
\end{center}
\end{figure}
\noindent
%%%%%%%%%%%%%%%%%%%%%%%%%%%%%%%%%%%%%%%

Comparing this result with  the Lagrangian  whose equations of motion
are the same as the equations of motion of a test particle in the PPN
metric (\ref{g00})--(\ref{gij}) (see eq.~(6.80) of ref.~\cite{Will_TEGP})
we find that,  to 1PN order and as far as the conservative dynamics is concerned, the introduction of $\b_3$ gives rise to a PPN theory with $\b=1+\b_3$, i.e. $\bar{\b}=\b_3$, while  $\g=1$ as in GR.  
Therefore the bound on $\bar{\b}$ from the perihelion of Mercury translates into $|\b_3|<3\cdot 10^{-3}$, while \eq{Cassini44} translates into the bound (at $68\%$ c.l.)
\be\label{lunar}
|\b_3|<2\cdot 10^{-4}\, .
\ee
This bound reflects the fact that $\b_3$, just as the PPN parameter $\bar{\beta}$, violates the weak equivalence principle.
The introduction of $\b_3$, however, also affects the radiative sector
of the theory, something that is not modeled in the phenomenological  PPN
framework since the latter by definition is only concerned with the
motion of test masses in a deformed metric, and therefore
only modifies the conservative part of the dynamics. 
Note however that, in the framework of multiscalar-tensor
theories, the extension of the PPN formalism introduced
in Ref~\cite{Dam_Far-ST_CQG}  allows for a consistent treatment of
both the conservative dynamics (including  the case of strongly
self-gravitating bodies) and of radiative effects.

It is clearly interesting to see what bounds on $\b_3$ can
be obtained from experiments that probe the radiative sector of GR, 
such as the timing of binary pulsars or the observation of the coalescence
of compact binaries at interferometers. 
The effective Lagrangian describing the interaction of the binary 
system with radiation gravitons is obtained by computing
the three graphs in  Fig.~\ref{figGR6} (corresponding to 
Fig.~6 of ref.~\cite{NRGR_paper}), and the introduction
of $\b_3$ affects the $HHh$ vertex in Fig.~\ref{figGR6}c. 

Computing these graphs as in ref.~\cite{NRGR_paper}, but with 
our modified three-graviton vertex, we find
\be\label{lquad}
{\cal L}_{\rm rad}=
\frac{1}{2\mpl}[Q_{ij}R_{0i0j}+qR_{0i0i} + \beta_3 
(3 V h_{00} +Z^{ij} h_{ij})]\, ,
\ee
where $Q_{ij}$ is the quadrupole moment of the source and we define
\bees
\label{eq:qVZ}
q&=& \frac{1}{3} \sum_a m_a x_a^2\, , \\
V(r)&=& \frac{G_N m_1m_2}{r}\, ,\\
Z^{ij}(r)&=&\frac{G_Nm_1m_2r^ir^j}{r^3}\, ,
\ees
where ${\bf r}={\bf x}_1-{\bf x}_2$.
The term $Q_{ij}R_{0i0j}$
in \eq{lquad} is the usual quadrupole interaction. 
The second term, $qR_{0i0i}$, is
non-radiating when $\b_3=0$, but we will see that
for $\b_3\neq 0$ it contributes to the radiated power when the orbit
is non-circular. The last two terms in \eq{lquad}
are the explicit
$\beta_3$-dependent terms induced by the modification of the
three-graviton vertex in Fig.~\ref{figGR6}c. 

Finally, we have omitted a term where $h_{00}$ is coupled to the conserved energy (at this order) and $h_{0i}$ is coupled to the conserved angular momentum, since these terms do not  generate gravitational radiation.

%%%%%%%%%%%%%%%%%%%%%%%%%%%%%%%%%%%%%%%
\begin{figure}
\begin{center}
\includegraphics[width=0.7\textwidth]{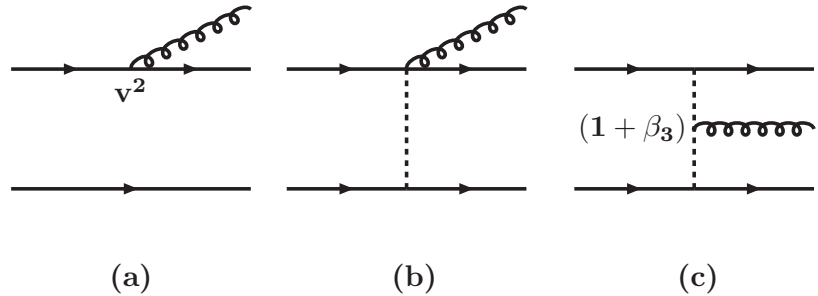}
\caption{\label{figGR6}
The diagrams that contribute to the matter-radiation Lagrangian.
Dashed lines denote potential gravitons, wiggly lines radiation
gravitons, and solid lines the point-like sources.}
\end{center}
\end{figure}
%%%%%%%%%%%%%%%%%%%%%%%%%%%%%%%%%%%%%%%

The back-reaction of GW emission on the source can be computed as usual
from the energy balance equation  
\be\label{balance}
P_{\rm GW}=-\dot{E}\, , 
\ee
where
$P_{\rm GW}$ is the power radiated in GWs and $E$ the orbital energy of
the system. To obtain 
the expression for the radiated power 
for $\b_3\neq 0$ we cannot simply use the quadrupole
formula of GR, since the introduction of 
$\b_3$ generates new contributions. To take them into
account we proceed as in ref.~\cite{NRGR_paper}, and we compute
the imaginary part of the graph shown in 
Fig.~\ref{figGR3}. The vertices of the graph can be read from
\eq{lquad}. When $\beta_3=0$ the only relevant vertex comes from
$Q_{ij}R_{0i0j}$, and the computation of the imaginary part gives back
the Einstein quadrupole formula \cite{NRGR_paper}. In our case we
have various possible vertices, and we must compute the imaginary part of
\be \label{imagine}
\frac{-i}{8\mpl^2}
\sum_{a,b=1}^4 \int dt_1dt_2\, I^a_{ij}(t_1)I^b_{kl}(t_2) 
\langle S^a_{ij}(t_1)S^b_{kl}(t_2)\rangle\, ,
\ee
where $I^a_{ij}=(Q_{ij},q\d_{ij},\beta_3 V\d_{ij},\beta_3  Z_{ij})$ 
depends on the matter variables
and $S^a_{ij}=(R_{0i0j}, \\ R_{0i0j}, \d_{ij}h_{00}, h_{ij})$ on the
gravitational field.
When both vertices of the diagram in 
Fig.~\ref{figGR3} are proportional to the quadrupole, one obtains the
usual GR result 
\be
\label{emqq}
P_{QQ}=\frac{G_N}{5}\langle \dddot{Q}_{ij}\dddot{Q}_{ij}\rangle\, ,
\ee
as already found in~\cite{NRGR_paper}. Computing the other
contributions we find that
the terms $P_{Qq}$ and $P_{qq}$ vanish identically. In fact,
the  $Qq$ and $qq$ graphs 
vanish because $Q_{ij}$ is traceless, while the $qq$ graph vanishes
because $\delta_{ij}\delta_{kl}$ gives zero when contracted
$\delta_{ik}\delta_{jl}+\delta_{il}\delta_{jk}-\frac 23\delta_{ij}\delta_{kl}$,
which is the tensor that comes out from the two-point function 
$\langle R_{0i0j}R_{0k0l}\rangle$.
The $QV$, $qZ$ and $VZ$ graphs vanish for similar reasons, so
the only relevant contributions come
from the $QZ$ and $qV$ graphs, and we find
\be
\label{exem1}
P_{QZ}=-2\beta_3G_N\langle\dddot Q_{ij}\dot Z_{ij}\rangle\, ,
\ee 
and
\be
\label{exem2}
P_{qV}=-6\beta_3G_N\langle\dddot q\dot V\rangle\, . 
\ee
As for the $VV$ and $ZZ$ graphs, they give a
contribution that, from the point of view of the multipole expansion,
is of the same order as the quadrupole radiation but  proportional
to $\b_3^2$, and can be neglected.

We can now use these results to perform the comparison with binary
pulsars and with interferometers.

%%%%%%%%%%%%%%%%%%%%%%%%%%%%%%%%%%%%%%%
\section{Comparison with experiments}\label{sec:exp}
%%%%%%%%%%%%%%%%%%%%%%%%%%%%%%%%%%%%%%%

As we already saw in \eq{lunar}, solar system experiments, and 
in particular  lunar laser ranging, give the bound
$|\b_3|<2\cdot 10^{-4}$. In this section we study the bounds on $\b_3$
that can be obtained from binary pulsars and from the detection of
coalescing binaries at interferometers.

%%%%%%%%%%%%%%%%%%%%%%%%%%%%%%%%%%%%%%%
\subsection{Binary pulsars}
%%%%%%%%%%%%%%%%%%%%%%%%%%%%%%%%%%%%%%%

Since  $\beta_3$ modifies the emitted power already at Newtonian
order, the energy of the orbit in
\eq{balance} can now be directly
computed using the Keplerian equations of motion. 
We see that this test of GR is conceptually different from the tests based
on solar system experiments. The latter only probe the conservative
part of the Lagrangian, i.e. the $\b_3$-dependent term given in
\eq{lb1}, 
while binary pulsars are sensitive to the
$\b_3$ dependence given in the radiation Lagrangian (\ref{lquad})
(even if the effect of $\beta_3$ on the conservative dynamics will
also enter, 
through the determination of the masses of the stars from the
periastron shift, see below).

Using the Keplerian equations of motion for an elliptic orbit of
eccentricity $e$ we get
\begin{eqnarray}
\label{pellips}
P_{QQ}&=&\frac {32 G_N^4 \mu^2 M^3}{5 a^5(1-e^2)^{7/2}}
\left(1+\frac{73}{24} e^2 +\frac{37}{96}e^4\right)\,,\\
P_{QZ}&=&\beta_3 \frac {32 G_N^4 \mu^2 M^3}{5 a^5(1-e^2)^{7/2}}
\left(\frac{5}{2}+\frac{175}{24}e^2+\frac{85}{96}e^4\right),\\
P_{qV}&=&-\beta_3 \frac {32 G_N^4 \mu^2 M^3}{5 a^5(1-e^2)^{7/2}} 
\left(\frac{5}{16}e^2+\frac{5}{64}e^4\right)\, ,
\end{eqnarray}
where $M=m_1+m_2$ is the total mass, $\mu=m_1m_2/M$ is the reduced mass,
and we will also use the notation
$\nu =m_1m_2/M^2$ for the symmetric mass ratio.
From the energy balance equation we then get the evolution of the
orbital period $P_b$,
\begin{eqnarray}
\label{dotT}
\frac{\dot{P}_b}{P_b}=-\frac{96}{5}G_N^{5/3}\nu\, M^{5/3}
\left(\frac{P_b}{2\pi}\right)^{-8/3}[f(e)+\beta_3 g(e)]\,,
\end{eqnarray}

%%%%%%%%%%%%%%%%%%%%%%%%%%%%%%%%%%%
\begin{figure}
\begin{center}
\includegraphics[width=0.4\textwidth]{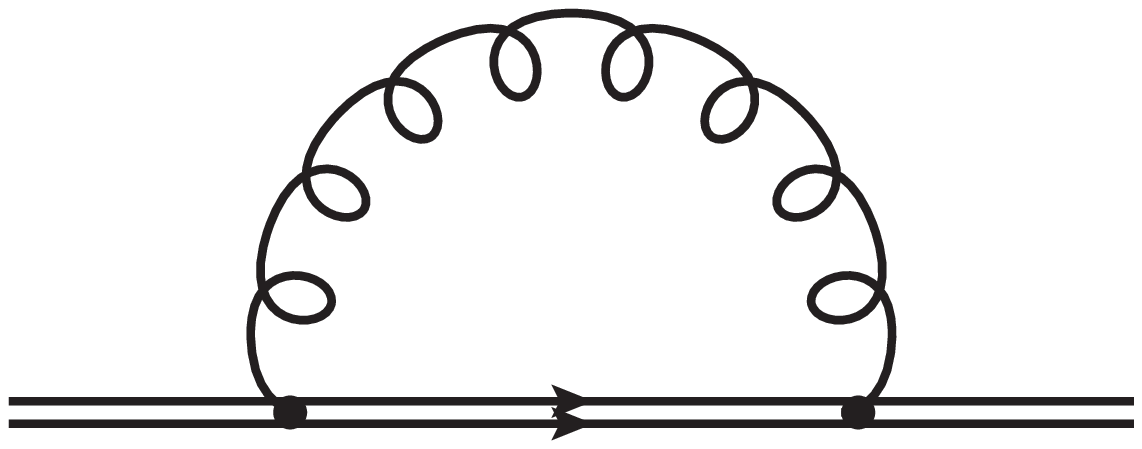}
\caption{The self-energy diagram whose imaginary part gives the 
radiated power. The wiggly line can refer either to $h_{ij}$ or to 
$h_{00}$, and the vertices to any of the four $I^a_{ij}S^a_{ij}$ 
with $a=1,\ldots,4$, see the text.} \label{figGR3} 
\end{center}
\end{figure}
%%%%%%%%%%%%%%%%%%%%%%%%%%%%%%%%%%%

\noindent where
\begin{eqnarray}
\label{fege}
&&f(e)=\frac{1}{(1-e^2)^{7/2}}
\left(1+\frac{73}{24} e^2 +\frac{37}{96}e^4\right)\,,\\
&&g(e)=\frac{1}{(1-e^2)^{7/2}}\left(\frac{5}{2}+\frac{335}{48}e^2+
\frac{155}{192}e^4\right)\,.
\end{eqnarray}
The term proportional to $f(e)$ is the standard GR
result~\cite{Peters:1963ux}, while the term proportional to $g(e)$ 
is the extra contribution due to $\b_3$.

In order to measure $\beta_3$ from the dynamics of
binary pulsars, we must also determine the  
dependence on $\b_3$ of the  periastron shift $\omega$ and of the 
Einstein time delay $\gamma_E$, since these two observables are used to
determine the masses of the two compact stars. 
In particular the periastron shift 
fixes the total mass $M$ of the system, while the Einstein
time delay $\g_E$ measures a different combination of masses,
see e.g. eqs.~(6.56) and (6.93) of ref.~\cite{Maggiorebook}.
Using the conservative Lagrangian with the modification (\ref{lb1})
and repeating the standard textbook computation of the 
periastron shift $\omega$,
we find that the value $\omega_{\b_3}$
computed in a theory with $\b_3\neq 0$ is related to the GR value 
$\omega_{\rm GR}$ by
\be
\omega_{\beta_3}=\(1-\frac{\beta_3}{3}\)\omega_{\rm GR}\, , 
\ee
while
the Einstein time delay is unchanged because it is
not affected by the post-Keplerian parameters. 
So, if $\b_3\neq 0$, the true
value of the total mass $M$ of the binary system, that enters in 
\eq{dotT}, is not the one that would be inferred from the  periastron
shift using the predictions of GR, but rather we get
\be
M_{\beta_3}=\(1+\frac{\beta_3}{2}\)M_{\rm GR}\, . 
\ee
Similarly, using  eq.~(6.93) of ref.~\cite{Maggiorebook},
for the symmetric mass ratio $\nu$ we get 
\be
\nu_{\beta_3}=(1+w\beta_3)\nu_{\rm GR}\, ,
\ee 
where 
\be
\label{nubeta}
w=\frac{\kappa}{3}\frac{\sqrt{1+4\kappa}-2}{\sqrt{1+4\kappa}}
\frac 1{(1+4\kappa)^{1/2}-(1+\kappa)}
\ee
and 
\be
\kappa=\frac{\gamma}{e}\, \(\frac{2\pi}{P_b}\)^{1/3}\,
(G_NM_{\rm GR})^{-2/3}\, . 
\ee
Putting everything together and keeping only the
linear order in $\beta_3$ we finally find that the ratio between the
value of $\dot{P}_b$ computed at $\b_3\neq 0$ and the value of
$\dot{P}_b$ computed in GR is
\be
\frac{\dot{P}_b^{(\beta_3)}}{\dot{P}_b^{\rm GR}}=1+\beta_3 \tilde{g}(e)\, ,
\ee
where 
\be
\tilde{g}(e)=\frac{g(e)}{f(e)}+ \frac{5}{6}+w\, .
\ee
Observe that the term ${g(e)}/{f(e)}$ comes from the  effect of $\b_3$
on the radiative sector of the theory, while the term $(5/6)+w$ comes
from the effect of $\b_3$ on the conservative sector, i.e. on the mass
determination. 
Inserting the numerical values for the Hulse-Taylor pulsar, we
get $\tilde{g}(e)\simeq 3.21$. Note that
${g(e)}/{f(e)}\simeq 2.38$, so $\tilde{g}(e)$
is  dominated by the  effect of $\b_3$
on the radiative sector of the theory.
For this binary pulsar, after correcting for
the Doppler shift due to the relative velocity between us and the
pulsar induced by the differential rotation of the Galaxy, the ratio
between the observed value   $\dot{P}_b^{\rm obs}$ and the GR prediction
$\dot{P}_b^{\rm GR}$ is 
$\dot{P}_b^{\rm obs}/\dot{P}_b^{\rm GR}=1.0013(21)$. Interpreting this as
a measurement of $\b_3$ we finally get 
$3.21\b_3=0.0013(21)$, i.e.
\be\label{boundb3}
\b_3=(4.0\pm 6.4)\cdot 10^{-4}\, ,
\ee
so the three-graviton vertex is consistent with the GR prediction  at the
0.1\% level. 
This bound is slightly worse, but comparable, to the one from lunar laser ranging, \eq{lunar}. It should be
stressed, however, that \eq{boundb3} is really a test involving the
radiative sector of GR, while \eq{lunar} only tests the conservative sector.
For comparison, observe that in the
Standard Model the triple gauge boson couplings are measured to an
accuracy of about 3\%~\cite{Schael:2004tq}.

For the double pulsar we find
$\tilde{g}(e)\simeq 3.3$. Since $\dot{P}_b$ for the double pulsar 
is presently measured at the $1.4\%$ level \cite{Kramer:2006nb}, 
we get a larger bound compared to \eq{boundb3}. 
However, further monitoring of this system is expected to bring 
the error on $\dot{P}_b$  down to the $0.1\%$ level.

%%%%%%%%%%%%%%%%%%%%%%%%%%%%%%%
\subsection{Binary coalescences at interferometers} \label{sec:coal}
%%%%%%%%%%%%%%%%%%%%%%%%%%%%%%%

We now compare these results with what can be expected from 
the detection of a binary coalescence at GW interferometers. 
In this case one can determine the physical parameters of 
the inspiraling bodies, by performing matched filtering of theoretical 
waveform templates.
In the matched filtering method any difference in the time behavior 
between the actual signal and the theoretical template model will 
eventually cause the two to go out of phase, with a consequent 
drop in the signal-to-noise ratio (SNR). 
The introduction of $\b_3$ and $\b_4$ affects the template, 
in particular the accumulated phase 
\be
\phi=2\pi \int_{t_{\rm min}}^{t_{\rm max}} \!\!\!\!f(t) dt\,,
\ee
where $f(t)$ is the time-varying frequency of the source, and the 
subscript min (max) denotes the values when the signal enters (leaves)
the detector band-width.
Thus in principle a detection of a GW signal from coalescing binaries
could be translated into a measurement of the three- and four-graviton 
vertices. 
In this section we investigate the accuracy of such a determination.

With respect to the timing of binary pulsars, there are at least three
important qualitative differences that affect the  accuracy at which
these systems can test the non-linearities of GR. First,
coalescing compact binaries in the last stage of the coalescence reach
values of $v/c\sim 1/3$, and are therefore much more relativistic than
binary pulsars, which rather  have $v/c\sim 10^{-3}$. Second,
the leading Newtonian result  for $\phi$
is  of order $(v/c)^{-5}$ so it is much larger than one, and to get the phase
with a precision $\D\phi\ll 1$, as needed by interferometers, all
the corrections at least up to $O(v^6/c^6)$ to the Newtonian result
must be included, so higher-order 
corrections are important even if they are numerically small
relative to the leading term.  In other words, even if PN corrections are
suppressed by powers of $v/c$ with respect to the leading term, they
can be probed up to high order
because what matters for GW interferometers is the overall
value of the PN corrections to the phase, 
and not their value relative to the large
Newtonian term. 
These two considerations should suggest that
interferometers are much more sensitive than pulsar timing to the
non-linearities of GR. 

On the other hand, for binary pulsars we can
measure not only the decay of the orbital period due to GW emission,
but also several other Keplerian observables,
that provide a determination of the geometry of the orbit, as well as
post-Keplerian obser\-vables, 
such as the periastron shift and the Einstein
time delay, which fix
the masses of
the stars in the binary system. This is not the case for the detection
of coalescences at interferometers. With interferometers the
parameters that determine the waveform, such as the
masses and spins of the stars, must be
determined from the phase of the GW itself, and one must then 
carefully investigate the
degeneracies between the determination of $\b_3$ (or of $\b_4$) and
the determination of the masses and spins of the stars. 
This effect clearly goes in the
direction of degrading the accuracy of parameter reconstruction
at GW interferometers, with
respect to binary pulsar timing, so in the end it is not obvious a
priori which  of the two,  GW interferometers or binary pulsar timing,
is more sensitive
to the non-linearities of GR. This question
is answered in what follows.

Repeating with $\b_3\neq 0$
the standard computation of the orbital
phase for a circularized orbit,
we find that  $\b_3$  
modifies the orbital phase $\phi(t)$ 
already at 0PN (i.e. Newtonian) level, where to linear order in
$\b_3$ we get
\be\label{phi0PN}
\phi^{\rm 0PN}=-\frac{\Theta^{5/8}}{\nu} (1+b_0\beta_3)\, ,
\ee
with $b_0=-5/2$, and $\Theta$ is defined as
\be\label{defTheta}
\Theta =\frac{\nu (t_c-t)}{5GM}\, (1-b_0\beta_3)\, ,
\ee 
where $t_c$ is the time of coalescence.
Combining the factors $\nu$ and $M$ which enter in the definition of $\Theta$
with the explicit factor $1/\nu$ in \eq{phi0PN} we recover the well-know result
that the Newtonian phase depends on the masses of the stars only
through the chirp mass $M_c=\nu^{3/5}M$. From \eq{phi0PN} we
immediately understand the crucial role that degeneracies have for
interferometers. In fact, since $M_c$ is determined from \eq{phi0PN}
itself, using only the 0PN phase (\ref{phi0PN}) it is impossible to
detect the deviation from the prediction of GR induced by $\b_3$. A
non-zero value of $\b_3$ would simply induce an error in the
determination of $M_c$. 

The same happens at 1PN level. In fact, at 1PN order and with
$\b_3\neq 0$ 
the phase has the general form
\be
\label{1PN}
\phi^{\rm 1PN}=-\frac{\Theta^{5/8}}{\nu}\[ (1+b_0\beta_3)
+a_1(\nu)(1+b_1\b_3 ) \Theta^{-1/4}\]
\, ,
\ee
where, as before, 
$b_0=-5/2$ is the 0PN correction proportional to $\b_3$, while
\be\label{a1nu}
a_1(\nu)=\frac{3715}{8064} + \frac{55}{96}\nu
\ee 
is the 1PN GR prediction \cite{Blan-LRR:06,Maggiorebook},
and $b_1$  (which is possibly $\nu$-dependent) parametrizes 
the 1PN correction due to $\b_3$. 
(For simplicity, we only wrote explicitly the term linear in $\b_3$ 
since $|\b_3|$ is much smaller than one, but all our considerations 
below can be trivially generalized to terms quadratic in $\b_3$, 
just by allowing the function $b_1(\nu)$ to depend also on $\b_3$). 
In general we expect $b_1$ to be $O(1)$, and we will see below 
that for our purposes this estimate is sufficient.

Using $M_c$ and $\nu$ as independent mass
variables, in place of $m_1$ and $m_2$, we see that while
the effect of $\b_3$ on the 0PN phase can be reabsorbed into $M_c$,
its effect on the 1PN phase can be reabsorbed into a rescaling of
$\nu$. Observe that, in the detection of a single coalescence event,
GW interferometers do not measure the functional dependence of $\nu$ of
the 1PN phase, but only its numerical value for the actual value of
$\nu$ of that binary system, so we cannot infer  the presence of a term
proportional to $\b_3$ from the fact that it changes the functional
form of the $\nu$-dependence from the one obtained by \eq{a1nu}.
Thus, even at 1PN order, 
it is impossible to detect the deviations from GR
induced by $\b_3$. A non-zero $\b_3$ would simply induce an error
on the determination of $M_c$ and $\nu$, i.e. on the masses of the two
stars. 

We then examine the situation at 1.5PN
order. Let us at first neglect the 
spin of the two stars. Then
the 1.5PN phase with $\b_3\neq 0$ has the generic form
\bees
\label{1.5PN}
\phi^{\rm 1.5PN}&=&-\frac{\Theta^{5/8}}{\nu}\[ (1+b_0\beta_3)
+a_1(\nu)(1+b_1\b_3 ) \Theta^{-1/4}\right.\nn\\
&&\left.\phantom{-\frac{\Theta^{5/8}}{\nu}} 
+a_2(1+b_2\b_3) \Theta^{-3/8}
\]
\, ,
\ees
where $a_2=-3\pi/4$ is the 1.5PN GR prediction and $b_2$ is the
(possibly $\nu$-dependent) 1.5PN correction
due to $\b_3$. Again, we will not need its exact value, and we will
simply make the natural assumption that it is $O(1)$.

However, for the purpose of determining $\b_3$,
neglecting  spin is not correct. Indeed, 
at 1.5PN order  the spin of the bodies enters 
through the spin-orbit coupling, and
the evolution of the GW frequency $f$ with time is given 
by~\cite{Kidder:1992fr}
\be\label{dfdt15PN}
\frac{df}{dt}=\frac{96}{5}\pi^{8/3} M_c^{5/3} f^{11/3}
\[ 1-\frac{24}{5} a_1(\nu) x +(4\pi -\b_{\rm LS})x^{3/2}\],
\ee
where $x=(\pi Mf)^{2/3}$, while $\b_{\rm LS}$ describes the spin-orbit
coupling  and is given by
\be
\label{eq:beta_ls}
\b_{\rm LS} =\frac{1}{12}\sum_{a=1}^2\[
113\frac{m_a^2}{M^2}+75\nu\]{\bf \hat{L}}\bdot\vchi_a\, ,
\ee
where ${\bf L}$ is the orbital angular momentum, 
$\vchi_a={\bf S}_a/m_a^2$ and
${\bf S}_a$ is the  spin of the $a$-th body.
In principle $\b_{\rm LS}$ evolves
with time because of the precession of ${\bf L}$, ${\bf S}_1$ and
${\bf S}_2$. However, it turns out that in practice it is almost
conserved, and can be treated as a
constant~\cite{Cutler:1994ys}. Integration of \eq{dfdt15PN} then
shows that, in the 1.5PN phase, $\b_{\rm LS}$ is exactly degenerate
with $\b_3$ in \eq{1.5PN}. Furthermore, observe that 
$\b_{\rm LS}$, depending on the spin configuration, can reach a
maximum value of about 8.5~\cite{Cutler:1994ys} (and its maximum value
remains large
even in the limit $\nu\ra 0$), while $\b_3$ is
already bound by laser ranging at the level of $2\times 10^{-4}$ and by
pulsar timing at the level of $10^{-3}$ 
(which tests the radiative sector, as do GW interferometers). Thus, 
the effect of $\b_3$ at 1.5PN is simply reabsorbed into a (very small)
shift of $\b_{\rm LS}$.

At 2PN order $\b_3$ is
degenerate with the parameter $\s$ that describes the spin-spin
interaction
\bees
\label{eq:spin}
\s &=&\frac{\nu}{48}
[721(\hat{\bf L}\bdot\vchi_1)-247\vchi_1\bdot\vchi_2
(\hat{\bf L}\bdot\vchi_2) ]\nn\\
&&+\frac{1}{96}\sum_{a=1}^2\frac{m_a^2}{M^2}
\[719 (\hat{\bf L}\bdot\vchi_a)^2-233\vchi_a^2\]\, .
\ees
The first term is the one which is usually quoted in the literature,
first computed in \cite{Kidder:1992fr} (see also
\cite{Poisson:1995ef, Blanchet:1995ez}). The term in the second line,
computed recently in \cite{Racine:2008kj},
is however of the same order, and must be included.

The first term is proportional to $\nu$, and reaches a maximum value
$\s_{\rm max}(\nu)\simeq 10\nu$. In a coalescence with
very small value of $\nu$, this term is therefore suppressed; e.g. in  
an extreme mass-ratio inspiral (EMRI) event at LISA~\cite{LISA} 
where a BH of mass $m_1=10\msun$ falls into a supermassive BH with
$m_2=10^6\msun$, one has $\nu=10^{-5}$ and   the term in the first line
has a maximum value $\sim 10^{-4}$.
If this standard term gave  the full answer, 
a value of $\b_3$ in excess of this value could therefore
give an effect that cannot be ascribed to $\s$. However, the
presence of the new term recently computed in 
\cite{Racine:2008kj} spoils this reasoning, since it is not
proportional to $\nu$. The conclusion is  that, just as with
$\b_{\rm SL}$ at 1.5PN order, the effect of $\b_3$ at 2PN order is
just reabsorbed into a small redefiniton of $\s$, and therefore simply
induces an error in the reconstruction of the spin configuration
(observe also that fixing $\b_{\rm LS}$ does not allow us to fix the
spin combinations that appear in $\s$.) 

%%%%%%%%%%%%%%%%%%%%%%%%%%%%%%%%%%%%
\begin{figure}[t]
\begin{center}
\includegraphics[width=0.6\textwidth]{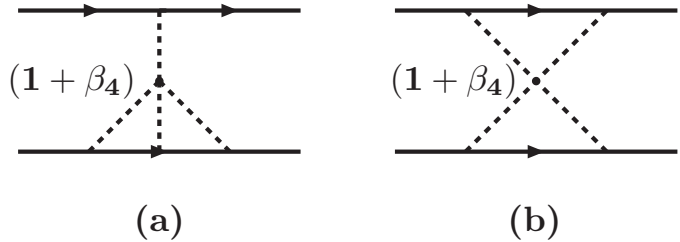}
\caption{\label{fig4}
The diagrams that contribute to the conservative dynamics, which are
affected by a modification of the four-graviton vertex.}
\end{center}
\end{figure}
%%%%%%%%%%%%%%%%%%%%%%%%%%%%%%%%%%%%

One could in principle investigate the effect of $\b_3$ on 
higher-order coefficients, such as the 2.5PN term $\psi_5$ and the 3PN
term $\psi_6$ in \eq{Psipsik}, which at LISA
can be measured with a precision of order 
$10^{-2}$~\cite{Arun:2006hn}. Unfortunately, it is difficult to
translate them into bounds on $\b_3$ because
at 2.5PN order one  finds that $\b_3$ is degenerate with a
different combination of spin and orbital variables, which is not
fixed by the 1.5PN spin-orbit term (see  Table~II of
ref.~\cite{Blanchet:2006gy}), while  
at 3PN order  the spin contribution is not yet known.
The conclusion is therefore that
interferometers cannot measure the three- and higher-order graviton
vertex, since the effect of a modified vertex is simply reabsorbed
into the determination of the masses and spin of the binary system.
The conclusion that interferometers are not competitive with pulsar 
timing for measuring deviations from GR was also reached in 
ref.~\cite{Dam_Far_GW_Tests:98}, although in a different context. 
In fact, ref.~\cite{Dam_Far_GW_Tests:98} was  concerned
with multiscalar-tensor theories, whose leading-order effect
is the  introduction of a term corresponding to dipole radiation (a
``minus one''-PN term).

We now examine what can be said about the four-graviton vertex,
parametrized by $\b_4$.
In the conservative part of the
Lagrangian $\b_4$ contributes through the diagrams of Fig.~\ref{fig4}.
However, these contributions  only affect the equations of motion at
2PN order, so there is no hope to see them in solar system
experiments, where the velocities at play are very small.
For the same reason, no significant bound can be obtained from 
binary pulsars; from a simple order of magnitude estimate we find
that the Hulse-Taylor pulsar can only give a limit  $\b_4<O(10)$,
which is not significant. 

%%%%%%%%%%%%%%%%%%%%%%%%%%%%%%%%%%%%
\begin{figure}[t]
\begin{center}
\includegraphics[width=0.3\textwidth]{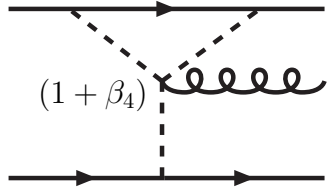}
\caption{\label{fig5}
The diagram that contributes to the radiative dynamics, which is
affected by a modification of the four-graviton vertex.}
\end{center}
\end{figure}
%%%%%%%%%%%%%%%%%%%%%%%%%%%%%%%%%%%%

At GW interferometers, $\b_4$ enters into the phase for the first time 
at 1PN order, through the diagram in Fig.~\ref{fig5}. 
However, it suffers exactly of the same degeneracy
issues as $\b_3$, so it cannot be measured to any interesting accuracy
at present or future interferometers, at least
with the technique discussed here.
Note however that in this paper we have worked in the restricted PN
approximation, in which only the harmonic at twice the source
frequency is retained. Higher-order harmonics 
however  break degeneracies between various parameters in the 
template~\cite{VanDenBroeck:2006ar}, and it would be interesting
to investigate  whether their inclusion  in the analysis allows one to 
put a bound on  $\b_3$ and $\b_4$ from binary coalescences.

%%%%%%%%%%%%%%%%%%%%%%%%%%%%%%%%%%
\section{Conclusions}\label{sec:group_concls}
%%%%%%%%%%%%%%%%%%%%%%%%%%%%%%%%%%

We have proposed to quantify the accuracy by which various experiments probe the non-linearities of GR, by translating their results  
into measurements of the non-Abelian vertices of the theory, such as the three-graviton vertex and the four-graviton vertex. 
This is similar in spirit to tests of the Standard Model of particle physics, where the non-Abelian vertices involving three and four gauge bosons have been measured at LEP and at the Tevatron.

We have shown that, at a phenomenological level, this can be done in a
consistent and gauge-invariant manner, by introducing
parameters $\b_3$ and $\b_4$ that quantify the deviations from the GR
prediction of the
three- and four-graviton vertices, respectively. 
We have found that, in the conservative sector of the
theory, i.e. as long as one neglects the emission of gravitational
radiation at infinity, the introduction of $\b_3$ at 1PN order is phenomenologically
equivalent to the
introduction of a parameter $\bppn=1+\b_3$ in the parametrized PN
formalism. Strong bounds on $\b_3$ therefore come from solar system
experiments, and most notably from lunar laser ranging, that provides
a measurement at the $0.02\%$ level.

The modification of the three-graviton vertex however also affects the
radiative sector of the theory, and we have found that the timing of
the Hulse-Taylor pulsar gives a bound on $\b_3$ at the 0.1\% level, 
not far from the one
obtained from lunar laser ranging. Conceptually, however, the two
bounds have different meanings, since lunar laser ranging 
only probes  the conservative sector of the theory, while pulsar timing
is also sensitive  to the radiative sector. 

We have then studied the
results that could be obtained from the detection of coalescences at
interferometers, and we have found that, even if $\b_3$ already
modifies the GW phase at the Newtonian level and $\b_4$ at 1PN order,
their effect can always be
reabsorbed into other parameters in
the template, such as the mass and spin of the two bodies so,
rather than detecting a deviation from the GR prediction, one would
simply make a small error in the estimation of these parameters.

%\vspace{5mm}\noindent
\subsubsection{Acknowledgments} 

The work of UC, SF, MM and HS is supported by the Fonds National Suisse.
We thank Alvaro de Rujula for a stimulating discussion and Thibault Damour
and the referees for useful comments. 
HS would like to thank Steven Carlip and Bei-Lok~Hu, and UC would like to thank Yi-Zen~Chu, for helpful discussions.

%% file: Chap_Berti/Chap_Berti.tex
% BBW commands
\newcommand{\hL}{\hat{\mathbf{L}}}
\newcommand{\hS}{\hat{\mathbf{S}}}
\newcommand{\vS}{\mathbf{S}}
\newcommand{\vL}{\mathbf{L}}
% LIGO rates commands
\newcommand{\Nup}{$\dot{N}_{\rm max}$}
\newcommand{\Npl}{$\dot{N}_{\rm high}$}
\newcommand{\Nre}{$\dot{N}_{\rm re}$}
\newcommand{\Nlow}{$\dot{N}_{\rm low}$}

%%%%%%%%%%%%%%%%%%%%%%%%%%%%%%%%
\chapter{The effects of modified graviton vertices in the templates \\ 
for binary coalescences} 
\chaptermark{The effects of modified graviton vertices in GW templates}
\label{chap:Berti}
%%%%%%%%%%%%%%%%%%%%%%%%%%%%%%%%

%%%%%%%%%%%%%%%%%%%%%%%%%%%%%%%%%
\section[The possibility of measuring the non-Abelian vertices of \\ General Relativity with \ifos]
    {The possibility of measuring the non-Abelian vertices of General Relativity with \ifos %
  \sectionmark{Measuring the non-Abelian vertices of GR at \ifos} } 
    \sectionmark{Measuring the non-Abelian vertices of GR at \ifos}
\label{sec:vert_meas}

% where:    
% "[]" is for the table of contents
% the subsequent "{}" is for where the section actually is
% 1st "sectionmark" is for the first page containing the section
% 2nd "sectionmark" is for the other pages containing the section

%%%%%%%%%%%%%%%%%%%%%%%%%%%%%%%%%

In the previous chapter I have described a parametrized framework to constrain gravity non-linearities from an effective field theory point of view~\cite{UC:09}. 
In this work, performed with the rest of my group, we tagged the three- and four-graviton vertices of GR with some parameters $\b_3$ and $\b_4$, respectively, and we investigated the bounds that could be put on them by means of experiments of relativistic gravity. 
One of the conclusion of our study is that the timing of binary pulsars can provide better tests of GR than what is possible with GW observations because the latter are plagued by degeneracies among parameters. 
A similar conclusion in favor of binary pulsar tests has been derived by DEF in their study of \sts, which I have discussed in \Sec{sec:Dam_Far}. 
As pointed out in that section, the validity of this type of conclusions is limited to the PN orders probed by pulsar timing, i.e. up to terms of ${\cal O}(v/c)^5$: these comprise only the lowest orders of the radiative regime, while GW observations aim at probing the PN structure at least until the 3.5 next-to-leading order order, i.e. up to terms of ${\cal O}(v/c)^{12}$.
Moreover, in confronting the probing power of different testing grounds, one should also consider how the extraction of physical parameters is done in the various experiments. 
Through the timing of binary pulsars one can measure the decay of the orbital period, which is a consequence of GW emission; on top of the decay, one has access to other post-Keplerian parameters, like the periastron shift and Einstein time delay, which are used to fix the unknown masses of the stars in a binary system.  
The situation is radically different for what concerns the detection of coalescences at interferometers: the entire physical information on the emitting system has to be extracted from the phase of the GW signal that has spent many cycles in the detector. 
Therefore, every type of parameters, be them the constituent masses or the coupling constants of alternative theories of gravity, must be determined from the a unique measured quantity. 
These differences between binary pulsar timing and coalescences at interferometers are at the heart of the conclusions of the work with my group~\cite{UC:09}. 
In fact, on one hand we were able to constrain the three- and four-graviton vertices to interesting accuracies using the timing of Hulse-Taylor pulsar; 
on the other, by looking at the GW phasing formula we could not derive any bound: possible deviations of the non-Abelian vertices from their GR values were found to be degenerate with numerically larger phasing coefficients of the GW template, notably those due to the spins of the objects.
Because we worked in the restricted PN framework, we concluded that these degeneracies could be broken by the inclusion of amplitude corrections in the template, as higher harmonics enrich the functional dependence of the parameters~\cite{VanDenBroeck:2006ar}.
Improvements in parameter estimation brought in by the use of full waveforms have been discussed in \Sec{sec:PPN_GW}, based on the study of ref.~\cite{Mishra:2010tp}.
However, a richer functional dependence could offer the possibility of bounding small parameters like $\b_3$ and $\b_4$ even in the restricted PN framework.
An example along this direction is the work of ref.~\cite{Berti:2004bd}, where Berti, Buonanno and Will (BBW) investigated the bounds on some alternative theories that could be put with LISA~\cite{LISA} observations if templates include spin-orbit and spin-spin effects. 
I will now describe the analysis of BBW by focusing on the aspects which are more salient to the problem of estimating \btf. 
In doing so, I will partially follow the presentation given in ref.~\cite{Berti:2004bd}.

%%%%%%%%%%%%%%%%%%%%%%%%%%%%%%%%%
\section[Constraining alternative theories at \ifos\ with spin effects \\ in the template -  
The case of the space-borne experiment LISA]
    {Constraining alternative theories at \ifos\ with spin effects in the template - \\ 
The case of the space-borne experiment LISA %
  \sectionmark{Constraining alternative theories with spin effects in the template} } 
    \sectionmark{Constraining alternative theories with spin effects in the template}
\label{sec:BBW}

% where:    
% "[]" is for the table of contents
% the subsequent "{}" is for where the section actually is
% 1st "sectionmark" is for the first page containing the section
% 2nd "sectionmark" is for the other pages containing the section

%%%%%%%%%%%%%%%%%%%%%%%%%%%%%%%%%

The focus of BBW was on Brans-Dicke (BD) \st\ and on the phenomenological effect of a massive graviton.
In the course of this thesis we have already encountered the extra contributions due to these alternatives to GR: 
%~\footnote{I call the phenomenological effect of a massive graviton a theory just to be able to refer to it at the same time as Brans-Dicke's.} bring with respect to GR: 
for a \st\ the relevant term is the dipolar flux~(\ref{eq:flux_dip}), for a massive graviton it is the shift in the 1PN phasing coefficient~(\ref{eq:a2shift}). 
Within the restricted post-Newtonian approximation, BBW used the following expression for the waveform in the frequency domain~\footnote{Other assumptions behind \eq{eq:h_restr} will be discussed in what follows.} 
\bees
\label{eq:h_restr}
\tilde h(f) &=&\frac{\sqrt{3}}{2}\,{\cal A}\,f^{-7/6}\,e^{i\psi(f)} 
%\quad \quad \alpha = {\rm I,II}\,, 
\\
{\cal A} &=& \frac{1}{\sqrt{30}\pi^{2/3}} \frac{{\cal M}^{5/6}}{D_{\rm L}} \nn \,,
\ees
where $D_{\rm L}$ is the luminosity distance given by
\be
\label{eq:DL}
D_L=\frac{1+z}{H_0}\int_0^z \frac{dz'}{\left[\Omega_M(1+z')^3
+\Omega_\Lambda\right]^{1/2}}\,.
\ee
in which one has assumed a zero--spatial-curvature Universe with $\Omega_M$ the energy density of matter and $\Omega_\Lambda$ the one of a cosmological constant. 

Concerning the phase $\psi$ in \eq{eq:h_restr}, BBW adopted an expression valid up to order 2PN: this is indeed the accuracy with which the spin terms of the phasing were known.
With a slight change of notation with respect to BBW~\footnote{Here I use $(\pi Mf)$ as an expansion parameter for consistency with \Sec{sec:PPN_GW}; BBW adopted the combination $(\pi {\cal M}f)$ where the chirp mass is used instead of the total mass $M$: this choice brings a dependence of the phasing coefficients on the mass ratio $\nu$ which is not present in \Sec{sec:PPN_GW}.\label{foot:M}}, the 2PN phasing formula reads
\bees 
\label{eq:phase_f}
\Psi(f)&=&2\pi f t_c-\Phi_c+ \frac{3}{128\,\nu}\,(\pi Mf)^{-5/3}\,
\left\{ 1- \frac{5 {\cal S}^2}{84 \omega_{\rm BD}}\,(\pi M f)^{-2/3} 
\right. \nn\\
 \nn\\
&& -\frac{128\,\nu}{3}\frac{\pi^2 D\,M}{\lambda_g^2\,(1+z)}\,(\pi Mf)^{2/3}  
+\left(\frac{3715}{756} +\frac{55}{9}\nu\right)\,(\pi M f)^{2/3} \nn\\
 \nn\\
&& \left. -16\pi\,(\pi M f) + 4 \b_{LS}\,(\pi M f) \right.\nn\\
 \nn\\
&&+\left. \left(\frac{15293365}{508032}+\frac{27145}{504}\nu+\frac{3085}{72}\nu^2\right)\, (\pi M f)^{4/3} 
-10\,\sigma\,(\pi M f)^{4/3} \right\} \,  
\ees
where:
\bd 

\item The first two terms are related to the time $t_c$ and phase $\Phi_c$ of coalescence, so they basically establish where the waveform begins or ends; 
$M$ is the total mass of the binary, $\nu$ its mass-ratio, $f$ the frequency variable. 
Because LISA will be able to probe sources at cosmological distances, $M$ includes a redshift factor, i.e. $M$ is the observed mass and it is related to the mass $M_s$ measured in the rest frame of the source by $M=M_s (1+z)$.

\item The combination $(\pi M f)$ sets the PN order of the expansion as $v =(\pi M f)^{1/3}$.

\item The first term inside the braces, the "1", together with the prefactor of the expression, comes from the GR quadrupole. 

\item The second term is the contribution of dipole gravitational radiation in BD theory, 
where ${\cal S}$ is proportional to the difference of the scalar charges of the bodies. 
From the discussion of \Sec{sec:Dam_Far}, we remember that a dipole term is larger than the quadrupole by a factor $v^{-2}$; however, $\cal S$ can be at most 1, while $\omega_{\rm BD}$ is at least $4\times 10^4$ (as we know from Solar-System tests~\cite{Cassini}): therefore, the dipole phasing coefficient is numerically small. 

\item The third term is the 1PN phasing coefficient that phenomenologically parametrizes a massive graviton by means of the graviton Compton wavelength $\lambda_g$. 
I have discussed it in \Sec{sec:PPN_GW}, where the distance $D$ was given for small redshift only. The full expression of $D$ reads 
\be
\label{defD}
D=\frac{1+z}{H_0}\int_0^z \frac{dz'}{(1+z')^2\left[\Omega_M(1+z')^3+\Omega_\Lambda\right]^{1/2}}\,,
\ee
so it is not a conventional cosmological distance measure as the luminosity distance~(\ref{eq:DL}). 
It emerges when studying how a massive graviton propagates in a cosmological background~\cite{Will:1997bb}.
For the Hubble constant one can assume $H_0=72$ km~s$^{-1}$~Mpc$^{-1}$,
according to the present observational estimates; %~\cite{cosmology}.
however, as we will see, the analysis of BBW reveals that the choice of a specific cosmological model affects the bound on $\lambda_g$ only weakly.

From the discussion of \Sec{sec:PPN_GW} we remember that the effect of a massive graviton is to alter the arrival time of waves of a given frequency. 
Moreover, let us mention also the graviton Compton wavelength is already bound from Solar-System tests~\cite{SS_Yukawa}: $\lambda_g\,\gsim\,10^{12}\,$km\,$\simeq 3\times10^{-2}\,$pc, which corresponds to a bound on the graviton mass $m_g\,\lsim \,2\times 10^{-22}$eV 
(cfr. the discussion of~\eq{lambdabound} later on). 
 
\item All the remaining terms are those predicted by GR.
Notably, the quantities $\b_{\rm LS}$ and $\sigma$ represent the spin-orbit and spin-spin contributions to the phasing, respectively~\footnote{The expression for the spin-spin contribution differs from the one reported in \eq{eq:spin} of \chap{chap:Group} because BBW was published in 2005 while the extra term in \eq{eq:spin} has been calculated in 2009~\cite{Racine:2008kj}.}: 
%The expressions used in BBW then read
\bees
\b_{\rm LS} &=& \frac{1}{12} \sum_{i=1}^2 \chi_i \left [113 \frac{m_i^2}{M^2} + 75
\nu \right ] \hL \cdot \hS_i \,,
\\
\sigma &=& \frac{\nu}{48} \chi_1 \chi_2 \left (-247 \hS_1 \cdot \hS_2 + 721 \hL
\cdot \hS_1 \hL \cdot \hS_2 \right )\,,
\ees
where $\hS_i$ and $\hL$ are unit vectors in the direction of the spins
and of the orbital angular momentum, respectively, and $\vS_i=\chi_i
m_i^2 \hS_i$.  
For black holes, the dimensionless spin parameters $\chi_i$ must be smaller than unity, while for neutron stars they are generally much smaller than unity.  
It follows that $|\b_{\rm LS}|\lesssim 9.4$ and $|\sigma|\lesssim 2.5$\,, which makes the phasing coefficients due to spin effects numerically large. 
\ed

\noindent
 
As one can see, the phasing formula (\ref{eq:phase_f}) contains two different types of parameters: those which are intrinsic to the source, like the masses, the distance $D$ and the spins of the objects, and those which are extrinsic, like the parameters $\omega_{\rm BD}$ and $\lambda_g$ which are characteristic of the theory of gravity. 
Denoting the full set of parameters by $\{\theta_a\}$, one can describe a waveform in the time domain as $h_i = h_i [t; \{\theta_a\}]$\, where the index~$i$ labels the waveform and $t$ is the time variable. 
To detect a GW signal and estimate the parameters of the emitting source, it is convenient to adopt the matched filtering technique, i.e. to maximize the convolution of a signal with a putative waveform; this convolution is expressed by the {\it inner} product 
\begin{equation}
\label{eq:inner}
(h_1|h_2) \equiv 2 \int_0^{\infty} \frac{ {\tilde{h}_1}^*\tilde{h}_2 +
{\tilde{h}_2}^*\tilde{h}_1 }{S_n(f)}\,df \,,
\end{equation}
where $\tilde h(f) \equiv \int_{-\infty}^{\infty} h(t) \exp(2\pi i f t) dt$ is the Fourier transform of the GW waveform, ${\tilde{h}_i}^*(f)$ its complex conjugate and $S_n(f)$ is the noise spectral density of the detector, a quantity that describes the sensitivity of an \ifo\ as a function of the frequency. 
By means of the inner product (\ref{eq:inner}), one can define the signal-to-noise ratio (SNR) for a given $h_i$ by 
\begin{equation}
\rho[h_i] \equiv (h_i | h_i)^{1/2} 
\label{rho}
\end{equation}
and the so-called {\it Fisher matrix} $\Gamma_{ab}$ %, with components given by
\begin{equation}
\label{eq:fisher}
\Gamma_{ab} \equiv \left( \frac{\partial h_i}{\partial\theta_a} \mid
\frac{\partial h_i}{\partial\theta_b} \right) \,,
\end{equation}
which can be used to estimate the root mean square error $\Delta\theta_a$ in measuring the parameter~$\theta_a$. 
In fact, in the limit of large SNR, these errors can be obtained by taking the square root of the diagonal elements of the inverse of the Fisher matrix,
\begin{equation}
\label{eq:errors}
\Delta\theta_a = \sqrt{\Sigma_{aa}} \,, \qquad  \Sigma = \Gamma^{-1} \,;
\end{equation}
analogously, the coefficients quantifying the correlation between two parameters $\theta_a$ and $\theta_b$ are given by 
\begin{equation}
\label{eq:corr}
c_{ab} = \Sigma_{ab}/\sqrt{\Sigma_{aa}\Sigma_{bb}} \,.
\end{equation}
%
\begin{comment}%%%%%%%%%%%%%%%%%%%%%%%%%%%%%%%%%%%
In this context, BBW used a compact notation to refer to the parameters of the alternative theories they examined~\footnote{I adapt BBW notation according to my choice of using the total mass $M$ instead of the chirp mass ${\cal M}$ (cfr. footnote~\ref{foot:M}).}
%
\be
\label{eq:compar}
\varpi \equiv \frac{1}{\omega_{\rm BD}} \;, \q
\beta_{g} \equiv \frac{\pi^2 D\,M}{\lambda_g^2\,(1+z)}\,; 
\ee
%
with these definitions, the corresponding derivatives of $\tilde{h}$ that appear in \eq{eq:fisher} through \eq{eq:inner} read
\end{comment}%%%%%%%%%%%%%%%%%%%%%%%%%%%%%%%%%%%

By making use of the Fisher matrix, BBW derive the errors and correlations coefficients on the various parameters to assess the role that spin terms have in estimating parameters with detections at LISA. 
The candidate events considered by BBW are from binary systems in circular orbits with spins aligned or anti-aligned with the orbital angular momentum, i.e. the spins of the objects are non-precessing. 
The choice of neglecting eccentricity is to simplify the treatment. 
Other simplifications stem from modeling LISA as a single \ifo\ and not taking into account the orbital motion of the spacecrafts. 
The latter assumptions are justified by the fact that the angles describing the position of the source and the orientation of LISA are parameters that modify the amplitude of the signal and not its phase: for this reason, there is no correlation among the angles and the intrinsic or extrinsic parameters of the phase (\ref{eq:phase_f}). 
%Within the restricted post-Newtonian approximation, form for $\tilde{h}$ in Eq. (\ref{hrestrict}),
In the part of the work where they ignore LISA's orbital motion, BBW find that the SNR takes the form
\be
\label{eq:snr}
\sqrt{\langle \rho^2 \rangle} = 6.245 \times 10^{-23} \,\left (\frac{M}{M_\odot}\right )^{5/6} \nu^{1/2} \left (\frac{1 {\rm Gpc}}{D_L}\right )\, 
\sqrt{\int_{f_{\rm in}}^{f_{\rm end}}\frac{3}{4} \frac{f^{-7/3}}{S_h(f)} df}\,,
\ee
where angular braces indicate the average over LISA's angular response and where the limits of integration in the frequency are set as follows.
The upper limit is chosen to be $f_{\rm fin}= {\rm min}(f_{\rm ISCO},f_{\rm end})$. 
Here $f_{\rm ISCO}$ is twice the conventional (Schwarzschild) frequency of the innermost stable circular orbit for a point mass, namely $f_{\rm ISCO} = (6^{3/2}\pi M)^{-1}$. 
On the other hand, $f_{\rm end}=1~$Hz is a conventional upper cutoff on the LISA noise curve. 
The initial frequency~$f_{\rm in}$ is determined by assuming that we observe the inspiral over a time $T_{\rm obs}$ before the ISCO and by selecting a cutoff frequency below which the LISA noise curve is not well characterized. 
BBW's default cutoff is $f_{\rm low}=10^{-5}$ Hz. 
The choice of these limits of integration also applies to the frequency integrals that are needed to compute the Fisher matrix (see \eqs{eq:fisher}{eq:inner}).
The expression of the SNR (\ref{eq:snr}) will be useful to understand the bounds on the parameters.

\smallskip 

When placing bounds on alternative theories of gravity, BBW treated only spin-orbit terms. 
The reason is due to the dimensionality of the Fisher matrix, which gets too large in presence of two types of spin contributions at the same time as gravity modifications.
Concerning the role of spin terms, the main conclusions of the work of BBW are the following ones.
%\ben 
%\item 

\subsubsection{How the presence of spin terms in the template generally affects parameter estimation}

There is a degradation in parameter estimation caused by spin contributions even in the absence of any modification to GR. 
This has been known since the work of ref.~\cite{Poisson:1995ef} and is due to the fact that if the spins are non-precessing, they are not modulated by the orbital motion: as a consequence, the spins are correlated with other intrinsic parameters, like the masses and distances. 
Indeed, ref.~\cite{Vecchio_BBW} has shown that spin modulation allows a better estimation of masses and distances with respect to the case when spins are aligned or anti-aligned with the orbital angular momentum. 
Therefore, as a general rule, the presence of a higher number of parameters dilutes the available information.

\subsubsection{Constraining Brans-Dicke theory} % \item

To constrain BD theory the best sources are asymmetric systems, which maximize the difference of the scalar charges of the bodies in the parameter ${\cal S}$ of \eq{eq:phase_f}. 
In fact, black holes have vanishing coupling to the scalar field because of the no-hair theorem, so black hole binaries are not useful to bound $\omega_{\rm BD}$. Neutron star binaries are not optimal either because their scalar charge is only weakly dependent on the equation of state (cfr. the discussion of \fig{fig:ST_Tests_Full} in \Sec{sec:Dam_Far}).
Finally, binaries comprising white dwarves can be excluded from consideration because these objects are not so compact, so they introduce tidal effects that complicate the analysis. 
BBW then focus their treatment on the inspiral of a neutron star into a black hole with the following masses: $M_{NS}=1.4\msun$ and $M_{BH}:(4\times10^2$--\,$1\times10^4)\msun$. 
They set an SNR=10, which, according to \eq{eq:snr}, implies that a higher total mass $M$ corresponds to a larger luminosity distance $D_L$ probed by LISA: notably, one has $D_L\simeq3\times10^2$\,Mpc for $M \simeq4\times10^4\msun$\,. 
However, a high total mass worsens the bound on $\omega_{\rm BD}$ for the following reason. 
The derivative of the waveform (\ref{eq:h_restr}) with respect to $\omega_{\rm BD}$ is 
\bees
\label{eq:derBD}
\frac{\pa \tilde h}{\pa \omega_{\rm BD}} &=& 
\frac{5 i}{3584} \, \frac{{\cal S}^2}{\omega_{\rm BD}^2} \nu^{-1}(\pi M f)^{-7/3}\,\tilde h \\ \nn
&\propto& \frac{1}{\omega_{\rm BD}^2} M^{-3} \nu^{-2/5} \\ \nn
&\sim& \frac{1}{\omega_{\rm BD}^2} M_{NS}^{-2/5} M_{BH}^{-13/5}
\ees
where in the second passage we have included the dependence of the amplitude of the waveform~(\ref{eq:h_restr}) on the mass parameters and in the last passage we have expanded $M$ and $\nu$ with the assumption $M_{NS}\ll M_{BH}$. 
With this behavior one can roughly estimate the error $\Delta\theta^\omega = \sqrt{\Sigma^{\omega\omega}}$ on the BD parameter.  
In fact the diagonal element of the Fisher matrix $\Gamma_{\omega\omega}$ is quadratic in the derivative~(\ref{eq:derBD}). 
If we invert only this element we obtain a bound on $\omega_{\rm BD}$ that BBW call {\it un-correlated} because this is the bound that could be put in the ideal case in which all the parameters were known and not correlated among themselves.  
The bound from $\Gamma_{\omega\omega}$ is explicitly quoted by BBW in their tables: it is seen to be always two orders of magnitude higher than the one comprising correlations. 
Here I will only use $\Gamma_{\omega\omega}$ to show the functional dependence of the error on the parameters. 
A part from the frequency dependence in the first line of \eq{eq:derBD} we have
\bees 
\label{eq:sigma_om}
{\Sigma_{\omega\omega}}^{1/2} &\sim& {\Gamma_{\omega\omega}}^{-1/2}
\sim \left(\frac{\pa \tilde h}{\pa \omega_{\rm BD}}\right)^{-1} \nn \\ 
&\sim& \omega_{\rm BD}^2 M_{BH}^{13/5}\,,
\ees
where we have ignored the dependence on $M_{NS}$ because this mass is kept constant in the analysis of BBW.
The lower bound on a parameter like $\omega_{\rm BD}$ can be chosen to be the error committed in estimating the parameter itself; denoting by $\bar\omega_{\rm BD}$ the lower value corresponding to the error (\ref{eq:sigma_om}) one has
$$
\bar\omega_{\rm BD} \sim \omega_{\rm BD}^2 M_{BH}^{13/5} \Longleftrightarrow 
\bar\omega_{\rm BD} \sim M_{BH}^{-13/5}\,,
$$
which explains what found by BBW concerning the behavior of the bound $\bar\omega_{\rm BD}$ with $M_{BH}$, i.e. why the most constraining system is a $1.4\,M_\odot$ neutron star inspiraling into a $400 \, M_\odot$ black hole. 

Concerning the dilution of information in the presence of spin, BBW find that the spin-orbit term reduces the bound on $\omega_{\rm BD}$ by factors of order 10 -- 20: for example, in the case of the most constraining system, the bound is degraded from $\omega_{\rm BD}\,\gsim\,8\times 10^5$ to $\omega_{\rm BD}\,\gsim\,3.9\times 10^4$.  
This last constraint is almost identical to the Solar-System bound from Cassini measurements of the Shapiro time delay~\cite{Cassini}. 
However, a bound coming from binary coalescences detected at LISA is of  greatest importance even if numerically equivalent to the constraint derived in a different testing ground. 
This situation is similar to the one concerning the bounds on \sts\ derived by DEF: as I have discussed in \Sec{sec:Dam_Far}, the dynamical regimes probed by \ifos\ are very different from those characterizing the Solar-System and binary pulsars. 
First and foremost, the LISA bound obtained by BBW comes from a 2PN template, i.e. from considering GR non-linearities that are two orders higher than those probed by binary pulsars. 
In a second place, if one of the binary constituents of a LISA source is a neutron star, the companion object will be a black hole of mass $M_{BH}\gsim 100\msun$: %black holes have scalar charges that vanish exactly due to the no-hair theorem but 
therefore, the LISA bound would come from the strongest gravitational fields possible.

\subsubsection{Constraining the presence of an effective mass term in graviton propagation} 
% \item  

To place bounds on the graviton Compton wavelength, BBW do not fix a thre\-shold on the SNR. 
In fact, to investigate this non GR behavior one is not limited to consider binary systems which comprise neutron stars, as it was needed to constrain BD theory. 
Therefore, one can take advantage of the high sensitivity of LISA to sources at cosmological distances; probing the Universe at such a deeper level, one awaits to detect black hole-binaries with constituent masses from $10^4\msun$ to $10^7\msun$. 
This is indeed the mass range investigated by BBW, who considered these sources at a luminosity distance $D_L=3$\,Gpc. 

The derivative of the waveform (\ref{eq:h_restr}) with respect to $\lambda_g$ is 
\bees
\label{eq:lambda_g}
\frac{\pa \tilde h}{\pa \lambda_g} &=& 
+2(\pi M f)^{-3/3}\, \frac{\pi^2 D M}{\lambda_g^3(1+z)} \, \tilde h \\ \nn
&\propto& \frac{\pi^2 D}{\lambda_g^3(1+z)} \frac{M^{5/6}}{D_L}\, \nu^{1/2} \\ \nn
&\sim& \frac{D}{(1+z)D_L} \frac{M^{5/6}}{\lambda_g^3}\, \nu^{1/2}\,,
\ees
where in the second passage we have included the dependence of the amplitude of the waveform~(\ref{eq:h_restr}) on the mass parameters. 
The behavior of \eq{eq:lambda_g} is different with respect to \eq{eq:derBD}, notably the derivative is now directly proportional to the total mass.
For this reason, BBW obtained the best bound on $\lambda_g$ from the system with highest total mass, notably an equal mass binary with $M=2\times10^7\msun$\,, whose SNR is 2063. 

Now I repeat the analysis based on the diagonal element of the Fisher matrix in order to estimate the error and understand its dependence on the other parameters.
From \eq{eq:lambda_g} we have 
\bees 
\label{eq:sigma_lam}
{\Sigma_{\lambda \lambda}}^{1/2} &\sim& {\Gamma_{\lambda \lambda}}^{-1/2}
\sim \left(\frac{\pa \tilde h}{\pa \lambda_g}\right)^{-1} \nn \\ 
&\sim& \left(\frac{D}{(1+z)D_L}\right )^{-1} \lambda_g^3 \, M^{-5/6} \;, \\ \nn 
\ees
in which we have omitted the dependence on the mass ratio because we are considering equal mass binaries. 
The corresponding lower bound on $\lambda_g$ reproduces the result of an earlier inve\-stigation by Will~\cite{Will:1997bb}:
\begin{equation}
\lambda_g \sim {\cal N} \left( {D \over (1+z)D_L} \right )^{1/2} M^{+11/12} \,, 
\label{lambdabound}
\end{equation}
where ${\cal N}$ encodes various contributions from the experimental noise and where the dependence on the mass accounts for that of the SNR~(cfr. \eq{eq:snr}).
As noted in ref.~\cite{Will:1997bb}, the bound on $\lambda_g$ depends only weakly on distance, via the redshift dependence of the factor $[D/(1+Z)D_L]^{1/2}$: in fact, this ratio varies from unity at $z=0$ to 0.45 at $z=1.5$.  
The mild dependence of the bound~(\ref{lambdabound}) on redshift is due to the fact that while the signal strength and the accuracy {\it decrease} with distance, the size of the arrival-time effect {\it increases} with distance. 
This is connected to the distortion effects caused by the presence of an effective mass term in the dispersion relation. 

For all the systems investigated the degradation factor due to the spin-orbit term never exceeds~5. 
The best constraint on the graviton Compton wavelength in the presence of spin-orbit is $\lambda_g \geq 2.2\times10^{16}$km, which is four orders of magnitude larger than the Solar-System one~\cite{SS_Yukawa} and corresponds to $\lambda_g \geq 7\times10^{2}\,$pc.
It is interesting to convert this bound in a constrain on the graviton mass as we did in \Sec{sec:PPN_GW}. 
In units $\hbar=c=1$ one has $1=\hbar c\simeq200$~MeV\,fm$^{-1}$, where fm=fermi=$10^{-15}$m: with a conversion factor of 1m=($2\times10^{-7}$ev)$^{-1}$, one can finally obtain  $m_g\,\lsim\,10^{-26}$eV.

%\een

%%%%%%%%%%%%%%%%%%%%%%%%%%%%%%%%%
\subsection[The study of gravitational-wave constraints on \btf\ and 
\\ its comparison with other analysis]
    {The study of gravitational-wave constraints on \btf\ and its comparison with other analysis %
  \sectionmark{GW constraints on \btf\ in comparison with other analysis} } 
    \sectionmark{GW constraints on \btf\ in comparison with other analysis }
\label{sec:connections}

% where:    
% "[]" is for the table of contents
% the subsequent "{}" is for where the section actually is
% 1st "sectionmark" is for the first page containing the section
% 2nd "sectionmark" is for the other pages containing the section

%%%%%%%%%%%%%%%%%%%%%%%%%%%%%%%%%

The conclusions of BBW imply that the phasing coefficients corresponding to $\omega_{\rm BD}$ and $\lambda_g$ are both numerically small. 
In fact, from the bounds discussed just above one has
\be
\frac{5\,S^2}{84\,\omega_{\rm BD}} \,\lsim\, \frac{5\,(0.6)^2}{84\,(4\times 10^{4})}
\,\sim\, 5.4 \times 10^{-7} 
\ee
while the phasing coefficient due to a massive graviton in the case of LISA reads 
\bees
\frac{D\,M}{(\lambda_g)^2} &\simeq& \frac{(3\times10^9\,pc)\cdot(3\times10^7\,km)}{(2 \times 10^{16}\,km)^2} \nn \\
\nn \\
&\simeq&\frac{(3\times10^{22}\,km)\cdot(3\times10^7\,km)}{4 \times 10^{32}\,km^2} 
\simeq 2 \times 10^{-3} \,.
\ees
This characteristic is especially relevant to assess if one can possibly estimate small parameters like $\b_3$ and $\b_4$ with GW templates comprising spin terms. 
In this respect, another result of BBW is of crucial importance.
Both $\omega_{\rm BD}$ and $\lambda_g$ contribute a significant number of cycles to the template: from 10 to 1000. 
This is what really matters for GW astronomy: it is the individual value of the PN corrections to the phase, not their value relative to a numerically larger term.
Therefore, the conclusions of BBW suggest that it would be possible to constrain $\b_3$ and $\b_4$ even in the presence of spin terms.

\begin{figure}[t!]
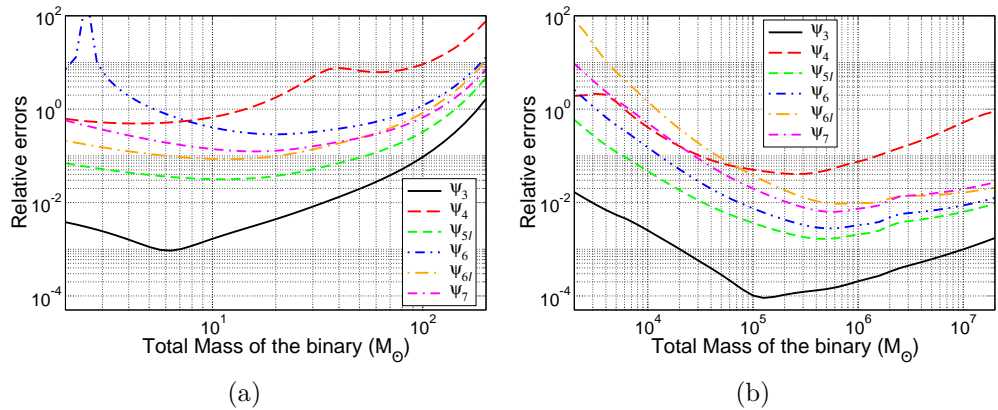

	\centering 
	\subfigure[]{
\includegraphics[width=2.5in]{How_To_Test/Sathya_Tests/EGO-errors.eps}
%		\label{fig:} 
} 
	\subfigure[]{
\includegraphics[width=2.5in]{How_To_Test/Sathya_Tests/LISA-errors.eps}
%		\label{fig:} 
}
\caption{Plot showing the relative errors in the test parameters $\psi_T=\psi_3, \psi_4, \psi_{5l}, \psi_6, \psi_{6l}, \psi_7$ as a function of the total mass $M$. 
The left panel displays the case of the third generation ground-based \ifo\ ET~\cite{ET} for a stellar mass compact binary at luminosity distance $D_L=200$ Mpc.
The right panel shows the situation for a supermassive black hole binary at a redshift of $z=1$ as observed for a year by LISA~\cite{LISA}.
The test parameters are analyzed one at a time.
Figure taken from ref.~\cite{Arun:2006hn}. \smallskip}
\label{fig:lisa-ego}
\end{figure}

\smallskip

It is interesting to compare the work of BBW with the studies of Arun, Iyer, Sathya\-prakash and collaborators (AISC)~\cite{Arun:2006yw,Arun:2006hn,Mishra:2010tp} that I have discussed in \Sec{sec:PPN_GW}. 
AISC did not include spin in their treatment so that a direct comparison between their analysis and the one from BBW can not be drawn. 
However, the results of the methodologies can be compared for what concerns the issues of degeneracies and the estimation accuracy achievable on individual parameters.
In the most powerful version of the test, AISC consider a reduced set of PN phasing coefficients and use two of them as a basis to express all the others: in so doing, AISC "take advantage" of the correlations among the coefficients.
One of the conclusions of AISC is that it will be possible to estimate the various PN phasing coefficients at future \ifos\ with the accuracies reported in \fig{fig:lisa-ego}. 
As we can see, the best accuracies are awaited with LISA~\cite{LISA}: even in this case the estimation precision is worse than $10^{-3}$, apart from one parameter. 
One could wonder how such accuracies might help in distinguishing an error on a GR coefficient from the actual presence of, say, $\b_3$ that is already known to be smaller than $10^{-4}$.
In answering this question, one should note that the focus of AISC is on estimating the PN phasing coefficients $\psi_j$'s; on the other hand, in BBW one tries to estimate the physical parameters of the template, both the intrinsic and the extrinsic ones. 
The difference lies in the fact that a single physical parameter like the mass ratio enters almost all the PN coefficients (apart from those that are constant, like $\psi_3$): therefore, the very aims of the two studies are different. 
As a consequence, for the purpose of estimating $\b_3$ and $\b_4$, the analysis implemented in BBW is the optimal one. 
On the other hand, for the sake of testing gravity in general, the best strategy is to implement more than one approach: on top of those I have confronted here, one could employ the ones suggested by AISC (see footnote \ref{foot:EOB} in \Sec{sec:PPN_GW}) and the parametrized post-Einstein framework proposed by Yunes and Pretorius~\cite{Yunes_PPE1}.

\begin{table}[t]\renewcommand{\arraystretch}{1.5}
\centering
\begin{tabular}{c@{\quad}c@{\quad\vline\quad}c@{\quad}c@{\quad}c@{\quad}c@{\quad}c@{\quad}c}
\hline
Interferometer & Source & \Nlow & \Nre & \Npl & \Nup\\
& & yr$^{-1}$ & yr$^{-1}$ & yr$^{-1}$ & yr$^{-1}$ & \phantom{bla} \\
\hline
& NS-NS 
& $2\times10^{-4}$ & $2\times 10^{-2}$ &$0.2$ &$0.6$ \\
Initial & NS-BH & $7\times 10^{-5}$ &$4\times10^{-3}$& $0.1$& \\
& BH-BH & $2\times 10^{-4}$ &$7\times10^{-3}$& 0.5  & \\
%& IMRI into IMBH &  &  & $<0.001$ & 0.01 \\
%& IMBH-IMBH & &  & $10^{-4}$  & $10^{-3}$\\
\hline
& NS-NS 
& 0.4 & 40 & 400 & 1000\\
Advanced & NS-BH  & 0.2  & 10 & 300 & \\
& BH-BH & 0.4 & 20 & 1000 & \\
%& IMRI into IMBH &  & & 10 & 300 \\
%& IMBH-IMBH & & & 0.1 & 1 \\
\hline
\end{tabular}
\caption{Detection rates for compact binaries: quantities are quoted in number of events per year; the physical significance of the various estimates in the columns is explained in Table~\ref{table_terminology}. 
Table adapted from ref.~\cite{Adv_Rates:09}. 
\label{table:det_rates} \smallskip}
\end{table}

\smallskip

So far, I have not discussed how the constraints could be affected by a joint analysis of many systems. 
Detection rates for ground-based \ifos\ will definitely improve in passing from 
the initial configuration to the advanced one, whose implementation is already underway. 
The most up-to-date estimates for detection rates have been reported in ref.~\cite{Adv_Rates:09}, from which Table~\ref{table:det_rates} is adapted. 
The work of ref.~\cite{Adv_Rates:09} included results appeared until September 2009 concerning various types of compact binary coalescences, among which I focus on the following ones: 
\bd
%\centering
\item neutron star-binaries (NS-NS);
\item neutron star-black hole binaries (NS-BH);
\item black hole-binaries (BH-BH).
\ed 
%The masses of the constituents objects are chosen as follows: $M_{NS}=1.4\msun$ and $M_{BH}=10\msun$.
For a given binary type in a LIGO-Virgo search, the table contains different estimates of the detection rate $\dot N = R \times N_G$, where $R$ is the coalescence rate of that type of binary per galaxy and $N_G$ is the number of galaxies accessible with a search for the relevant binary type.
The physical significance of the various estimates is explained in Table~\ref{table_terminology}. % which is adapted from ref.~\cite{Adv_Rates:09}.

\begin{table}[t]\renewcommand{\arraystretch}{1.5}
\centering
\begin{tabular}{c@{\quad\vline\quad}l@{\quad\vline\quad}l}
\hline
Symbol & Rate statement & Physical significance\\
\hline
\Nlow & Plausible pessimistic estimate & Rates could be as low as...\\
\Nre & Realistic estimate & Rates are likely to be...\\
\Npl & Plausible optimistic estimate & Rates could be as high as...\\
\Nup & Upper limit & Rates should be no higher than...\\
\hline
\end{tabular}
\caption{Rate statement terminology. 
Table adapted from ref.~\cite{Adv_Rates:09}. 
\label{table_terminology} \smallskip}
\end{table}
%
\begin{comment}%%%%%%%%%%%%%%%%%%%%%%%%%%%%%%%%%%
The coalescence rates for various coalescences are given: per Milky Way Equivalent Galaxy, per $L_{10}^1$ or per Mpc$^3$ in the case of NS-NS, NS-BH and BH-BH binaries respectively. 
The actual detection threshold for a network of interferometers will depend on a number of factors, including:
\bd 
\item the network conﬁguration (the relative locations, orientations, and noise power spectral densities of the detectors);
\item the characteristics of the detector noise (its Gaussianity and stationarity);  
\item the search strategy used (coincident vs. coherent search). 
\ed 
A full treatment of these effects was beyond the scope of ref.~\cite{Adv_Rates:09}. 
Instead, event rates detectable by the LIGO-Virgo network are estimated by scaling to an average volume within which a single detector is sensitive to coalescences above a ﬁducial SNR threshold of 8. 
This is a conservative choice if the detector noise is Gaussian and stationary and if there are two or more detectors operating in coincidence. 
\end{comment}%%%%%%%%%%%%%%%%%%%%%%%%%%%%%%%%%%
For the purpose of calculating the rates in Table \ref{table:det_rates}, ref.~\cite{Adv_Rates:09}  assumed that all neutron stars have a mass of $1.4 M_\odot$ and all black holes have a mass of $10 M_\odot$. 
Although both types of objects will rather cover a range of masses (see, e.g., \cite{OShaughnessy:2006wh,Mandel:2009nx}), the present knowledge of the mass distribution is not sufficient to provide detailed models; moreover, the uncertainties in the coalescence rates dominate errors from the simplified assumptions about component masses. 
Within these assumptions, ref.~\cite{Adv_Rates:09} estimated the event rates detectable by the LIGO-Virgo network by scaling to an average volume within which a single detector is sensitive to coalescences above a ﬁducial SNR threshold of 8. 
For a single interferometer, this volume probed is characterized by the so-called {\it horizon distance}: according to the type of source and the \ifo\ involved in the search, the horizon distance assumes the values reported in Table~\ref{table:hor_dist}.

\smallskip

As it is evident from Table~\ref{table:det_rates}, with the advanced versions of LIGO~\cite{ALIGO} and Virgo~\cite{AVirgo} one can realistically await as many as 70 detections per year, i.e. more than one per week! 
This large statistics will constitute a major improvement in estimating parameters like \btf: in fact, the values of extrinsic parameters depend only on the theory of gravity and not on the system detected; on the other hand, each source will be unique for what concerns the intrinsic parameters like the spin orientation.  
Therefore, if $N$ events are detected, each one can be interpreted as providing an independent measurement of an extrinsic parameter, for example $\b_3$: the total error on $\b_3$ would then be a fraction $1/\sqrt{N}$ of the error coming from a single observation. 
Detection of many systems has a further advantage in measuring extrinsic parameters: as for PPK tests with binary pulsars, there might be systems that are better suited for measuring one parameter instead of another. 
For example, the scalar charge of a neutron star is better measurable if one can detect a dipolar flux from an asymmetric system like a pulsar-white dwarf binary (see \eq{eq:flux_dip} in \Sec{sec:Dam_Far}).
Moreover, the bound derived on a given parameter from one system might be used as a prior when testing a different parameter with another source: in this way, one should be able to disentangle the effects of various extrinsic parameters. %if they happen to be present in the GW template at the same PN order. 
A final note concerns the possibility to further increase the accuracy of GW experiments by cross-correlating the outputs of many \ifos: this enables one to dig signals lower than the sensitivity of an individual instrument and has proven crucial to obtain the bound on the GW background of stochastic origin with first-generation instruments~\cite{GW_bck-Nat}. 

\begin{table}[t]\renewcommand{\arraystretch}{1.5}
\centering
\begin{tabular}{c@{\quad}c@{\quad\vline\quad}c@{\quad}}
\hline
Interferometer & Source & $D_{horizon}$(Mpc) \\
\hline
& NS-NS & 33  \\
Initial LIGO & NS-BH & 70  \\
& BH-BH & 161  \\
\hline
& NS-NS & 445 \\
Advanced LIGO & NS-BH  & 927  \\
& BH-BH & 2187 \\
\hline
\end{tabular}
\caption{According to the source type, a given \ifo\ can probe different volumes of space, characterized by the so-called {\it horizon distance}. 
With a fiducial signal-to-noise threshold of 8, Initial and Advanced LIGO are sensitive to the indicated values of the horizon distance~\cite{Adv_Rates:09}. 
\label{table:hor_dist} %\smallskip
}
\end{table}

%\smallskip

The considerations discussed in the previous sections motivate a dedicated analysis to extend the study of \btf\ to PN orders in the GW phasing.
I have started this investigation thanks to a collaboration with Emanuele Berti, one of the authors of BBW. % , I have started the follow-up of the work conducted with my group~\cite{UC:09}: the aim is to see which bounds can be put on $\b_3$ and $\b_4$ by GW detection at \ifos, when one considers this as a parameter estimation problem.
In what follows I describe part of the ongoing work in this direction. 
We expect that the analysis presented in this chapter will be finalized soon~\cite{withBerti}.

%%%%%%%%%%%%%%%%%%%%%%%%%%%%%%%%%
\section[Feynman diagrams for the phase shift induced by the modified \\ three- and four-graviton vertices up to first post-Newtonian order]
    {Feynman diagrams for the phase shift induced by the modified three- and four-graviton vertices up to first post-Newtonian order %
  \sectionmark{Diagrams for the modified non-Abelian vertices of GR up to 1PN} } 
    \sectionmark{Diagrams for the modified non-Abelian vertices of GR up to 1PN}
\label{sec:vert_diags}

% where:    
% "[]" is for the table of contents
% the subsequent "{}" is for where the section actually is
% 1st "sectionmark" is for the first page containing the section
% 2nd "sectionmark" is for the other pages containing the section

%%%%%%%%%%%%%%%%%%%%%%%%%%%%%%%%%

As we saw in \Sec{sec:PPN_GW}, the starting point to compute the evolution of the orbital phase $\phi(t)$ is the energy balance equation (\ref{eq:balance}) that we re-write here as 
\bees
\label{eq:bal}
P=-\dot{E}\,,
\ees
where $P$ is the power emitted in GWs and $E$ is the mechanical energy of the system.
The PN expansions of $P$ and $E$ that we met in \Sec{sec:PPN_GW} \eq{eq:EF} can be conveniently dealt with by means of a gauge invariant adimensional quantity 
\be
x \equiv \left(G_N M\, \omega_s\right)^{2/3} \sim {\cal O} (v^2)\,,
\ee
where $\omega_s$ is the frequency of the source bulk movement, for example that of its orbital motion;
%$G_N$ is the Newton constant and $M$ the total mass of the bodies constituting the binary system, 
up to 1PN, the expressions of $P$ and $E$ in terms of $x$ read 
\begin{eqnarray}
\label{eq:EPx}
E=E_0 x (1+ \eps_1 x)\,,\quad P=P_0 x^5 (\pi_0+\pi_1x)\,,
\end{eqnarray}
where $\eps_1$ and $\pi_1$ indicate the 1PN corrections.
The energy balance equation (\ref{eq:bal}) then translates as the time evolution of the variable~$x$, which is given by
\begin{eqnarray}
\label{eq:dotx}
\dot{x}=-\frac{P_0 \pi_0}{E_0}x^5\left[1+\frac{\pi_1-2\pi_0 \eps_1}{\pi_0}x\right]
\end{eqnarray}
and can be easily solved perturbatively ($E_0<0,P_0>0$).
By redefining the time parameter as 
\be
\label{eq:Theta}
\Theta=-\frac{P_0\pi_0}{64 E_0} (t_c-t) \q \text{with}\;
 \(\Theta\)^{-1/4} \sim {\cal O}(v^2) \; ,
\ee
the solution of \eq{eq:dotx} reads:
\begin{eqnarray}
x=\frac{\Theta^{-1/4}}{4}\left[1+\frac{\pi_1-2\eps_1\pi_0}{12\pi_0}
\Theta^{-1/4}\right]\,.
\end{eqnarray}
Finally, defining the source phase as ${\rm d}\phi/{\rm d}t=\omega_s$, one gets the time evolution of the orbital phase in terms of the parameter $\Theta$
\begin{eqnarray}
\label{eq:phi_EP}
\phi &=& -\frac{\Theta^{5/8}}{\pi_0 \nu}\left[1+5\frac{2\eps_1\pi_0-\pi_1}{24 \pi_0} \Theta^{-1/4}\right] +\text{const} \nn \\
\nn \\
&\equiv& -\frac{\Theta^{5/8}}{\nu}\left[ \phi_0 + \; \phi_2 \;\Theta^{-1/4}\right] +\text{const} \,
\end{eqnarray}
where I have introduced the compact notation for the phasing coefficients: $\phi_0$ for the leading order term and $\phi_2$ for the ${\cal O}(v^2)$ 1PN correction.
As I have discussed in \Sec{sec:PPN_GW}, every theory of gravity has its own prediction for these coefficients: in the case of GR, this prediction reads 
\begin{eqnarray}
&&E_0=-\frac{\mu}{2}\,,\quad \eps_1=-\frac{3}{4}-\frac{\nu}{12}\,,\nn\\
\nn \\
&&P_0=\frac{32\nu^2}{5G}\,,\quad \pi_0=1\,,\quad \pi_1=
-\frac{1247}{336}-\frac{35}{12}\nu 
\end{eqnarray}
where $\mu$ is the reduced mass of the system $\mu\equiv m_1 m_2 /M$ and we remind that $E_0$ and $P_0$ contribute to the phase through the time parameter $\Theta$~(see \eq{eq:Theta}). 
Plugging this prediction in \eq{eq:phi_EP} gives the GR result for the phase as a function of time
\begin{eqnarray}
\phi=-\frac{\Theta^{5/8}}{\nu}
\left[1+\left(\frac{3715}{8064}+\frac{55}{96}\nu\right)
\Theta^{-1/4}\right]+\text{const} \; ;
\end{eqnarray}
therefore, in GR one has 
\be
\phi_0 =1 \q , \q
\phi_2 =\left(\frac{3715}{8064}+\frac{55}{96}\nu\right) 
\ee
which coincide, respectively, with the coefficients $\a_0$ and $\a_2$ in \eq{eq:alpha}.

\begin{figure*}[t]
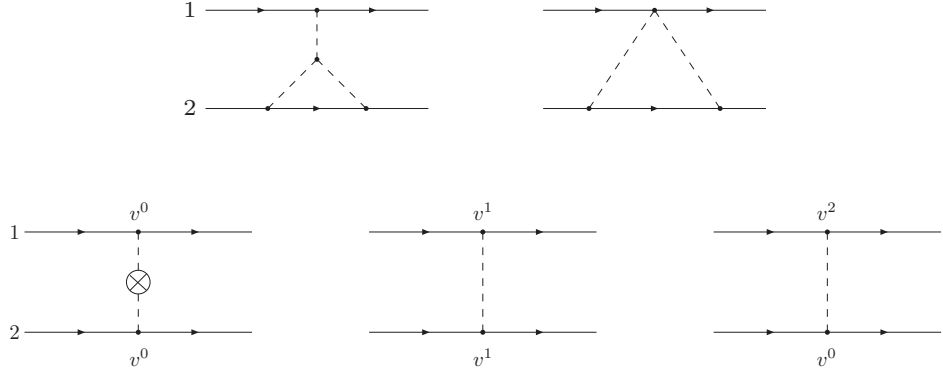

\centering
\includegraphics[width=0.50\textwidth,angle=0]
{Chap_EFT/EIH_G2.eps}
\vskip 1.0cm
\includegraphics[width=0.80\textwidth,angle=0]
{Chap_EFT/EIH_v2.eps}
\caption{The Feynman diagrams that, in NRGR, represent the next-to-leading 
order contributions to the conservative dynamics, i.e. corrections to the 
Newtonian potential which are ${\cal O}({m\,G_N}/r)$, upper row, and 
${\cal O}(v^2)$, lower row, with respect to it. 
Figure taken from ref.~\cite{NRGR_paper}. }
\label{fig:EIH_b}
\end{figure*}
\begin{figure}
    \centering
    \includegraphics[width=6cm]{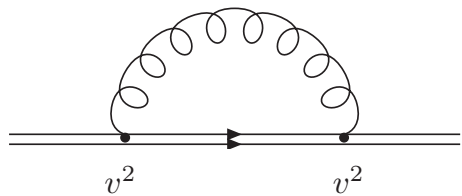}
\caption{NRGR diagram whose imaginary part gives rise to the quadrupolar gravitational radiation power spectrum (see text and \Sec{sec:eft_bin} for discussion).
%Feynman diagram representing the emission and re-absorption of a radiation graviton: the amplitude of this process, where no quantum propagates to $\infty$, enables one to calculate the power emitted in GWs (see \chap{chap:EFT})
Figure taken from ref.~\cite{NRGR_paper}. }
\label{fig:selfrad}
\end{figure}

Looking at \eq{eq:phi_EP}, one can see that the phasing coefficients originate from both the conservative and the radiative dynamics of the system; notably, the 1PN correction to the phase $\phi_2$ is made of $\pi_0$ and $\pi_1$, coming from the expansion of $P$, and $\eps_1$ coming from the expansion of $E$ (see \eq{eq:EPx}).
In the EFT terminology, these expansions can be obtained by means of Feynman diagrams. 
For what concerns the energy function, the 1PN correction~$\eps_1$ is calculable from the diagrams of \fig{fig:EIH_b} that represent the EIH Lagrangian, like in \Sec{sec:eft_iso}. 
On the other hand, the EFT prescription to calculate the power emitted in GWs is to take the imaginary part of the self-interaction diagram of \fig{fig:selfrad}.
We remind from \Sec{sec:eft_bin} that this Feynman diagram describes a process in the effective theory which is valid at scales $\ell$ larger than the orbital radius~$r$\,: for this reason, the binary system is un-resolved in the diagram and indicated with a double line. 
In the figure, the individual matter-gravity vertex corresponds to many vertices of the theory valid at scales $\ell < r$. 
These vertices are described by the diagrams of \fig{fig:leading_rad}: 
%, one of which contains a three-graviton vertex: 
they are ${\cal O}(v^2)$ with respect to the coupling of a radiation graviton with the mass monopole of the system (cfr. discussion in \Sec{sec:eft_bin}). 
In GR both this coupling and the coupling to the dipole vanish: therefore, the leading order in the radiation dynamics of GR is represented by the ${\cal O}(v^2)$ terms of \fig{fig:leading_rad}. 
Looking at these diagrams, one can see that the three-graviton vertex enters the power at leading order. 
A possible deviation of this vertex from its GR value is reflected in a shift in the power coefficient~$\pi_0$\,: tagging by $\b_3$ the \tgv, we can keep track of this deviation and of its contribution to the phase.
The shift produced by $\b_3$ at leading order in the power is what enabled my group and me to put the bound $\b_3\leq10^{-4}$ from the data on Hulse-Taylor pulsar. 
This calculation in the strong-field/radiative regime did not need to take into account the shift produced by $\b_3$ in the conservative dynamics. 
In fact, as one can see from \fig{fig:EIH_b}, this last shift affects the EIH Lagrangian, which is related to the 1PN correction to the energy~$\eps_1$. 
In other words, the three-graviton vertex enters the energy at next-to-leading order: %through a modification of the EIH Lagrangian: 
this produces a shift in $\eps_1$ which affects the phase (\ref{eq:phi_EP}) at 1PN order.
In ref.~\cite{UC:09}, we rather used the 1PN change in the conservative part to constrain~$\b_3$ by means of Solar-System tests.
 
For the purpose of estimating \btf\ through GW observations, it is necessary to see how the phase is affected by the modified \tfgv\ at least to order 1PN, i.e. one has to calculate the modification of the coefficient~$\phi_2$ (\ref{eq:phi_EP}). 
For this one needs the shifts in $\pi_1$ and $\eps_1$: because I already have the latter, I only have to calculate the former.
The shift in $\pi_1$ can be obtained from the Feynman diagrams that contain \tfgvs\ in the radiative dynamics at 1PN. 
I display these diagrams in the following sections.

\begin{figure*}[t!]
    \centering 
    \includegraphics[width=12cm]{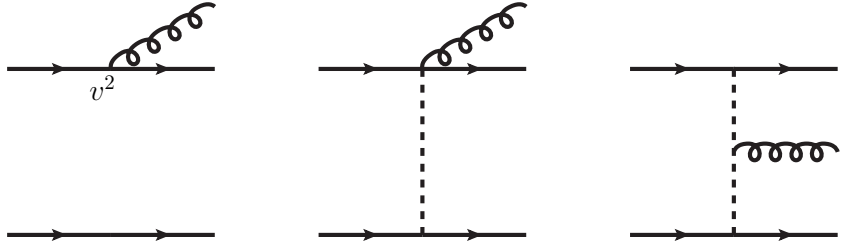}
    \caption{Feynman diagrams describing the leading order (0PN) radiation Lagrangian in GR (see text for discussion).}
    \label{fig:leading_rad}
\end{figure*}
%

%%%%%%%%%%%%%%%%%%%%%%%%%%%%%%%%%
\subsection[The phase shift induced by the modified three-graviton vertex \\ 
at first post-Newtonian order]
    {The phase shift induced by the modified three-graviton vertex 
    at first post-Newtonian order %
  \sectionmark{1PN phase shift induced by the modified three-graviton vertex} } 
    \sectionmark{1PN phase shift induced by the modified three-graviton vertex }
\label{sec:rad_3}

% where:    
% "[]" is for the table of contents
% the subsequent "{}" is for where the section actually is
% 1st "sectionmark" is for the first page containing the section
% 2nd "sectionmark" is for the other pages containing the section

%%%%%%%%%%%%%%%%%%%%%%%%%%%%%%%%%

We start the investigation of the radiative sector from the contributions due to the three-graviton vertex. 
We have to consider which are the possible modifications of ${\cal O}(v^2)$ 
or of ${\cal O}(G_N)$ to the \tgv\ diagram of \fig{fig:leading_rad}.
%This is spirit we will use in what follows to refer to the modifications.

A first type of modifications comes from the PN corrections of the matter-graviton vertices with factors of the velocity: these come from the expansion of the point particle action 
\be
\label{eq:Spp}
{\cal S}_{pp} = {m\over \mpl}\int dt \left[-{1\over 2} H_{00} - 
H_{0i} v_i - {1\over 4} H_{00} {\bf v}^2 -{1\over 2} H_{ij} v^i v^j\right] \,.
\ee
and produce the diagrams of \fig{fig:beta3_v2}. 
\begin{figure*}[htbp]
    \centering 
    \includegraphics[width=8cm]{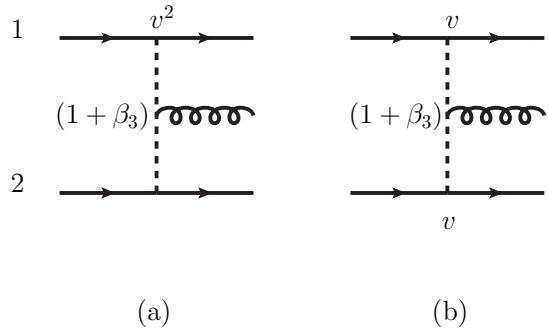}
    \caption{1PN modifications of the three-graviton vertex coming from the velocity expansion in the point-particle action \eq{eq:Spp}.}
\label{fig:beta3_v2}
\end{figure*}

In the spirit of the PN expansion, one treats time derivatives as perturbations. 
For what concerns the EFT approach we met an example of this attitude in \Sec{sec:EIH} when we were dealing with the propagator of a potential graviton (see the derivation of \eqs{eq:Hprop}{eq:fullprop}). 
We report excerpts from that discussion here for a more complete treatment. 
By considering potential modes in Fourier space one can expand their propagator as: 
\begin{equation}
{1\over k^2_0 - {\bf k}^2} =-{1\over {\bf k}^2}\left[1 + {k^2_0\over {\bf k}^2}+\cdots\right] =-{1\over {\bf k}^2}\left[1 + {\cal O}(v^2)\right] \,,
\end{equation}
whose leading order term ${\bf k}^{-2}$ gives the $1/r$ dependence of Newtonian potential once Fourier-transformed according to  
\begin{equation}
\label{eq:Four_New}
\int {d^4 k\over (2\pi)^4} e^{-i k\cdot x} {1\over {\bf k}^2} = {1\over 4\pi |{\bf x}|} \delta(x^0) \, .
\end{equation}

Therefore, to leading order, the propagator for $H_{{\bf k}\mu\nu}$ reads 
\begin{equation}
\label{eq:Hprop_b}
\langle T H_{\bf k\mu\nu}(x^0) H_{\bf q\alpha\beta}(0)\rangle = -{i\over{\bf k}^2} (2\pi)^3 \delta^3({\bf k}+{\bf q})\delta(x^0) P_{\mu\nu;\alpha\beta} \,,
\end{equation}
where $T$ stands for {\it time-ordering} and the tensor structure is given by \eq{eq:Pprop}. 
On the other hand the ${\cal O}(v^2)$ terms give a propagator which reads
\be 
\label{eq:fullprop_b}
\langle H_{\vk\mu\nu}(x^0) H_{{\bf q}\alpha\beta}(0)\rangle_{\otimes} = 
-(2\pi)^3\delta^3(\vk + {\vq}){i\pa_t^2\over \vk^4}\delta(x^0)
P_{\mu\nu;\alpha\beta} \,,
\ee
where the subscript $\otimes$ is to distinguish this propagator from the one in \eq{eq:Hprop_b} and where we stress that
\be
{\pa_t^2\over \vk^4} \sim {v^2\pa_i^2\over \vk^4} 
\sim {v^2\vk^2\over \vk^4} \sim {v^2\over \vk^2} \,,
\ee 
so that \eq{eq:fullprop_b} is indeed suppressed by $v^2$ with respect to \eq{eq:Hprop_b}.

In the present context, time derivatives are to be considered also in graviton self-interactions like the $hHH$ vertex of \eq{eq:hHH} in \Sec{sec:eft_bin}.
The resulting diagrams are those of Fig.~(\ref{fig:beta3_deriv}), where the blobs and the acronyms $1D$, $2D$ indicate corrections due to the inclusion of one or two time derivatives, respectively.
\begin{figure*}[t]
    \centering 
    \includegraphics[width=14cm]{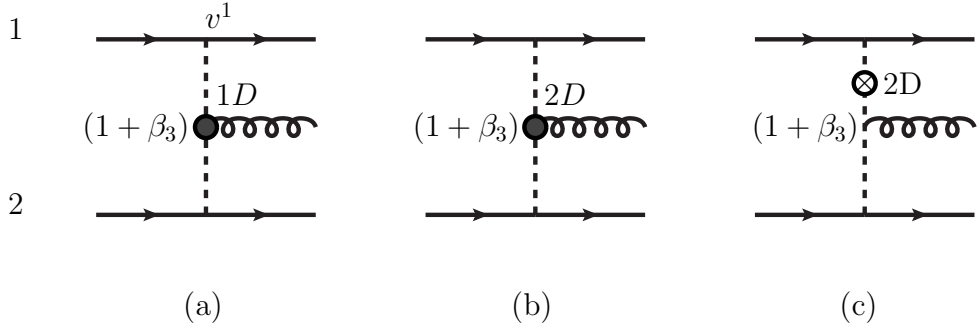}
    \caption{1PN modifications of the three-graviton vertex coming from the inclusion of time derivatives.}
\label{fig:beta3_deriv}
\end{figure*}

\smallskip 

There is another way of getting PN modifications from derivatives that is peculiar to radiative dynamics: it is the multipole-expansion of the radiative field at the level of the action, a necessary ingredient of the EFT treatment~\cite{QCD_mult}.  
Condensing the discussion of \Sec{sec:eft_bin}, one can take the origin of the reference frame to be the center of mass of the system and write the radiative field as 
\be
\label{eq:multi}
h_{\mu\nu} (\vx,t) \simeq h_{\mu\nu}(0,t) + x^i \pa_i h_{\mu\nu}(0,t) 
+ \frac 12 x^i x^j \pa_i \pa_j h_{\mu\nu} (0,t) \,,
\ee
where $x^i$'s are coordinates with respect to the center of mass and where the (spatial) derivative scales as $\pa_i \sim v / r$ so that $x^i \pa_i \sim v$. 
To have a correction of ${\cal O}(v^2)$ we need to consider the last term of \eq{eq:multi}, as it is the case for the diagram in Fig.~(\ref{fig:beta3_multi2}).

\begin{figure*}[t]
    \centering 
    \includegraphics[width=5cm]{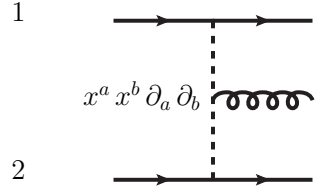}
    \caption{1PN modification of the three-graviton vertex coming from the multipole-expansion of the radiative field, a necessary ingredient of the EFT treatment.}
\label{fig:beta3_multi2}
\end{figure*}

Finally, we have the ${\cal O}(G_N)$ modifications due to graviton self-interactions. 
A first example is the one represented in Fig.~(\ref{fig:seagull_rad}). 
%({\bf cfr. EIH Lagrangian's diagrams w.r.t. Newton potential's diagram}).
%
\begin{figure*}[t]
    \centering 
    \includegraphics[width=4.5cm]{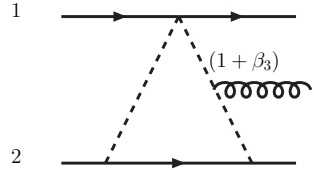}
    \caption{1PN contribution to the radiative sector coming from ${\cal O}(G_N)$ modifications of the diagrams in Fig.~(\ref{fig:leading_rad}).}
\label{fig:seagull_rad}
\end{figure*}

After Fig.~(\ref{fig:seagull_rad}) it is easy to build the diagrams of Fig.~(\ref{fig:3H_1h}) and Fig.~(\ref{fig:two_3grav_rad}).
\begin{figure*}[p]
    \centering 
    \includegraphics[width=10cm]{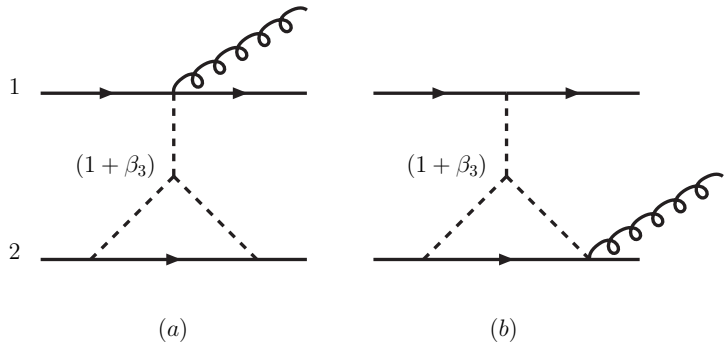}
    \caption{1PN contributions to the radiative sector coming from ${\cal O}(G_N)$ modifications of the diagrams in Fig.~(\ref{fig:leading_rad}).}
\label{fig:3H_1h}
\end{figure*}
\begin{figure*}[p]
    \centering 
    \includegraphics[width=10cm]{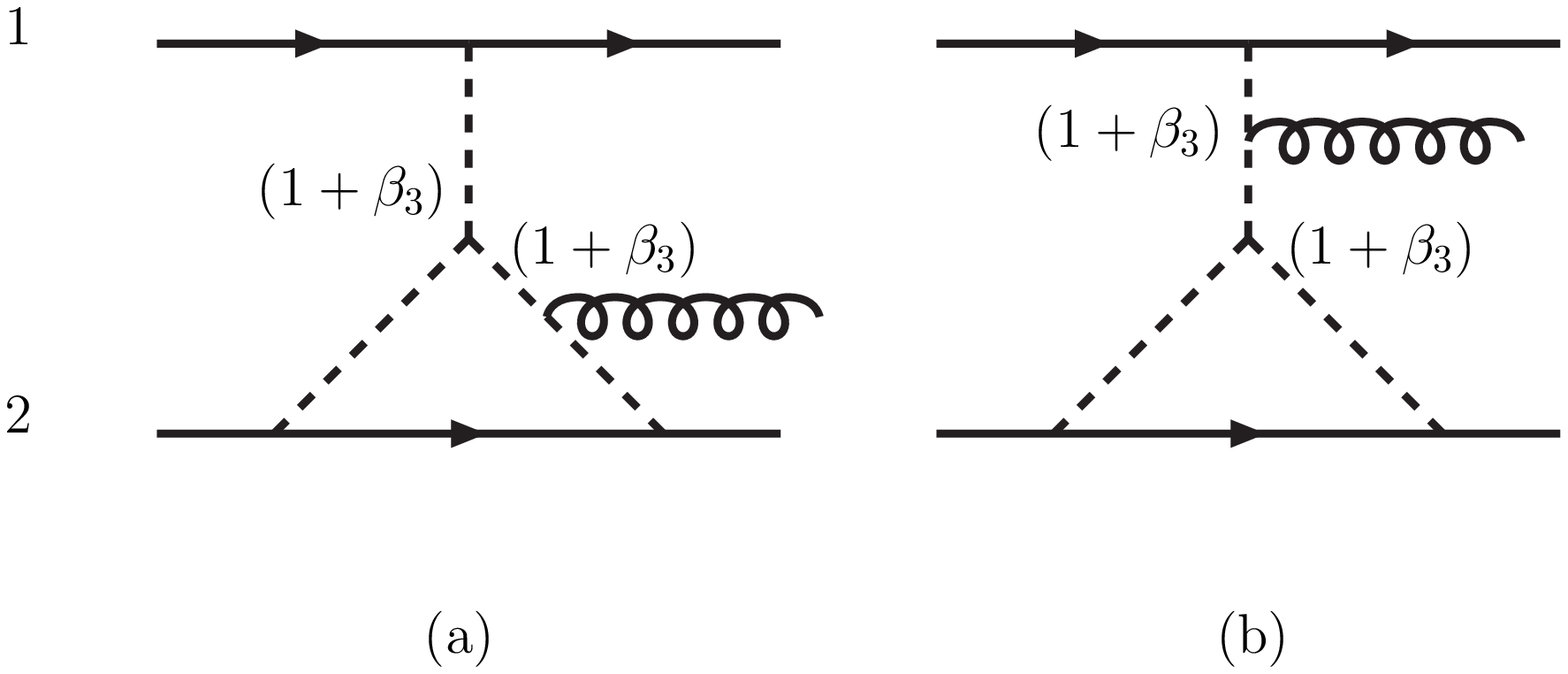}
    \caption{1PN contributions to the radiative sector coming from ${\cal O}(G_N)$ modifications of the diagrams in Fig.~(\ref{fig:leading_rad}).}
\label{fig:two_3grav_rad}
\end{figure*}

\clearpage

%%%%%%%%%%%%%%%%%%%%%%%%%%%%%%%%%
\subsection[The phase shift induced by the modified four-graviton vertex \\ 
at first post-Newtonian order]
    {The phase shift induced by the modified four-graviton vertex 
    at first post-Newtonian order %
  \sectionmark{1PN phase shift induced by the modified four-graviton vertex} } 
    \sectionmark{1PN phase shift induced by the modified four-graviton vertex }
\label{sec:rad_4}

% where:    
% "[]" is for the table of contents
% the subsequent "{}" is for where the section actually is
% 1st "sectionmark" is for the first page containing the section
% 2nd "sectionmark" is for the other pages containing the section

%%%%%%%%%%%%%%%%%%%%%%%%%%%%%%%%%

Because of the scalings of graviton self-interactions, a diagram containing a four-graviton vertex is of order 1PN with respect to diagrams where a three-graviton vertex is present.
Therefore, the four-graviton vertex contributes for the first time to the phasing directly at 1PN order through the diagram of Fig.~(\ref{fig:beta_4_rad}).
\begin{figure*}[htbp]
    \centering 
    \includegraphics[width=5cm]{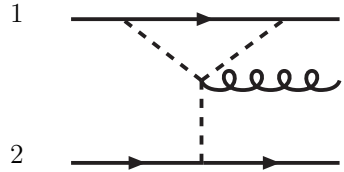}
    \caption{The only diagram containing a four-graviton vertex that contributes to the radiative sector at order 1PN.}
\label{fig:beta_4_rad}
\end{figure*}

%\clearpage

%%%%%%%%%%%%%%%%%%%%%%%%%%%%%%%%%
\subsection[The phase shift induced by the modified three-graviton vertex \\ 
at half post-Newtonian order]
    {The phase shift induced by the modified three-graviton vertex 
    at half post-Newtonian order %
  \sectionmark{0.5PN phase shift induced by the modified three-graviton vertex} } 
    \sectionmark{0.5PN phase shift induced by the modified three-graviton vertex }
\label{sec:beta3_half}

% where:    
% "[]" is for the table of contents
% the subsequent "{}" is for where the section actually is
% 1st "sectionmark" is for the first page containing the section
% 2nd "sectionmark" is for the other pages containing the section

%%%%%%%%%%%%%%%%%%%%%%%%%%%%%%%%%

%%%%%%%%% 
       % 0.5 PN
%%%%%%%%%
\begin{figure*}[htbp]
    \centering 
    \includegraphics[width=8cm]{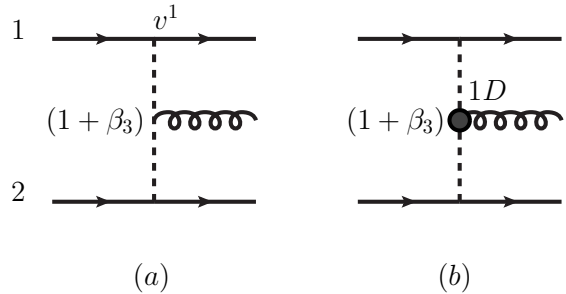}
    \caption{Feynman diagrams representing possible modifications of a \tgv\ at half PN order in the phase. (See text for complete discussion.)}
\label{fig:beta3_half}
\end{figure*}

With scaling arguments in mind, one can also build the diagrams of \fig{fig:beta3_half}, which are ${\cal O}(v)$ with respect to the quadrupole.
In GR both these diagrams do no contribute. 
Let us see in more detail why this is so.

We start from the diagram in panel (a). 
The velocity factor attached to the worldline is contracted with a field $H_{0i}$, which in turn requires a couple of indices $0j$ in the three-graviton vertex,  otherwise the propagator vanishes.
There are two possible ways to further contract these indices.
The first possibility is to have $k_0 k_j$ i.e. one time derivative in the vertex;  however, this would bring another factor of order $v$. 
The diagram in Fig.\ref{fig:beta3_half}(a) would then be incomplete as it stands
and would not belong to a 0.5PN order: if one draws the 1D symbol for the derivative, the result is one of the 1PN diagrams we met in the preceding section.
A second possible way of contracting the indices $0j$ in the vertex is offered by the polarization of the radiative graviton: this mode is non-radiating in GR to obey a tenet of the theory, so it makes sense to consider it in presence of non-vanishing $\b_3$ and $\b_4$.

Now we consider Fig.~\ref{fig:beta3_half}(b). 
A time derivative in the vertex amounts to having a term $k_0 k_j$ which we can contract with a radiative $h_{0j}$ mode: this mode is forbidden in GR but not necessarily in alternative theories. 
Incidentally, we remind here that a similar case of terms which are  forbidden in GR allowed us to put a constrain on $\b_3$ (see the discussion at the end of \Sec{sec:comp}, starting from \eq{lquad}). % (see the power for circular orbits).
 
%One can then conclude that both diagrams in Fig.~(\ref{fig:beta3_half}) can contribute to a 0.5PN order for the case $\b_3\neq 0$ and should be retained when studying the phasing.

At the same 0.5PN order there is also a term from the multipole expansion (cfr. \fig{fig:beta3_multi2}). 
The bottom line is that the presence of a shift $\b_3$ in the \tgv\ contributes terms of order 0.5 PN in the phase on top of those of order 1PN described in the previous sections. 
Therefore, the full set of diagrams that should be investigated at order 0.5 PN in the phase are those of \fig{fig:05PN}. 
\begin{figure*}[!h]
    \centering 
    \includegraphics[width=12cm]{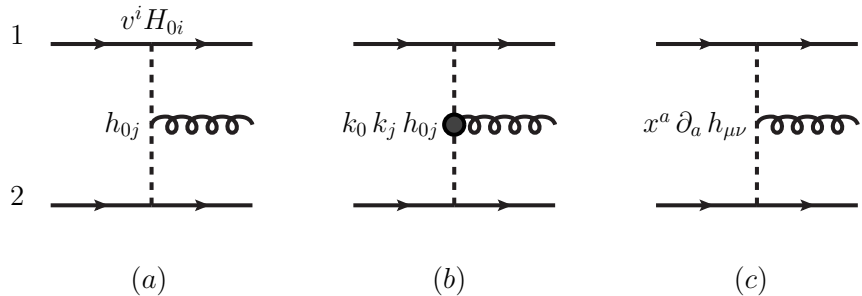}
    \caption{Terms of half-PN order in the phase excited by the \tgv\ modification $\b_3$. The 0.5PN order is absent in GR.}
\label{fig:05PN}
\end{figure*}
%

%\clearpage

%%%%%%%%%%%%%%%%%%%%%%%%%%%%%%%%%%%%%%%% 
\section{Final remarks} 
%%%%%%%%%%%%%%%%%%%%%%%%%%%%%%%%%%%%%%%%

In this chapter I have presented an extension of the work I performed with the rest of the gravity group at the University of Geneva~\cite{UC:09}: there we proposed a field-theoretical extension of the PPN, where the parameters are used to tag the strength of graviton vertices, notably the deviation of the \tfgv\ from the GR predictions. 
%described the possibility of testing gravity with GWs by means of a new \ph\ approach, the one originally proposed in ref.~\cite{UC:09} by the gravity group of the University of Geneva to which I belong. 
With available data from experiments in the Solar System and from the timing of binary pulsars, we could constrain these deviations to interesting accuracies. 
Concerning the bounds that could be put with future detections at \ifos, we qualitatively concluded that the numerical smallness of $\beta_3$ and $\beta_4$ with respect to other parameters in the waveform at the same PN order could have prevented us from getting significant constraints. 
For this reason we decided not to go further in calculating the PN modifications that $\beta_3$ and $\beta_4$ introduce in the phase of GWs and to publish our results. 
Because we worked in the restricted PN framework, we suggested that these degeneracies could be broken by the inclusion of amplitude corrections in the template, as higher harmonics enrich the functional dependence of the parameters~\cite{VanDenBroeck:2006ar}.
%Improvements in parameter estimation brought in by the use of full waveforms have been discussed in \Sec{sec:PPN_GW}, based on the study of ref.~\cite{Mishra:2010tp}.
However, a richer functional dependence could offer the possibility of binding small parameters like \btf\ even in the restricted PN framework.
An example along this direction is the work of ref.~\cite{Berti:2004bd}, where Berti, Buonanno and Will investigated the bounds on some alternative theories that could be put with LISA observations if templates include spin-orbit and spin-spin effects. 
I have described this analysis in \Sec{sec:BBW}, focusing on the aspects which are more salient to the problem of estimating \btf; in this context, the most notable conclusion is the following. 
Deriving the Fisher matrix, Berti {\it et al.} found that the inclusion of spin-orbit contributions in the waveform degrades the accuracy in estimating the parameters of alternative theories; however, the possibility of deriving meaningful constraints on these parameters in the realistic case of spinning objects does not get spoilt. 
On these grounds, it seems likely that one can constrain $\b_3$ and $\b_4$ even in the presence of spin terms.
In collaboration with Emanuele Berti I am conducting this follow-up of the work of ref.~\cite{UC:09}. 
The details of this investigation have been discussed in \Sec{sec:vert_diags}, where I have presented the Feynman diagrams needed to calculate how the modified \tfgvs\ affect the phase up to 1PN order. 
Once the phase modification is obtained, an analysis of the type of ref.~\cite{Berti:2004bd} %Berti {\it et al.}
should enable Berti and me to derive the constraints that future GW observations will put on the parameters \btf~\cite{withBerti}. 

The field-theoretical extension of the PPN that characterizes refs.~\cite{UC:09,withBerti} sets itself in the active research domain of testing gravity with GWs. 
In sections \ref{sec:PPN_GW} and \ref{sec:connections} I have discussed two other phenomenological approaches that have appeared in the literature to parametrize the gravitational waveform: the series of studies by Arun {\it et al.}~\cite{Arun:2006yw,Arun:2006hn,Mishra:2010tp} and the parametrized post-Einstein framework put forward by Yunes and Pretorius in ref.~\cite{Yunes_PPE1}. 
These studies and the work discussed in the present chapter constitute relevant phenomenological frameworks to test gravity in the strong-field/radiative regime represented by GWs.

%% file: Conclusions.tex
\chapter*{Conclusions}
\addcontentsline{toc}{chapter}{Conclusions}
\chaptermark{Conclusions}

This thesis is based on the effective field theory approach that has been recently proposed to tackle the two-body problem in General Relativity. 
Such a framework has been borrowed from particle physics studies of non-relativistic bound states: for this reason it has been dubbed Non-Relativistic General Relativity (NRGR) by their authors, Goldberger and Rothstein.
Through NRGR one can reformulate the post-Newtonian expansion of binary dynamics in a streamlined fashion. 
In fact, an effective field theory approach is best suited to treat a problem like the inspiral of compact binaries, where the disparity of physical scales allows one to separate the relevant degrees of freedom in subsets with decoupled dynamics.
At each scale, by means of power counting rules, one can assess the number of terms to be included in the computation of a given observable: in a field-theoretical context, this perturbative  calculation is conveniently performed by means of Feynman diagrams. 

The use of these diagrammatic techniques is quite natural when discussing alternative theories of gravity with respect to General Relativity, as first shown by Damour and Esposito-Farese. 
A major example in this sense is represented by \sts\ where the dynamics of the helicity-2 gravitons characterizing General Relativity is supplemented by a sector of helicity-0 modes. 
In collaboration with Riccardo Sturani of the gravity group at the University of Geneva, I applied NRGR to both point-like and string-like sources in a \st\ of gravity. 
Notably, we treated the issue of the renormalization of the energy-momentum tensor and reproduced two types of results which had been obtained previously: with the insight of NRGR we were able to re-derive and confirm the resolution of an apparent discrepancy between those results. 

Beside representing a convenient computational tool, diagrammatic techniques enable one to shed a different light on the information extracted from experiments of relativistic gravity. 
This field-theoretical perspective is the standpoint taken in the present thesis and it has been exposed in Chapters 4 and 5. 
There I have reported investigations which I developed in two steps: in a first stage I worked with the entire gravity group at the University of Geneva composed of Stefano Foffa, Michele Maggiore, Hillary Sanctuary, Riccardo Sturani and me; afterwards, I started a collaboration with Emanuele Berti at the University of Mississippi. 
In these works we adopt NRGR description of non-linearities in General Relativity: these are represented by a perturbative series of Feynman diagrams in which classical gravitons interact with matter sources and among themselves. 
Focusing on the purely gravitational sector, we tag self-interaction vertices with parameters: in such a way, my group and I were able, for example, to translate the measure of the period decay of Hulse-Taylor pulsar in a constraint on the \tgv\ at the 0.1\% level.
What emerges from this attitude is an extension of the parametrized post-Newtonian framework of Will and collaborators to the strong-field/radiative regime of gravity. 
Such a formulation is very useful in view of gravitational wave detection, which will constitute a unique source of knowledge in the domains of gravity, in particular, and physics, in general. 
In this context, our parametrized field-theoretical framework constitutes a complementary alternative to other approaches in the field: the reference example is the series of studies put forward by Arun, Iyer, Sathyaprakash and collaborators to test General Relativity by measuring the coefficients of the gravitational wave phase.

%% file: abbreviations.tex
\chapter*{Abbreviations List}
\addcontentsline{toc}{chapter}{Abbreviations List}

\bd

\item BD = 
%in \chap{chap:EFT}: Blanchet \& Damour, to refer to the approach they pursued with their collaborators in building the post-Newtonian expansion of the two-body problem (see ref.~\cite{Blan-LRR:06} and references therein); 
in \chap{chap:Ric}: Buonanno \& Damour, to refer to their work on the effective action of cosmic strings~\cite{Buonanno:1998kx},
in \chap{chap:Berti}: Brans \& Dicke, to refer to their scalar-tensor theory~\cite{Brans:1961sx} 

\item BH = Black Hole

\item BBH = Binary Black Hole

\item DEF = Damour and Esposito-Farese, as in their treatment of \sts, described in \chap{chap:How_To}

\item DH = Dabholkar and Harvey, to indicate their study~\cite{Dabholkar:1989jt} discussed in \chap{chap:Ric}

\item EEP = Einstein Equivalence Principle, i.e. universality of free-fall together with local Lorentz invariance and local position invariance (see Will’s review~\cite{Will_LRR:2006} for a comprehensive discussion)

\item EFT = Effective Field Theory

\item EH = Einstein-Hilbert (referred to the action)

\item EIH = Einstein-Infeld-Hoffmann (referred to the action)

\item EMT = Energy-Momentum Tensor

\item ET = Einstein Telescope~\cite{ET}

\item GR = General Relativity

\item GW(s) = Gravitational Wave(s) 

\item IR = Infra Red

\item LIGO = Laser Interferometer Gravitational-wave Observatory~\cite{LIGO} 

\item LISA = Laser Interferometer Space Antenna~\cite{LISA}  

\item LLR = Lunar Laser Ranging, referred to the experiment which uses retroflecting mirrors left by the first men on the Moon to monitor the distance of the Earth from its natural satellite (see for example ref.~\cite{LLR:04})

%\item LO = Leading Order\item NLO = Next-to-Leading Order\item NNLO = Next-to-next-to-Leading Order

\item NRGR = Non Relativistic General Relativity, which is how the EFT of ref.~\cite{NRGR_paper} has been coined by the authors in analogy with the treatment of two-body problems in QED and QCD

\item NS = Neutron Star

\item PK = Post-Keplerian

\item PM = Post-Minkowskian

\item PN = Post-Newtonian

\item PPK = Parametrized Post-Keplerian, referring to the phenomenological approach proposed in ref.~\cite{Dam-PPK} to analyze pulsar data in a theory-indipendent way

\item PPN = Parametrized Post-Newtonian, referring to the original phenomenological framework used to interpret Solar-System experiments on gravity (see ref.~\cite{Will_TEGP} for a comprehensive treatment)

\item PSR = Pulsar 

\item SEP = Strong Equivalence Principle, i.e. the version of the principle which is valid for self-gravitating objects

\item UV = Ultra Violet

\ed